\documentclass[imslayout,preprint]{imsart}
\def\journal@name{}
\usepackage[margin=1.5in]{geometry}

\usepackage{import, subfiles}
\usepackage{url}
\usepackage[dvipsnames]{xcolor}
\usepackage{tikz, pgf}
\usetikzlibrary{calc}
\usetikzlibrary{decorations.pathreplacing}
\usepackage{amsmath, amsthm, bm, bbm, amssymb}
\usepackage{mathtools}
\usepackage[inline]{enumitem}

\usepackage{amsfonts}
\newenvironment{proofof}[1]{\renewcommand{\proofname}{Proof of #1}\proof}{\endproof}
\usepackage{algpseudocode, algorithm}
\algtext*{EndWhile}%
\algtext*{EndIf}%
\algtext*{EndFor}%
\usepackage{multirow}
\usepackage{wrapfig}
\usepackage{graphicx} %
\usepackage{natbib}
\usepackage[hypertexnames=false,colorlinks,citecolor=blue,urlcolor=black,linkcolor=blue,pdfborder={0 0 0}]{hyperref}
\usepackage{todonotes}
\usepackage[capitalize,sort]{cleveref}  

\usepackage{caption}
\usepackage{subcaption}
\usepackage{comment}

\crefname{nlem}{Lemma}{Lemmas}
\crefname{nprop}{Proposition}{Propositions}
\crefname{ncor}{Corollary}{Corollaries}
\crefname{nthm}{Theorem}{Theorems}
\crefname{assumption}{Assumption}{Assumptions}
\crefname{condition}{Condition}{Conditions}
\crefname{exa}{Example}{Examples}
\usepackage{jhh-misc2}
\makeatletter
\newcommand{\inlineitem}[1][]{%
\ifnum\enit@type=\tw@
    {\descriptionlabel{#1}}
  \hspace{\labelsep}%
\else
  \ifnum\enit@type=\z@
       \refstepcounter{\@listctr}\fi
    \quad\@itemlabel\hspace{\labelsep}%
\fi}
\makeatother
\newcommand{\tdim}{D'}

\newcommand{\genpar}{GenPar} %

\newcommand{\thirdpass}[1]{\textcolor{black}{#1}}
\newcommand{\secondpass}[1]{\textcolor{black}{#1}}
\newcommand{\firstpass}[1]{\textcolor{black}{#1}}

\newcommand{\normZ}{Z}
\newcommand{\traitLL}{\ell}
\newcommand{\targetcondLL}{h}
\newcommand{\approxcondLL}{\widetilde{h}}
\newcommand{\targetLatent}{X}
\newcommand{\targetObs}{Y}
\newcommand{\approxLatent}{Z}
\newcommand{\approxObs}{W}
\newcommand{\approxlev}{K}
\newcommand{\support}{U}
\newcommand{\CRMtypical}{(\beta+1)\max(C(K,C_1),C(N,C_1))}
\newcommand\AIFA[1]{\mathrm{AIFA}_{#1}}

\DeclarePairedDelimiter{\floor}{\lfloor}{\rfloor}
\newcommand{\euler}{e}
\newcommand\indict[1]{\mathbf{1}\{#1\}}
 
\newcommand{\distBNB}{\mathrm{BNB}}
\newcommand{\distPoisson}{\mathrm{Poisson}}

\newcommand{\distBer}{\mathrm{Ber}}
\newcommand{\distGamma}{\mathrm{Gamma}}
\newcommand{\distNB}{\mathrm{NB}}
\newcommand{\distBetaP}{\mathrm{BP}}
\newcommand{\distIFA}{\mathrm{IFA}}

\newcommand{\distCMP}{\mathrm{CMP}}

\newcommand{\distXGamma}{\mathrm{XGamma}}
\DeclarePairedDelimiter\inner{\langle}{\rangle}%
\newcommand{\approxrate}{\rho}
\newcommand{\dTV}{d_{\text{TV}}}
\newcommand\zerovec[1]{0_{#1-1}}
\newcommand\BPproc[2]{P^{\text{BP}}_{#1,#2}}
\newcommand{\distGenGamma}{\mathrm{GenGamma}}
\newcommand{\EPPFblocks}{b}

\newcommand{\naturals}{\mathbb{N}} %

\allowdisplaybreaks[3]
\boldshortcuts

\begin{document}
	
\title{Independent finite approximations for Bayesian nonparametric inference}	
\runtitle{Independent finite approximations}	
	
\begin{aug}
	\author[A]{\fnms{Tin D.} \snm{Nguyen}\ead[label=e1]{tdn@mit.edu}},
		\author[B]{\fnms{Jonathan} \snm{Huggins}\ead[label=e2]{huggins@bu.edu}},
		\author[A]{\fnms{Lorenzo} \snm{Masoero}\ead[label=e3]{lom@mit.edu}},
		\author[C]{\fnms{Lester} \snm{Mackey}\ead[label=e4]{lmackey@microsoft.com}},
		\author[A]{\fnms{Tamara} \snm{Broderick}\ead[label=e5]{tamarab@mit.edu}}
		
\runauthor{T. Nguyen et al.}

\address[A]{LIDS, MIT, \printead{e1,e3,e5}}
\address[B]{Department of Mathematics \& Statistics, Boston University, \printead{e2}}
\address[C]{Microsoft Research New England, \printead{e4}}

\end{aug}

\begin{abstract}

\thirdpass{
Completely random measures (CRMs) and their normalizations (NCRMs)
offer flexible models in Bayesian nonparametrics.
But their infinite dimensionality presents challenges for inference.
Two popular finite approximations are truncated finite approximations (TFAs)
and independent finite approximations (IFAs). While the former have been
well-studied, IFAs lack similarly general bounds on approximation error, and
there has been no systematic comparison between the two options.
In the present work, we propose a general recipe to construct practical
finite-dimensional approximations for homogeneous CRMs and NCRMs,
in the presence or absence of power laws. We call our construction the
\emph{automated independent finite approximation} (AIFA). Relative to
TFAs, we show that AIFAs facilitate more straightforward derivations and use
of parallel computing in approximate inference. We upper bound the
approximation error of AIFAs for a wide class of common CRMs and NCRMs
--- and thereby develop guidelines for choosing the approximation level. Our
lower bounds in key cases suggest that our upper bounds are tight. We prove
that, for worst-case choices of observation likelihoods, TFAs are more efficient
than AIFAs. Conversely, we find that in real-data experiments with standard
likelihoods, AIFAs and TFAs perform similarly. Moreover, we demonstrate that
AIFAs can be used for hyperparameter estimation even when other potential
IFA options struggle or do not apply.
}

\end{abstract}

\maketitle

\section{Introduction}
Many data analysis problems can be seen as discovering a latent set of traits in a population
--- for example, recovering topics or themes from scientific papers,
ancestral populations from genetic data, interest groups from social network data, or unique speakers 
across audio recordings of many meetings \citep{Palla:2012tv,Blei:2010je,Fox:2010a}. 
In all of these cases, we might reasonably expect the number of latent traits present in a data
set to grow with the number of observations. 
One might choose a prior for different data set sizes, but then model construction potentially becomes inconvenient and unwieldy.
A simpler approach is to choose a single prior that naturally yields different expected numbers of traits for different numbers of data points.
In theory, Bayesian nonparametric (BNP) priors have exactly this desirable property due to a countable infinity of traits, 
so that there are always more traits to reveal through the accumulation of more data. 

\secondpass{However, the infinite-dimensional parameter presents a practical challenge;
namely, it}
is impossible to store an infinity of random variables in memory
or learn the distribution over an infinite number of variables in finite time. 
\secondpass{Some authors have developed conjugate priors and likelihoods \citep{Orbanz:2010ul,James:2017,broderick2018posteriors}
to circumvent the infinite representation; in particular, these models allow marginalization of the infinite collection of latent traits.
These models will typically be part of a more complex generative model where the remaining components are all finite.
Therefore, users can apply approximate inference schemes
such as Gibbs sampling. However, these marginal forms typically limit the user to a constrained family of models; 
are not amenable to parallelization;
would require substantial new development to use with modern inference
engines like NIMBLE \citep{devalpine2017nimble};
and are not straightforward to use with variational Bayes.
}

An alternative approach %
is to approximate the infinite-dimensional prior
with a finite-dimensional prior that essentially replaces the infinite collection of random traits by a finite subset of ``likely'' traits. 
Unlike a fixed finite-dimensional prior across all data set sizes, this finite-dimensional prior is
an approximation to the BNP prior. 
Therefore, its cardinality \secondpass{can be} informed directly by the BNP prior and the size of the observed data. 
\secondpass{Any moderately complex model will necessitate approximate inference, such as \secondpass{Markov chain Monte Carlo (MCMC) or variational Bayes (VB)}. Therefore,}
as long as the error due to the finite-dimensional prior approximation is small compared to the 
error due to using approximate inference, inferential quality is not affected.
\secondpass{Unlike marginal representations, probabilistic programming languages like NIMBLE \citep{devalpine2017nimble} natively support such finite approximations.
}

Much of the previous work on finite approximations developed and analyzed truncations of series representations of the random measures underlying 
the nonparametric prior; we call these \emph{truncated finite approximations} (TFAs) 
and refer to \citet{campbell2019truncated} for a thorough study. TFAs start from a sequential ordering of population traits
in a random measure.
The TFA retains a finite set of approximating traits; these match the population traits until a finite point and do not include terms beyond that \citep{doshi-velez2009variational,paisley2012stick,roychowdhury2015gamma,arbel2017moment,campbell2019truncated}.
\secondpass{However, we show in \cref{sec:conceptual} that the sequential nature of TFAs makes it difficult to derive update steps in an approximate
inference algorithm (either MCMC or VB) and is not amenable to parallelization.}

Here, we instead develop and analyze a general-purpose finite approximation consisting of independent
and identically distributed (\iid)\ representations of the traits together with their rates within the population;
we call these \emph{independent finite approximations} (IFAs). 
At the time of writing, we are aware of two \secondpass{alternative lines of work} on generic constructions of finite approximations using i.i.d.\ random variables, 
namely \citet{lijoi2023finite} and \citet{lee2022unified,lee2016finite}.
\citet{lijoi2023finite} design approximations for clustering models, characterize the posterior predictive distribution, and derive tractable inference schemes. 
\secondpass{However, the authors have not developed their method for trait allocations, where data points can potentially belong to multiple traits and can potentially exhibit traits in different amounts. And in particular it would require additional development to perform inference in trait allocation models using their approximations.}\footnote{\secondpass{We also note that, without modification, their approximation is not suitable for use in statistical models where the unnormalized atom sizes of the CRM are bounded, as arise when modeling the frequencies (in $[0,1]$) of traits. While model reparameterization may help, it requires (at least) additional steps.}}
\citet{lee2022unified,lee2016finite} construct finite approximations through a novel augmentation scheme. %
However, \secondpass{\citet{lee2022unified,lee2016finite}} lack explicit constructions in important situations, such as exponential-family rate measures, 
because the functions involved in the augmentation are, in general, only implicitly defined.
When the augmentation is implicit, \secondpass{there is not currently} a way to evaluate (up to proportionality constant) the probability density of the finite-dimensional distribution\thirdpass{; therefore standard Markov chain Monte Carlo and variational approaches for approximate inference are unavailable.}

\textbf{Our contributions.}
We propose a general-purpose construction for IFAs that subsumes a number of special cases that have already been successfully
used in applications \secondpass{(\cref{subsec:general-IFA}). We call our construction the \emph{automated independent finite approximation}, or AIFA.
We show that AIFAs can handle a wide variety of models --- including homogeneous completely random measures (CRMs) and normalized CRMs (NCRMs) (\cref{subsec:NCRM}).\footnote{\secondpass{NCRMs are also called \emph{normalized random measures with independent increments }(NRMIs) \citep{regazzini2003distributional,James:2009ux}}.} \thirdpass{Our construction} can handle (N)CRMs exhibiting power laws and has an especially convenient \thirdpass{form} for exponential family CRMs (\cref{subsec:eCRM-IFA}). We show that our construction works for useful CRMs not previously seen in the BNP literature (\cref{exa:eG-CMP}).
Unlike marginal representations, AIFAs do not require conditional conjugacy and can be used with VB.
We show that, unlike TFAs, AIFAs facilitate straightforward derivations within approximate inference schemes such as MCMC or VB
and are amenable to parallelization during inference (\cref{sec:conceptual}). In existing special cases, practitioners report similar \thirdpass{predictive} performance between AIFAs and TFAs~\citep{Kurihara:2007} and that AIFAs are also simpler to use compared to TFAs~\citep{Fox:2010a,Johnson:2013}. 
In contrast to the methods of \citet{lee2022unified,lee2016finite}, one can always evaluate the probability density (up to a proportionality constant) of AIFAs; furthermore,
in \cref{subsec:discount-estimation}, AIFAs accurately learn model hyperparameters by maximizing the marginal likelihood where the methods of \citet{lee2022unified,lee2016finite} struggle.}

\secondpass{In \cref{sec:bounds}, we bound the error induced by approximating an exact infinite-dimensional prior with an AIFA.}
Our analysis provides interpretable error bounds with explicit dependence on the size of the approximation and the data cardinality; our bounds can be used to set the size of the approximation in practice.
Our error bounds reveal that for the worst-case choice of observation likelihood, to approximate the target to a desired accuracy, it is necessary to use a large IFA model while a small TFA model would suffice. 
\secondpass{However, in practical experiments with standard observations likelihoods, we find that AIFAs and TFAs of equal sizes have similar performance.
Likewise, we find that, when both apply, AIFAs and alternative IFAs \citep{lee2022unified,lee2016finite} exhibit similar predictive performance (\cref{subsec:synthetic}). But AIFAs apply more broadly and are amenable to hyperparameter learning via optimizing the marginal likelihood, unlike \citet{lee2022unified,lee2016finite} (\cref{subsec:discount-estimation}). As a further illustration, we show that we are able to learn whether a model is over- or underdispersed, and by how much, using an AIFA approximating a novel BNP prior in \cref{subsec:dispersion-estimation}.}
\section{Background}
\label{sec:bkg}

Our work will approximate nonparametric
priors, so we first review construction of these priors from completely random measures (CRMs). Then we cover existing work on the construction of
truncated and independent finite approximations for these CRM priors. 
\secondpass{For some space $\Psi$, let} $\psi_i \in \Psi$ represent the $i$-th trait of interest, and let $\theta_{i} > 0$ represent the corresponding rate or frequency of this trait in the population.
\secondpass{If the set of traits is finite, we let $I$ equal its cardinality; if the set of traits is countably infinite, we let $I = \infty$.}
Collect the pairs of traits and frequencies in a measure $\Theta$ that places non-negative mass $\theta_i$ at location $\psi_i$: $\Theta := \sum_{i=1}^{I} \theta_i \delta_{\psi_i}$, where $\delta_{\psi_i}$ is a Dirac measure placing mass 1 at location $\psi_i$.
To perform Bayesian inference, we need to choose a prior distribution on $\Theta$ and a likelihood for the observed
data $\targetObs_{1:N} := \{\targetObs_{n}\}_{n=1}^{N}$ given $\Theta$. 
Then, applying a disintegration, we can obtain the posterior on $\Theta$ given
the observed data.

\textbf{Homogeneous completely random measures.} 
\secondpass{Many common BNP priors can be formulated as completely random measures \citep{Kingman:1967,lijoi2010models}.}\footnote{
\secondpass{
Conversely, some important priors, such as Pitman-Yor processes, are not CRMs or their normalizations
and are outside the scope of the present paper \citep{pitman1997two,arbel2019stochastic,lijoi2020pitman}}.
}
CRMs are constructed from Poisson point processes,\footnote{For brevity, we do not consider the fixed-location and deterministic components of a
CRM~\citep{Kingman:1967}. When these are purely atomic, they can be added to our analysis without undue effort.}
which are straightforward to manipulate analytically \citep{Kingman:1992}. 
Consider a Poisson point process on $\mathbb{R}_+ \defined \left[0, \infty\right)$ with rate measure $\nu(\dee\theta)$ such that 
$
  \nu(\mathbb{R}_+)=\infty
$
and
$
  \int \min(1, \theta) \nu(\dee\theta) < \infty. 
$
\secondpass{Such a process generates a countably infinite set of rates
$\left(\theta_i\right)_{i=1}^\infty$ with $\theta_i \in \mathbb{R}_+$ and $0 < \sum_{i=1}^{\infty} \theta_i < \infty$ almost surely.}
We assume throughout that $\psi_i \distiid H$ for some diffuse
distribution $H$. 
The distribution $H$, called the ground measure, serves as a prior on the traits in the space $\Psi$. 
For example, consider a common topic model. Each trait $\psi_i$ represents a latent topic, modeled as a probability vector in the simplex of vocabulary words. And $\theta_i$ represents the frequency with which the topic $\psi_i$ appears across documents in a corpus. $H$ is a Dirichlet distribution over the probability simplex, with dimension given by the number of words in the vocabulary.

By pairing the rates from the Poisson process with traits drawn from the ground measure, we obtain a completely random measure and use the shorthand $\distCRM(H, \nu)$ for its law: \secondpass{
$
\Theta = \sum_i \theta_i \delta_{\psi_i} \dist \distCRM(H, \nu).
$}
Since the traits $\psi_i$ and the rates $\theta_i$ are independent, the CRM is \emph{homogeneous}.
\secondpass{When the total mass $\Theta(\Psi)$ is strictly positive and finite, the corresponding \emph{normalized CRM} (NCRM) is $\Xi \defined \Theta / \Theta(\Psi)$, which is a 
discrete probability measure:
} $\Xi = \sum_{i} \xi_i \delta_{\psi_i}$, where $\xi_i = \theta_i/(\sum_{j} \theta_j)$
\secondpass{\citep{regazzini2003distributional,James:2009ux}}.

The CRM prior on $\Theta$ is typically combined with a likelihood that generates trait counts for each data point. 
Let $\traitLL(\cdot \given \theta)$ be a proper probability mass function on $\nats \cup \{0\}$ for all 
$\theta$ in the support of $\nu$. 
The process $\targetLatent_{n} \defined \sum_i x_{ni}\delta_{\psi_i}$ collects the trait counts,
where $x_{ni} \given \Theta \sim \traitLL(\cdot \given \theta_i)$ independently across atom index $i$ and \iid\ across data index $n$. 
We denote the distribution of $\targetLatent_n$ as $\distLP(\traitLL, \Theta)$, which we call the \emph{likelihood process}.
\secondpass{Together, the prior on $\Theta$ and likelihood on $\targetLatent$ given $\Theta$ form a generative model for allocation of data points to traits; hence, this generative model is a special case of a \emph{trait allocation model} \citep{campbell2018exchangeable}. Analogously, when the trait counts are restricted to $\{0,1\}$, this generative model represents a special case of a \emph{feature allocation model}.}

Since the trait counts are typically just a latent component in a full generative model
specification, we define the observed data to be $\targetObs_{n} \given \targetLatent_n \distind f(\cdot \given \targetLatent_n)$ for a probability kernel $f(\dee Y \given X)$. 
Consider the topic modeling example: $\theta_i$ represents the rate of topic $\psi_i$ in a document corpus; $\Theta$ captures the rates of all topics; $X_{n}$ captures how many words in document $n$ are generated from each topic; and $Y_{n}$ gives the 
observed collection of words for that document.

\textbf{Finite approximations.}
Since the set $\{\theta_i\}_{i=1}^{\infty}$ is countably infinite, it is not possible to simulate or perform posterior
inference for every $\theta_i$. One approximation scheme uses a \emph{finite approximation}
$\Theta_{\approxlev} \defined \sum_{i=1}^\approxlev \approxrate_i \delta_{\psi_i}$. 
The atom sizes $\{\approxrate_i\}_{i=1}^{K}$ are designed so that $\Theta_{\approxlev}$ is a good approximation of $\Theta$ in a suitable sense. 
\secondpass{Since it involves a finite number of parameters unlike $\Theta$, $\Theta_{\approxlev}$ can be used directly in standard posterior approximation schemes such as Markov chain Monte Carlo or variational Bayes. But not using the full CRM $\Theta$ introduces approximation error.}

A \emph{truncated finite approximation} \secondpass{\citep[TFA;][]{doshi-velez2009variational,paisley2012stick,roychowdhury2015gamma,arbel2017moment,campbell2019truncated}} requires constructing an ordering on the set of rates from the Poisson process; let $(\theta_i)_{i=1}^{\infty}$ be the corresponding \emph{sequence} of rates.
The approximation uses $\approxrate_i = \theta_i$ for $i$ up to some $K$; i.e.\ one \secondpass{keeps} the first $K$ rates in the sequence and \secondpass{ignores} the remaining ones. 
\secondpass{We refer} to the number of instantiated atoms $K$ as the \emph{approximation level.}
\secondpass{\citet{campbell2019truncated}} categorizes and analyzes TFAs. TFAs offer an attractive nested structure:
to refine an existing truncation, it suffices to generate the additional terms in the sequence. 
However, the complex dependencies between the \secondpass{rates $(\theta_i)_{i=1}^{K}$} potentially make inference more challenging. 

We instead develop a family of \emph{independent finite approximations} (IFAs). 
An IFA is defined by a sequence of probability measures $\nu_{1}, \nu_{2}, \dots$ such that 
at approximation level $\approxlev$, there are $\approxlev$ atoms whose weights are given by $\approxrate_{1},\dots,\approxrate_{\approxlev} \distiid \nu_{\approxlev}$.
The probability measures are chosen so that the sequence of approximations converges in distribution to the target CRM: $\Theta_{\approxlev} \convD \Theta$ as $K \to \infty$. 
For random measures, convergence in distribution can also be characterized by convergence of integrals under the measures \citep[Lemma 12.1 and Theorem 16.16]{Kallenberg:2002}.  
The advantages and disadvantages of IFAs reverse those of TFAs: the atoms are now i.i.d., potentially making inference easier, but a completely new approximation
must be constructed if $\approxlev$ changes. 

Next consider approximating \secondpass{an NCRM} $\Xi = \sum_{i} \xi_i \delta_{\psi_i}$, where $\xi_i =  \theta_i/(\sum_{j} \theta_j)$\secondpass{,} with a finite approximation. 
A normalized TFA might be defined in one of two ways. 
In the first approach, the rates $\{\approxrate_i\}_{i=1}^{K}$ that target the CRM rates $\{\theta_i\}_{i=1}^{\infty}$ are normalized to form the NCRM approximation; i.e.\ the approximation has atom sizes $\approxrate_i/ \sum_{j=1}^{K} \approxrate_j$ \citep{campbell2019truncated}. 
The second approach directly constructs an ordering over the sequence of normalized rates $\xi_i$
and truncates this representation.\footnote{\secondpass{In this case, $\sum_{i=1}^{K} \xi_i < 1$. Therefore, setting the final atom size in the NCRM approximation to be $1 - \sum_{i=1}^{K} \xi_i$ ensures the approximation is a probability measure.}} We construct normalized IFAs in a similar manner to the first TFA approach: the \secondpass{NCRM} approximation has atom sizes $\approxrate_i/ \sum_{j=1}^{K} \approxrate_j$ where $\{\approxrate_i\}_{i=1}^{K}$ are the IFA rates.

In the past, independent finite approximations have largely been developed on a case-by-case basis \citep{paisley2009nonparametric,Broderick:2015,Acharya:2015,lee2016finite}. 
Our goal is to provide a general-purpose mechanism. 
\citet{lijoi2023finite} and \citet{lee2022unified} have also recently pursued a more general construction, but we believe there remains room for improvement.
\secondpass{\citet{lijoi2023finite} focus on NCRMs for clustering; it is not immediately clear how to adapt this work for inference in trait allocation models.}
Also, \citet[Theorem 1]{lijoi2023finite} employ infinitely divisible random variables. 
Since infinitely divisible distributions that are not Dirac measures cannot have bounded support, the approximate rates $\{\approxrate_i\}_{i=1}^{K}$ are not naturally compatible with the trait likelihood $\traitLL(\cdot \given \theta)$ if the support of the rate measure $\nu$ is bounded. But the support of $\nu$ is often bounded in applications to \secondpass{trait allocation} models; e.g., $\theta_i$ may represent a feature frequency, taking values in $[0,1]$, and $\traitLL(\cdot \given \theta)$ may take the form of a Bernoulli, binomial, or negative binomial distribution.
Therefore, applications of the finite approximations of \citet[Theorem 1]{lijoi2023finite} to these models \secondpass{may require some additional work.}
The construction in \citet[Proposition 3.2]{lee2022unified} yields $\{\approxrate_i\}_{i=1}^{K}$ that are compatible with $\traitLL(\cdot \given \theta)$ and recovers important cases in the literature.
However, outside these special cases, it is unknown if the i.i.d.\ distributions are tractable because \secondpass{the} densities $\nu_{\approxlev}$ are not explicitly defined\secondpass{;} see the discussion around \cref{eq:arrival-time-approx} for more details.

\bexa[Running example: beta process] \label{exa:bp_intro}
For concreteness, we consider the \emph{(three-parameter) beta process}\footnote{\secondpass{Also known as the \emph{stable beta process} \citep{Teh:2009}}} \citep{Teh:2009,broderick2012beta} as a running example of a CRM. 
The process $\distBP(\gamma, \alpha, d)$ is defined by a mass parameter $\gamma > 0$, discount parameter $d \in [0,1)$, and \secondpass{concentration} parameter $\alpha > -d$. It has rate measure
\[
\nu(\dee \theta) = \gamma \frac{\Gamma(\alpha+1)}{\Gamma(1-d) \Gamma(\alpha+d)} \indict{0 \le \theta \le 1} \theta^{-d-1}(1-\theta)^{\alpha+d-1}\dee \theta. 
\]
The $d = 0$ case yields the standard beta process \citep{hjort1990nonparametric,Thibaux:2007}. 
The beta process is typically paired with the Bernoulli likelihood process
with conditional distribution $\traitLL(x \given \theta) = \theta^{x}(1-\theta)^{1-x} \indict{x \in \{0,1\}}$.
The \secondpass{resulting} \emph{beta--Bernoulli process} has been used in factor analysis models \citep{doshi-velez2009variational,paisley2012stick} and for dictionary learning \citep{zhou2009nonparametric}.
\eexa

\section{\secondpass{Automated independent finite approximations}}%
\label{sec:construction}
In this section we introduce \secondpass{\emph{automated independent finite approximations,}}
a practical construction of independent finite approximations (IFAs) for a broad class of CRMs. 
We highlight a useful special case of our construction for exponential family CRMs~\citep{broderick2018posteriors} without power laws
and \secondpass{apply our construction to approximate NCRMs}. 
In all of these cases, we prove that as the approximation size increases, the distribution of the approximation converges (in some relevant sense)
to that of the exact infinite-dimensional model.

\subsection{\secondpass{Applying our approximation to CRMs}} \label{subsec:general-IFA}

Formally, we define IFAs in terms of a fixed, diffuse probability measure $H$ and a sequence of probability measures $\nu_{1}, \nu_{2}, \dots$.
The $\approxlev$-atom IFA $\Theta_{\approxlev}$ is
\begin{align*}
	\Theta_{\approxlev} &\defined {\textstyle\sum_{i=1}^{\approxlev}} \approxrate_{i}\delta_{\psi_{i}}, & 
	\approxrate_{i} &\distiid \nu_{\approxlev},  &
	\psi_{i} \distiid H,
\end{align*}
which we write as $\Theta_{\approxlev} \sim  \distIFA_{\approxlev}(H, \nu_{\approxlev})$.
We consider CRM rate measures $\nu$ with densities that, near zero, are (roughly) proportional to $\theta^{-1-d}$, where $d \in [0,1)$ is the \emph{discount} parameter.
We will propose a general construction for IFAs \secondpass{given a target random measure}
and prove that it converges to the \secondpass{target} (\cref{thm:default-CRM-convergence}). 
We first summarize our requirements for which CRMs we approximate in \cref{assume:rate-measure-near-0}.
We show in \cref{app:more-examples} that popular BNP priors satisfy \cref{assume:rate-measure-near-0};
specifically, \secondpass{we check} the beta, \thirdpass{gamma \citep{Ferguson:1972,Kingman:1975,Titsias:2008},
generalized gamma \citep{Brix:1999},
beta prime \citep{Broderick:2015}, and PG($\alpha,\zeta$)-generalized gamma \citep{James:2013} processes.}

\bassume \label{assume:rate-measure-near-0}
For $d \in [0,1 )$ and $\eta \in V \subseteq \reals^{d}$, we take $\Theta \dist \distCRM(H, \nu(\cdot; \secondpass{\gamma}, d, \eta))$ for 
\begin{equation*}
	\nu(\dee\theta; \secondpass{\gamma}, d, \eta) \defined \gamma\theta^{-1-d}g(\theta)^{-d} \frac{h(\theta; \eta)}{Z(1-d, \eta)}\dee\theta
\end{equation*}
such that
\benum[leftmargin=*]
\item for $\xi > 0$ and $\eta \in V$, $Z(\xi, \eta) \defined \int \theta^{\xi-1}g(\theta)^{\xi}h(\theta; \eta)\dee\theta < \infty$;
\item $g$ is continuous, $g(0) = 1$, and there exist constants $0 < c_{*} \le c^{*} < \infty$ such that $c_{*} \le g(\theta)^{-1} \le c^{*}(1 + \theta)$; 
\item there exists $\eps > 0$ such that for all $\eta \in V$, the map $\theta \mapsto h(\theta; \eta)$ is continuous and bounded on $[0,\eps]$. 
\eenum
\eassume

Other than the discount $d$ and mass $\gamma$, the rate measure $\nu$ potentially depends on additional hyperparameters $\eta$. 
The finiteness of the normalizer $Z$ is necessary in defining finite-dimensional distributions whose densities are similar in form to $\nu$. 
The conditions on the behaviors of $g(\theta)$ and $h(\theta;\eta)$ ensure that the overall rate measure's behavior near $\theta = 0$ is dominated by the $\theta^{-1-d}$ term. 
The support of the rate measure is implicitly determined by $h(\theta; \eta)$.

Given a CRM satisfying \cref{assume:rate-measure-near-0}, we can construct a sequence of IFAs that converge in distribution to that CRM. 

\bnthm \label{thm:default-CRM-convergence}

Suppose \cref{assume:rate-measure-near-0} holds. Let 
\begin{equation} \label{eq:default-Sb}
	S_{b}(\theta) =
	\begin{cases}
		\exp\left(\frac{-1}{1-(\theta-b)^{2}/b^2}+1\right) & \text{if } \theta \in (0,b) \\
		\indict{\theta > 0} & \text{otherwise.}
	\end{cases} 
\end{equation}
For $c \defined \gamma {h(0; \eta)}/{Z(1-d,\eta)}$, let
\begin{equation*}
	\nu_{\approxlev}(\dee\theta) 
	\defined \theta^{-1+c\approxlev^{-1} - d S_{1/K}(\theta - 1/K)} g(\theta)^{c\approxlev^{-1} - d}h(\theta; \eta)Z_{\approxlev}^{-1}\dee\theta
\end{equation*}
be a family of probability densities, where $Z_{\approxlev}$ is chosen such that $\int \nu_{\approxlev}(\dee\theta) = 1$. If $\Theta_{\approxlev} \dist \distIFA_{\approxlev}(H, \nu_{\approxlev})$, then $\Theta_{\approxlev} \convD \Theta$ as $\approxlev \to \infty$. 
\enthm

See \cref{app:CRM-construction} for a proof of \cref{thm:default-CRM-convergence}.
We choose the particular form of $S_b(\theta)$ in \cref{eq:default-Sb} for concreteness and convenience. But our theory still holds for a more general class of $S_b$ forms, as we describe in more detail in the proof of \cref{thm:default-CRM-convergence}.

\bnumdefn
We call the $\approxlev$-atom IFA resulting from \cref{thm:default-CRM-convergence} the \emph{automated IFA} ($\AIFA{K}$). 
\enumdefn

Although the normalization constant $Z_K$ is not always available analytically, numerical implementation remains straightforward.
When $Z_K$ is a quantity of interest, such as in \cref{subsec:discount-estimation}, we estimate it using standard numerical integration schemes for a one-dimensional integral \citep{piessens2012quadpack,virtanen2020scipy}. 
\secondpass{
For other tasks, we need not access $Z_K$ directly. 
\thirdpass{In our experiments, we show that we can use either Markov chain Monte Carlo (\cref{subsec:image-denoising,subsec:dispersion-estimation}) or variational Bayes (\cref{subsec:topic-modelling,subsec:synthetic}) with the unnormalized density.}
}

\firstpass{
\secondpass{To illustrate our construction, we next apply \cref{thm:default-CRM-convergence} to $\distBP(\gamma, \alpha, d)$ from \cref{exa:bp_intro}.
In \cref{app:more-examples}, we show how to construct AIFAs for the beta prime, gamma, generalized gamma, and PG($\alpha,\zeta$)-generalized gamma processes.}
\bexa[\secondpass{Beta process AIFA}] \label{exa:bp_aifa}
\secondpass{
To apply \cref{assume:rate-measure-near-0}, let $\eta = \alpha + d$, $V = \reals_{+}$, $g(\theta) = 1$, $h(\theta; \eta) = (1- \theta)^{\eta - 1}\ind[\theta \le 1]$, and $Z(\xi, \eta)$ equal the beta function $B(\xi, \eta)$. Then the CRM rate measure $\nu$ in \cref{assume:rate-measure-near-0} corresponds to that of $\distBP(\gamma, \alpha, d)$ from \cref{exa:bp_intro}. Note that we make no additional restrictions on the hyperparameters $\gamma, \alpha, d$ beyond those in the original CRM (\cref{exa:bp_intro}). Observe that $h$ is continuous and bounded on $[0,1/2]$, and the normalization function $B(\xi, \eta)$ is finite for $\xi > 0, \eta \in V$; it follows that \cref{assume:rate-measure-near-0} holds. By \cref{thm:default-CRM-convergence}, then,} the AIFA density is 
\begin{equation*}
	\frac{1}{Z_K} \theta^{-1+c/K-dS_{\secondpass{1/K}}(\theta-1/K)}(1-\theta)^{\alpha+d-1} \mathbf{1} \{0 \leq \theta \leq 1\} d\theta,
\end{equation*}
where $c \coloneqq \gamma/B(\alpha+d,1-d)$ and $Z_K$ is the normalization constant. \secondpass{The density does not in general reduce to a beta distribution in $\theta$ due to the $\theta$ in the exponent.}
\eexa 
}

\textbf{Comparison to an alternative IFA construction.}
\secondpass{ \citet[Proposition 3.2]{lee2022unified} verify} the validity of a different IFA construction. 
Their construction requires two functions: (1) a bivariate function $\Lambda(\theta,t)$ such that for any $t > 0, \Delta(t) \coloneqq \int \Lambda(\theta,t) \nu(d\theta) < \infty$ and (2) a univariate function $f(n)$ such that $\Delta(f(n))$ is bounded from both above and below by $n$ as $n \to \infty.$
If these functions exist \secondpass{and}
\begin{equation} \label{eq:arrival-time-approx}
	\widetilde{\nu}_K(d\theta) \coloneqq \frac{\Lambda(\theta, f(K)) \nu(d\theta)}{\Delta(f(K))},
\end{equation} 
\citet[Proposition 3.2]{lee2022unified} show that $\distIFA_K(H, \widetilde{\nu}_K)$ converges in distribution to $\distCRM(H, \nu)$ as $\approxlev \to \infty$. 
The usability of \cref{eq:arrival-time-approx} in practice depends on the tractability of $\Lambda$ and $f$. 
There are typically many tractable $\Lambda(\theta, t)$ \citep[Section 4]{lee2022unified}. Proposition B.2 of
\citet{lee2022unified} lists tractable $f$ for the important cases of \thirdpass{the} beta \secondpass{process} and \secondpass{and generalized gamma process} with $d > 0$. 
However, the choice of $f$ provided there for general power-law processes is not tractable because its evaluation requires computing complicated inverses in the asymptotic regime. 
Furthermore, for processes without power laws, no general recipe for $f$ is known.
In contrast, the AIFA construction in \cref{thm:default-CRM-convergence} always yields densities that can be evaluated up to proportionality constants.

\bexa[\secondpass{Beta process: an IFA comparison}] \label{exa:bp_comparison}
\secondpass{We next compare our beta process AIFA to the two separate IFAs proposed by \citet{lee2022unified} and \citet{lee2016finite} for disjoint subcases within the case $d > 0$.
First consider the subcase where $\alpha = 0, d>0$. 
\citet{lee2016finite} derive\footnote{\firstpass{There is a typo in \citet[Theorem 2, item (iii)]{lee2016finite}: $\theta/K$ should be $(\theta/\Gamma(\alpha))/K$.}} 
what we call\footnote{\secondpass{\citet{devroye2014simulation} introduce the acronym BFRY to denote a distribution named for the authors \citet{bertoin2006particular}. 
We here use ``BFRY IFA'' to denote what \citet{lee2016finite} call the ``BFRY process'' and thereby emphasize that this process forms an IFA.}} the \emph{BFRY IFA}. The IFA density, denoted $\nu_{\text{BFRY}}(d\theta)$, is equal to }
\begin{equation} \label{eq:BFRY-nuK1}
	\frac{\gamma}{K} \frac{ \theta^{-d - 1} (1-\theta)^{d - 1} }{B(d,1-d)} \left[1  - \exp\left( - \left(\frac{K\Gamma(d)d}{\gamma}\right)^{1/d} \frac{\theta}{1-\theta}  \right) \right] \mathbf{1} \{0 \leq \theta \leq 1\}d\theta.
\end{equation}

\secondpass{
Second, consider the subcase where $\alpha > 0, d > 0$, \citet[Section 4.5]{lee2022unified} derive another $K$-atom IFA, which we call\footnote{\secondpass{We use the term ``generalized Pareto'' because \citet[Section 4.5]{lee2022unified} \secondpass{use} generalized Pareto variates to define $\Lambda(\theta, t)$ from \cref{eq:arrival-time-approx}.}} the \emph{generalized Pareto IFA} (\genpar{} IFA). The IFA density, denoted $\nu_{\text{\genpar{}}}(d\theta)$, is equal to }
\begin{equation} \label{eq:Pareto_IFA}
	\frac{\gamma}{K} \frac{\theta^{-d - 1}(1-\theta)^{\alpha + d - 1} }{B(1-d, \alpha + d)} \left(1  - \frac{1}{\left( \theta \left[\left( 1 + \frac{K d}{\gamma \alpha } \right)^{\frac{1}{d}} -1 \right] + 1   \right)^{\alpha} } \right)    \mathbf{1} \{0 \leq \theta \leq 1\} d\theta.
\end{equation}

\secondpass{Since the BFRY IFA and \genpar{} IFA apply to disjoint hyperparameter regimes, they are not directly comparable.
Since our AIFA applies to the whole domain $\alpha \geq -d$, we can separately compare it to each of these alternative IFAs; we also highlight that the AIFA still applies when $\alpha \in (-d,0)$, a case not covered by either the BFRY IFA or \genpar{} IFA.}

\secondpass{We find in \Cref{subsec:synthetic} that the AIFA and BFRY IFA have comparable predictive performance; the AIFA and \genpar{} IFA also have comparable predictive performance.
But in \Cref{subsec:discount-estimation}, we show that the AIFA is much more reliable than the BFRY IFA or the \genpar{} IFA for estimating the discount ($d$) hyperparameter by maximizing the marginal likelihood.
Conversely, sampling from a BFRY IFA or \genpar{} IFA prior is easier than sampling from an AIFA prior since the BFRY and \genpar{} IFA priors are formed from standard distributions.}
\eexa

\subsection{\secondpass{Applying our approximation to} exponential family CRMs} \label{subsec:eCRM-IFA}

\emph{Exponential family CRMs} with $d = 0$ comprise \secondpass{a widely used special case of CRMs. In what follows, we show how \cref{thm:default-CRM-convergence} simplifies in this special case.}

In common BNP models, the relationship between the likelihood $\traitLL(\cdot \given \theta)$ and the CRM prior is closely related to finite-dimensional exponential family conjugacy \cite[Section 4]{broderick2018posteriors}. In particular, the likelihood has an exponential family form\secondpass{,}
\begin{equation}
\label{bkg:expLPh}
\traitLL(x \mid \theta) \defined \kappa(x) \theta^{\phi(x)} \exp \left( \inner{\mu(\theta), t(x)}   - A(\theta)\right). 
\end{equation}
Here $x \in \mathbb{N} \cup \{0\}$, $\kappa(x) \in \mathbb{R}$ is the base density, $\phi(x) \in \mathbb{R}$ and $t(x) \in \mathbb{R}^{\tdim}$ (for some $\tdim$) form the vector of sufficient statistics $(t(x), \phi(x))^{T}$, $A(\theta) \in \mathbb{R}$ is the log partition function, $\mu(\theta) \in \mathbb{R}^{\tdim}$ and $\ln \theta$ form the vector of natural parameters $(\mu(\theta), \ln  \theta)^{T}$, and $\inner{\mu(\theta), t(x)}$ denotes the standard Euclidean inner product. 
The rate measure nearly matches the form of the conjugate prior, but behaves like $\theta^{-1}$ near $0$:
\begin{equation}
\label{bkg:expRateMeasure}
\nu(\dee\theta) \defined \gamma' \theta^{-1} \exp \left\{ \left\langle { \bmat \psi \\ \lambda \emat , \bmat \mu(\theta) \\ -A(\theta) \emat } \right\rangle \right\} \indict{\theta \in \support} \dee \theta,
\end{equation}
where $\gamma' > 0$, $\lambda > 0$, $\psi \in \mathbb{R}^{D'}$ and $\support \subseteq \mathbb{R}_{+}$ is the support of $\nu$.
\cref{bkg:expRateMeasure} leads to the suggestive terminology of \emph{exponential family} CRMs. The $\theta^{-1}$ dependence near $0$ means that these models lack power-law behavior. 
Models that can be cast in this form include the standard beta process with Bernoulli or negative binomial likelihood \citep{zhou2012beta,Broderick:2015}
and the gamma process with Poisson likelihood \citep{Acharya:2015,roychowdhury2015gamma}.
We refer to these models as, respectively, the beta--Bernoulli, beta--negative binomial, and gamma--Poisson processes. 

\secondpass{We now specialize \cref{assume:rate-measure-near-0} and \cref{thm:default-CRM-convergence} to exponential family CRMs in \cref{assume:eCRM-near-0} and \cref{cor:expCRM-d-zero}, respectively}.

\bassume  \label{assume:eCRM-near-0}
Let $\nu$ be of the form \secondpass{in} \cref{bkg:expRateMeasure} and assume that
\benum[leftmargin=*]
\item For any $\xi > - 1$, for any $\eta = (\psi, \lambda)^T$ where $\lambda > 0$, the normalizer defined as 
\begin{equation} \label{bkg:normalizer}
	\normZ(\xi, \eta) \defined \int_{\support} \theta^{\xi} \exp \left\{ \left\langle { \eta , \bmat \mu(\theta) \\ -A(\theta) \emat } \right\rangle \right\} d\theta
\end{equation}
is finite, and 
\item there exists $\epsilon > 0$ such that, for any $\eta = (\psi, \lambda)^T$ where $\lambda > 0$, the map 
\begin{equation*}
	\varsigma : \theta \mapsto  \exp \left\{ \left\langle \eta , \bmat \mu(\theta) \\ -A(\theta) \emat \right\rangle \right\} \indict{\theta \in \support}
\end{equation*}
is a continuous and bounded function of $\theta$ on $[0,\epsilon]$.
\eenum 
\eassume

\bncor  \label{cor:expCRM-d-zero}
Suppose \cref{assume:eCRM-near-0} holds.
For $c \defined \gamma' \varsigma(0)$, let
\begin{equation}
	\label{rev-eq:nu_K}
	\nu_K(\theta) \defined \frac{\theta^{c/K-1}\varsigma(\theta)}{\normZ \left( c/K-1, \eta\right)}.
\end{equation}
If $\Theta_{\approxlev} \dist \distIFA_{\approxlev}(H, \nu_{\approxlev})$, then $\Theta_{\approxlev} \convD \Theta$.
\encor

The density in \cref{rev-eq:nu_K} is almost the same as the rate measure of \cref{bkg:expRateMeasure}, except the $\theta^{-1}$ term has become $\theta^{c/K-1}$. 
As a result, \cref{rev-eq:nu_K} is a proper exponential-family distribution. 
\secondpass{In \cref{app:more-examples}, we detail the corresponding $d=0$ special cases of the AIFA for beta prime, \thirdpass{gamma}, generalized gamma, and PG($\alpha$,$\zeta$)-generalized gamma processes. We cover the beta process case next.}

\bexa[\secondpass{Beta process AIFA for $d=0$}] \label{exa:bp_aifa_nodiscount}
\cref{cor:expCRM-d-zero} is sufficient to recover known IFA results for $\distBP(\gamma,\alpha,0)$\secondpass{;} \firstpass{
when $d = 0$, the AIFA from \cref{exa:bp_aifa} simplifies to $\nu_K = \distBeta \left(\gamma \alpha/K, \alpha\right).$
}
\citet{doshi-velez2009variational} approximates $\distBP(\gamma,1,0)$  with $\nu_K=\distBeta \left(\gamma/K, 1\right)$. For $\distBP(\gamma,\alpha,0)$,
\citet{Griffiths:2011a} set $\nu_K=\distBeta \left(\gamma \alpha/K, \alpha\right)$, and \citet{paisley2009nonparametric} use $\nu_K=\distBeta \left(\gamma \alpha/K, \alpha(1-1/K)\right)$.
The difference between $\distBeta \left(\gamma \alpha/K, \alpha\right)$ and $\distBeta \left(\gamma \alpha/K, \alpha(1-1/K)\right)$ is negligible for moderately large $K$. 
\eexa

\secondpass{We can also use \cref{cor:expCRM-d-zero} to create a new finite approximation} for a nonparametric process so far not explored in the Bayesian nonparametric literature.
\bexa[CMP likelihood and extended gamma process] \label{exa:eG-CMP}
The \emph{CMP likelihood}\footnote{\secondpass{CMP stands for Conway-Maxwell-Poisson.}} \citep{shmueli2005useful} is given by 
\begin{equation} \label{eq:CMP}
	\traitLL(x \mid \theta) = \frac{\theta^x}{(x!)^{\tau}} \frac{1}{Z_{\tau}(\theta)},
	\quad\textrm{ where } 
	Z_\tau(\theta) \coloneqq \sum_{y=0}^{\infty} \frac{\theta^{y}}{(y!)^{\tau}}. 
\end{equation}
The conjugate CRM prior, which we call an \emph{extended gamma} (or \emph{Xgamma}) \emph{process}, has four hyperparameters: 
mass $\gamma$, concentration $c$, maximum $T$, and shape $\tau$\secondpass{:}
\begin{equation} \label{eq:eG}
	\nu(d\theta) = \gamma \theta^{-1} Z_{\tau}^{-c}(\theta) 1 \{ 0 \leq \theta \leq T\} d\theta.
\end{equation}
\eexa
Unlike existing BNP models, the model in \cref{eq:CMP,eq:eG}, which we call \emph{Xgamma--CMP \secondpass{process}}, is able to capture different dispersion regimes.
For $\tau < 1$, the variance of the counts from $\traitLL(x \mid \theta)$ is larger than the mean of the counts, corresponding to overdispersion.
For $\tau > 1$, the variance of the counts from $\traitLL(x \mid \theta)$ is smaller than the mean of the counts, corresponding to underdispersion.
As we show in \cref{subsec:dispersion-estimation}, the latent shape $\tau$ can be inferred using observed data.
\citet{zhou2012beta,Broderick:2015} provide BNP trait allocation models that handle overdispersion.  
\citet{canale2011bayesian} provide a BNP model that handles both underdispersion and overdispersion\secondpass{, but for clustering rather than traits}. 
We are not aware of trait allocation models that handle underdispersion, or any trait allocation models that handle both underdispersion and overdispersion.
Following the approach of \citet{broderick2018posteriors}, in \cref{app:admissible} we show that as long as
$\gamma > 0$, $c > 0$, $T  \geq 1$, and $\tau > 0$, the total mass of the rate measure is infinite and the number
 of active traits is almost surely finite.
Under these conditions, we show in \cref{app:more-examples} that \cref{cor:expCRM-d-zero} applies to the CRM in \cref{eq:eG}\secondpass{,
and we construct the resulting AIFA}.

\subsection{Normalized independent finite approximations} \label{subsec:NCRM}

Given that AIFAs are approximations that converge to the corresponding target CRM, it is natural to ask if normalizations of AIFAs converge to the corresponding normalization of the target CRM, i.e., the corresponding NCRM. 
Our next result shows that normalized AIFAs indeed converge, in the sense that the exchangeable partition probability functions, or EPPFs \citep{pitman1995exchangeable}, converge. 
Given a random sample of size $N$ from an NCRM $\Xi$, the EPPF gives the probability of the induced partition from such a sample. 
In particular, consider the model $\Xi \sim \text{NCRM}, \targetLatent_n \mid \Xi \distiid \Xi \text{ for } 1 \leq n \leq N$.\footnote{We reuse the $\targetLatent_n$ notation from the CRM description, even though $\targetLatent_n$ now is a scalar, because the role of the draws from $\Xi$ is the same as that of the draws from $\Theta$.} Grouping the indices $n$ with the same value of $X_n$ induces a partition over the set $\{1,2,\ldots, N\}$. 
Let $\EPPFblocks$ represent the number of distinct values in the set $\{X_n\}_{n=1}^{N}$, so $\EPPFblocks \le N$. Let $n_i$ be the number of indices $n$ with $X_n$ equal to the $i$-th distinct value of $X_n$, for some ordering of the values. So $\sum_{i=1}^{\EPPFblocks}n_i = N$ and $\forall i, n_i \geq 1$. 
With this notation in hand, we can write the EPPF, which gives the probability of the induced partition under the model, as a symmetric function $p(n_1,n_2,\ldots,n_\EPPFblocks)$ that depends only on the counts $n_i$. 
Similarly, we let $p_\approxlev(n_1,n_2,\ldots,n_\EPPFblocks)$ be the EPPF for the normalized $\AIFA{\approxlev}$. Note that $p_\approxlev(n_1,n_2,\ldots,n_\EPPFblocks) = 0$ when $\approxlev < \EPPFblocks$ since the normalized $\AIFA{\approxlev}$ at approximation level $\approxlev$ generates at most $\approxlev$ blocks.

\bnthm \label{thm:EPPF-convergence}
Suppose \cref{assume:rate-measure-near-0} holds. Take any positive integers $N,b,\{n_i\}_{i=1}^b$ such that $\EPPFblocks \leq N$, $n_i \geq 1$, and $\sum_{i=1}^{\EPPFblocks}n_i = N$. 
Let $p$ be the EPPF of the NCRM $\Xi \defined \Theta/\Theta(\Psi)$. 
If $\Theta_{\approxlev}$ is the AIFA for $\Theta$ at approximation level $\approxlev$, %
and $p_K$ is the EPPF for the corresponding NCRM approximation $\Theta_{\approxlev}/\Theta_{\approxlev}(\Psi)$, then
\begin{equation*}
	\lim_{K \to \infty} p_{K}(n_1,n_2,\ldots,n_\EPPFblocks) = p(n_1,n_2,\ldots,n_\EPPFblocks).
\end{equation*}
\enthm

See \cref{app:NCRM-construction} for the proof. 
Since the EPPF gives the probability of each partition, the point-wise convergence in \cref{thm:EPPF-convergence} certifies that the distribution over partitions induced by sampling from the normalized $\AIFA{\approxlev}$ converges to that induced by sampling from the target NCRM, for any finite sample size $N$.

\section{Non-asymptotic error bounds}
\label{sec:bounds}

\cref{thm:default-CRM-convergence,thm:EPPF-convergence} justify the use of our proposed AIFA construction in the limit $\approxlev \to \infty$ 
but do not provide guidance on how to choose the approximation level $\approxlev$ when $N$ observations are available. 
In \cref{subsec:CRM}, we quantify the error introduced by replacing \secondpass{an exponential family} $\distCRM$ with the AIFA.
In \cref{subsec:mHDP}, we quantify the error introduced by replacing a Dirichlet process ($\distDP$) \citep{Ferguson:1973,Sethuraman:1994} with the corresponding normalized AIFA.
We derive error bounds that are simple to manipulate and yield recommendations for the appropriate $\approxlev$ for a given $N$ and a desired accuracy level. 

\subsection{\secondpass{Bounds when approximating an exponential family CRM}} \label{subsec:CRM}
Recall from \cref{sec:bkg} that the CRM prior $\Theta$ is typically paired with a likelihood process $\distLP$\secondpass{, which} manifests features $X_n$\secondpass{,} and a probability kernel $f$ relating active features to observations $Y_n$. 
The target nonparametric model can be summarized as 
\[
\begin{split}
	\Theta &\sim \distCRM(H, \nu), \\
	X_n \mid \Theta &\distiid \distLP(\traitLL, \Theta), \quad n = 1, 2, \ldots, N, \\
	Y_n \mid X_n &\distind f(\cdot \given X_n), \quad n = 1, 2, \ldots, N.
\end{split} 	\label{eq:target-model}
\]

The approximating model, with $\nu_K$ as in \cref{thm:default-CRM-convergence} (or \cref{cor:expCRM-d-zero}), is
\[
	\begin{split}
	\Theta_{\approxlev} &\sim \secondpass{\AIFA{\approxlev}}(H, \nu_\approxlev),  \\
	Z_n \mid \Theta_{\approxlev} &\distiid \distLP(\traitLL, \Theta_\approxlev), \quad n = 1, 2, \ldots, N, \\
	W_n \mid Z_n &\distind f(\cdot \mid Z_n),  \,\,~\quad n = 1, 2, \ldots, N. \\
	\end{split}	
		\label{thm-eq:approx-model}
\]
Active traits in the approximate model are collected in $Z_n$ and observations are $W_n.$
Let $P_{N,\infty}$ be the marginal distribution of the observations $Y_{1:N}$ and $P_{N,K}$ be the marginal distribution of the observations $W_{1:N}$. 
The \emph{approximation error} we analyze is the total variation distance  $\dTV(P_{N,K}, P_{N,\infty}) \defined \sup_{0 \le g \le 1}|\int g \dee P_{N,K} - \int g \dee P_{N,\infty}|$
between the two observational processes, 
one using the $\distCRM$ and the other one using the approximate $\AIFA{K}$ as the prior. 
Total variation is a standard choice of error when analyzing CRM approximations \citep{Ishwaran:2002,doshi-velez2009variational,paisley2012stick,campbell2019truncated}.
Small total variation distance implies small differences in expectations of bounded functions.

\textbf{Conditions.}
\secondpass{In our analysis, we focus on exponential family CRMs and conjugate likelihood processes. We will suppose \cref{assume:eCRM-near-0} holds.}
Our analysis guarantees that $\dTV(P_{N,K}, P_{N,\infty})$ is small whenever \secondpass{a conjugate exponential family CRM--likelihood pair and the corresponding AIFA model satisfy certain conditions, beyond those already stated in \cref{assume:eCRM-near-0}.}
In the proof of the error bound, these conditions serve as intermediate results that ultimately lead to small approximation error.
Because we can verify the conditions for common models, we have error bounds in the most prevalent use cases of CRMs. 
To express these conditions, we use the \emph{marginal process} representation of the target and the approximate model, i.e., the series of conditional distributions of $X_n \given X_{1:(n-1)}$ (or $Z_n \given Z_{1:(n-1)}$) with $\Theta$ (or $\Theta_K$) integrated out. \secondpass{Corollary 6.2 of \citet{broderick2018posteriors}
guarantees that the marginal $X_n \given X_{1:(n-1)}$ is a random measure with finite support and with a convenient form. Since we will use this form to write our conditions (\cref{condition:marginal-process} below), we first review the requisite notation --- and establish analogous notation for $Z_n \given Z_{1:(n-1)}$.}

We \secondpass{start by defining} $\targetcondLL$ and $M$ to describe the conditional distribution $X_n \given X_{1:(n-1)}$.
Let $K_{n-1}$ be the number of unique atom locations in $X_1,X_2,\ldots,X_{n-1}$, and let $\{\zeta_i\}_{i=1}^{K_{n-1}}$ be the collection of unique atom locations in $X_1,X_2,\ldots,X_{n-1}$. 
Fix an atom location $\zeta_j$ (the choice of $j$ does not matter). 
For $m$ with $1 \leq m \le n$, let $x_{m}$ be the atom size of $X_m$ at atom location $\zeta_j$; $x_m$ may be zero if there is no atom at $\zeta_j$ in $X_m$.
The distribution of $x_n$ depends \emph{only} on the $x_{1:(n-1)}$ values, which are the atom sizes of previous measures $X_m$ at $\zeta_j$.
We use $\targetcondLL(x \mid x_{1:(n-1)})$ to denote the probability mass function (p.m.f.)\ of $x_{n}$ at value $x$.
Furthermore, $X_n$ has a finite number of new atoms, which can be grouped together by atom size.
Consider any potential atom size $x \in \nats$. 
Define $p_{n,x}$ to be the number of atoms of size $x$.
Regardless of atom size, each atom location is a fresh draw from the ground measure $H$ and
$p_{n,x}$ is Poisson-distributed; we use $M_{n,x}$ to denote the mean of $p_{n,x}$.

Next, we define $\approxcondLL$, which governs the conditional distribution of $Z_n \given Z_{1:(n-1)}$.
Let $\zerovec{n}$ be the zero vector with $n-1$ components. Although $\targetcondLL(x \mid x_{1:(n-1)})$ is  defined only for count vectors $x_{1:(n-1)}$ that are not identically zero, we will see that $\approxcondLL(x \mid \zerovec{n})$ is well-defined.
In particular, let $\{\zeta_i\}_{i=1}^{K_{n-1}}$ be the union of atom locations in $Z_1,Z_2,\ldots,Z_{n-1}$.
Fix an atom location $\zeta_j$. 
For $1 \leq m \leq n$, let $x_{m}$ be the atom size of $Z_m$ at atom location $\zeta_j$. 
We write the p.m.f.\ of $x_{n}$ at $x$ as $\approxcondLL(x \mid x_{1:(n-1)})$. 
In addition, $Z_n$ also has a maximum of $K-K_{n-1}$ new atoms with locations disjoint from $\{\zeta_i\}_{i=1}^{K_{n-1}}$, and the distribution of atom sizes is governed by $\approxcondLL(x \mid \zerovec{n})$.
Note that we reuse the $x_n$ and $\zeta_j$ notation from $X_n \given X_{1:(n-1)}$ without risk of confusion, since $x_n$ and $\zeta_j$ are dummy variables whose meanings are clear given the context of $\targetcondLL$ or $\approxcondLL$.

In \cref{app:marginal}, we describe the marginal processes in more detail and give formulas for $\targetcondLL$, $\approxcondLL$, and $M_{n,x}$ in terms of the functions that parametrize \cref{bkg:expRateMeasure,bkg:expLPh} and the normalizer \cref{bkg:normalizer}. 
For the beta--Bernoulli process with $d = 0$, the functions have particularly \secondpass{convenient} forms. 

\bexa \label{exa:beta-Ber}
For the beta--Bernoulli model with $d = 0$, we have
\begin{equation*}
	\targetcondLL(x \mid x_{1:(n-1)}) = \frac{\sum_{i=1}^{n-1}x_i}{\alpha-1+n} \indict{x = 1} + \frac{\alpha +  \sum_{i=1}^{n-1}(1-x_i)}{\alpha-1+n} \indict{x = 0}.
\end{equation*}
\begin{equation*}
	\approxcondLL(x \mid x_{1:(n-1)}) = \frac{\sum_{i=1}^{n-1}x_i+\gamma\alpha/K}{\alpha-1+n+\gamma\alpha/K} \indict{x = 1}  + \frac{\alpha +  \sum_{i=1}^{n-1}(1-x_i)}{\alpha-1+n+\gamma\alpha/K} \indict{x = 0},
\end{equation*}
\begin{equation*}
	M_{n,1} = \frac{\gamma \alpha}{\alpha - 1 + n}, \hspace{10pt} M_{n,x} = 0 \text{ for } x > 1.
\end{equation*}
\eexa 

We now formulate conditions on $\targetcondLL$, $\approxcondLL$, and $M_{n,x}$ that will yield small $\dTV(P_{N,K}, P_{N,\infty})$.

\bcond \label{condition:marginal-process}
There exist constants $\{C_i\}_{i=1}^{5}$ such that
\benum 
\item for all $n \in \nats$, 
\begin{equation}
	\label{thm-ass:eCRM-tot}
	\sum_{x = 1}^{\infty} M_{n,x} \leq \frac{C_1}{n-1+C_1};
\end{equation}
\item for all $n \in \nats$,
\begin{equation}
	\label{thm-ass:IFA-tot}
	\sum_{x=1}^{\infty} \approxcondLL(x \mid x_{1:(n-1)}=\zerovec{n}) \leq \frac{1}{K}\frac{C_1}{n-1+C_1};
\end{equation}
\item for any $n \in \nats$, for any $\{x_i\}_{i=1}^{n-1} \neq \zerovec{n}$,
\begin{equation}
	\label{thm-ass:old-location}
	\sum_{x=0}^{\infty} \left| \targetcondLL(x\mid x_{1:(n-1)}) - \approxcondLL(x \mid x_{1:(n-1)}) \right| \leq   \frac{1}{K} \frac{C_1}{n-1+C_1}; \textrm{ and }
\end{equation}
\item 	for all $n \in \nats$, for any $K \geq C_2(\ln n + C_3)$,
\begin{equation}
	\label{thm-ass:new-location}
	\sum_{x=1}^{\infty} \left| M_{n,x} - K \approxcondLL(x \mid x_{1:(n-1)}=\zerovec{n})\right| \leq \frac{1}{K} \frac{C_4\ln n+C_5}{n-1+C_1}.
\end{equation}
\eenum 
\econd

Note that the conditions depend only on the functions governing the \secondpass{exponential family CRM prior and its conjugate likelihood process} --- and not on the observation likelihood $f$.
\cref{thm-ass:eCRM-tot} constrains the growth rate of the target model since
$\sum_{n=1}^{N} \sum_{x=1}^{\infty} M_{n,x}$ is the expected number of components for data cardinality $N$.
Because each $\sum_{x = 1}^{\infty} M_{n,x}$ is at most $O(1/n)$, the total number of components after $N$ samples is $O(\ln N)$. 
Similarly, \cref{thm-ass:IFA-tot} constrains the growth rate of the approximate model. 
The third condition (\cref{thm-ass:old-location}) ensures that $\approxcondLL$ is a good approximation of $\targetcondLL$ in total variation distance
and that there is also a reduction in the error as $n$ increases. 
Finally, \cref{thm-ass:new-location} implies that $K\approxcondLL(x \given \zerovec{n})$ is an accurate approximation of $M_{n,x}$, and there is also a reduction in the error as $n$ increases. 

We show that \cref{condition:marginal-process} holds for the most commonly used non-power-law CRM models; see \cref{exa:beta--Bernoulli-marginal} for the case of the beta--Bernoulli model with discount $d=0$ and \cref{app:more-verification} for the beta--negative binomial and gamma--Poisson models with $d = 0$.
As we detail next, we believe \cref{condition:marginal-process} is also reasonable beyond these common models. 
The $O(1/n)$ quantity in \cref{thm-ass:eCRM-tot} is the typical expected number of new features after observing $n$ observations in non-power-law BNP models.
\cref{thm-ass:IFA-tot,thm-ass:old-location,thm-ass:new-location} are likely to hold when $\approxcondLL$ is a small perturbation of $\targetcondLL$ and $K \approxcondLL$ is a small perturbation of $M_{n,x}$.
For instance, in \cref{exa:beta-Ber}, the functional form of $\approxcondLL$ is very similar to that of $\targetcondLL$, except that $\approxcondLL$ has the additional $\gamma \alpha/K$ factor in both numerator and denominator.
The functional form of $K\approxcondLL$ is very similar to that of $M_{n,x}$, except that $K\approxcondLL$ has an additional $\gamma \alpha/K$ factor in the denominator.
\bexa [{Beta--Bernoulli with $d = 0$, continued}] \label{exa:beta--Bernoulli-marginal}
The growth rate of the target model is
\begin{equation*}
	\sum_{x=1}^{\infty} M_{n,x} = M_{n,1} =  \frac{\gamma \alpha}{n-1 + \alpha}.
\end{equation*}
Since $\approxcondLL$ is supported on $\{0,1\}$, the growth rate of the approximate model is 
\begin{equation*}
	\approxcondLL(1 \mid x_{1:(n-1)}=\zerovec{n}) = \frac{\gamma \alpha/K}{\alpha - 1 + n + \gamma \alpha/K} \leq \frac{1}{K} \frac{\gamma \alpha}{n-1+\alpha}.
\end{equation*}
Since both $\targetcondLL$ and $\approxcondLL$ are supported on $\{0,1\}$, \cref{thm-ass:old-location} becomes 
\begin{equation*}
	\left| \targetcondLL(1 \mid x_{1:(n-1)}) - \approxcondLL(1 \mid x_{1:(n-1)}) \right| = \left| \frac{\sum_{i=1}^{n-1}x_i+\gamma\alpha/K}{\alpha-1+n+\gamma\alpha/K} - \frac{\sum_{i=1}^{n-1}x_i}{\alpha-1+n}  \right| \leq \frac{\gamma \alpha}{K} \frac{1}{n - 1 + \alpha}.
\end{equation*}
And because $M_{n,x} = 0 = \approxcondLL(x \mid \cdot \;)$ for $x > 1$, \cref{thm-ass:new-location} becomes 
\begin{equation*}
	\left|M_{n,1} - K \approxcondLL(1 \mid x_{1:(n-1)}=\zerovec{n})\right| = \left| \frac{\gamma \alpha}{\alpha - 1 + n} - \frac{\gamma \alpha}{\alpha - 1 + n + \frac{\gamma \alpha}{K}} \right| \leq \frac{\gamma^2 \alpha}{K} \frac{1}{n - 1 + \alpha}.
\end{equation*}
Calibrating $\{C_i\}$ based on these inequalities is straightforward.
\eexa 

\textbf{Upper bound.}
We now make use of \cref{condition:marginal-process} to derive an upper bound on the approximation error induced by AIFAs.

\bnthm [Upper bound for exponential family CRMs] \label{thm:CRM-upperbound}
Recall that $P_{N,\infty}$ is the distribution of $\targetObs_{1:N}$ from \cref{eq:target-model} while $P_{N,K}$ is the distribution of $\approxObs_{1:N}$ from \cref{thm-eq:approx-model}.
If \secondpass{\cref{assume:eCRM-near-0}} and \cref{condition:marginal-process} hold, then there exist positive constants \firstpass{$C',C'',C''',C''''$} depending only on $\{C_i\}_{i=1}^{5}$ such that 
\firstpass{
\begin{equation*}
	\dTV \left( P_{N,\infty}, P_{N,K} \right) \leq \frac{C' + C''\ln^2 N + C''' \ln N \ln K +  C'''' \ln K}{K}.
\end{equation*}
}
\enthm

\firstpass{See \cref{app:upperbound-proof} for explicit values of the constants as well as the proof.} 
\cref{thm:CRM-upperbound} states that the AIFA approximation error grows as $O(\ln^2 N)$ with fixed $K$, and decreases as $O \left( \ln K/K \right)$ for fixed $N$. 
The bound accords with our intuition that, for fixed $K$, the error should increase as $N$ increases: with more data, the expected number of latent components in the data increases, demanding finite approximations of increasingly larger sizes. 
In particular, $O(\ln N)$ is the standard Bayesian nonparametric growth rate for non-power law models. It is likely that the $O(\ln^2 N)$ factor can be improved to $O(\ln N)$ due to $O(\ln N)$ being the natural growth rate; more generally, we {conjecture} that the error directly depends on the expected number of latent components in a model for $N$ observations. 
On the other hand, for fixed $N$, we expect that error should decrease as $K$ increases and the approximation thus has greater capacity. This behavior also matches \cref{thm:default-CRM-convergence}, which guarantees that sufficiently large finite models have small error.

\secondpass{We highlight that \cref{thm:CRM-upperbound} provides upper bounds both (i) for approximations that were already known in the literature but where bounds were not already known, as in the case of the beta–negative binomial process, and (ii) for processes and approximations not previously studied in the literature in any form.}

\textbf{Lower bounds.}
From the upper bound in \cref{thm:CRM-upperbound}, we know how to set a sufficient number of atoms for accurate approximations: for the total variation to be less than some $\epsilon$, we solve for the smallest $K$ such that the right hand side of \cref{thm:CRM-upperbound} is smaller than $\epsilon$.
We now derive lower bounds on the AIFA approximation error to characterize a \emph{necessary} number of atoms for accurate approximations, by looking at worst-case observational likelihoods $f$. \secondpass{In particular, \cref{thm:CRM-upperbound} implies that an AIFA with $K = O \left(\text{poly}(\ln N)/\epsilon\right)$ atoms suffices in approximating the target model to less than $\epsilon$ error. In \cref{thm:betaBer-lnN-necessary} below, we establish that $K$ must grow at least at a  $\ln N$ rate in the worst case. In \cref{thm:betaBer-1/K-lowerbound} below, we establish that the $ 1/\epsilon$ term is necessary.
To the best of our knowledge, \cref{thm:betaBer-lnN-necessary,thm:betaBer-1/K-lowerbound} are the \emph{first} lower bounds on IFA approximation error for any process.}

Our lower bounds apply to the beta--Bernoulli process with $d = 0$. 
Recall that $P_{N,\infty}$ is the distribution of $\targetObs_{1:N}$ from \cref{eq:target-model} while $P_{N,K}$ is the distribution of $\approxObs_{1:N}$ from \cref{thm-eq:approx-model}.
\secondpass{In what follows, $\BPproc{N}{\infty}$ refers to the marginal distribution of the observations that arises when we use the prior $\distBP(\gamma, \alpha, 0)$. Analogously, $\BPproc{N}{K}$ is the observational distribution that arises when we use the $\AIFA{K}$ approximation in \cref{exa:bp_aifa}. T}he observational likelihood $f$ will be clear from context.
The worst-case observational likelihoods $f$ are pathological.
We leave to future work to lower bound the approximation error when more common likelihoods $f$, such as Gaussian or Dirichlet, are used. 

\secondpass{For the first result, it will be useful to define the \emph{growth function} for any $N \in \nats$, $\alpha > 0$:}
\begin{equation} \label{eq:growth-function}
	C(N, \alpha)  \defined \sum_{n=1}^{N} \frac{\alpha}{n-1+\alpha}.
\end{equation}
\secondpass{$C(N, \alpha)$ satisfies $\lim_{N \rightarrow \infty} C(N,\alpha) / (\alpha \ln N) = 1$; this asymptotic equivalence is a corollary of \cref{lem:harmonic-like-sum} or Theorem 2.3 from \citet{korwar1972contributions}.}
Our next result shows that our AIFA \secondpass{approximation can be poor} if the approximation level \secondpass{$K$} is too small compared to the growth function $C(N,\alpha)$. 

\bnthm [$\ln N$ is necessary] \label{thm:betaBer-lnN-necessary}
For the beta--Bernoulli \secondpass{process} model with $d = 0$, 
there exists an observation likelihood $f$, independent of $K$ and $N$, such that for any $N$, if $K \leq 0.5\gamma C(N,\alpha)$, then 
\begin{equation*}
	\dTV(\BPproc{N}{\infty}, \BPproc{N}{K}) \geq 1 - \frac{C}{N^{\gamma \alpha/8}},
\end{equation*}
where $C$ is a constant depending only on $\gamma$ and $\alpha$.
\enthm

See \cref{crm-proof:betaBer-lnN-necessary} for the proof. 
The intuition is that, with high probability, the number of features that manifest in the target $\targetLatent_{1:N}$ is greater than $0.5\gamma C(N,\alpha)$. However, the finite model $\approxLatent_{1:N}$ has fewer than $0.5\gamma C(N,\alpha)$ components. Hence, there is an event where the target and approximation assign drastically different probability masses. 
\cref{thm:betaBer-lnN-necessary} implies that as $N$ grows, if the approximation level $K$ fails to surpass the $0.5\gamma C(N,\alpha)$ threshold, then the total variation between the approximate and the target model remains bounded from zero; in fact, the error tends to one. 

\secondpass{We next show that the ${1}/{K}$ factor in the upper bound from \cref{thm:CRM-upperbound}} is \emph{tight} (up to logarithmic factors). 

\bnthm [Lower bound of $1/K$] \label{thm:betaBer-1/K-lowerbound}
For the beta--Bernoulli \secondpass{process} model with $d = 0$, there exists an observation likelihood $f$, independent of $K$ and $N$, such that for any $N$,
\begin{equation*}
	\dTV(\BPproc{N}{\infty}, \BPproc{N}{K}) \geq C\frac{1}{(1+\gamma/K)^2} \frac{1}{K} ,
\end{equation*}
where $C$ is a constant depending only on $\gamma$.
\enthm

See \cref{crm-proof:betaBer-1/K-lowerbound} for the proof. 
The intuition is that, under the pathological likelihood $f$, analyzing the AIFA approximation error is the same as analyzing the binomial--Poisson approximation error \citep{lecam1960approximation}. 
We then show that ${1}/{K}$ is a lower bound using the techniques from  \cite{barbour1984rate}. 
\cref{thm:betaBer-1/K-lowerbound} implies that an AIFA with $K = \Omega \left( 1/\epsilon\right)$ atoms is necessary in the worst case. 

\secondpass{Our} lower bounds (which apply specifically to the beta--Bernoulli process) are much less general than our upper bounds\secondpass{. However,} as a practical matter\secondpass{, generality in the lower bounds} is 
not so crucial due to \secondpass{the} different roles played by upper and lower bounds. 
Upper bounds give control over the approximation error; this control is what is needed to trust the approximation and to set
the approximation level. Whether or not we have access to lower bounds, general-purpose upper bounds give us this control. 
Lower bounds, on the other hand, serve as a helpful check that the upper bounds are not too loose --- and reassure us that we are not inefficiently
using too many atoms in a too-large approximation. From that standpoint, the need for general-purpose lower bounds is not as pressing.

The dependence on the accuracy level in the $d=0$ beta--Bernoulli process is worse for AIFAs than for TFAs.
For example, consider the Bondesson approximation \citep{Bondesson:1982,campbell2019truncated} of $\distBP(\gamma, \alpha, 0)$; we will see next that this approximation is a TFA with excellent error bounds.

\bexa [Bondesson approximation \citep{Bondesson:1982}] \label{exa:Bondesson-of-beta}
Fix $\alpha \geq 1$, let $E_l \distiid \distExp(1)$, and and let $\Gamma_k \defined \sum_{l=1}^{k} E_l$. The $K$-atom Bondesson approximation of $\distBP(\gamma, \alpha, 0)$ is a TFA $\sum_{k=1}^{K} \theta_k \delta_{\psi_{k}}$, where $\theta_k \defined V_k \exp (-\Gamma_k/\gamma \alpha), V_k \distiid \distBeta(1,\alpha-1)$, and $\psi_{k} \distiid H$.
\eexa 

The following result gives a bound on the error of the Bondesson approximation\secondpass{.}

\bnprop {\citep[Appendix A.1]{campbell2019truncated}} \label{prop:TFA-upperbound}
For $\gamma > 0, \alpha \geq 1$, let $\Theta_K$ be distributed according to a level-$K$ Bondesson approximation of $\distBP(\gamma, \alpha, 0)$, $R_n \given \Theta_K \distiid \distLP(\traitLL; \Theta_K), T_n \given R_n \distind f(\cdot \given R_n)$ with $N$ observations. Let $Q_{N,K}$ be the distribution of the observations $T_{1:N}$. Then:
$
\dTV \left(\BPproc{N}{\infty}, Q_{N,K}\right) \leq N \gamma \left( \frac{\gamma \alpha}{1+\gamma \alpha} \right)^{K}.
$
\enprop 
\cref{prop:TFA-upperbound} implies that a TFA with $K = O \left( \ln\{ N/\epsilon \} \right)$ atoms suffices in approximating the target model to less than $\epsilon$ error. Up to log factors in $N$, comparing the necessary ${1}/{\epsilon}$ level for an AIFA and the sufficient $\ln \left({1}/{\epsilon}\right)$ level for a TFA,  we conclude that the necessary size for an AIFA is exponentially larger than the sufficient size for a TFA, in the worst-case observational likelihood $f.$

\subsection{Approximating a (hierarchical) Dirichlet process} \label{subsec:mHDP}
So far we have analyzed AIFA error for CRM-based models. 
In this section, we analyze the error \secondpass{that arises from} using a normalized AIFA as an approximation for an NCRM; here, we focus on a 
Dirichlet process --- i.e., a normalized gamma process without power-law behavior. 
We first consider a generative model with the same number of layers as in previous sections. 
But we also consider a more complex generative model, with an additional layer --- as is common in, e.g., text analysis. 
Indeed, one of the strengths of Bayesian modeling is the flexibility facilitated by hierarchical modeling, 
and a goal of probabilistic programming is to provide fast, automated inference for these more complex models.

\textbf{Dirichlet process.}
The Dirichlet process is one of the most widely used nonparametric priors \secondpass{and arises as a normalized gamma process}.
The \secondpass{generalized} gamma process CRM is characterized by the rate measure 
$
\nu(\dee \theta) = \gamma\frac{\lambda^{1-d}}{\Gamma(1-d)}
\theta^{-d-1}e^{-\lambda \theta}\dee \theta. 
$ We denote its distribution as $\distGammaP(\gamma, \lambda, d)$. A normalized draw from $\distGammaP(\gamma, 1, 0)$ is Dirichlet-process distributed with mass parameter $\gamma$ \citep{Kingman:1975,Ferguson:1973}. By \cref{cor:expCRM-d-zero}, $\distIFA_{\approxlev}(H, \nu_{\approxlev})$ with 
$
\nu_{\approxlev} = \distGam(\gamma /\approxlev, 1)
$
converges to $\distGammaP(\gamma,1,0)$. Because the normalization of independent gamma random variables is a Dirichlet random variable, a normalized draw from $\distIFA_\approxlev(H, \nu_\approxlev)$ is equal in distribution to $\sum_{i=1}^{\approxlev} p_i \delta_{\psi_i}$ where $\psi_i \distiid H$ and $\{p_i\}_{i=1}^{\approxlev} \sim \distDir(\{{\gamma}/{\approxlev}\}\mathbf{1}_{\approxlev}).$ We call this distribution the \emph{finite symmetric Dirichlet} ($\distFSD$), and denote 
it as $\distFSD_{\approxlev}(\gamma,H)$.\footnote{The name ``finite symmetric Dirichlet'' comes from \citet{Kurihara:2007}. See \citet[Section 2.2]{Ishwaran:2001} for other names this distribution has had in the literature.}

In the simplest use case, the Dirichlet process is used as the de Finetti measure for observations $X_n$; i.e., $\Xi \sim \distDP, X_{n} \mid \Xi \distiid \Xi \text{ for } 1 \leq n \leq N$. 
In \cref{app:dp-results}, we state error bounds when $\distFSD_K$ replaces the Dirichlet process as the mixing measure that are analogous to the results in \cref{subsec:CRM}.
The upper bound is similar to \cref{thm:CRM-upperbound} in that the error grows as $O(\ln^2 N)$ with fixed $K$, and decreases as $O \left( \ln K/K \right)$ for fixed $N$. 
The lower bounds, which are the analogues of \cref{thm:betaBer-lnN-necessary,thm:betaBer-1/K-lowerbound}, state that $K = \Omega(\ln N)$ is necessary for accurate approximations, and that truncation-based approximations are better than $\distFSD_K$, in the worst case. 
In comparison to existing results \citep{ishwaran2000markov,Ishwaran:2002}, Theorem 1 of \citet{ishwaran2000markov} does not bound the distance between observational processes, so it is not directly comparable to our error bound.  
We improve upon Theorem 4 of \citet{Ishwaran:2002}, whose upper bound on the $\distFSD$ approximation error lacks an explicit dependence on $K$ or $N$. 
So, unlike our bounds, that bound cannot be inverted to determine a sufficient approximation level $K$.

\textbf{Hierarchical Dirichlet process.}
In modern applications such as text analysis, practitioners use additional hierarchical levels to capture group structure in observed data. 
In text, we might have $D$ documents with $N$ words in each. More, generally, we might have $D$ groups (each indexed by $d$) with $N$ observations (each indexed by $n$) each.
We target the influential model of \citet{wang2011online,Hoffman:2013}, which is a variant of the hierarchical Dirichlet process \citep[HDP;][]{Teh:2006b} and which we refer to as the \emph{modified HDP}. 
In the HDP, $G$ is a population measure with $G \sim \distDP(\omega,H)$. The measure for the $d$-th subpopulation is $G_d \given G \sim \distDP(\alpha, G)$; the concentrations $\omega$ and $\alpha$ are potentially different from each other. 
The modified HDP is defined in terms of the \emph{truncated stick-breaking (TSB) approximation}:

\bnumdefn [Stick-breaking approximation \citep{Sethuraman:1994}] \label{exa:TSB}
For $i = 1,2,\ldots, K - 1$, let $v_i \distiid \text{Beta}(1,\alpha)$. Set $v_K = 1$. Let $\xi_i = v_i \prod_{j=1}^{i-1}(1-v_j)$. Let $\psi_k \distiid H$, and $\Xi_K = \sum_{k=1}^{K} \xi_k \delta_{\psi_{k}}.$ We denote the distribution of $\Xi_K$ as $\distTSB_K(\alpha, H)$.
\enumdefn

In the modified HDP, the sub-population measure is distributed as $G_d \given G \sim \distTSB_T(\alpha, G)$.
\citet{wang2011online} and \citet{Hoffman:2013} set $T$ to be small so that inference in the
modified HDP is more efficient than in the HDP, since the number of parameters per group is greatly reduced.
From a modeling standpoint, small $T$ is a reasonable assumption since documents typically manifest a small number of topics from the corpus,
with the total number depending on the document length and independent of corpus size.
For completeness, the generative process of the modified HDP is
\begin{equation} \label{eq:mHDP}
	\begin{aligned}
		G &\sim \distDP(\omega, H), \\
		H_{d} \mid G &\distiid \distTSB_T(\alpha, G) &\text{ across } d, \\
		\beta_{dn} \mid H_d &\distind H_d(\cdot) & \text{ across } d,n \\
		W_{dn} \mid \beta_{dn} &\distind f(\cdot \mid \beta_{dn}) & \text{ across } d,n.
	\end{aligned}
\end{equation}
$H_d$ contains at most $T$ distinct atom locations, all shared with the base measure $G$. 

The finite approximation we consider replaces the population-level Dirichlet process with $\distFSD_K$, keeping the other conditionals intact:\footnote{Our construction in \cref{eq:IFA-mHDP} is slightly different from Eqs.\ 5.5 and 5.6 in \citet{Fox:2010a}.
Our document-level process $F_d$ contains at most $T$ topics from the underlying corpus; by contrast, the \citet{Fox:2010a} document-level process contains as many topics as the corpus-level process.
However, the novelty of \cref{eq:IFA-mHDP} is incidental since the replacement of the population-level DP with the $\distFSD$ in the modified HDP is analogous to the DP case.}
\begin{equation} \label{eq:IFA-mHDP}
	\begin{aligned}
		G_K &\sim \distFSD_K(\omega, H), \\
		F_d \mid G_K &\distiid \distTSB_T(\alpha, G_K) & \text{across } d, \\
		\psi_{dn} \mid F_d &\distind  F_d(\cdot) &\text{across } d,n, \\
		 Z_{dn} \mid \psi_{dn} &\distind f(\cdot \mid\psi_{dn}) &\text{across } d,n.
	\end{aligned}
\end{equation}
Our contribution is analyzing the error of \cref{eq:IFA-mHDP}.

Let $P_{(N,D),\infty}$ be the distribution of the observations $\{W_{dn}\}$. Let $P_{(N,D),K}$ be the distribution of the observations $\{Z_{dn}\}$.
We have the following bound on the total variation distance between $P_{(N,D),\infty}$ and $P_{(N,D),K}$. 

\bnthm [Upper bound for modified HDP] \label{thm:mHDP-upperbound}
For some constants \firstpass{$C',C'',C'''$, $C'''$} that depend only on $\omega$, 
\secondpass{
\begin{equation*}
	\dTV  \left(P_{(N,D),\infty}, P_{(N,D),K}\right) \leq\frac{C' + C'' \ln^2(DT) + C''' \ln (DT) \ln K + C'''' \ln K}{K}.
\end{equation*}
}
\enthm

\firstpass{See \cref{app-proof:mHDP-upperbound} for explicit values of the constants as well as the theorem's proof. }
For fixed $K$, \cref{thm:mHDP-upperbound} is independent of $N$, the number of observations in each group, but scales with the number of groups $D$ like $O(\text{poly}(\ln D))$. 
For fixed $D$, the approximation error decreases to zero at rate no slower that $O\left( \ln K/K \right)$.
The $O(\ln(DT))$ factor is related to the expected logarithmic growth rate of Dirichlet process mixture models \citep[Section 5.2]{arratia2003logarithmic} in the following way. Since there are $D$ groups, each manifesting at most $T$ distinct atom locations from an underlying Dirichlet process prior, the situation is akin to generating $DT$ samples from a common Dirichlet process prior. 
Hence, the expected number of unique samples is $O(\ln(DT))$. 
Similar to \cref{thm:CRM-upperbound}, we speculate that the $O(\ln^2(DT))$ factor can be improved to $O(\ln(DT))$. 
For error bounds of truncation-based approximations of hierarchical processes, such as the HDP, we refer to \citet[Theorem 1]{lijoi2020sampling}.

\section{Conceptual benefits of finite approximations}
\label{sec:conceptual}

Though approximation error lends itself more readily to analysis, ease-of-use considerations are often at the forefront of users' choice of finite approximation in practice. Therefore, we next compare AIFAs to TFAs in this dimension. We see that AIFAs offer more straightforward updates in approximate inference algorithms and easier implementation of parallelism.

To reduce notation in this section, we let a term without subscripts represent the collection of all subscripted terms: $\approxrate \defined (\approxrate_k)_{k=1}^{K}$ denotes the collection of atom sizes, 
$\psi \defined (\psi_k)_{k=1}^{K}$ denotes the collection of atom locations, 
$x \defined (x_{n,k})_{k=1,n=1}^{K,N}$ denotes the latent trait counts of each observation\secondpass{,}\footnote{The usage of $x$ in this section is different from the usage in the remaining sections: in \cref{bkg:expLPh}, $x$ is a single observation from the likelihood process.} 
and $y \defined (y_n)_{n=1}^{N}$ denotes the observed data. 
\secondpass{We use a dot to collect terms across the corresponding} subscript: $x_{.,k} \defined (x_{n,k})_{n=1}^{N}$ denotes trait counts across observations of the $k$-th trait.
We next consider algorithms to approximate the posterior distribution $\Pr(\approxrate,\psi,x\given y)$ of the finite approximation.

\textbf{Gibbs sampling.} When all latent parameters are continuous, Hamiltonian Monte Carlo methods are increasingly standard for performing Markov chain Monte Carlo (MCMC) posterior approximation \citep{hoffman2014nuts,Carpenter:2017}. 
However, due to the discreteness of the trait counts $x$, successful MCMC algorithms for CRMs or their approximations have been based largely on Gibbs sampling \citep{Geman:1984hh}. 
In particular, blocked Gibbs sampling utilizing the natural Markov blanket structure is straightforward to implement when the complete conditionals 
$\Pr(\approxrate\given x,\psi,y),\Pr(x\given \psi,\approxrate,y)$, and $\Pr(\psi\given x,\approxrate,y)$ are easy to simulate from.\footnote{
Because of the factorization $\Pr(x\given \psi,\approxrate,y) = \prod_{n=1}^{N} \Pr(x_{n,.} \given \psi,\approxrate,y_n)$, Gibbs sampling over the finite approximation can be an appealing technique even when Gibbs sampling over the marginal process is not.
In particular, the wall-time of a Gibbs iteration for the finite approximation can be small by drawing $\Pr(x_{n,.} \given \psi,\approxrate,y_n)$ in parallel.
Meanwhile, any iteration to update the trait counts with the marginal process representation needs to sequentially process the data points, prohibiting speed up through parallelism.
}

Different finite approximations with the same number of atoms $K$ change only $\Pr(\approxrate)$ in the generative model. So, of the conditionals, we expect only $\Pr(\approxrate\given x,\psi,y)$ to differ across finite approximations. We next show in \cref{lem:IFA-conditional-conjugacy} that the form of $\Pr(\approxrate\given x,\psi,y)$ is particularly tractable for AIFAs. Then we will discuss how Gibbs derivations are substantially more involved for TFAs.
\bnprop [{Conditional conjugacy of AIFA}] \label{lem:IFA-conditional-conjugacy}
Suppose the likelihood is an exponential family (\cref{bkg:expLPh}) and the AIFA prior $\nu_K$ is as in \cref{cor:expCRM-d-zero}. 
Then the complete conditional of the atom sizes factorizes across atoms as:
\begin{equation*}
	\Pr(\approxrate\mid x,\psi,y) = \prod_{k=1}^{K} \Pr(\approxrate_k \mid x_{.,k}).
\end{equation*}
Furthermore, each $\Pr(\approxrate_k\given x_{.,k})$ is in the same exponential family as the AIFA prior, with density proportional to
\begin{equation}
	\label{eq:tractable-expectation}
	\begin{aligned}
	&\indict{\approxrate \in \support} \approxrate^{c/K+\sum_{n=1}^{N}\phi(x_{n,k})-1}\exp \left( \inner{\psi+ \sum_{n=1}^{N}t(x_{n,k}) , \mu(\approxrate)} + (\lambda+N) [-A(\approxrate)] \right).
	\end{aligned}
\end{equation}
\enprop

See \cref{app-proof:conjugacy} for the proof of \cref{lem:IFA-conditional-conjugacy}.
For common models --- such as beta--Bernoulli, gamma--Poisson, and beta--negative binomial --- we see that the complete conditionals over AIFA atom sizes are in forms that are well known and easy to simulate.

There are many different types of TFAs, but typical TFA Gibbs updates pose additional challenges. Even when $\Pr(\approxrate)$ is easy to sample from, $\Pr(\approxrate\given x)$ can be intractable, as we see in the following example.

\bexa [Stick-breaking approximation \citep{broderick2012beta,paisley2011variational}] \label{exa:stick-breaking-beta}
Consider the TFA for $\distBP(\gamma,\alpha,0)$ given by
\begin{equation*}
\Theta_K = \sum_{i=1}^{K} \sum_{j=1}^{C_i} V_{i,j}^{(i)} \prod_{l=1}^{i-1} (1-V_{i,j}^{(l)}) \delta_{\psi_{ij}},
\end{equation*}
where $C_i \distiid \text{Poisson}(\gamma)$, $ V_{i,j}^{(l)} \distiid  \text{Beta}(1,\alpha)$ and $\psi_{i,j} \distiid H$. 
One can sample the atom sizes $V_{i,j}^{(i)} \prod_{l=1}^{i-1}(1-V_{i,j}^{(l)})$. But there is no tractable way to sample from the conditional distribution $\Pr(\approxrate\given x)$ because of the dependence on $C_i$ as well as the entangled form of each $\approxrate.$ 
Strategies to make sampling more tractable include introducing auxiliary round indicator variables $r_k$ and marginalizing out the stick-breaking proportions \citep{broderick2012beta}.
However, the final model still contains one Gibbs conditional that is difficult to sample from \citep[Equation 37]{broderick2012beta}.
\eexa 

Other superposition-based approximations, like decoupled Bondesson or power-law \citep{campbell2019truncated}, 
present similar challenges due to the number of atoms per round variables $C_i$ and the dependence among the atom sizes.

\textbf{Mean-field variational inference (MFVI).} Analogous to Hamiltonian Monte Carlo for MCMC, black-box variational methods are increasingly used for variational inference when the latent parameters are continuous \citep{Ranganath:2014,Kingma:2014,Rezende:2014,Burda:2016:IWAE,kucukelbir2017automatic,bingham2018pyro}. Mean-field coordinate ascent updates \citep[Section 6.3]{Wainwright:2008} remain popular for cases with discrete variables, including the present trait counts $x$.\footnote{
When discrete latent variables are present, black-box variational methods typically utilize enumeration strategies to marginalize out the discrete variables.
There exists a tradeoff between user time and wall time. 
The user time is small since there is no need to derive update equations, but the wall time can \secondpass{be} large depending on the enumeration strategy.}

MFVI posits a factorized distribution $q$ to approximate the exact posterior. In our case, we approximate $\Pr(\approxrate,\psi,x\given y)$ with $q(\approxrate,\psi,x) = q_\approxrate(\approxrate)q_\psi(\psi)q_x(x)$.
We focus on $q_{\approxrate}(\approxrate)$.
For fixed $q_\psi(\psi)$ and $q_x(x)$, the optimal $q_\approxrate^*$ minimizes the (reverse) Kullback-Leibler divergence between the posterior and $q_\approxrate^*q_\psi q_x$: 
\begin{equation} \label{eq:q1-as-optim}
	q_{\approxrate}^*  := \underset{q_\approxrate}{\mathrm{argmin}} \text{ KL} \left( q_\approxrate(\cdot)q_\psi(\cdot)q_x(\cdot) \mid \mid  \Pr(\cdot,\cdot,\cdot\given y) \right).
\end{equation}

Our next result shows that $q^*_\approxrate$ takes a convenient form when using AIFAs.
\bncor [{AIFA optimal distribution is in exponential family}] \label{cor:AIFA-q1}
Suppose the likelihood is an exponential family (\cref{bkg:expLPh}) and the AIFA prior $\nu_K$ is as in \cref{cor:expCRM-d-zero}. 
Then, the density of $q_\approxrate^*$ is given by
\begin{equation} \label{eq:AIFA-q1}
	q_\approxrate^*(\rho) = \prod_{k} \widetilde{p}_k(\rho_k),
\end{equation}
where each $\widetilde{p}_k$ has density at $\rho_k$ proportional to
\begin{equation} \label{eq:ptilde}
\indict{\approxrate_k \in \support} \rho_k^{c/K + \sum_n \EE_{x_{n,k} \sim q_x} \phi (x_{n,k})  - 1} \exp \left\langle  \begin{bmatrix}
		\psi + \sum_{n} \EE_{x_{n,k} \sim q_x} t(x_{n,k}) \\
		\lambda + N
	\end{bmatrix},  \begin{bmatrix}
		\mu(\rho_k) \\ -A(\rho_k)
	\end{bmatrix}  \right\rangle
\end{equation}
where $x_{n,k} \sim q_x$ denotes the marginal distribution of $x_{n,k}$ under $q_x(x)$.
\encor 
That is, when using the AIFA, the optimal $q_\approxrate^*$ factorizes across the $K$ atoms, and each distribution is in the conjugate exponential family for the likelihood $\traitLL(x_{n,k} \given \rho_k)$.
Typically users will report summary statistics like means or variances of the variational approximations $q_{\rho}^*$. These are typically straightforward from the exponential family form.

The TFA case is much more complex and requires both more steps in the inference scheme as well as additional approximations. See \cref{app:conceptual-proofs} for two illustrative examples.

\textbf{Parallelization.} We end with a brief discussion on parallelization. 
In both \cref{lem:IFA-conditional-conjugacy} and \cref{cor:AIFA-q1}, the update distribution for $\rho$ factorizes across the $K$ atoms.
Hence, AIFA updates can be done in parallel across atoms, yielding speed-ups in wall-clock time, with the gains being greatest when there are many instantiated atoms.
For TFAs, due to the complicating coupling among the atom rates, there is no such benefit from parallelization.

\section{Empirical evaluation}
\label{sec:experiments}

\secondpass{In our experiments, we compare our AIFA constructions to TFAs and to other IFA constructions \citep{lee2016finite,lee2022unified} on a variety of synthetic and real-data examples. Even though our theory suggests better performance of TFAs than AIFAs for worst-case likelihoods, we find comparable performance of TFAs and AIFAs in predictive tasks (\cref{subsec:image-denoising,subsec:topic-modelling}). Likewise, we find comparable performance of AIFAs and alternative IFAs in predictive tasks (\cref{subsec:synthetic}). However, we find that AIFAs can be used to learn model hyperparameters where alternative IFA approximations fail (\cref{subsec:discount-estimation}). And we show that AIFAs can be used to learn model hyperparameters for new models, not previously explored in the BNP literature (\cref{subsec:dispersion-estimation}).
}

In relation to prior studies, existing empirical work has compared IFAs and TFAs \secondpass{only} for simpler models and smaller data sets (e.g., \citet[Table 1,2]{doshi-velez2009variational} and \citet[Figure 4]{Kurihara:2007}). 
Our comparison is grounded in models with more levels and analyzes datasets of much larger sizes. 
For instance, in our topic modeling application, we analyze nearly $1$ million documents, while the comparison in \citet{Kurihara:2007} utilizes only $200$ synthetic data points.

\subsection{Image denoising with the beta--Bernoulli process}
\label{subsec:image-denoising}

\secondpass{Our first experiments show comparable performance of the AIFA and TFA at an image denoising task with a CRM-based target model.}
We use MCMC for image denoising through dictionary learning because it is an application where finite approximations of BNP models --- in particular \secondpass{the} beta--Bernoulli \secondpass{process} with $d = 0$ --- have proven useful \citep{zhou2009nonparametric}. 
The observation likelihood in this dictionary learning model is not one of the worst cases in \cref{subsec:CRM}. 
We find that the performance of AIFAs and TFAs is comparable across $\approxlev$, and the posterior modes across TFA and AIFA models are similar to each other. 

\begin{figure}[!t]
	\begin{subfigure}[b]{.24\linewidth}
		\centering
		\includegraphics[width=\linewidth]{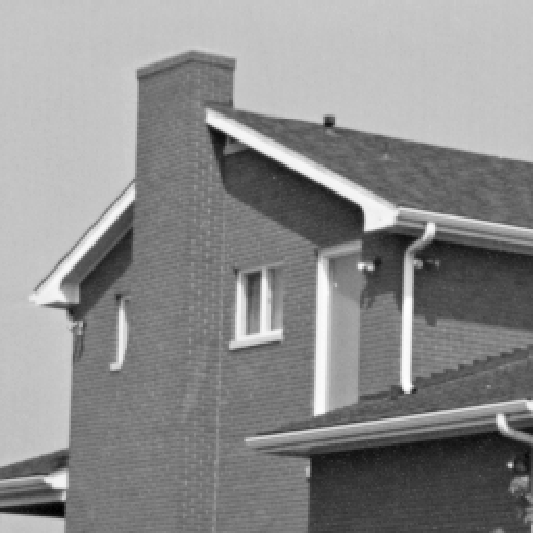}
		\caption{Original}\label{sub-fig:original}
	\end{subfigure}%
	\begin{subfigure}[b]{.24\linewidth}
		\centering
		\includegraphics[width=\linewidth]{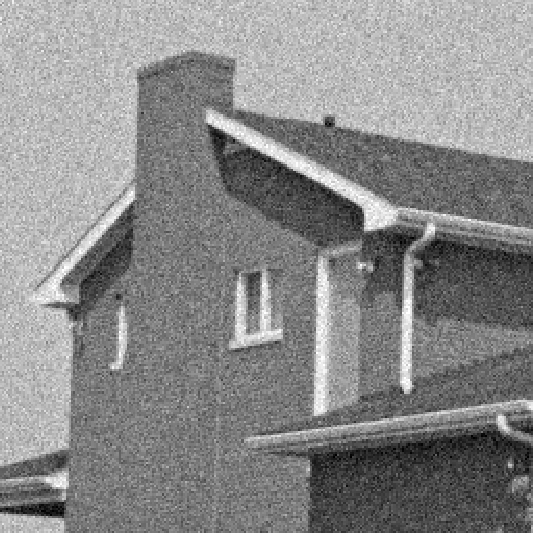}
		\caption{Input, $24.64$ dB}\label{sub-fig:corrupted}
	\end{subfigure}%
	\begin{subfigure}[b]{.24\linewidth}
		\includegraphics[width=\linewidth]{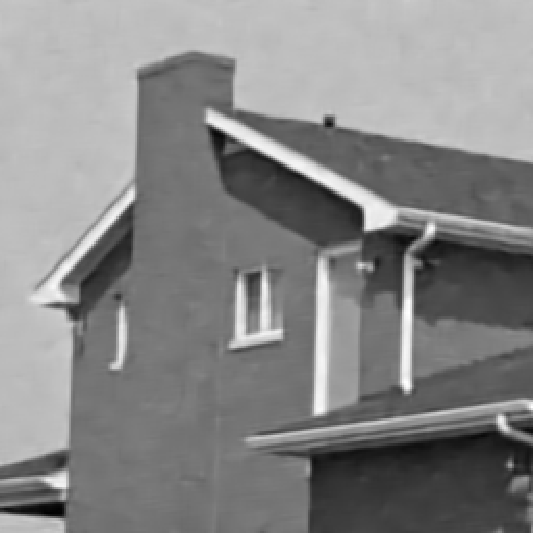}
		\caption{AIFA, $33.81$ dB}\label{sub-fig:IFA}
	\end{subfigure}%
	\begin{subfigure}[b]{.24\linewidth}
		\includegraphics[width=\linewidth]{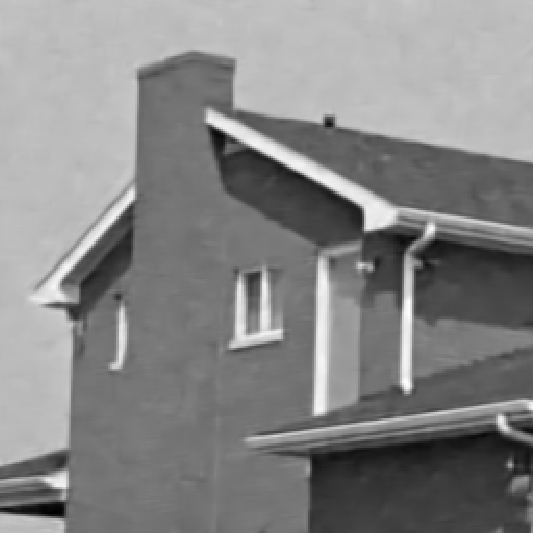}
		\caption{TFA, $34.03$ dB}\label{sub-fig:TFA}
	\end{subfigure}
	\caption{ AIFA and TFA denoised images have comparable quality. \textbf{(a)} The noiseless image. \textbf{(b)} The corrupted image. \textbf{(c,d)} Sample denoised images from finite models with $K = 60$. We report PSNR (in dB) with respect to the noiseless image.}
	\label{exp-fig:example-denoising}
\end{figure}

The goal of image denoising is to recover the original, noiseless image (e.g., \cref{sub-fig:original}) from a corrupted one (e.g., \cref{sub-fig:corrupted}). 
The input image is first decomposed into small contiguous patches.
The model assumes that each patch is a combination of latent \emph{basis elements}. 
By estimating the coefficients expressing the combination, one can denoise the individual patches and ultimately the overall image. 
The beta--Bernoulli process allows simultaneous estimation of both basis elements and basis assignments. 
The number of extracted patches depends on both the patch size and the input image size. So even on the same input image, the analysis might process a varying number of ``observations.''
The nonparametric nature of the beta--Bernoulli process sidesteps the cumbersome problem of calibrating the number of basis elements for these different data set sizes,
which can be large even for a relatively small image; for a $256 \times 256$ image like \cref{sub-fig:corrupted}, 
the number of extracted patches, $N$, is about 60,000. 
We quantify denoising quality by computing the peak signal-to-noise ratio (PNSR) between the original and the denoised image \citep{hore2010image}. The higher the PNSR, the more similar the images. 

We use Gibbs sampling to approximate the posterior distributions. 
To ensure stability and accuracy of the sampler, patches (i.e., observations) are gradually introduced in epochs, and the sampler modifies only the latent variables of the current epoch's observations.
See \cref{app:image-setup} for more details about the finite approximations, the hyperparameter settings, and the inference algorithm.

\begin{figure}[!t]
	\begin{subfigure}[b]{.33\linewidth}
		\centering
		\includegraphics[width=\linewidth]{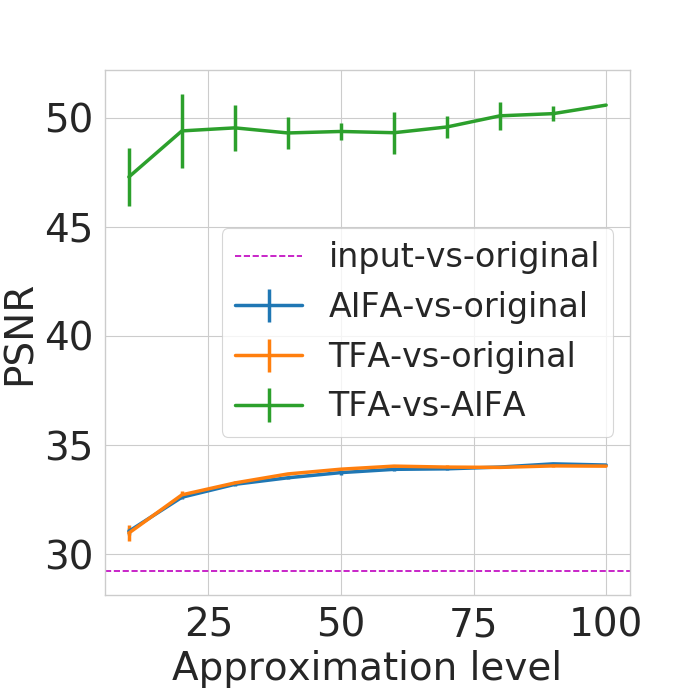}
		\caption{Performance across $K$}\label{sub-fig:house256-similar-metric}
	\end{subfigure}%
	\begin{subfigure}[b]{.33\linewidth}
		\centering
		\includegraphics[width=\linewidth]{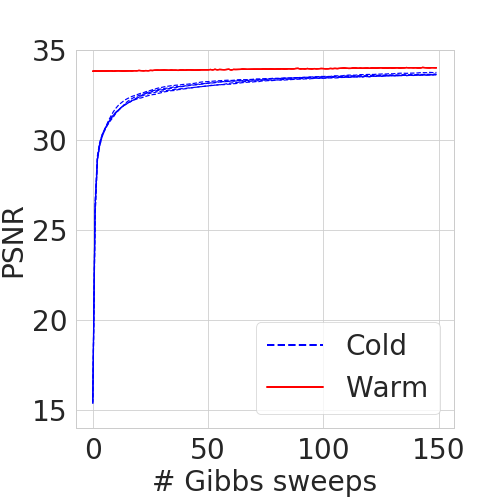}
		\caption{TFA training}\label{sub-fig:house256-TFA-similar-posterior}
	\end{subfigure}%
	\begin{subfigure}[b]{.33\linewidth}
		\includegraphics[width=\linewidth]{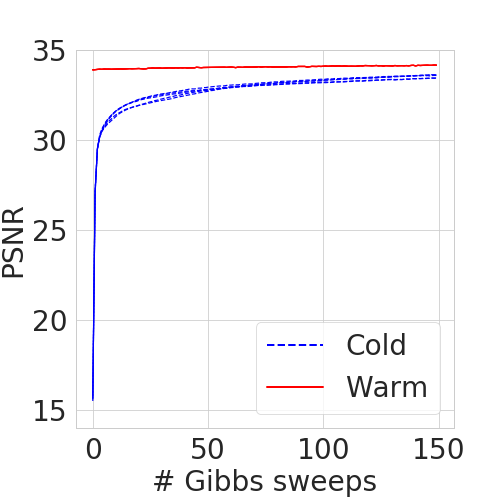}
		\caption{AIFA training}\label{sub-fig:house256-IFA-similar-posterior}
	\end{subfigure}
	\caption{\textbf{(a)} Peak signal-to-noise ratio (PNSR) as a function of approximation level $K$.
		Error bars depict 1-standard-deviation ranges across $5$ trials. 
		\textbf{(b,c)} How PSNR evolves during inference \secondpass{across 10 trials, with 5 each starting from respectively cold or warm starts.} 
		}
	\label{fig:house256-results}
\end{figure}

\cref{sub-fig:IFA,sub-fig:TFA} visually summarize the results of posterior inference for a particular image.
We report experiments with other images in \cref{app:more-denoising}.
Our results across all images
indicate that the AIFA and TFA perform similarly, and 
both approximations perform much better than the baseline (i.e., the noisy input image).
\cref{fig:house256-results} quantitatively confirms these qualitative findings; \cref{sub-fig:house256-similar-metric} shows that, for approximation levels we considered, the PSNR between either the TFA or AIFA output image and the original image are always very similar and substantially higher (between 30 and 35) than the PSNR between the original and corrupted image (below 30).
In fact, each TFA denoised image is more similar to the AIFA denoised image than to the original image;
the PSNR between the TFA and AIFA outputs is about 50.
We also see from \cref{sub-fig:house256-similar-metric} that the quality of denoised images improves with increasing $K$. 
The improvement with $K$ is largest for small $K$, and plateaus for larger values of $K$. 
In addition to randomly initializing the latent variables at the beginning of the Gibbs sampler of one model (``cold start''), we can use the last configuration of latent variables visited in the other model as the initial state of the Gibbs sampler (``warm start''). 
In \cref{sub-fig:house256-TFA-similar-posterior}, the warm-start curve uses the output of inference with the AIFA as an initial value for inference with the TFA; similarly, the warm-start curve of \cref{sub-fig:house256-IFA-similar-posterior} uses the output with the TFA to initialize inference with the AIFA.
For both approximations, $K = 60$. 
At the end of training, all latent variables for all patches have been assigned, so for the warm start experiment, we make all patches available from the start instead of gradually introducing patches.  
For both approximations, the Gibbs sampler initialized at the warm start visits candidate images that essentially have the same PSNR as the starting configuration; the PSNR values never deviate from the initial PSNR by more than $1\%.$
The early iterates of the cold-start Gibbs sampler are noticeably lower in quality compared to the warm-start iterates, and the quality at the plateau is still lower than that of the warm start.\footnote{Because \secondpass{the warm start represents the end of the training from the cold start} with gradually introduced patches, the gap in final PSNR is due to the gradual patch introduction.}
Each PSNR trace corresponds to a different set of initial values and simulation of the conditionals.
The variation across \secondpass{the 5 warm-start trials} is small; the variation across \secondpass{the 5 cold-start trials} is larger but still quite small. 
In all, the modes of TFA posterior are good initializations for inference with the AIFA model, and vice versa.

\subsection{Topic modelling with the modified hierarchical Dirichlet process}
\label{subsec:topic-modelling}
We next compare the performance of normalized AIFAs (namely, $\distFSD_K$) and TFAs (namely, $\distTSB_K$) in a $\distDP$-based model with additional hierarchy: the modified HDP from \cref{subsec:mHDP}. 
As in \cref{subsec:image-denoising}, we find that the approximations perform similarly. 

We use the modified HDP for topic modeling. 
We apply stochastic variational inference with mean-field factorization \citep{Hoffman:2013} to approximate the posterior over the latent topics. 
The training corpus consists of nearly one million documents from Wikipedia.
We measure the quality of inferred topics via predictive log-likelihood on a set of $10{,}000$ held-out documents. 
See \cref{app:topic-setup} for complete experimental details. 

\begin{figure}[!t]
	\begin{subfigure}[b]{.32\linewidth}
		\centering
		\includegraphics[width=\linewidth]{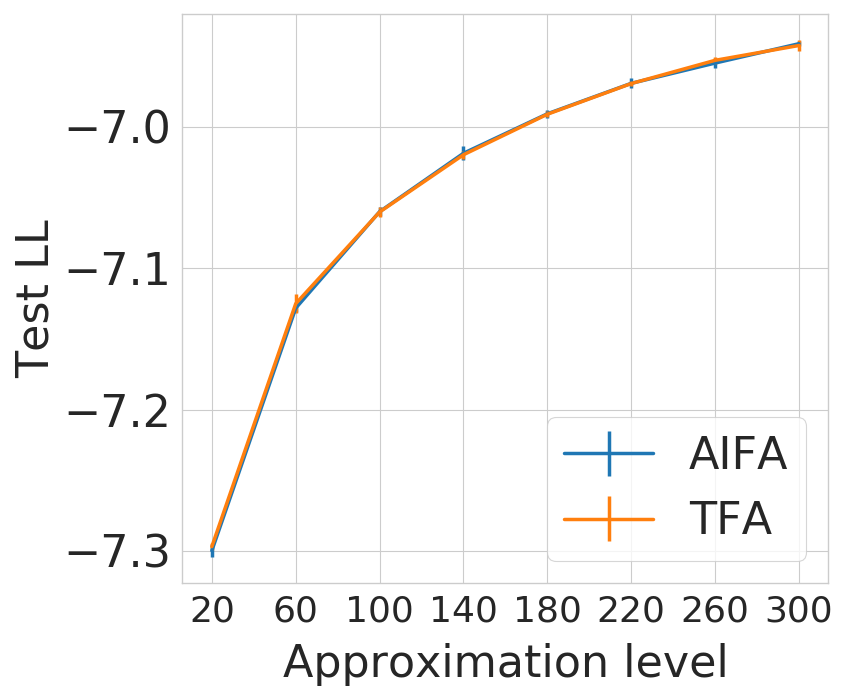}
		\caption{Performance across $K$}\label{sub-fig:wiki1m-similar-metric}
	\end{subfigure}%
	\begin{subfigure}[b]{.345\linewidth}
		\centering
		\includegraphics[width=\linewidth]{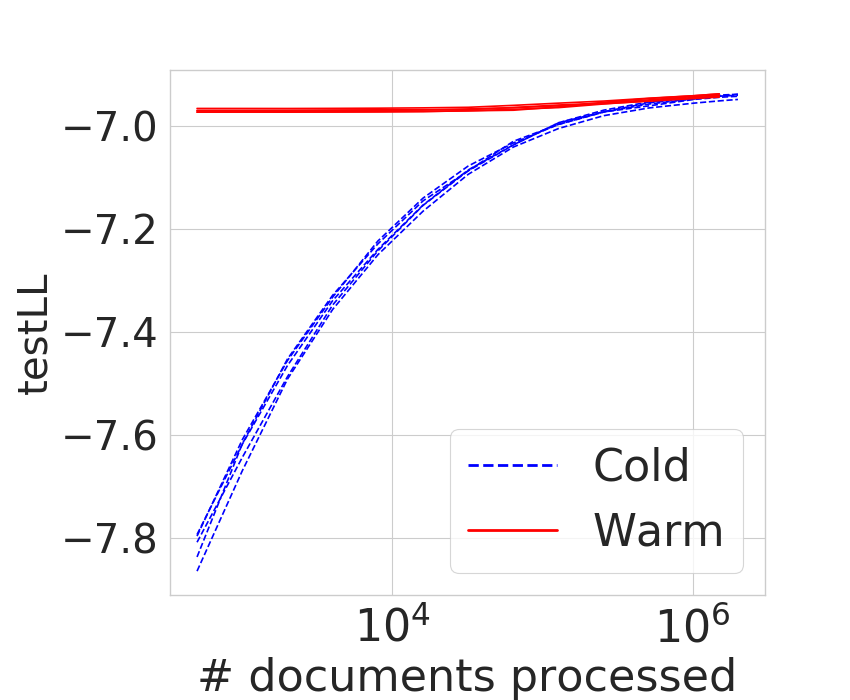}
		\caption{TFA training}\label{sub-fig:wiki1m-TFA-similar-posterior}
	\end{subfigure}%
	\begin{subfigure}[b]{0.345\linewidth}
		\includegraphics[width=\linewidth]{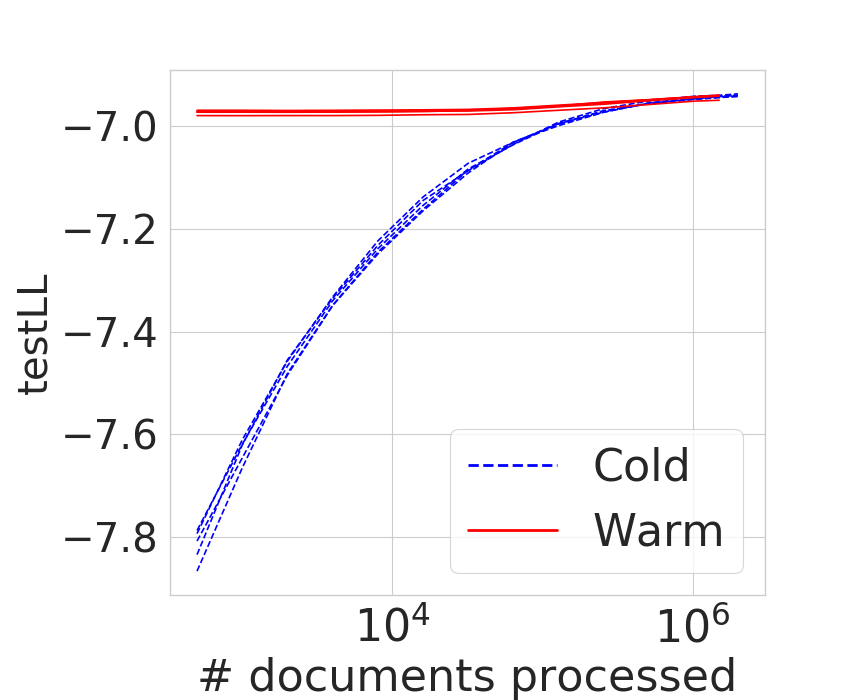}
		\caption{AIFA training}\label{sub-fig:wiki1m-IFA-similar-posterior}
	\end{subfigure}%
	\\
	\caption{
		\textbf{(a)} Test log-likelihood (testLL) as a function of approximation level $K$. Error bars show 1 standard deviation across 5 trials. 
		\textbf{(b,c)} TestLL change during inference. }
	\label{fig:wiki1m-results}
\end{figure}

\cref{sub-fig:wiki1m-similar-metric} shows that, as expected,
the quality of the inferred topics improves as the approximation level grows.
For a given approximation level, the quality of the topics learned using the TFA and the normalized AIFA are almost the same. 

The warm start in this case corresponds to using variational parameters at the end of the other model's training.
\cref{sub-fig:wiki1m-TFA-similar-posterior} uses the outputs of inference with the normalized AIFA approximation as initial values for inference with the normalized TFA; similarly \cref{sub-fig:wiki1m-IFA-similar-posterior} uses the TFA to initialize inference with the AIFA. 
We fix the number of topics to $K = 300$ \secondpass{and run 5 trials each with the cold start and warm start, respectively.}
For both approximations, the test log-likelihood stays nearly the same for warm-start training iterates; the test log-likelihood for the iterates never deviate more than $0.5\%$ from the initial value.
The early iterates after the cold start are noticeably lower in quality compared to the warm iterates; however at the end of training, the test log-likelihoods are nearly the same. 
Each trace corresponds to a different set of initial values and ordering of data batches processed.
The variation across either cold starts or warm starts is small. 
So, in sum, the modes of the TFA posterior are good initializations for inference with the AIFA model, and vice versa.

\subsection{Comparing \secondpass{predictions across} independent finite approximations}
\label{subsec:synthetic}

\secondpass{We next show that AIFAs have comparable predictive performance with other IFAs, namely the BFRY IFA and \genpar{} IFA.}
We consider a linear--Gaussian factor analysis model with the power-law beta--Bernoulli process \citep{Griffiths:2011a}, where \secondpass{the AIFA, BFRY IFA, or \genpar{} IFA} can be used directly. 

\begin{figure}[t]
	\begin{subfigure}[b]{0.5\linewidth}
		\centering
		\includegraphics[width=\linewidth]{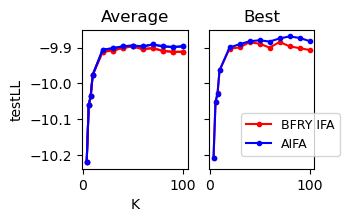}
		\caption{BFRY IFA versus AIFA}\label{sub-fig:bfry-v-aifa}
	\end{subfigure}%
	\begin{subfigure}[b]{0.5\linewidth}
		\centering
		\includegraphics[width=\linewidth]{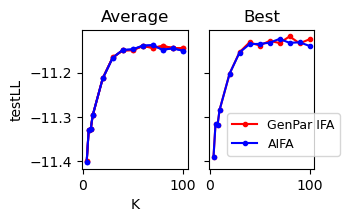}
		\caption{\secondpass{\genpar{}} IFA vs AIFA}\label{sub-fig:genpar-v-aifa}
	\end{subfigure}%
	\caption{ \firstpass{
	\textbf{(a)} The left panel shows \secondpass{the average predictive log-likelihood of the AIFA (blue) and BFRY IFA (red) as a function of the approximation level $K$; the average is across $10$ trials with different random seeds for the stochastic optimizer.
	 The right panel shows highest predictive log-likelihood across the same $10$ trials.}
	 \textbf{(b)} The panels are analogous to \textbf{(a)}, except \secondpass{the \genpar{}} IFA is in red.}}
	\label{fig:aifa-vs-other-ifa}
\end{figure}

\secondpass{Recall that the BFRY IFA applies only when the concentration hyperparameter is zero, and the \genpar{} IFA applies
only when the concentration parameter is positive. We consider it a strength of the AIFA that it applies to both cases (and the negative range of the concentration
hyperparameter) simultaneously. Nonetheless, we here generate two separate synthetic datasets: one to compare the BFRY IFA with the AIFA and one to compare the \genpar{} IFA with the AIFA. In each case, } \firstpass{ 
we generate $2{,}000$ data points from the full CRM model with a discount of $d = 0.6$.}
We use $1{,}500$ for training and report predictive log-likelihood on the $500$ held-out data points.
For posterior approximation, we use automatic differentiation variational inference as implemented in Pyro \citep{bingham2018pyro}.
To isolate the effect of the approximation type, we use ``ideal'' initialization conditions: we initialize the variational parameters
using the latent features, assignments, and variances that generated the training set.
See \cref{app:synthetic-setup} for more details about the BRFY IFA, \secondpass{\genpar{} IFA,} and the approximate inference scheme.
\cref{sub-fig:bfry-v-aifa} shows that across approximation levels $K$, the predictive performances of the AIFA and BFRY IFA are similar. 
\secondpass{Likewise, \cref{sub-fig:genpar-v-aifa} shows that the predictive performance of \secondpass{the AIFA and \genpar{}} IFA are similar.
}

\subsection{Discount estimation}
\label{subsec:discount-estimation}

\secondpass{We next show that AIFAs can reliably recover the beta process discount hyperparameter $d$, which governs the power law growth in the number of features. By contrast, we show that the BFRY IFA or \genpar{} IFA struggle at this task. In \cref{app:mass-concentration}, we show that the AIFA can also reliably estimate the mass and concentration hyperparameters.}

\secondpass{We generate a synthetic dataset so that the ground truth hyperparameter values are known. The data takes the form of a binary matrix $X$, with $N$ rows and $\tilde{K}$ columns.
We} generate $X$ from an Indian buffet process prior; recall that the Indian buffet process is the marginal process of a beta process CRM paired with Bernoulli likelihood.
\secondpass{To learn the hyperparameter values with an AIFA, we maximize the marginal likelihood of the observed matrix $X$ implied by the AIFA.}
\secondpass{In particular, we} compute the marginal likelihood by integrating the Bernoulli likelihood $\Pr(x_{n,k} \given \theta_k)$ over $\theta_k$ distributed as the $K$-atom AIFA $\nu_K$.
To quantify the variability of the estimation procedure, we generate $50$ feature matrices and compute the maximum likelihood estimate for each of these $50$ trials.
See \cref{app:discount-setup} for more experimental details.

\begin{figure}[t]
	\begin{subfigure}[b]{0.5\linewidth}
		\centering
		\includegraphics[width=\linewidth]{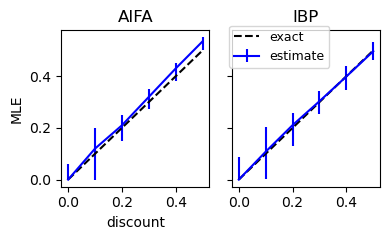}
		\caption{Maximum likelihood estimates}
		\label{sub-fig:discountEstimation}
	\end{subfigure}%
	\begin{subfigure}[b]{0.5\linewidth}
		\centering
		\includegraphics[width=\linewidth]{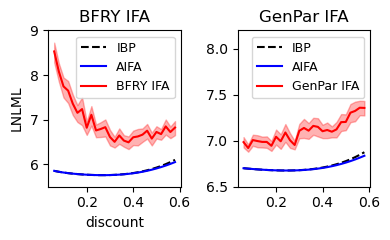}
		\caption{Log negative log marginal likelihood}
		\label{sub-fig:likelihoodCurve}
	\end{subfigure}
	\caption{\textbf{(a)} We estimate the discount by maximizing the marginal likelihood of the AIFA (left) or the full process (right).
	The solid blue line is the median of the estimated discounts, while the lower and upper bounds of the error bars are the $20\%$ and $80\%$ quantiles. 
	The black dashed line is the ideal value of the estimated discount, equal to the ground-truth discount. 
	\firstpass{\textbf{(b)} In each panel, the solid red line is the average log of negative log marginal likelihood (LNLML) across batches. 
	\secondpass{The light red region depicts two standard errors in either direction from the mean.}
	}}
\end{figure}

\cref{sub-fig:discountEstimation} shows that we can use an AIFA to estimate the underlying discount for a variety of ground-truth discounts\secondpass{.
Since the estimates and error bars are similar whether we use the AIFA (left) or full nonparametric process (right), we conclude that using the AIFA yields comparable inference to using the full process.}

In theory, the marginal likelihood of the BFRY IFA can also be used to estimate the discount\secondpass{, but in practice we find that this approach is not straightforward and can yield unreliable estimates.}
\secondpass{At the time of writing,} such an experiment \secondpass{had not yet} been \secondpass{attempted; \citet{lee2016finite} focus} on clustering models \secondpass{and do} not discuss strategies \secondpass{to estimate any hyperparameter in a feature allocation model with a BFRY IFA.}
\firstpass{We are not aware of a closed-form formula for the marginal likelihood.
Default schemes to numerically integrate $\Pr(0 \mid \theta_k)$ against the BFRY prior for $\theta_k$ fail because of overflow issues. 
$(K \Gamma(d) d/\gamma)^{1/d}$ is typically very large, especially for small $d$.
Due to finite precision, $1-\exp\left( - (Kd/\gamma)^{1/d} \frac{\theta}{1-\theta} \right)$ evaluates to $1$ on the quadrature grid used by numerical integrators \citep{piessens2012quadpack}}.
\thirdpass{In this case, \cref{eq:BFRY-nuK1} behaves as $\theta^{-d - 1}$ near $0$, and thus the integral over $\theta$ diverges.}
\firstpass{To create the left panel of \cref{sub-fig:likelihoodCurve}, we view the marginal likelihood as an expectation and construct Monte Carlo estimates\secondpass{;} we draw $10^5$ BFRY samples to estimate the marginal likelihood, and \secondpass{we} take the estimate's logarithm as an approximation to the log marginal likelihood \secondpass{(red line)}.
To quantify the uncertainty, we draw $100$ batches of $10^5$ samples \secondpass{(light red region)}}.
Even for \secondpass{this large number of Monte Carlo samples}, the \secondpass{estimated} log marginal likelihood curve is too noisy to be useful for \secondpass{hyperparameter} estimation.
\secondpass{By comparison, we can compute the log marginal likelihood analytically for the IBP (dashed black line); it is much smoother and features a clear minimum.
Moreover, we can compute the AIFA log marginal likelihood via numerical integration (solid blue line); it is also very smooth and features a clear minimum.}

\secondpass{We again consider the BFRY IFA and \genpar{} IFA separately and generate separate simulated data for each case due to their disjoint assumptions; we generate date with concentration $\alpha=0$ for the BFRY IFA and with $\alpha > 0$ for the \genpar{} IFA.
An experiment to recover a discount hyperparameter with the \genpar{} IFA, analogous to the experiment above with the BFRY IFA, has also not previously been attempted.}
\secondpass{There is no analytical formula for the \genpar{} IFA marginal likelihood, 
and we again encounter overflow when trying numerical integration. 
Therefore, we resort to Monte Carlo; we find that estimates of the log marginal likelihood are too noisy} for practical use in recovering the discount (the right panel of \Cref{sub-fig:likelihoodCurve}).

\subsection{Dispersion estimation}
\label{subsec:dispersion-estimation}

\secondpass{Finally, we show that the AIFA can straightforwardly be adapted to estimate hyperparameters in other BNP processes, not just the beta process. In particular we show that AIFAs can be used to learn the dispersion parameter $\tau$ in the novel Xgamma--CMP process that we introduced in \cref{exa:eG-CMP}.} 
We consider a well-known application of BNP trait-allocation models to matrix-factorization--based topic modeling \citep{roychowdhury2015gamma}. 
The observed data is a count matrix $X$, with $N$ rows, representing documents, and $V$ columns, representing vocabulary words.
We adjust the model of \citet{roychowdhury2015gamma} to use the Xgamma--CMP process of \cref{exa:eG-CMP} instead of a gamma--Poisson process.
The added flexibility of $\tau$ allows modeling trait count distributions that are over- or under-dispersed, which cannot be done with the gamma-Poisson process.

To have a notion of ground truth, we generate synthetic data (with $N = 600$) from a large AIFA (with $K = 500$) of the Xgamma--CMP process, which is a good approximation of the BNP limit.\footnote{For the chosen number of documents $N$, let the number of traits with positive count be $\widehat{K}$. There is no noticeable difference in the distribution of $\widehat{K}$ between $K = 500$ and $K > 500$. The rates of the inactive (zero count) traits are smaller than $1/N$. %
}
\secondpass{In each set of experiments, the data are overdispersed ($\tau < 1$) or underdispersed ($\tau > 1$). 
In this case, we take a Bayesian approach to estimating $\tau$, and put a uniform prior on $\tau \in (0,100]$ since $\tau$ must be strictly positive.
For smaller values} of $K$ ($K = 50$ to $K = 150$), we approximate the posterior for \secondpass{the} $K$-atom AIFA using Gibbs sampling\secondpass{.}
See \cref{app:dispersion-setup} for more \secondpass{details about the} experimental setup.

\begin{figure}[t]
	\centering
	\includegraphics[width=\linewidth]{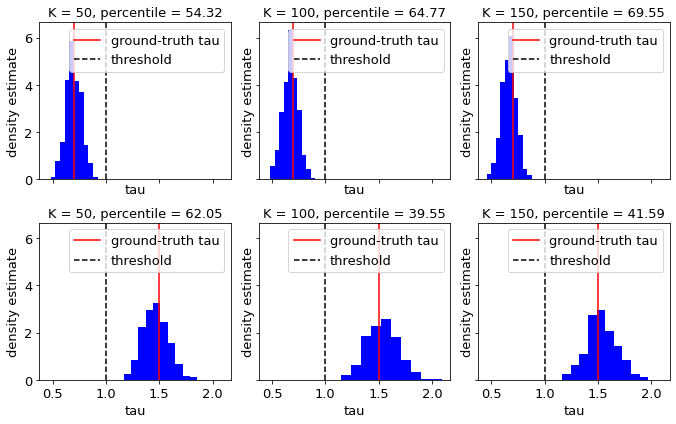}
	\caption{\secondpass{Blue histograms show posterior density} estimates for $\tau$ from MCMC draws. The ground-truth $\tau$ \secondpass{(solid red line) is $0.7$ in the overdispersed case (upper row) and $1.5$ in the underdispersed case (lower row).} The \secondpass{threshold $\tau=1$ (dashed black line)} marks the transition from overdispersion ($\tau < 1.0$) to underdispersion ($\tau > 1.0$). The percentile in each panel's title is the percentile \secondpass{where} the ground truth $\tau$ \secondpass{falls} in the posterior draws. \secondpass{The approximation size $K$ of the AIFA increases in the plots from left to right.}}
	\label{fig:dispersion-estimation}
\end{figure}

\cref{fig:dispersion-estimation} shows that \secondpass{the posterior approximation agrees with the ground truth on the dispersion type (over or under) in each case.}
\secondpass{We also see from the figures that} the $95\%$ credible intervals contain the ground-truth $\tau$ \secondpass{value in each case}.

\section{Discussion} 
\label{sec:discussion}

We have provided a general construction of automated independent finite approximations (AIFAs) for completely random measures and their normalizations. 
Our construction provides novel finite approximations not previously seen in the literature.
For processes without power-law behavior, we provide approximation error bounds; our bounds show that we can ensure accurate approximation by setting the number of atoms $K$ to be (1) logarithmic in the number of observations $N$ and (2) inverse to the error tolerance $\epsilon$.
We have discussed how the independence and automatic construction of AIFA atom sizes lead to convenient inference schemes. 
A natural competitor for AIFAs is a truncated finite approximation (TFA). We show that, for the worst case choice of observational likelihood and the same $K$, AIFAs can incur larger error than the corresponding TFAs.
However, in our experiments, we find that the two methods have essentially the same performance in practice.
Meanwhile, AIFAs are overall easier to work with than TFAs, whose coupled atoms complicate the development of inference schemes.
Future work might extend our error bound analysis to conjugate exponential family CRMs with power-law behavior. 
An obstacle to upper bounds for the positive-discount case is the verification of the clauses in \cref{condition:marginal-process}.
In the positive-discount case, the functions $h$ and $M_{n,x}$, which describe the marginal representation of the nonparametric process, take forms that are straightforwardly amenable to analysis. But the function $\widetilde{h}$, which describes the finite approximations, is complex.
In general, $\widetilde{h}$ is equal to the ratio of two normalization constants of different AIFAs.
The normalization constants can be computed numerically. However, to make theoretical statements such as the clauses in \cref{condition:marginal-process}, we need to prove their smoothness properties.
Another direction is to tighten the error upper bound by focusing on specific, commonly-used observational likelihoods --- in contrast to the worst-case analysis we provide here.
Finally, more work is required to directly compare the size of error in the finite approximation to the size of error due to approximate inference algorithms such as Markov chain Monte Carlo or variational inference.

\section*{Acknowledgments} 
\label{sec:acknowledgments}

Tin D. Nguyen, Jonathan Huggins, Lorenzo Masoero, and Tamara Broderick were supported in part by ONR grant N00014-17-1-2072, 
NSF grant CCF-2029016, ONR MURI grant N00014-11-1-0688, and a Google Faculty Research Award.
Jonathan Huggins was also supported by the National Institute of General Medical Sciences of the National Institutes of Health 
under grant number R01GM144963 as part of the Joint NSF/NIGMS Mathematical Biology Program. 
The content is solely the responsibility of the authors and does not necessarily represent the official views of the National Institutes of Health.

\bibliographystyle{imsart-nameyear}
\bibliography{library,refs}
\newpage
\appendix

\numberwithin{equation}{section}
\numberwithin{figure}{section}

\section{Additional examples of AIFA construction} \label{app:more-examples}
Let $B(\alpha, \beta) = \frac{\Gamma(\alpha)\Gamma(\beta)}{\Gamma(\alpha + \beta)}$ denote the beta function.

\bexa[Beta prime process]
Taking $V = \reals_{+}$, $g(\theta) = (1 + \theta)^{-1}$, $h(\theta; \eta) = (1 + \theta)^{-\eta}$,
and $Z(\xi, \eta) = B(\xi, \eta)$ in \cref{thm:default-CRM-convergence} yields the beta prime process of \citet{Broderick:2015}, 
which has rate measure
\begin{equation*}
	\nu(\dee\theta) = \frac{\gamma}{B(\eta, 1 - d)}\theta^{-1-d}(1 + \theta)^{- d - \eta}\dee\theta.
\end{equation*}
Since $g$ is continuous, $g(0) = 1$, $1 \le g(\theta) \le 1 + \theta$, and $h(\theta; \eta)$ 
is continuous and bounded on $[0,1]$, \cref{assume:rate-measure-near-0} holds.

In the case of $d = 0$, the corresponding exponential family distribution is beta prime.
With two placeholder parameters $\alpha$ and $\beta$, the beta prime density at $\theta > 0$ is
\begin{equation*}
	\distBetaPrime(\theta; \alpha, \beta) =  \frac{\theta^{\alpha-1}(1+\theta)^{-\alpha-\beta}}{B \left( \alpha,\beta \right)}.
\end{equation*}
To construct AIFA using \cref{cor:expCRM-d-zero}, we set $c = \gamma \eta$ and 
\begin{equation*}
\nu_{\approxlev}(\theta) = \distBetaPrime(\theta; \gamma\eta/\approxlev, \eta).
\end{equation*}
\eexa

\bexa[Generalized gamma process] \label{exa:gamma}
Taking $V = \reals_{+}$, $g(\theta) = 1$, $h(\theta; \lambda) = e^{-\lambda\theta}$, and 
$Z(\xi, \lambda) = \Gamma(\xi)\lambda^{-\xi}$ in \cref{thm:default-CRM-convergence} yields the generalized gamma process, with rate measure
\begin{equation*}
\nu(\dee\theta) = \gamma \frac{\lambda^{1-d}}{\Gamma(1-d)}\theta^{-d-1}e^{-\lambda \theta}\dee\theta.
\end{equation*}
Since $h(\theta;\eta)$ is continuous and bounded on $[0,1]$, \cref{assume:rate-measure-near-0} holds. 

In the case of $d = 0$ i.e.\ the gamma process, the corresponding exponential family distribution is gamma.
To construct AIFA using \cref{cor:expCRM-d-zero}, we set $c = \gamma \lambda$ and 
\begin{equation*}
	\nu_{\approxlev}(\theta) = \distGamma(\theta; \gamma\lambda/\approxlev, \lambda).
\end{equation*}
\eexa 

\bexa[PG($\alpha$,$\zeta$)-generalized gamma process] \label{exa:generalized-gamma}
Taking $V = \reals_{+}^{2}$, $g(\theta) = 1$, $h(\theta; \eta) = e^{-(\eta_{1}\theta)^{\eta_{2}}}$, 
and $Z(\xi, \eta) = \Gamma(\xi/\eta_{2})(\eta_{1}\eta_{2})^{-\xi}$
in \cref{thm:default-CRM-convergence} yields the PG($\alpha$,$\zeta$)-generalized gamma process whose rate measure is 
\begin{equation*}
\nu(\dee\theta) = \frac{\gamma (\eta_{1}\eta_{2})^{1-d}}{\Gamma((1-d)/\eta_{2})} \theta^{-d-1} e^{- (\eta_{1} \theta)^{\eta_{2}}}\dee\theta.
\end{equation*}
Since $h(\theta; \eta)$ is continuous and bounded on $[0,1]$, \cref{assume:rate-measure-near-0} holds.

In the positive discount case, let $c = \frac{\gamma (\eta_{1}\eta_{2})^{1-d}}{\Gamma((1-d)/\eta_{2})}$, and the finite-dimensional distribution has density equalling
\begin{equation*}
	\frac{1}{Z_K} \theta^{c/K - 1 - dS_{1/K}(1-1/K)}e^{- (\eta_{1} \theta)^{\eta_{2}}}\dee\theta,
\end{equation*}
where $Z_K := \int_0^{\infty} \theta^{c/K - 1 - dS_{1/K}(1-1/K)}e^{- (\eta_{1} \theta)^{\eta_{2}}}\dee\theta.$

In the case of $d = 0$, the corresponding exponential family distribution is generalized gamma. 
With three placeholder parameters $\xi'$, $\eta_1'$, and $\eta_2'$, the generalized gamma density at $\theta> 0$ is
\begin{equation*}
	\distGenGamma(\theta; \xi', \eta_{1}', \eta_{2}') = \frac{(\eta_2'/\xi'^{\eta_1'}) \theta^{\eta_1'-1} e^{-(\theta/\xi')^{\eta_2'}} }{\Gamma(\eta_1'/\eta_2')}
\end{equation*}
To construct AIFA using \cref{cor:expCRM-d-zero}, we set $c = \frac{\gamma \eta_{1}\eta_{2}}{\Gamma(\eta_{2}^{-1})}$ and 
\begin{equation*}
\nu_{\approxlev}(\theta) = \distGenGamma\left(\theta ; \frac{1}{\eta_{1}}, \frac{\gamma \eta_{1}\eta_{2}}{\approxlev \Gamma(\eta_{2}^{-1})}, \eta_{2}\right).
\end{equation*}
\eexa 

\bexa[Extended gamma process]
Taking $V = (0, \infty) \times (1,\infty)$, $g(\theta)  = 1$, $h(\theta; \eta) = Z_{\tau}^{-c}(\theta)$, $U = [0,T]$,
and $Z(\xi, \eta) = \int_{0}^{\infty} \theta^{\xi - 1} Z_{\tau}^{-c}(\theta) d\theta$ in \cref{thm:default-CRM-convergence} yields the extended gamma process from \cref{eq:eG}.
Since $g(\theta) = 1$, the second condition in \cref{assume:rate-measure-near-0} holds.
For any $\tau$ and $c$, $Z_{\tau}^{-c}(\theta)$ is continuous and bounded on $[0,1]$, so the third condition in \cref{assume:rate-measure-near-0} holds.
As for the first condition, we note that $ Z_{\tau}^{-c}(\theta) \leq (1+ \theta)^{-c}$, since the minimum of $Z_{\tau}(\theta)$ with respect to $\tau$ is $1 + \theta$, attained at $\tau = \infty$.
Therefore, $Z(\xi, \eta)$ is finite if 
\begin{equation*}
	\int_{0}^{T} \theta^{\xi - 1}(1 + \theta)^{-c}d\theta
\end{equation*}
is finite.
Since $(1 + \theta)^{-c} \leq 1$, the last integral is at most
\begin{equation*}
	\int_{0}^{T} \theta^{\xi - 1} d\theta = \frac{T^{\xi}}{\xi},
\end{equation*}
which is finite.
Hence, all three conditions of \cref{assume:rate-measure-near-0} hold, and we can apply \cref{cor:expCRM-d-zero}. 
The AIFA is
\begin{equation*}
	\nu_{\approxlev}(\theta) = \frac{1}{Z_K} \theta^{\gamma / K - 1} Z_{\tau}^{-c}(\theta) 1 \{ 0 \leq \theta \leq T \} d\theta,
\end{equation*}
where $Z_K$ is the normalization constant  $Z_K = \int_{0}^{T}\theta^{\gamma / K - 1} Z_{\tau}^{-c}(\theta) d\theta$.
More generally, for $\gamma, c, \tau > 0$ and $T \geq 1$, we use the notation $\distXGamma(\gamma, c, \tau, T)$ to denote the real-valued distribution with density at $\theta$ equal to:
\begin{equation} \label{def:Xgamma}
	\distXGamma(\theta; \gamma, c, \tau, T) := \frac{\theta^{\gamma - 1} Z_{\tau}^{-c}(\theta) 1 \{ 0 \leq \theta \leq T \}  }{\int_{0}^{T}\theta^{\gamma - 1} Z_{\tau}^{-c}(\theta) d\theta}.
\end{equation}

\eexa

\section{Proofs of AIFA convergence} \label{app:construction-proofs}
In this appendix, to highlight the fact that the i.i.d.\ distributions are different across $K$, we use $\approxrate_{K,i}$ to denote the $i$-th atom size in the approximation of level $K$ i.e. the $K$-atom AIFA is
\begin{align*}
	\Theta_{\approxlev} &\defined {\textstyle\sum_{i=1}^{\approxlev}} \approxrate_{K,i}\delta_{\psi_{K,i}}, & 
	\approxrate_{K,i} &\distiid \nu_{\approxlev},  &
	\psi_{K,i} \distiid H.
\end{align*}

\subsection{AIFA converges to CRM in distribution} \label{app:CRM-construction}

We first state a more general construction than \cref{thm:default-CRM-convergence}, and proceed to prove that result, as a proof of \cref{thm:default-CRM-convergence}.

For the more general construction, we first generalize the $S_b(\theta)$ as in \cref{thm:default-CRM-convergence} with so-called approximate indicators.
\bnumdefn
The parameterized function family $\theset{S_{b}}_{b \in \reals_{+}}$ is composed of 
\emph{approximate indicators} if, for any $b \in \reals_{+}$, 
$S_{b}(\theta)$ is a real, non-decreasing function such that $S_{b}(\theta) = 0$ for $\theta \le 0$
and $S_{b}(\theta) = 1$ for $\theta \ge b$. 
\enumdefn
Valid examples of approximate indicators are the indicator function $S_{b}(\theta) = \indict{\theta > 0}$
and the smoothed indicator function from \cref{thm:default-CRM-convergence}.
Some approximate indicators have a point of discontinuity; e.g., $S_{b}(\theta) = \indict{\theta > 0}$. But the smoothed indicator is both continuous and differentiable; see \cref{app:diff-smooth-indicator}.

\bnthm \label{thm:general-CRM-convergence} 
Suppose \cref{assume:rate-measure-near-0} holds, and let $\theset{S_{b}}_{b \in \reals_{+}}$ be a family of approximate indicators. 
Fix $a > 0$, and let $(b_{\approxlev})_{\approxlev \in \nats}$ be a decreasing sequence such that $b_{\approxlev} \to 0$. 
For $c \defined \gamma {h(0; \eta)}/{Z(1-d,\eta)}$, let
\begin{equation*}
	\nu_{\approxlev}(\dee\theta) 
	\defined \theta^{-1+c\approxlev^{-1} - d S_{b_{\approxlev}}(\theta - a\approxlev^{-1})} g(\theta)^{c\approxlev^{-1} - d}h(\theta; \eta)Z_{\approxlev}^{-1}\dee\theta
\end{equation*}
be a family of probability densities, where $Z_{\approxlev}$ is chosen such that $\int \nu_{\approxlev}(\dee\theta) = 1$. If $\Theta_{\approxlev} \dist \distIFA_{\approxlev}(H, \nu_{\approxlev})$, then $\Theta_{\approxlev} \convD \Theta$ as $\approxlev \to \infty$. 
\enthm

\cref{thm:general-CRM-convergence} recovers \cref{thm:default-CRM-convergence} by setting $S_b$ equaling the smoothed indicator, $a = 1$, and $b_K = 1/K$. 
See \cref{app:ab-impact} for discussions on the impact of the tuning hyperaparameters on the performance of our IFA.

In order to prove \cref{thm:general-CRM-convergence} , we require a few auxiliary results. 

\bnlem[{\citet[Lemma 12.1, Lemma 12.2 and Theorem 16.16]{Kallenberg:2002}}] \label{lem:CRM-convergence}
Let $\Theta$ be a random measure and $\Theta_{1},\Theta_{2},\dots$ a sequence of random measures. 
If for all measurable sets $A$ and $t > 0$, 
\begin{equation*}
	\lim_{\approxlev \to \infty} \EE[e^{-t \Theta_{\approxlev}(A)}] = \EE[e^{-t \Theta(A)}],
\end{equation*}
then $\Theta_{\approxlev} \convD \Theta$. 
\enlem

For a density $f$, let $\mu(t, f): \theta \mapsto (1 - e^{-t \theta})f(\theta)$. 
In results that follow we assume all measures on $\reals_{+}$ have densities with respect to Lebesgue measure.
We abuse notation and use the same symbol to denote the measure and the density. 

\bnprop
Let $\Theta \dist \distCRM(H, \nu)$ and for $\approxlev = 1,2,\dots$, let $\Theta_{\approxlev} \dist \distIFA_{\approxlev}(H, \nu_{\approxlev})$
where $\nu$ is a measure and $\nu_{1}, \nu_{2},\dots$ are probability measures on $\reals_{+}$, all absolutely
continuous with respect to Lebesgue measure. 
If $\|\mu(1, n\nu_{\approxlev}) - \mu(1, \nu)\|_{1} \to 0$, then $\Theta_{\approxlev} \convD \Theta$. 
\enprop
\bprf
Let $t > 0$ and $A$ a measurable set. 
First, recall that the Laplace functional of the CRM $\Theta$ is
\begin{equation*}
	\EE[e^{-t \Theta(A)}] = \exp\left\{ -H(A) \int_{0}^{\infty} \mu(t, \nu)(\theta)\,\dee\theta\right\}.
\end{equation*}
We have
\(
\EE[e^{-t \approxrate_{\approxlev,1} \ind(\psi_{\approxlev,1} \in A)}]
&= \Pr(\psi_{\approxlev,1} \in A)\EE[e^{-t \approxrate_{\approxlev,1}}] + \Pr(\psi_{\approxlev,1} \notin A) \\
&= H(A)\EE[e^{-t \approxrate_{\approxlev,1}}] + 1 - H(A) \\
&= 1 - H(A)(1 - \EE[e^{-t \approxrate_{\approxlev,1}}]) \\
&= 1 - \frac{H(A)}{\approxlev}\int_{0}^{\infty} \mu(t, \approxlev\nu_{\approxlev})(\theta)\,\dee\theta.
\)
Since $\frac{|1 - e^{-t \theta}|}{|1 - e^{-\theta}|} \le \max(1, t)$, it follows
by hypothesis that $\|\mu(t, \approxlev\nu_{\approxlev}) - \mu(t, \nu)\|_{1} \to 0$. 
Thus, by dominated convergence and the standard exponential limit,
\(
\lim_{\approxlev \to \infty} \EE[e^{-t \approxrate_{\approxlev,1} \ind(\psi_{\approxlev,1} \in A)}]^{\approxlev}
&=  \lim_{\approxlev \to \infty} \left(1 - \frac{H(A)}{\approxlev}\int_{0}^{\infty} \mu(t, \approxlev\nu_{\approxlev})(\theta)\,\dee\theta\right)^{\approxlev} \\
&= \exp\left\{ -  \lim_{\approxlev \to \infty}  H(A)\int_{0}^{\infty} \mu(t, \approxlev\nu_{\approxlev})(\theta)\,\dee\theta\right\} \\
&= \exp\left\{ - H(A)\int_{0}^{\infty} \mu(t, \nu)(\theta)\,\dee\theta\right\}.
\)
Finally, by the independence of the random variables $\theset{\theta_{\approxlev,i}}_{i=1}^{\approxlev}$ and $\theset{\psi_{\approxlev,i}}_{i=1}^{\approxlev}$,
\begin{equation*}
	\lim_{\approxlev \to \infty} \EE[e^{-t \Theta_{\approxlev}(A)}] 
	= \lim_{\approxlev \to \infty} \EE[e^{-t \approxrate_{\approxlev,1} \ind(\psi_{\approxlev,1} \in A)}]^{\approxlev},
\end{equation*}
so the result follows from \cref{lem:CRM-convergence}. 
\eprf

\bnlem \label{lem:TV-convergence-2}
If there exist measures $\pi(\theta)\,\dee\theta$ and $\pi'(\theta)\,\dee\theta$ on $\reals_{+}$ such that 
for some $\kappa > 0$ and $c,c'$,
\benum
\item the measures $\mu, \mu_{1}, \mu_{2},\dots$ have densities $f, f_{1}, f_{2}, \dots$ with respect to $\pi$
and densities $f', f_{1}', f_{2}', \dots$ with respect to $\pi'$,
\item $\int_{0}^{\kappa}|f'(\theta) - f_{\approxlev}'(\theta)|\dee\theta \xrightarrow{K \to \infty} 0$,
\item $\sup_{\theta \in [\kappa,\infty)}|f(\theta) - f_{\approxlev}(\theta)| \xrightarrow{K \to \infty} 0$,
\item $\sup_{\theta \in [0,\kappa]}\pi'(\theta) \le c' < \infty$, and
\item $\int_{\kappa}^{\infty} \pi(\theta)\,\dee\theta \le c < \infty$,
\eenum
then 
\begin{equation*}
\|\mu - \mu_{\approxlev}\|_{1} \xrightarrow{K \to \infty} 0.
\end{equation*} 
\enlem

\bprf
We have, using the assumptions and H\"older's inequality,
\(
\|\mu - \mu_{\approxlev}\|_{1} 
&= \int_{0}^{\kappa} |f'(\theta)-f_{\approxlev}'(\theta)|\pi'(\dee\theta) +  \int_{\kappa}^{\infty}|f(\theta)-f_{\approxlev}(\theta)|\pi(\dee\theta) \\
\begin{split}
&\le \left(\sup_{\theta \in [0,\kappa]}\pi'(\theta) \right) \int_{0}^{\kappa} |f'(\theta)-f_{\approxlev}'(\theta)|\dee\theta \\
&\phantom{\le~} +  \left(\sup_{\theta \in [\kappa,\infty)}|f(\theta)-f_{\approxlev}(\theta)|\right)\int_{\kappa}^{\infty}\pi(\dee\theta) 
\end{split} \\
&\le c' \int_{0}^{\kappa}|f'(\theta)-f_{\approxlev}'(\theta)|\dee\theta + c\sup_{\theta \in [\kappa,\infty)}|f(\theta)-f_{\approxlev}(\theta)|.
\)
The conclusion follows by the assumptions. 
\eprf

\begin{proofof}{\cref{thm:general-CRM-convergence}} \label{app:proof-CRM-construction}
Note that since $h$ is continuous and bounded on $[0,\eps]$, $c$ as given in the theorem statement is finite. 
We will apply \cref{lem:TV-convergence-2} with $\kappa = \min(1, \eps)$, 
$\mu = \mu(1, \nu)$, $\mu_{\approxlev} = \mu(1, n \nu_{\approxlev})$, 
\begin{equation*}
\pi(\theta) = \frac{\theta^{-d}g(\theta)^{1-d}h(\theta; \eta)}{Z(1-d, \eta)},
\end{equation*}
and $\pi'(\theta) \defined (\theta g(\theta))^{d}\pi(\theta)$.
Because of the finiteness of $Z(\xi, \eta)$, item $5$ of \cref{lem:TV-convergence-2}, which asks for $\int_{\kappa}^{\infty} \pi(\theta)\,\dee\theta < \infty$, is satisfied.
Thus, $f(\theta) = \gamma(1- e^{-\theta})(\theta g(\theta))^{-1}$,
\begin{equation*}
f_{\approxlev}(\theta) = n Z_{\approxlev}^{-1}(1-e^{-\theta})\theta^{-1+c \approxlev^{-1}+d-dS_{b_{\approxlev}}(\theta - a\approxlev^{-1})} g(\theta)^{-1+c\approxlev^{-1}},
\end{equation*}
and $f'(\theta) = (\theta g(\theta))^{-d}f(\theta)$, and $f'_{\approxlev}(\theta) = (\theta g(\theta))^{-d}f_{\approxlev}(\theta)$.

We now note a few useful properties that we will use repeatedly in the proof.
Observe that $(a/\approxlev)^{c\approxlev^{-1}} = 1 + o(1)$.
The assumption that $h$ is bounded and continuous 
implies that on $[0,a/\approxlev]$, $h(\theta; \eta) = h(0; \eta) + o(1)$. 
Similarly, for any $\delta > 0$, $g(\theta)$ is bounded and continuous for $\theta \in [0,\delta]$ and therefore, 
together with the fact that $g(0) = 1$, we can conclude that on $[0,a/\approxlev]$, $g(\theta) = 1 + o(1)$. 

For the remainder of the proof we will consider $\approxlev$ large enough that $a\approxlev^{-1} + 2b_{\approxlev}$ and $c\approxlev^{-1}$ are less than $\kappa$. 
The normalizing constant $Z_{\approxlev}$ can be written as 
\(
\begin{split}
Z_{\approxlev} 
&= \int_{0}^{a/\approxlev}(\theta g(\theta))^{-1+ c\approxlev^{-1}}\pi'(\dee\theta) \\
&\phantom{=~} + \int_{a/\approxlev}^{\kappa}\theta^{-1+ c\approxlev^{-1} - dS_{b_{\approxlev}}(\theta - a\approxlev^{-1})}g(\theta)^{-1+c\approxlev^{-1}}\pi'(\dee\theta) \\
&\phantom{=~} + \int_{\kappa}^{\infty}(\theta g(\theta))^{-1+ c\approxlev^{-1} - d}\pi'(\dee\theta).
\end{split}
\)
We rewrite each term in turn. 
For the first term,
\(
\int_{0}^{a/\approxlev}\theta^{-1+ c\approxlev^{-1}}g(\theta)^{-1+c\approxlev^{-1}}\pi'(\dee\theta) 
&= (c/\gamma + o(1))\int_{0}^{a/\approxlev} \theta^{-1 + c\approxlev^{-1}}\dee\theta \\
&= (c/\gamma + o(1))\frac{\approxlev}{c}\left(\frac{a}{\approxlev}\right)^{c\approxlev^{-1}} \\
&= \frac{\approxlev}{\gamma} + o(\approxlev).
\)
Since $\kappa \le 1$ and $S_{b_{\approxlev}} \in [0,1]$, for 
$\theta \in [a/\approxlev, \kappa]$, $\theta^{-dS_{b_{\approxlev}}(\theta - a\approxlev^{-1})} \le \theta^{-d}$. 
Since $g(0) = 1$, $c_{*} \le 1$ and therefore 
$g(\theta)^{-1+c\approxlev^{-1}} \le c_{*}^{-1+c}$. 
Hence the second term is upper bounded by
\(
c_{*}^{-1+c}\int_{a/\approxlev}^{\kappa}\theta^{-1+ c\approxlev^{-1} - d}\pi'(\dee\theta)
&\le c_{*}^{-1}(c/\gamma + O(1))\frac{\approxlev^{d}}{a^{d}} \frac{\approxlev}{c} (\kappa^{c\approxlev^{-1}} - (a/\approxlev)^{c\approxlev^{-1}}) \\
&= O(\approxlev^{d}) \times O(\ln  \approxlev) \\
&= o(\approxlev). 
\)
For the third term, 
\(
\int_{\kappa}^{\infty}(\theta g(\theta))^{-1+ c\approxlev^{-1} - d}\pi'(\dee\theta)
&= \int_{\kappa}^{\infty}(\theta g(\theta))^{-1+ c\approxlev^{-1}}\pi(\dee\theta) \\
&\le (\kappa c_{*})^{-1 + c \approxlev^{-1}}\int_{\kappa}^{\infty}\pi(\dee\theta) \\
&\le (\kappa c_{*})^{-1}.
\)
Hence, $Z_{\approxlev} = \frac{\approxlev}{\gamma} + o(\approxlev)$ and $\approxlev Z_{\approxlev}^{-1} = \gamma(1 + e_{\approxlev})$, 
where $e_{\approxlev} = o(1)$. 

Next, we have
\begin{align} \label{eq:partial-f-sup}
	\lefteqn{\sup_{\theta \in [\kappa, \infty)} |f(\theta) - f_{\approxlev}(\theta)|} \nonumber \\
	&= \sup_{\theta \in [\kappa, \infty)} (1-e^{-\theta})(\theta g(\theta))^{-1}|\gamma - \approxlev Z_{\approxlev}^{-1}(\theta g(\theta))^{c\approxlev^{-1}}|\nonumber \\
	&\le \sup_{\theta \in [\kappa, \infty)} \gamma(\theta g(\theta))^{-1}|1 - (1 + e_{\approxlev})(\theta g(\theta))^{c\approxlev^{-1}}|\nonumber \\
	\begin{split}
		&\le \gamma\sup_{\theta \in [\kappa, \infty)} (\theta g(\theta))^{-1}|1 - (\theta g(\theta))^{c\approxlev^{-1}}|  \\
		&\phantom{\le~} +  \gamma e_{\approxlev} \sup_{\theta \in [\kappa, \infty)}(\theta g(\theta))^{-1+c\approxlev^{-1}}. 
	\end{split}
\end{align}
To bound the two terms we will use the fact that if $\theta \ge \kappa$, then 
\begin{equation*}
\theta g(\theta) \ge   \frac{\theta}{c^{*}(1 + \theta)} \ge \frac{\kappa}{c^{*}(1 + \kappa)} \defines \tkappa
\end{equation*}
and if $\theta \le 1$ then $\theta g(\theta) \le c_{*} \le 1$. 
Hence, letting $\psi \defined \theta g(\theta)$, for the first term in \cref{eq:partial-f-sup} we have 
\(
\lefteqn{\gamma\sup_{\theta \in [\kappa, \infty)} (\theta g(\theta))^{-1}|1 - (\theta g(\theta))^{c\approxlev^{-1}}|} \\
&\le \gamma\sup_{\psi \in [\tkappa, \infty)} \psi^{-1}|1 - \psi^{c\approxlev^{-1}}| \\
&\le \gamma  \sup_{\psi \in [\tkappa, 1]}  \psi^{-1} |1 -  \psi^{c\approxlev^{-1}}|  + \gamma \sup_{\psi \in [1, \infty)} \psi^{-1}|1 - \psi^{c\approxlev^{-1}}|\\
&\le \gamma \tkappa^{-1} \sup_{\psi \in [\tkappa, 1]}|1 - \psi^{c\approxlev^{-1}}| + \gamma \left(\frac{\approxlev-c}{\approxlev}\right)^{\approxlev c^{-1}}\left|1 - \frac{\approxlev}{\approxlev - c}\right|  \\
&\le \gamma \tkappa^{-1} (1 - \tkappa^{c\approxlev^{-1}}) +O(1) \times \frac{c}{\approxlev - c}\\
&= \gamma \tkappa^{-1} \times o(1) + O(\approxlev^{-1})  \\
&\to 0. 
\)
Similarly, for the second term in \cref{eq:partial-f-sup} we have 
\(
\gamma e_{\approxlev} \sup_{\theta \in [\kappa, \infty)}(\theta g(\theta))^{-1+c\approxlev^{-1}}
&\le \gamma e_{\approxlev} \sup_{\psi \in [\tkappa, \infty)} \psi^{-1 + c\approxlev^{-1}} \\
&\le \gamma \tkappa^{-1} e_{\approxlev} \\
&\to 0. 
\)
Since $g(\theta)$ is bounded on $[0,\kappa]$, $g(\theta)^{c\approxlev^{-1}} = 1 + o(1)$ and therefore 
$(1 + e_{\approxlev})g(\theta)^{c\approxlev^{-1}} = 1 + e_{\approxlev}'$, where $e_{\approxlev}' = o(1)$. 
Using this observation together with the bound ${(1 - e^{-\theta})\theta^{-1} \le 1}$, we have
\begin{align}\label{eq:partial-f-integral}
\lefteqn{\int_{0}^{\kappa} |f'(\theta) - f_{\approxlev}'(\theta)|\dee\theta = \int_{0}^{\kappa} (\theta g(\theta))^{-d}|f(\theta) - f_{\approxlev}(\theta)|\dee\theta} \nonumber \\
&= \int_{0}^{\kappa} (1-e^{-\theta}) (\theta g(\theta))^{-1-d}|\gamma - \approxlev Z_{\approxlev}^{-1} \theta^{c\approxlev^{-1}+d - d S_{b_{\approxlev}}(\theta - a\approxlev^{-1})}g(\theta)^{c\approxlev^{-1}}| \dee\theta \nonumber \\
&\le \gamma [c^{*}(1 + \kappa)]^{1+d} \int_{0}^{\kappa}\theta^{-d} |1 - (1+e_{\approxlev}')\theta^{c\approxlev^{-1}+d - d S_{b_{\approxlev}}(\theta - a\approxlev^{-1})}| \dee\theta \nonumber \\
\begin{split} 
&\le \gamma \int_{0}^{\kappa}\theta^{-d} |1 -  \theta^{c\approxlev^{-1}+d - d S_{b_{\approxlev}}(\theta - a\approxlev^{-1})}| \dee\theta + \gamma e_{\approxlev}'\int_{0}^{\kappa} \theta^{c\approxlev^{-1}+d - d S_{b_{\approxlev}}(\theta - a\approxlev^{-1})}\dee\theta. 
\end{split} 
\end{align}
We bound the first integral in \cref{eq:partial-f-integral} in four parts: from 0 to $a\approxlev^{-1}$, 
from $a\approxlev^{-1}$ to $a\approxlev^{-1} + b_{\approxlev}$, from $a\approxlev^{-1} + b_{\approxlev}$ to $\kappa - b_{\approxlev}$, and
from $\kappa - b_{\approxlev}$ to $\kappa$. 
The first part is equal to 
\(
\int_{0}^{a\approxlev^{-1}}\theta^{-d} |1 -  \theta^{d + c\approxlev^{-1}}| \dee\theta 
&\le \int_{0}^{a\approxlev^{-1}}\theta^{-d} + \theta^{c\approxlev^{-1}} \dee\theta \\
&= \left. \frac{\theta^{1-d}}{1-d} +  \frac{\approxlev}{c + \approxlev} \theta^{1 + c\approxlev^{-1}} \right|_{0}^{a\approxlev^{-1}} \\
&= \frac{1}{1-d}(a\approxlev^{-1})^{1-d} +  \frac{\approxlev}{c + \approxlev} (a\approxlev^{-1})^{1+c\approxlev^{-1}} \\
&\to 0.
\)
The second part is equal to
\(
\int_{a\approxlev^{-1}}^{a\approxlev^{-1} + b_{\approxlev}}\theta^{-d} |1 -  \theta^{c\approxlev^{-1}+d - d S_{b_{\approxlev}}(\theta - a\approxlev^{-1})}| \dee\theta
&\le \int_{a\approxlev^{-1}}^{a\approxlev^{-1} + b_{\approxlev}}\theta^{-d} +  \theta^{c\approxlev^{-1}-d} \dee\theta \\
&\le2\int_{a\approxlev^{-1}}^{a\approxlev^{-1} + b_{\approxlev}}\theta^{-d} \dee\theta \\
&= \left. \frac{2}{1-d}\theta^{1-d} \right|_{a\approxlev^{-1}}^{a\approxlev^{-1} + b_{\approxlev}} \\
&=  \frac{2}{1-d}\left[(\frac{a}{\approxlev} + b_{\approxlev})^{1-d} - \left(\frac{a}{\approxlev}\right)^{1-d}\right] \\
&\to 0.
\)
The third part is equal to 
\(
\int_{a\approxlev^{-1} + b_{\approxlev}}^{\kappa - b_{\approxlev}}\theta^{-d} |1 - \theta^{c\approxlev^{-1}}| \dee\theta
&= \int_{a\approxlev^{-1} + b_{\approxlev}}^{\kappa - b_{\approxlev}}\theta^{-d} - \theta^{c\approxlev^{-1}-d } \dee\theta \\
&= \left. \frac{1}{1-d}\theta^{1-d} -  \frac{\approxlev}{c + \approxlev(1-d)} \theta^{1 - d + c\approxlev^{-1}} \right|_{a\approxlev^{-1} + b_{\approxlev}}^{\kappa - b_{\approxlev}} \\
\begin{split}
&= \frac{(\kappa - b_{\approxlev})^{1-d}}{1-d} -  \frac{\approxlev}{c + \approxlev(1-d)}(\kappa - b_{\approxlev})^{1 - d + c\approxlev^{-1}} \\
&\phantom{=~} -  \frac{(a\approxlev^{-1} + b_{\approxlev})^{1-d}}{1-d}  + \frac{\approxlev}{c + \approxlev} (a\approxlev^{-1} + b_{\approxlev})^{1 - d + c\approxlev^{-1}}  
\end{split} \\
&\to 0.
\)
The fourth part is equal to
\(
\int_{\kappa - b_{\approxlev}}^{\kappa}\theta^{-d} |1 -  \theta^{c\approxlev^{-1}}| \dee\theta 
&\le \int_{\kappa - b_{\approxlev}}^{\kappa}\theta^{-d} + \theta^{c\approxlev^{-1}-d} \dee\theta \\
& \to 0
\)
using the same argument as the second part. 
The second integral in \cref{eq:partial-f-integral} is upper bounded by
\(
\gamma e_{\approxlev}'\int_{0}^{\kappa}\theta^{c\approxlev^{-1}-d S_{b_{\approxlev}}(\theta - a\approxlev^{-1})}  \dee\theta
\le \gamma  e_{\approxlev}'\int_{0}^{\kappa}\theta^{-d}\dee\theta 
= \gamma e_{\approxlev}'  \frac{\kappa^{1-d}}{1-d} = o(\approxlev). 
\)
Since $\sup_{\theta \in [0,\kappa]}\pi'(\theta) < \infty$ by the boundedness of $g$ and $h$ and $\pi$ is a probability density 
by construction, conclude using \cref{lem:TV-convergence-2} that $\|\mu-\mu_{\approxlev}\|_{1} \to 0$.
It then follows from \cref{lem:CRM-convergence} that $\Theta_{\approxlev} \convD \Theta$.
\end{proofof}

\subsection{Differentiability of smoothed indicator} \label{app:diff-smooth-indicator}
We show that 
\begin{equation*}
	S_{b}(\theta) =
	\begin{cases}
		\exp\left(\frac{-1}{1-(\theta-b)^{2}/b^2}+1\right) & \text{if } \theta \in (0,b) \\
		\ind[\theta > 0] & \text{otherwise.}
	\end{cases} 
\end{equation*}
is differentiable over the whole real line. 
Since on the separate domains $(-\infty, 0)$, $(0,b)$, and $(b, \infty)$, the derivative exists and is continuous, we only need to show that the values of the derivative at $\theta = 0$ and $\theta = b$ from either side match. 

To start, we show that $S_{b}(\theta)$ is continuous at $\theta = 0$ and $\theta = b$.
\begin{equation*}
	\begin{aligned}
		\lim_{\theta \to b^{-}} S_b(\theta) &= \exp \left(1 - \frac{1}{1-0}\right) = 1, \\
		\lim_{\theta \to 0^{+}} S_b(\theta) &= \exp \left(1 - \frac{1}{\infty}\right) = 0. 
	\end{aligned}
\end{equation*}

For $\theta = b$, the derivative from the right ($\theta \xrightarrow{} b^{+}$) is $0$ since constant function. 
The derivative on the interval $(0,b)$ equals
\begin{equation} \label{eq:dS-in-0b}
	\frac{dS_{b}}{d\theta} = S_b(\theta) \frac{-1}{[(\theta-b)^2/b^2-1]^2} \frac{2(\theta-b)}{b^2}.
\end{equation}
The limit as we approach $b$ from the left is $0$ since $\lim_{\theta \to b^{-}} S_b(\theta) = 1$ and the term $(\theta-b)$ vanishes.
So the one-sided derivative is continuous at $\theta = b$. 

For $\theta = 0$, the derivative from the left ($\theta \xrightarrow{} 0^{-}$) is $0$ since also constant function. 
The limit of \cref{eq:dS-in-0b} as we approach $0$ from the right is also $0$. 
It suffices to show
\begin{equation*}
	\lim_{\theta \to 0^{+}}	S_b(\theta) \frac{-1}{[(\theta-b)^2/b^2-1]^2} = 0. 
\end{equation*}
Reparametrizing $x = \frac{1}{1-(\theta - b)^2/b^2}$, we have that $x \to \infty$ and $\theta \to 0^{+}$. 
The last limit becomes
\begin{equation*}
	\lim_{x \to \infty} \frac{\exp(-x)}{x^2} = 0,
\end{equation*}
which is true because the decay of the exponential function is faster than any polynomial. 

The derivative defined over disjoint intervals are continuous at the boundary points, so the overall approximate indicator is differentiable.

\subsection{Normalized AIFA EPPF converges to NCRM EPPF} \label{app:NCRM-construction}
\begin{proofof}{\cref{thm:EPPF-convergence}}
	First, we show that the total mass of AIFA converges in distribution to the total mass of CRM. 
	It suffices to consider $K \geq \EPPFblocks$ so that the AIFA EPPF is non-zero since we only care about the asymptotic behavior of $p_{K}(n_1,n_2,\ldots,n_\EPPFblocks).$
	Through \cref{app:proof-CRM-construction}, we have shown that for all measurable sets $A$ and $t > 0$, the Laplace functionals converge:
	\begin{equation*}
	\lim_{\approxlev \to \infty} \EE[e^{-t \Theta_{\approxlev}(A)}] = \EE[e^{-t \Theta(A)}],
	\end{equation*}
	By choosing $A = \Psi$ i.e. the ground space, we have that $\Theta_{\approxlev}(\Psi)$ is the total mass of AIFA and $\Theta(\Psi)$ is the total mass of CRM
	\begin{equation*}
	\Theta_{\approxlev}(\Psi) = \sum_{i=1}^{K}\approxrate_{K,i}, \hspace{10pt} \Theta(\Psi) = \sum_{i=1}^{\infty}\theta_i.
	\end{equation*}
	Since for any $t>0$, the Laplace transform of $\Theta_{\approxlev}(\Psi)$ converges to that of $\Theta(\Psi)$, we conclude that $\Theta_{\approxlev}(\Psi)$ converges to $\Theta(\Psi)$ in distribution \citep[Theorem 5.3]{Kallenberg:2002}:
	\begin{equation} \label{proof:totIFAconvtotCRM}
	\sum_{i=1}^{K}\approxrate_{K,i} \convD \Theta(\Psi).
	\end{equation}
	
	Second, we show that the decreasing order statistics of AIFA atom sizes converges (in finite-dimensional distributions i.e., in f.d.d) to the decreasing order statistics of CRM atom sizes. For each $K$, the decreasing order statistics of AIFA atoms is denoted by $\{\approxrate_{K,(i)} \}_{i=1}^{K}$: 
	\begin{equation*}
	\approxrate_{K,(1)} \geq \approxrate_{K,(2)} \geq \cdots \geq \approxrate_{K,(K)}.
	\end{equation*}
	We will leverage \citet[Theorem 4 and page 191]{loeve1956ranking} to find the limiting distribution $\{\approxrate_{K,(i)}\}_{i=1}^{K}$ as $K \to \infty$. It is easy to verify the conditions to use the theorem: because the sums $\sum_{i=1}^{K}\approxrate_{K,i}$ converge in distribution to a limit, we know that all the $\approxrate_{K,i}$'s are uniformly asymptotically negligible \cite[Lemma 15.13]{Kallenberg:2002}. Now, we discuss what the limits are. It is well-known that $\Theta(\Psi)$ is an infinitely divisible positive random variable with no drift component and Levy measure exactly $\nu(d\theta)$ \citep{Perman:1992}. In the terminology of \citet[Equation 2]{loeve1956ranking}, the characteristics of $\Theta(\Psi)$ are $a = b = 0$ (no drift or Gaussian parts), $L(x) = 0$, and
	\begin{equation*}
	M(x) = -\nu([x,\infty)).
	\end{equation*}
	Let $I$ be a counting process \emph{in reverse} over $(0,\infty)$ defined based on the Poisson point process $\{\theta_i\}_{i=1}^{\infty}$ in the following way. For any $x$, $I(x)$ is the number of points $\theta_i$ exceeding the threshold $x$: 
	\begin{equation*}
	I(x) \defined |\{i: \theta_i \geq x\}|.
	\end{equation*}
	We augment $I(0) = \infty$ and $I(\infty) = 0$. As a stochastic process, $I$ has independent increments, in that for all $0 = t_0 < t_1 < \cdots < t_k$ , the increments $I(t_i)-I(t_{i-1})$ are independent, furthermore the law of the increments is $I(t_{i-1})-I(t_i) \sim \distPoisson(M(t_i)-M(t_{i-1}))$. These properties are simple consequences of the counting measure induced by the Poisson point process. According to \citet[Page 191]{loeve1956ranking}, the limiting distribution of $\{\approxrate_{K,(i)} \}_{i=1}^{K}$ is governed by $I$, in the sense that for any fixed $t \in \nats$, for any $x_1,x_2,\ldots,x_t \in [0,\infty)$:
	\begin{equation} \label{proof:IFAordered-haslimit}
	\begin{aligned}
	\lim_{K \to \infty} \Pr(\approxrate_{K,(1)} < x_1 &, \approxrate_{K,(2)} < x_2, \ldots, \approxrate_{K,(t)} < x_t) \\
	&= \Pr(I(x_1) < 1, I(x_2) < 2, \ldots, I(x_t) < t).
	\end{aligned}
	\end{equation}
	Because the $\theta_i$'s induce $I$, we can relate the left hand side to the order statistics of the Poisson point process. We denote the decreasing order statistic of the $\{\theta_i\}_{i=1}^{\infty}$ as:
	\begin{equation*}
	\theta_{(1)} \geq \theta_{(2)} \geq \cdots \geq \theta_{(n)} \geq \cdots 
	\end{equation*}
	Clearly, for any $t \in \nats$, the event that $I(x)$ exceeds $t$ is the same as the top $t$ jumps among the $\{\theta_i\}_{i=1}^{\infty}$ exceed x: $I(x) \geq t \iff  \theta_{(t)} \geq x$. Therefore \cref{proof:IFAordered-haslimit} can be rewritten as, for any fixed $t \in \nats$, for any $x_1,x_2,\ldots,x_t \in [0,\infty)$:
	\begin{equation} \label{proof:IFAordered-conv-PPPorderd}
	\lim_{K \to \infty} \Pr(\approxrate_{K,(1)} < x_1, \approxrate_{K,(2)} < x_2, \ldots, \approxrate_{K,(t)} < x_t) = \Pr(\theta_{(1)} < x_1, \theta_{(2)} < x_2, \ldots, \theta_{(t)} < x_t).
	\end{equation}
	It is well-known that convergence of the distribution function imply weak convergence --- for instance, see \citet[Chapter III, Problem 1]{pollard2012convergence}. 
	Actually, from \citet[Theorem 5 and page 194]{loeve1956ranking}, for any fixed $t \in \nats$, the convergence in distribution of $\{\approxrate_{K,(i)}\}_{i=1}^{t}$ to $\{\theta_i\}_{i=1}^{t}$ holds jointly with the convergence of $\sum_{i=1}^{K} \approxrate_{K,(i)}$ to $\sum_{i=1}^{\infty}\theta_i$: the two conditions of the theorem, which are continuity of the distribution function of each $\approxrate_{K,i}$ and $M(0)=-\infty$\footnote{There is a typo in \citet{loeve1956ranking}.}, are easily verified. 
	Therefore, by the continuous mapping theorem, if we define the normalized atom sizes:
	\begin{equation*}
	p_{K,(s)}  \defined \frac{\approxrate_{K,(s)}}{\sum_{i=1}^{K} \approxrate_{K,i}}, \hspace{10pt}
	p_{(s)} \defined \frac{\theta_{(s)}}{\sum_{i=1}^{\infty} \theta_{i}},
	\end{equation*}
	we also have that the normalized decreasing order statistics converge:
	\begin{equation*} 
	(p_{K,{i}})_{i=1}^{K} \stackrel{f.d.d.}{\to} (p_{K,(i)})_{i=1}^{\infty}.
	\end{equation*}
	
	Finally we show that the EPPFs converge. In addition, if we define the \emph{size-biased permutation} (in the sense of \citet[Section 2]{gnedin1998convergence}) of the normalized atom sizes:
	\begin{equation*}
	\{ \widetilde{p}_{K,i} \} \sim \text{SBP}(p_{K,(s)}), \hspace{10pt} \{ \widetilde{p}_{i} \}  \sim \text{SBP}(p_{(s)}),
	\end{equation*}
	then by \citet[Theorem 1]{gnedin1998convergence}, the finite-dimensional distributions of the size-biased permutation also converges:
	\begin{equation} \label{proof:normeIFAsbp-conv-normedPPPsbp}
	(\widetilde{p}_{K,i})_{i=1}^{K} \stackrel{f.d.d.}{\to} (\widetilde{p}_{i})_{i=1}^{\infty}.
	\end{equation}
	 \citet[Equation 45]{pitman1996some} gives the EPPF of $\Xi = \Theta/\Theta(\Psi)$:
	\begin{equation*}
	p(n_1,n_2,\ldots,n_\EPPFblocks) = \mathbb{E} \left( \prod_{i=1}^{\EPPFblocks} \widetilde{p}_i^{n_i-1} \prod_{i=1}^{\EPPFblocks-1} \left(1-\sum_{j=1}^{i} \widetilde{p}_j \right) \right),
	\end{equation*}
	Likewise, the EPPF of $\Xi_K = \Theta_K/\Theta_K(\Psi)$ is:
	\begin{equation*}
	p_K(n_1,n_2,\ldots,n_t) = \mathbb{E} \left( \prod_{i=1}^{\EPPFblocks} \widetilde{p}_{K,i}^{n_i-1} \prod_{i=1}^{\EPPFblocks-1} \left(1-\sum_{j=1}^{i} \widetilde{p}_{K,j} \right) \right).
	\end{equation*}
	Since $\EPPFblocks$ is fixed, and each $p_j$ is $[0,1]$ valued, the mapping from the $\EPPFblocks$-dimensional vector $p$ to the product $\prod_{i=1}^{\EPPFblocks} p_i^{n_i-1} \prod_{i=1}^{\EPPFblocks-1} \left(1-\sum_{j=1}^{i} p_j \right)$  is continuous and bounded. 
	The choice of $N$, $\EPPFblocks$, $n_i$ have been fixed but arbitrary. Hence, the convergence in finite-dimensional distributions of in \cref{proof:normeIFAsbp-conv-normedPPPsbp} imply that the EPPFs converge. 
\end{proofof}

\section{Marginal processes of exponential CRMs} \label{app:marginal}
The marginal process characterization describes the probabilistic model not through the two-stage sampling $\Theta \sim \distCRM(H, \nu)$ and $\targetLatent_{n} \given \Theta  \distiid \distLP(\traitLL; \Theta)$, but through the conditional distributions $X_{n} \given X_{n-1}, X_{n-2}, \ldots, X_1$ i.e. the underlying $\Theta$ has been \emph{marginalized out}. This perspective removes the need to infer a countably infinite set of target variables. In addition, the \emph{exchangeability} between $X_1,X_2,\ldots,X_N$ i.e. the joint distribution's invariance with respect to ordering of observations \citep{Aldous:1985}, often enables the development of inference algorithms, namely Gibbs samplers.

\citet[Corollary 6.2]{broderick2018posteriors} derive the conditional distributions $X_{n} \given  X_{n-1}, X_{n-2}, \ldots, X_1$ for general exponential family CRMs \cref{bkg:expLPh,bkg:expRateMeasure}. 

\bnprop [{Target's marginal process \citep[Corollary 6.2]{broderick2018posteriors}}] \label{prop:target-marginal}
For any $n$, $X_{n} \given X_{n-1}, \ldots, X_1$ is a random measure with finite support.
\benum[leftmargin=*]
\item Let $\{\zeta_i\}_{i=1}^{K_{n-1}}$ be the union of atom locations in $X_1,X_2,\ldots,X_{n-1}$. For $1 \leq m \leq n-1$, let $x_{m,j}$ be the atom size of $X_m$ at atom location $\zeta_j$. Denote $x_{n,i}$ to be the atom size of $X_n$ at atom location $\zeta_i$. The $x_{n,i}$'s are independent across $i$ and the p.m.f.\ of $x_{n,i}$ at $x$ is
\begin{equation*}
	\begin{aligned}
	\targetcondLL(x\given x_{1:(n-1)}) &= \\
	&\kappa(x) \frac{\normZ \left(-1+\sum_{m=1}^{n-1} \phi(x_{m,i}) + \phi(x), \eta + \bmat \sum_{m=1}^{n-1}t(x_{m,i}) + t(x)\\ n \emat \right)}{\normZ \left(-1+\sum_{m=1}^{n-1} \phi(x_{m,i}), \eta + \bmat \sum_{m=1}^{n-1}t(x_{m,i}) \\ n-1\emat \right)}.
	\end{aligned}
\end{equation*}
\item For each $x \in \mathbb{N}$, $X_n$ has $p_{n,x}$ atoms whose atom size is exactly $x$. The locations of each atom are iid $H$: as $H$ is diffuse, they are disjoint from the existing union of atoms $\{\zeta_i\}_{i=1}^{K_n-1}$.  $p_{n,x}$ is Poisson-distributed, independently across $x$, with mean:
\begin{equation*}
\begin{aligned}
	M_{n,x} &= \\
	&\gamma' \kappa(0)^{n-1} \kappa(x)\normZ \left(-1+(n-1) \phi(0) + \phi(x), \eta + \bmat (n-1)t(0) + t(x))\\ n \emat \right).
\end{aligned}
\end{equation*}
\eenum
\enprop

In \cref{prop:approx-marginal}, we state a similar characterization of $Z_{n} \given Z_{n-1}, Z_{n-2}, \ldots, Z_1$ for the finite-dimensional model in \cref{thm-eq:approx-model} and give the proof.

\bnprop [Approximation's marginal process] \label{prop:approx-marginal}
For any $n$, $Z_{n} \given Z_{n-1}, \ldots, Z_1$ is a random measure with finite support.
\benum[leftmargin=*]
\item Let $\{\zeta_i\}_{i=1}^{K_{n-1}}$ be the union of atom locations in $Z_1,Z_2,\ldots,Z_{n-1}$. For $1 \leq m \leq n-1$, let $z_{m,j}$ be the atom size of $Z_m$ at atom location $\zeta_j$. Denote $z_{n,i}$ to be the atom size of $Z_n$ at atom location $\zeta_i$. $z_{n,i}$'s are independently across $i$ and the p.m.f. of $z_{n,i}$ at $x$ is:
\begin{equation*}
	\begin{aligned}
	&\approxcondLL(x\given z_{1:(n-1)}) = \\
	& \kappa(x) \frac{\normZ \left(c/K-1+\sum_{m=1}^{n-1} \phi(z_{m,i}) + \phi(x), \eta + \bmat \sum_{m=1}^{n-1}t(z_{m,i}) + t(x)\\ n \emat \right)}{\normZ \left(c/K-1+\sum_{m=1}^{n-1} \phi(z_{m,i}), \eta + \bmat \sum_{m=1}^{n-1}t(z_{m,i}) \\ n-1\emat \right)}.
	\end{aligned}
\end{equation*}
\item $K - K_{n-1}$ atom locations are generated iid from $H$. $Z_n$ has $p_{n,x}$ atoms whose size is exactly $x$ (for $x \in \mathbb{N} \cup \{0\}$) over these $K-K_{n-1}$ atom locations (the $p_{n,0}$ atoms whose atom size is $0$ can be interpreted as not present in $Z_n$). The joint distribution of $p_{n,x}$ is a multinomial with $K-K_{n-1}$ trials, with success of type $x$ having probability:
\begin{equation*}
	\begin{aligned}
		&\approxcondLL(x\given z_{1:(n-1)}=\zerovec{n}) = \\
		&\kappa(x) \frac{\normZ \left(c/K-1+(n-1) \phi(0) + \phi(x), \eta + \bmat (n-1)t(0)+ t(x)\\ n \emat \right)}{\normZ \left(c/K-1+(n-1)\phi(0), \eta + \bmat (n-1)t(0) \\ n-1\emat \right)}.
	\end{aligned}
\end{equation*}
\eenum
\enprop

\begin{proofof}{\cref{prop:approx-marginal}} \label{proof:approx-marginal}
	We only need to prove the conditional distributions for the atom sizes: that the $K$ distinct atom locations are generated iid from the base measure is clear. 
	
	First we consider $n = 1$. By construction in \cref{cor:expCRM-d-zero}, a priori, the trait frequencies $\{\approxrate_i\}_{i=1}^{K}$ are independent, each following the distribution:
	\begin{equation*}
	\Pr(\approxrate_i \in d\theta) = \frac{\indict{\theta \in \support}}{\normZ \left( c/K-1, \eta\right)} \theta^{c/K-1} \exp \left( \left\langle \eta , \bmat \mu(\theta) \\ -A(\theta) \emat \right\rangle \right).
	\end{equation*}
	Conditioned on $\{ \approxrate_i \}_{i=1}^{K}$, the atom sizes $z_{1,i}$ that $Z_1$ puts on the $i$-th atom location are independent across $i$ and each is distributed as:
	\begin{equation*}
	\Pr(z_{1,i} = x\given \approxrate_i) = \kappa(x) \approxrate^{\phi(x)} \exp \left( \inner{\mu(\approxrate_i), t(x)}  - A(\approxrate_i)\right). 
	\end{equation*}
	Integrating out $\approxrate_i$, the marginal distribution for $z_{1,i}$ is:
	\(
	\Pr(z_{1,i} = x) &= \int \Pr(z_{1,i} = x\given \approxrate_i = \theta) \Pr(\approxrate_i \in d\theta) \\
	&= \frac{\kappa(x)}{\normZ \left( c/K-1, \eta\right)} \int_{\support} \theta^{c/K-1+\phi(x)} \exp \left( \left\langle \eta + \bmat t(x) \\ 1 \emat , \bmat \mu(\theta) \\ -A(\theta) \emat \right\rangle \right) d\theta \\
	&= \kappa(x)\frac{\normZ \left( c/K-1+\phi(x), \eta + \bmat t(x) \\ 1 \emat \right)}{\normZ \left( c/K-1, \eta\right)},
	\)
	by definition of $\normZ$ as the normalizer \cref{bkg:normalizer}.
	
	Now we consider $n \geq 2$. The distribution of $z_{n,i}$ only depends on the distribution of $z_{n-1,i},z_{n-2,i},\ldots,z_{1,i}$ since the atom sizes across different atoms are independent of each other both a priori and a posteriori. The predictive distribution is an integral:
	\begin{equation*}
	\Pr(z_{n,i} = x\given z_{1:(n-1),i}) = \int \Pr(z_{n,i} = x\given \approxrate_i) \Pr(\approxrate_i \in d\theta\given z_{1:(n-1),i}).
	\end{equation*}
	Because the prior over $\approxrate_i$ is conjugate for the likelihood $z_{i,j}\given \approxrate_i$, and the observations $z_{i,j}$ are conditionally independent given $\approxrate_i$, the posterior $\Pr(\approxrate_i \in d\theta\given z_{1:(n-1),i})$ is in the same exponential family but with different natural parameters:
	\begin{equation*}
	\indict{\theta \in \support} \frac{\theta^{c/K-1+\sum_{m=1}^{n-1} \phi(z_{m,i})} \exp \left( \left\langle \eta + \bmat \sum_{m=1}^{n-1} t(z_{m,i}) \\ n-1 \emat , \bmat \mu(\theta) \\ -A(\theta) \emat \right\rangle \right) d\theta}{\normZ \left(c/K-1+\sum_{m=1}^{n-1} \phi(z_{m,i}), \eta +\bmat \sum_{m=1}^{n-1} t(z_{m,i}) \\ n - 1 \emat \right)}.
	\end{equation*}
	This means that the predictive distribution $\Pr(z_{n,i} = x\given z_{1:(n-1),i})$ equals:
	\(
	&\kappa(x)\frac{\int_{\support} \theta^{c/K-1+\sum_{m=1}^{n-1} \phi(z_{m,i}) + \phi(x)} \exp \left( \left\langle \eta + \bmat \sum_{m=1}^{n-1} t(z_{m,i}) + t(x) \\ n \emat , \bmat \mu(\theta) \\ -A(\theta) \emat \right\rangle \right) d\theta}{\normZ \left(c/K-1+\sum_{m=1}^{n-1} \phi(z_{m,i}), \eta +\bmat \sum_{m=1}^{n-1} t(z_{m,i}) \\ n - 1 \emat \right)}  \\
	&= \kappa(x) \frac{{\normZ \left(c/K-1+\sum_{m=1}^{n-1} \phi(z_{m,i}) + \phi(x), \eta +\bmat \sum_{m=1}^{n-1} t(z_{m,i}) + t(x) \\ n \emat \right)}}{\normZ \left(c/K-1+\sum_{m=1}^{n-1} \phi(z_{m,i}), \eta +\bmat \sum_{m=1}^{n-1} t(z_{m,i}) \\ n - 1 \emat \right)}.
	\)
	The predictive distribution $\Pr(z_{n,i} = x\given z_{1:(n-1),i})$ govern both the distribution of atom sizes for known atom locations and new atom locations.
\end{proofof}

\section{Admissible hyperparameters of extended gamma process} \label{app:admissible}

We first describe the two desidarata of a useful Bayesian nonparametric model in more detail.
The condition that the total mass of the rate measure needs to be infinite reads as 
\begin{equation*}
	\int_{0}^{\infty} \nu(d\theta) = \infty
\end{equation*}
This is \citet[A1]{broderick2018posteriors}.
To ensure that the number of active traits is almost surely finite, it suffices to ensure that the expected number of traits is finite. 
The condition that the expected number of active traits is finite reads as 
\begin{equation*}
	\int_{0}^{\infty} (1 - Z_{\tau}^{-1}(\theta)) \nu (d\theta) < \infty.
\end{equation*}
This is \citet[A2]{broderick2018posteriors}: note that $Z_{\tau}^{-1}(\theta)$ is exactly the probability that a trait with rate $\theta$ does not manifest.

\bnlem[{Hyperparameters for extended gamma rate measure}] \label{lem:params-CRM-eG}
	For any $\gamma > 0$, $c > 0$, $T \geq 1$, $\tau > 0$, for the rate measure $\nu(\d theta)$ from \cref{eq:eG}
	Then,
	\begin{itemize}
		\item $\int_{0}^{\infty} \nu(d\theta) = \infty.$
		\item  $\int_{0}^{\infty}[1 - Z^{-1}_{\tau}(\theta)] \nu(d\theta) < \infty.$
	\end{itemize}
\enlem
\begin{proof}[{Proof of \cref{lem:params-CRM-eG}}]
	
	We observe that it suffices to show the two conclusions for $\gamma = 1$, since any positive scaling of the rate measure will preserve the finiteness (or infiniteness) of the integrals.
	In addition, we can replace the upper limit of integration, $\infty$, by $T$, since the rate measure is zero for $\theta > T$.
	
	We begin with elementary observations about the monotonicity of $Z_{\tau}(\theta)$. 
	$Z_{\tau}(\theta)$ is increasing in $\theta$ but decreasing in $\tau$.
	In the limit of $\tau \to \infty$, $Z_{\tau}(\theta)$ approaches $1 + \theta$.
	
	To prove the first statement, we use a simple lower bound on $\int_{0}^{T} \nu(d\theta)$, which holds since $T \geq 1$:
	\begin{equation*}
		\begin{aligned}
			\int_{0}^{T} \theta^{-1} Z_{\tau}^{-c}(\theta) d\theta \geq  \int_{0}^{1} \theta^{-1} Z_{\tau}^{-c}(\theta) d\theta \\
			&\geq Z_{\tau}^{-c}(1) \int_{0}^{1} \theta^{-1}  d\theta = \infty.
		\end{aligned}
	\end{equation*}
	Since $Z_{\tau}(\theta)$ is increasing in $\theta$, for all $\theta \in [0,1]$, $Z_{\tau}^{-c}(\theta) \geq Z_{\tau}^{-c}(1) > 0.$ 
	There are many ways to show $\int_{0}^{1} \theta^{-1}  d\theta = \infty$ --- the connection with the harmonic series is one.
	
	To prove the second statement, we consider two cases separately.
	
	In the first case, $\tau \leq 1.0$. 
	We first show that, there exists a constant $\kappa > 0$ such that, for $\theta \in [0,1]$:
	\begin{equation} \label{eq:minusZ-order}
		1 -  Z_{\tau}^{-1}(\theta) \leq \theta + \kappa  \theta^2.
	\end{equation}
	Consider the Taylor series of $Z_{\tau}(\theta).$ By recursion, the $j$th derivative of $Z_{\tau}(\theta)$ equals
	\begin{equation} \label{eq:Ztau-d}
		\frac{d^i}{d\theta^i} Z_{\tau}(\theta) = \sum_{j=0}^{\infty} \left( \prod_{k=1}^{i} (j+k) \right)^{1-\tau} \frac{\theta^{j}}{(j!)^{\tau}}. 
	\end{equation}
	It is easy to check that the infinite sums in \cref{eq:Ztau-d} converge for any $\theta.$ By absolute convergence theorems\footnote{see, e.g.\ \url{https://www.whitman.edu/mathematics/calculus_online/section11.06.html}}, it suffices to inspect $\theta > 0.$ By the ratio test, subsequent terms have ratio
	\begin{equation*}
		\frac{\theta^{j+1}}{[(j+1)!]^{\tau}} \left( \prod_{k=1}^{i} (j+1+k) \right)^{1-\tau} / \frac{\theta^{j}}{[(j)!]^{\tau}} \left( \prod_{k=1}^{i} (j+k) \right)^{1-\tau} = \frac{\theta (j+1+i)^{1-\tau}}{j+1} \xrightarrow{j \to \infty} 0.
	\end{equation*}
	Clearly $Z_\tau(0) = 1.$ Hence, for all $\theta$ close enough to $0$, $Z_\tau(\theta)$ is strictly positive. Therefore, $Z_{\tau}^{-1}(\theta)$ also has derivatives of all orders in an open interval containing $[0,1]$. Note that $\frac{d}{d\theta} Z_{\tau}(\theta) \big|_{\theta = 0} = 1$. Therefore
	\begin{equation*}
		\frac{d}{d\theta} Z^{-1}_{\tau}(\theta) \big|_{\theta = 0} = \frac{-\frac{d}{d\theta} Z_{\tau}(\theta) \big|_{\theta = 0}}{Z_{\tau}^2(0)} = -1.
	\end{equation*}
	By Taylor's theorem \citet[Section 20.3]{kline1998calculus}, for any $\theta \in [0,1]$, there exists a $y$ between $0$ and $\theta$ such that
	\begin{equation*}
		Z^{-1}_\tau(\theta) = 1 - \theta + \frac{1}{2}\left(\frac{d^2}{d\theta^2} Z^{-1}_{\tau}(\theta)\big|_{\theta = y}\right) \theta^2. 
	\end{equation*}
	It is clear that the second derivative $\frac{d^2}{d\theta^2} Z^{-1}_{\tau}(\theta)\big|_{\theta = y}$ is bounded by a constant independent of $y$ for $y \in [0,1]$, since
	\begin{equation*}
		\frac{d^2}{d\theta^2} Z^{-1}_{\tau}(\theta) = \frac{\frac{d^2}{d\theta^2} Z_{\tau}(\theta)}{Z^{2}_{\tau}(\theta)} - 2 \left(\frac{d}{d\theta} Z_{\tau}(\theta) \right)^2 \frac{1}{Z^{3}_{\tau}(\theta)},
	\end{equation*}
	with the $Z_{\tau}(\theta)$ being at least $1$ and the derivatives being bounded. 
	This shows \cref{eq:minusZ-order}. 
	Therefore:
	\begin{equation*}
		\begin{aligned}
		\int_{0}^{T}[1 - Z^{-1}_{\tau}(\theta)] \nu(d\theta) &\leq \int_{0}^{1}[1 - Z^{-1}_{\tau}(\theta)] \theta^{-1} Z_{\tau}^{-c}(\theta) d\theta +  \int_{1}^{T} \theta^{-1} Z_{\tau}^{-c}(\theta) d\theta \\
		&= A + B.
			\end{aligned}
	\end{equation*}
	We use the estimate $1 -  Z_{\tau}^{-1}(\theta) \theta + \kappa  \theta^2$ in the first part ($A$):
	\begin{equation*}
		\int_{0}^{1}[1 - Z^{-1}_{\tau}(\theta)] \theta^{-1} Z_{\tau}^{-c}(\theta) d\theta \leq \int_{0}^{1} (1 + \kappa \theta) Z_{\tau}^{-c}(\theta) d\theta.
	\end{equation*}
	Since $Z_{\tau}^{-c}(\theta) \leq \exp(-c \theta)$, it is true that $A$ is finite.
	For the second part ($B$), we again use the upper bound $Z_{\tau}^{-c}(\theta) \leq \exp(-c \theta)$ and also $\theta^{-1} \leq 1$ to conclude that $B$ is finite.
	Overall $A + B$ is finite.
	
	In the second case, $\tau > 1.0$.
	Since $Z_{\tau}(\theta) \leq Z_1(\theta)$, $1 - Z_{\tau}^{-1}(\theta) \leq 1 - Z_{1}^{-1}(\theta) = 1 - \exp(-\theta)$.
	In addition, since $Z_{\tau}(\theta) \geq Z_{\infty} (\theta)$, we also have $Z_{\tau}^{-c}(\theta) \leq Z_{\infty}^{-c} = \frac{1}{(1+\theta)^c}$.
	Hence
	\begin{equation*}
		\int_{0}^{T}[1 - Z^{-1}_{\tau}(\theta)] \nu(d\theta) \leq \int_{0}^{T} (1- \exp(-\theta)) \theta^{-1} \frac{1}{(1+\theta)^c} d\theta.
	\end{equation*}
	Observe that for any positive $\theta$, $(1- \exp(-\theta)) \theta^{-1} \leq 1$.
	Therefore
	\begin{equation*}
		\int_{0}^{T}[1 - Z^{-1}_{\tau}(\theta)] \nu(d\theta) \leq \int_{0}^{T} \frac{1}{(1+\theta)^c} d\theta.
	\end{equation*}
	The integrand $\frac{1}{(1+\theta)^c}$ is continous and upper bounded on $[0,T]$, so the overall integral is finite.
	
\end{proof}

\section{Technical lemmas} \label{app:technical}
\subsection{Concentration}
\bnlem [Modified upper tail Chernoff bound] \label{lem:upper-Chernoff}
Let $X = \sum_{i=1}^{n} X_i$, where $X_i = 1$ with probability $p_i$ and $X_i = 0$ with
probability $1-p_i$, and all $X_i$ are independent. Let $\mu$ be an upper bound on $E(X) = \sum_{i=1}^{n} p_i$. Then for all $\delta > 0$:
\begin{equation*}
	\Pr(X \geq (1+\delta) \mu) \leq \exp \left( -\frac{\delta^2}{2+\delta} \mu \right).
\end{equation*}
\enlem 

\begin{proofof}{\cref{lem:upper-Chernoff}}
	The proof relies on the regular upper tail Chernoff bound \citep[Theorem 1.10.1]{doerr2019theory} and an argument using stochastic domination. 
	We pad the first $n$ Poisson trials that define $X$ with additional trials $X_{n+1},X_{n+2},\ldots,X_{n+m}$. $m$ is the smallest natural number such that $\frac{\mu-\mathbb{E}[X]}{m} \leq 1$. Each $X_{n+i}$ is a Bernoulli with probability $\frac{\mu-\mathbb{E}[X]}{m}$, and the trials are independent. Then $Y = X + \sum_{j=1}^{m}X_{n+j}$ is itself the sum of Poisson trials with mean exactly $\mu$, so the regular Chernoff bound applies:
	\begin{equation*}
	\Pr(Y \geq (1+\delta)\mu) \leq \exp \left( -\frac{\delta^2}{2+\delta} \mu \right),
	\end{equation*}
	where we used \citep[Equation 1.10.13]{doerr2019theory} and the simple observation that $2/3 \delta < \delta$. 
	By construction, $X$ is stochastically dominated by $Y$, so the tail probabilities of $X$ are upper bounded by the tail probabilities of $Y$.
\end{proofof}

\bnlem [{Lower tail Chernoff bound \citep[Theorem 1.10.5]{doerr2019theory}}] \label{lem:lower-Chernoff}
Let $X = \sum_{i=1}^{n} X_i$, where $X_i = 1$ with probability $p_i$ and $X_i = 0$ with
probability $1-p_i$, and all $X_i$ are independent. Let $\mu \defined E(X) = \sum_{i=1}^{n} p_i$. Then for all $\delta \in (0,1)$:
\begin{equation*}
\Pr(X \leq (1-\delta)\mu) \leq \exp(-\mu \delta^2/2).
\end{equation*} 
\enlem 

\bnlem [Tail bounds for Poisson distribution] \label{lem:poisson-tail}
    If $X \sim \distPoisson(\lambda)$ then for any $x > 0$:
        \begin{equation*}
        \Pr(X \geq \lambda + x) \leq \exp\left(-\frac{x^2}{2(\lambda+x)}\right),
        \end{equation*}
   and for any $0 < x < \lambda$:
   \begin{equation*}
   	\Pr(X \leq \lambda - x) \leq \exp \left(- \frac{x^2}{2\lambda} \right).
   \end{equation*}
\enlem 

\begin{proofof}{\cref{lem:poisson-tail}}
For $x \geq -1$, let $\psi(x) \defined 2((1+x)\ln(1+x)-x)/x^2$.

We first inspect the upper tail bound. If $X \sim \distPoisson(\lambda)$, for any $x > 0$, \citet[Exercise 3 p.272]{pollard2001user} implies that:
\begin{equation*}
\Pr (X \geq \lambda + x) \leq \exp \left( - \frac{x^2}{2\lambda} \psi \left( \frac{x}{\lambda} \right) \right).
\end{equation*}
To show the upper tail bound, it suffices to prove that $\frac{x^2}{2\lambda} \psi \left( \frac{x}{\lambda} \right)$ is greater than $\frac{x^2}{2(\lambda+x)}$. In general, we show that for $u \geq 0$:
\begin{equation}\label{proof:psi-ineq}
(u+1)\psi(u) - 1 \geq 0. 
\end{equation}
The denominator of $(u+1)\psi(u) - 1 $ is clearly positive. Consider the numerator of $(u+1)\psi(u) - 1$, which is $g(u) \defined 2((u+1)^2 \ln (u+1) - u(u+1) - u^2$. Its 1st and 2nd derivatives are:
\(
    g'(u) &= 4(u+1) \ln (u+1)  - 2u + 1 \\
    g''(u) &= 4\ln (u+1) + 2. 
\)
Since $g''(u) \geq 0$, $g'(u)$ is monotone increasing. Since $g'(0) = 1$, $g'(u) > 0$ for $u \geq 0$, hence $g(u)$ is monotone increasing. Because $g(0) = 0$, we conclude that $g(u) \geq 0$ for $u > 0$ and \cref{proof:psi-ineq} holds. Plugging in $u = x/\lambda$:
\begin{equation*}
\psi\left( \frac{x}{\lambda} \right) \geq \frac{1}{1+\frac{x}{\lambda}} = \frac{\lambda}{x+\lambda},
\end{equation*}
which shows $\frac{x^2}{2\lambda} \psi \left( \frac{x}{\lambda} \right) \geq \frac{x^2}{2(\lambda+x)}$. 

Now we inspect the lower tail bound. We follow the proof of \citet[Theorem 1]{canonne}. We first argue that:
\begin{equation} \label{proof:inter-lower-tail}
	\Pr(X \leq \lambda - x ) \leq \exp \left( - \frac{x^2}{2\lambda} \psi \left( -\frac{x}{\lambda}\right) \right).
\end{equation}
For any $\theta$, the moment generating function $\mathbb{E}[\exp(\theta X)]$ is well-defined and well-known:
\begin{equation*}
	\mathbb{E}[\exp(\theta X)] \defined \exp(\lambda(\exp(\theta)-1)).
\end{equation*}
Therefore:
\(
	\Pr(X \leq \lambda - x ) \leq \Pr(\exp(\theta X) \leq \exp(\theta(\lambda - x )) &\leq \Pr(\exp(\theta(\lambda-x-X)) \geq 1 ) \\
	&\leq \exp(\theta(\lambda-x)) \mathbb{E}[\exp(-\theta X)],
\)
where we have used Markov's inequality. 

We now aim to minimize $\exp(\theta(\lambda-x)) \mathbb{E}[\exp(-\theta X)]$ as a function of $\theta$. Its logarithm is:
\begin{equation*}
	\lambda (\exp(-\theta)-1) + \theta(\lambda-x).
\end{equation*}
This is a convex function, whose derivative vanishes at $\theta = - \ln \left(1-\frac{x}{\lambda} \right)$. Overall this means the best upper bound on $\Pr(X \leq \lambda - x)$ is:
\begin{equation*}
\exp \left( -\lambda \left( \frac{x}{\lambda} + (1-\frac{x}{\lambda})  \ln (1-\frac{x}{\lambda})\right) \right),
\end{equation*}
which is exactly the right hand side of \cref{proof:inter-lower-tail}. Hence to demonstrate the lower tail bound, it suffices to show that:
\begin{equation*}
	 \psi \left( -\frac{x}{\lambda}\right) \geq 1. 
\end{equation*}
More generally, we show that for $-1 \leq u \leq 0$, $\psi(u) - 1 \geq 0$. Consider the numerator of $\psi(u) - 1$, which is $h(u) \defined 2((1+u) \ln (1+u) - u) - u^2$. The first two derivatives are:
\(
h'(u) &= 2(1 + \ln(1+u)) - 2 u \\
h''(u) &= \frac{2}{1+u} - 2
\)
Since $h''(u) \geq 0$, $h(u)$ is convex on $[-1,0]$. Note that $h(0) = 0$. Also, by simple continuity argument, $h(-1) = 2$. Therefore, $h$ is non-negative on $[0,1]$, meaning that $\psi(u) \geq 1$.
\end{proofof}

\bnlem  [Multinomial-Poisson approximation] \label{lem:multi-poiss-approx}
Let $\{ p_i \}_{i=1}^{\infty}$, $p_i \geq 0$, $\sum_{i = 1}^{\infty}p_i < 1$. Suppose there are $n$ independent trials: in each trial, success of type $i$ has probability $p_i$.  Let $X = \{X_i\}_{i=1}^{\infty}$ be the number of type $i$ successes after $n$ trial. Let $Y = \{ Y_i \}_{i=1}^{\infty}$ be independent Poisson random variables, where $Y_i$ has mean $np_i$. Then, there exists a coupling $(\widehat{X}, \widehat{Y})$ of $P_X$ and $P_Y$ such that 
\begin{equation*}
	\Pr(\widehat{X} \neq \widehat{Y}) \leq n \left(\sum_{i=1}^{\infty}p_i\right)^2. 
\end{equation*}
Furthermore, the joint distribution $(\widehat{X}, \widehat{Y})$ naturally disintegrates i.e.\ the conditional distribution $\widehat{X} \given \widehat{Y}$ exists.
\enlem

\begin{proofof}{\cref{lem:multi-poiss-approx}}
	First, we recognize that both $X$ and $Y$ can be sampled in two steps. 
	\begin{itemize}
		\item Regarding $X$, first sample $N_1 \sim \distBinom \left( n, \sum_{i=1}^{\infty}p_i\right)$. Then, for each $1 \leq k \neq N_1$, independently sample $Z_k$ where $\Pr(Z_k = i) = \frac{p_i}{\sum_{j=1}^{\infty}p_j}$. Then, $X_i = \sum_{k=1}^{N_1} \indict{Z_k = i}$ for each $i$. 
		
		\item Regarding $Y$, first sample $N_2 \sim \distPoisson \left( n \sum_{i=1}^{\infty}p_i \right)$. Then, for each $1 \leq k \leq N_2$, independently sample $T_k$ where $\Pr(T_k = i) = \frac{p_i}{\sum_{j=1}^{\infty}p_j}$. Then, $Y_i = \sum_{k=1}^{N_2} \indict{T_k = i}$ for each $i$. 
	\end{itemize}
	The two-step sampling perspective for $X$ comes from rejection sampling: to generate a success of type $k$, we first generate some type of success, and then re-calibrate to get the right proportion for type $k$. The two-step perspective for $Y$ comes from the thinning property of Poisson distribution \citep[Exercise 1.5]{last2017lectures}. The thinning property implies that for any finite index set $\mathcal{K}$, all $\{ Y_i \}$ for $i \in \mathcal{K}$ are mutually independent and marginally, $Y_i \sim \distPoisson(n p_i)$. Hence the whole collection $\{Y_i\}_{i=1}^{\infty}$ are independent Poissons and the mean of $Y_i$ is $np_i$. 
	
	Observing that the conditional $X \given N_1=n$ is the same as $Y \given N_2 = n$, we propose the coupling that essentially proves propagation rule \cref{lem:tv-prop-rule}. The proposed coupling $(\widehat{X}, \widehat{Y})$ is that
	\begin{itemize}
		\item Sample $(\widehat{N_1}, \widehat{N_2})$ from the maximal coupling that attains $\dTV$ between the two distributions: $\distBinom \left( n, \sum_{i=1}^{\infty}p_i\right)$  and $\distPoisson \left( n \sum_{i=1}^{\infty}p_i \right)$. 
		\item If $\widehat{N_1} = \widehat{N_2}$, let the common value be $n$, sample $\widehat{X} \given \widehat{N_1}=n$ and set $\widehat{Y} = \widehat{X}.$ Else $\widehat{N_1} \neq \widehat{N_2}$, independently sample $\widehat{X} \given \widehat{N_1}$ and $\widehat{Y} \given \widehat{N_2}.$
	\end{itemize}
	From the classic binomial-Poisson approximation \citep{lecam1960approximation}, we know that
	\begin{equation*}
		\Pr(\widehat{N_1} \neq \widehat{N_2}) = \dTV(P_{N_1},P_{N_2}) \leq n \left(\sum_{i=1}^{\infty}p_i\right)^2,
	\end{equation*}
	which guarantees that
	\begin{equation*}
		\Pr(\widehat{X} \neq \widehat{Y}) \leq n \left(\sum_{i=1}^{\infty}p_i\right)^2.
	\end{equation*}
	
	Alternatively, we can sample from the conditional $\widehat{X} \given \widehat{Y}$ in the following way. From $\widehat{Y}$, compute $\widehat{N_2}$, which is just  $\sum_{x=1}^{\infty} \widehat{Y}_i.$ Sample $\widehat{N_1}$ from the conditional distribution $\widehat{N_1} \given \widehat{N_2}$ of the maximal coupling that attains the binomial-Poisson total variation. If $\widehat{N_1} = \widehat{N_2}$, set $\widehat{X} = \widehat{Y}.$ Else sample $\widehat{X}$ from the conditional $\widehat{X} \given \widehat{N_1}.$ It is straightforward to verify that this is the conditional $\widehat{X} \given \widehat{Y}$ of the joint  $(\widehat{X}, \widehat{Y})$ described above. 
\end{proofof}

\bnlem [Total variation between Poissons {\citep[Corrollary 3.1]{adell2005sharp}}] \label{lem:poisson-tv}
Let $P_1$ be the Poisson distribution with mean $s$, $P_2$ the Poisson distribution with mean $t$. Then:
\begin{equation*}
    \dTV(P_1, P_2) \leq 1-\exp(-|s-t|) \leq |s-t|. 
\end{equation*}
\enlem

\subsection{Total variation}
We will frequently use the following relationship between total variation and coupling.
For two distributions $P_X$ and $P_Y$ over the same measurable space, it is well-known that the total variation distance between $P_X$ and $P_Y$ is at most the infimum over joint distributions $(\widehat{X}, \widehat{Y})$ which are couplings of $P_X$ and $P_Y$:
\begin{equation*}
	\dTV(P_X, P_Y) \leq \inf_{ \widehat{X}, \widehat{Y} \text{ coupling of } P_{X},P_{Y}} \Pr(  \widehat{X} \neq \widehat{Y}).
\end{equation*}
When $P_X$ and $P_Y$ are discrete distributions, the inequality is actual equality, and there exists couplings that attain the equality \citep[Proposition 4.7]{levin2017markov}.

We first state the chain rule, which will be applied to compare joint distributions that admit densities. 
\bnlem [Chain rule] \label{lem:tv-chain-rule}
Suppose $P_{X_1,Y_1}$ and $P_{X_2,Y_2}$ are two distributions that have densities with respect to a common measure over the ground space $\mathcal{A} \times \mathcal{B}$. Then:
\begin{equation*}
    \dTV(P_{X_1,Y_1}, P_{X_2,Y_2}) \leq \dTV(P_{X_1}, P_{X_2}) + \sup_{a \in \mathcal{A}} \dTV(P_{Y_1\given X_1=a}, P_{Y_2\given X_2=a}). 
\end{equation*}
\enlem 

\begin{proofof}{\cref{lem:tv-chain-rule}}
Because both $P_{X_1,Y_1}$ and $P_{X_2,Y_2}$ have densities, total variation distance is half of $L_1$ distance between the densities:
\(
   & \dTV(P_{X_1,Y_1}, P_{X_2,Y_2}) = \frac{1}{2} \int_{\mathcal{A} \times \mathcal{B}} \given P_{X_1,Y_1}(a,b) - P_{X_2,Y_2}(a,b) \given  da db \\
    &= \frac{1}{2} \int_{\mathcal{A} \times \mathcal{B}} | P_{X_1,Y_1}(a,b) - P_{X_2}(a) P_{Y_1\given X_1}(b\given a) \\
    & \qquad\qquad+ P_{X_2}(a) P_{Y_1\given X_1}(b\given a) - P_{X_2,Y_2}(a,b) |  da db \\
    &\leq \frac{1}{2} \int_{\mathcal{A} \times \mathcal{B}}  P_{Y_1\given X_1}(b\given a)\given P_{X_1}(a)- P_{X_2}(a) \given  \\
    & \qquad\qquad+ P_{X_2}(a) \given P_{Y_1\given X_1}(b\given a) - P_{Y_2\given X_2}(b\given a)\given  da db\\
    &= \frac{1}{2} \int_{\mathcal{A} \times \mathcal{B}} P_{Y_1\given X_1}(b\given a)\given P_{X_1}(a)- P_{X_2}(a) \given da db \\
    &+  \frac{1}{2} \int_{\mathcal{A} \times \mathcal{B}} P_{X_2}(a) \given P_{Y_1\given X_1}(b\given a) - P_{Y_2\given X_2}(b\given a)| da db,
\)
where we have used triangle inequality. Regarding the first term, using Fubini: 
\(
    &\frac{1}{2} \int_{\mathcal{A} \times \mathcal{B}} P_{Y_1\given X_1}(b\given a)|P_{X_1}(a)- P_{X_2}(a) |da db \\
    &= \frac{1}{2} \int_{a \in \mathcal{A}} \left( \int_{b \in \mathcal{B}} P_{Y_1\given X_1}(b\given a) db\right) |P_{X_1}(a)- P_{X_2}(a) | da \\
    &= \frac{1}{2} \int_{a \in \mathcal{A}} |P_{X_1}(a)- P_{X_2}(a) |da \\
    &= \dTV(P_{X_1},P_{X_2}). 
\)
Regarding the second term:
\(
    &\frac{1}{2} \int_{\mathcal{A} \times \mathcal{B}} P_{X_2}(a) |P_{Y_1\given X_1}(b\given a) - P_{Y_2\given X_2}(b\given a)| da db \\
    &=  \int_{a \in \mathcal{A}} \left( \frac{1}{2}\int_{b \in \mathcal{B}} |P_{Y_1\given X_1}(b\given a) - P_{Y_2\given X_2}(b\given a)|  db \right) P_{X_2}(a) da \\
    &\leq \left( \sup_{a \in \mathcal{A}} \dTV(P_{Y_1\given X_1=a}, P_{Y_2\given X_2=a})  \right) \int_{a \in \mathcal{A}} P_{X_2}(a) da \\
    &= \sup_{a \in \mathcal{A}} \dTV(P_{Y_1\given X_1=a}, P_{Y_2\given X_2=a}).
\)
The sum between the first and second upper bound gives the total variation chain rule.
\end{proofof}

An important consequence of \cref{lem:tv-chain-rule} is when the distributions being compared have natural independence structures. 
\bnlem [Product rule] \label{lem:tv-prod-rule}
Let $P_{X_1,Y_1}$ and $P_{X_2,Y_2}$ be discrete distributions. In addition, suppose $P_{X_1,Y_1}$ factorizes into $P_{X_1} P_{Y_1}$ and similarly $P_{X_2,Y_2} = P_{X_2} P_{Y_2}$.
Then:
\begin{equation*}
	\dTV(P_{X_1,Y_1}, P_{X_2,Y_2}) \leq \dTV(P_{X_1}, P_{X_2}) + \dTV(P_{Y_1}, P_{Y_2}).
\end{equation*}
\enlem 

\begin{proofof}{\cref{lem:tv-prod-rule}}
	Since $P_{X_1,Y_1}$ and $P_{X_2,Y_2}$ are discrete distributions, we can apply \cref{lem:tv-chain-rule} (the common measure is the counting measure). Because each joint distribution $P_{X_i,Y_i}$ factorizes into $P_{X_i}P_{Y_i}$, for any $a \in \mathcal{A}$, the right most term in the inequality of \cref{lem:tv-chain-rule} simplifies into
	\begin{equation*}
		\sup_{a \in \mathcal{A}} \dTV(P_{Y_1\given X_1=a}, P_{Y_2\given X_2=a}) = \dTV(P_{Y_1}, P_{Y_2}),
	\end{equation*}
	since $P_{Y_1} = P_{Y_1 \given X_1 = a}$ and $P_{Y_2} = P_{Y_2 \given X_2 = a}$ for any $a.$
\end{proofof}

We call the next lemma the propagation rule, which applies even if distributions do not have densities.
\bnlem [Propagation rule] \label{lem:tv-prop-rule}
Suppose $P_{X_1,Y_1}$ and $P_{X_2,Y_2}$ are two distributions over the same measurable space. Suppose that the conditional $Y_2\given X_2=a$ is the same as the conditional $Y_1\given X_1=a$, which we just denote as $Y\given X=a$. Then:
\begin{equation*}
    \dTV(P_{Y_1},P_{Y_2}) \leq \inf_{ \widehat{X_1}, \widehat{X_2} \text{ coupling of } P_{X_1},P_{X_2}} \Pr(  \widehat{X_1} \neq \widehat{X_2}). 
\end{equation*}
If $P_{X_1}$ and $P_{X_2}$ are discrete distributions, we also have:
\begin{equation*}
	 \dTV(P_{Y_1},P_{Y_2}) \leq  \dTV(P_{X_1},P_{X_2}).
\end{equation*}
\enlem

\begin{proofof}{\cref{lem:tv-prop-rule}}
Let $(\widehat{X_1}, \widehat{X_2})$ be any coupling of $P_{X_1}$ and $P_{X_2}$. The following two-step process generates a coupling of $P_{Y_1}$ and $P_{Y_2}$:
\begin{itemize}
	\item Sample $(\widehat{X_1}, \widehat{X_2}).$
	\item If $\widehat{X_1} =  \widehat{X_2}$, let the common value be $x$. Sample $\widehat{Y_1}$ from the conditional distribution $Y \given X = x$, and set $\widehat{Y_2} = \widehat{Y_1}$. Else if $\widehat{X_1} \neq \widehat{X_2}$, independently sample $\widehat{Y_1}$ from $Y \given X =\widehat{X_1}$ and $\widehat{Y_2}$ from $Y \given X =\widehat{X_2}.$
\end{itemize}
It is easy to verify that the tuple $(\widehat{Y_1},\widehat{Y_2})$ is a coupling of $P_{Y_1}$ and $P_{Y_2}$. In addition, $(\widehat{Y_1},\widehat{Y_2})$ has the property that
\begin{equation*}
	\Pr(\widehat{Y_1} \neq \widehat{Y_2}, \widehat{X_1} = \widehat{X_2}) = 0,
\end{equation*}
since conditioned on $\widehat{X_1} = \widehat{X_2}$, the values of $\widehat{Y_1}$ and $\widehat{Y_2}$ always agree. Therefore:
\begin{equation*}
	\Pr(\widehat{Y_1} \neq \widehat{Y_2}) = \Pr(\widehat{Y_1} \neq \widehat{Y_2}, \widehat{X_1} \neq \widehat{X_2}) \leq \Pr(\widehat{X_1} \neq \widehat{X_2}).
\end{equation*}
This means that $\dTV(P_{Y_1},P_{Y_2})$ is small:
\begin{equation*}
	\dTV(P_{Y_1},P_{Y_2}) \leq \Pr(\widehat{X_1} \neq \widehat{X_2}).
\end{equation*}
So far $(\widehat{X_1}, \widehat{X_2})$ has been an arbitrary coupling between $P_{X_1}$ and $P_{X_2}$. The final step is taking the infimum on the right hand side over couplings.
When $P_{X_1}$ and $P_{X_2}$ are discrete distributions, the infimum over couplings is equal to the total variation distance.
\end{proofof}

The final lemma is the reduction rule, which says that the a larger collection of random variables, in general, has larger total variation distance than a smaller one. 
\bnlem [Reduction rule] \label{lem:tv-redux-rule}
	Suppose $P_{X_1,Y_1}$ and $P_{X_2,Y_2}$ are two distributions over the same measurable space $\mathcal{A} \times \mathcal{B}.$ Then:
	\begin{equation*}
		d_{TV} (P_{X_1,Y_1}, P_{X_2,Y_2}) \geq 	d_{TV} (P_{X_1}, P_{X_2}).
	\end{equation*}
\enlem

\begin{proofof}{\cref{lem:tv-redux-rule}}
	By definition,
	\begin{equation*}
		d_{TV} (P_{X_1}, P_{X_2}) = \sup_{\text{measurable }A} | P_{X_1}(A) - P_{X_2}(A)|.  
	\end{equation*}
	For any measurable $A$, the product $A \times \mathcal{B}$ is also measurable. In addition:
	\begin{equation*}
		P_{X_1}(A) - P_{X_2}(A) = P_{X_1,Y_1}(A,\mathcal{B}) - P_{X_2,Y_2}(A,\mathcal{B}).
	\end{equation*}
	Therefore, for any $A$,
	\begin{equation*}
		|P_{X_1}(A) - P_{X_2}(A)| \leq d_{TV} (P_{X_1,Y_1}, P_{X_2,Y_2}),
	\end{equation*}
	since $P_{X_1,Y_1}(A,\mathcal{B}) - P_{X_2,Y_2}(A,\mathcal{B})$ is the difference in probability mass for one measurable event. The final step is taking supremum of the left hand side.
\end{proofof}

\subsection{Miscellaneous}

\bnlem [Order of growth of harmonic-like sums]  \label{lem:harmonic-like-sum}
\begin{equation*}
	\alpha \left[ \ln N + \ln(\alpha + 1) - \psi(\alpha)  \right]   \geq \sum_{n=1}^{N} \frac{\alpha}{n-1+\alpha} \geq \alpha (\ln N - \psi(\alpha) - 1).
\end{equation*}
where $\psi$ is the digamma function.
\enlem 

\begin{proofof}{\cref{lem:harmonic-like-sum}}
	Because of the digamma function identity $\psi(z+1) = \psi(z) + 1/z$ for $z > 0$, we have:
	\begin{equation*}
		\sum_{n=1}^{N} \frac{\alpha}{n-1+\alpha} = \alpha[\psi(\alpha + N) - \psi(\alpha)]
	\end{equation*}
	\citet[Theorem 5]{gordon1994stochastic} says that
	\begin{equation*}
	\psi(\alpha + N) \geq \ln(\alpha + N) - \frac{1}{2(\alpha+N)} - \frac{1}{12(\alpha+N)^2} \geq \ln N - 1.
	\end{equation*}
	\citet[Theorem 5]{gordon1994stochastic} also says that
	\begin{equation*}
		\psi(\alpha + N) \leq \ln(\alpha + N) \leq \ln( (\alpha + 1) N ) = \ln(1 + \alpha) + \ln N,
	\end{equation*}
	where it's a simple proof that $\alpha + N \leq \alpha N + N = (\alpha + 1 )N$ when $\alpha > 0, N \geq 1$ 
\end{proofof}

We list a collection of technical lemmas that are used when verifying \cref{condition:marginal-process} for the recurring examples.

The first set assists in the beta--Bernoulli model. 

\begin{itemize}
    \item For $\alpha > 0$ and $i = 1,2,3,\ldots$:
    \begin{equation}
    	\label{ex-BetaBer:helper-ineq1}
    	\frac{1}{i + \alpha - 1} \leq 2 \left( \frac{1}{2\alpha} \indict{i=1} + \frac{1}{i} \indict{i > 1} \right).
    \end{equation}

    \item For $m,x,y > 0$, $m \leq y$:
    \begin{equation}\label{ex-BetaBer:helper-ineq2}
        \left| \frac{m+x}{y+x} - \frac{m}{y}  \right| \leq \frac{x}{y}.
    \end{equation}
\end{itemize}

\begin{proofof}{\cref{ex-BetaBer:helper-ineq1}}
If $i = 1$, $ \frac{1}{i+\alpha-1} = \frac{1}{\alpha}$. If $i \geq 1$, $\frac{1}{i+\alpha-1} \leq \frac{1}{i-1} \leq \frac{2}{i}$.
\end{proofof}

\begin{proofof}{\cref{ex-BetaBer:helper-ineq2}}
\begin{equation*}
     \left| \frac{m+x}{y+x} - \frac{m}{y}  \right| =  \left| \frac{(m+x)y-m(y+x)}{y(y+x)} \right| = \left| \frac{x(y-m)}{y(y+x)} \right| \leq \frac{x}{y}.
\end{equation*}
\end{proofof}

The second set aid in the gamma--Poisson model.

\begin{itemize}
    \item For $x \in [0,1)$;
    \begin{equation}\label{ex-GammaPoisson:helper-ineq0}
        (1-x)\ln(1-x) + x \geq 0. 
    \end{equation}

    \item For $x \in (0,1)$, for $p \geq 0$:
    \begin{equation}\label{ex-GammaPoisson:helper-ineq1}
        (1-x)^p + p \frac{x}{1-x} \geq 1.
     \end{equation}

    \item For $\lambda > 0$, for $m > 0, t > 1, x > 0$:
     \begin{equation}\label{ex-GammaPoisson:helper-ineq2}
        d_{TV} \left( \distNB (m, t^{-1}), \distNB(m+x,t^{-1}) \right) \leq x \frac{1/t}{1-1/t},
    \end{equation}
 	where $\distNB(r,\theta)$ is the negative binomial distribution. 
    
    \item For $y \in \naturals$, $K > m > 0$:
   \begin{equation}\label{ex-GammaPoisson:helper-ineq3}
        \left| \frac{m}{y} - K \frac{\Gamma(m/K+y)}{\Gamma(m/K) y!} \right| \leq \euler \frac{m^2}{K}.
   \end{equation}
    where $\euler$ is the Euler constant and $\Gamma(y)$ is the gamma function.
\end{itemize}

\begin{proofof}{\cref{ex-GammaPoisson:helper-ineq0}}
Set $g(x)$ to $(1-x)\ln(1-x) + x$. Then its derivative is $g'(x) = -\ln(1-x) \geq 0$, meaning the function is monotone increasing. Since $g(0) = 0$, it's true that $g(x) \geq 0$ over $[0,1)$. 
\end{proofof}

\begin{proofof}{\cref{ex-GammaPoisson:helper-ineq1}}
Let $f(p) =  (1-x)^p + p \frac{x}{1-x} - 1$. Then $f'(p) = \ln(1-x)(1-x)^p + \frac{x}{1-x}$. Also $f''(p) = (\ln(1-x))^2 (1-x)^p > 0$. So $f'(p)$ is monotone increasing. At $p = 0$, $f'(0) = \ln(1-x) + \frac{x}{1-x} \geq 0$. Therefore $f'(p) \geq 0$ for all $p$. So $f(p)$ is increasing. Since $f(0) = 0$, it's true that $f(p) \geq 0$ for all $p$.
\end{proofof}

\begin{proofof}{\cref{ex-GammaPoisson:helper-ineq2}}
It is known that $\distNB(r,\theta)$ is a Poisson stopped sum distribution \citep[Equation 5.15]{johnson2005univariate}:
\begin{itemize}
    \item $N \sim \distPoisson(-r \ln(1-\theta))$.
    \item $Y_i \distiid \text{Log}(\theta)$ where the  $\text{Log}(\theta)$ distribution's pmf at $k$ equals $\frac{-\theta^k}{k \ln(1-\theta)}$. 
    \item $\sum_{i=1}^{N} Y_i \sim \distNB(r,\theta)$. 
\end{itemize}

Therefore, by the propagation rule \cref{lem:tv-prop-rule}, to compare $\distNB(m,t^{-1})$ with $\distNB(m+x,t^{-1})$,  it suffices to compare the two generating Poissons. 
\(
&d_{TV}\left(\distNB(m,t^{-1}), \distNB(m+x,t^{-1}) \right) \\
&\leq d_{TV}(\distPoisson(-m\ln(1-t^{-1}), \distPoisson(-(m+x)\ln(1-t^{-1}))) \\
&\leq -\ln(1-t^{-1}) x \leq x \frac{t^{-1}}{1-t^{-1}}. 
\) 
We have used the fact that total variation distance between Poissons is dominated by their different in means \cref{lem:poisson-tv} and \cref{ex-GammaPoisson:helper-ineq0}. 

\end{proofof}

\begin{proofof}{\cref{ex-GammaPoisson:helper-ineq3}}
Since $\Gamma\left(\frac{m}{K}+y \right) = \left(\prod_{j=0}^{y-1} (\frac{m}{K} + j) \right) \Gamma \left(\frac{m}{K}\right) = \Gamma \left(\frac{m}{K}\right) \frac{m}{K} \prod_{j=1}^{y-1} (\frac{m}{K} + j)$, we have:
\begin{equation*}
    \left| \frac{m}{y} - K \frac{\Gamma(m/K+y)}{\Gamma(m/K)y!} \right| = \frac{m}{y} \left(\prod_{j=1}^{y-1} \frac{m/K+j}{j} - 1\right).
\end{equation*}
We inspect the product in more detail.
\(
\prod_{j=1}^{y-1} \frac{m/K+j}{j} &= \prod_{j=1}^{y-1} \left(1+\frac{m/K}{j}\right) \leq \prod_{j=1}^{y-1} \exp \left(\frac{m/K}{j}\right) \\
&= \exp \left(\frac{m}{K} \sum_{j=1}^{y-1}\frac{1}{j}\right) \leq \exp  \left(\frac{m}{K} \left( \ln y + 1 \right)\right) = (\euler y)^{m/K}. 
\)
where the $(y-1)$-th Harmonic sum is bounded by $\ln y + 1$. 
Therefore
\begin{equation*}
    \left| \frac{m}{y} - K \frac{\Gamma(m/K+y)}{\Gamma(m/K)y!} \right| \leq \frac{m}{y} \left( (\euler y)^{m/K} - 1 \right).
\end{equation*}
We quickly prove that for any $u \geq 1, 0 < a < 1$, we have
\begin{equation*}
	u^a - 1 \leq a(u-1).
\end{equation*}
Truly, consider the function $g(u) = a(u-1) - u^a + 1.$ The derivative is $g'(u) = a - au^{a-1} = a(1-u^{a-1}).$ Since $a \in (0,1)$ and $u \geq 1$, $g'(u) > 0.$ Therefore $g(u)$ is monotone increasing. Since $g(1) = 0$, we have reached the conclusion. Applying to our situation:
\begin{equation*}
	(\euler y)^{m/K} - 1 \leq \frac{m}{K} (\euler y - 1).
\end{equation*}
In all:
\begin{equation*}
	 \left| \frac{m}{y} - K \frac{\Gamma(m/K+y)}{\Gamma(m/K)y!} \right| \leq  \euler \frac{m^2}{K}.
\end{equation*}
\end{proofof}

The third set aid in the beta--negative binomial model.

\begin{itemize}
    \item For $x > 0$, $z \geq y > 1$:
    \begin{equation}\label{ex-BetaNegBin:helper-ineq2}
        B(x,y) - B(x,z) \leq (z-y)B(x+1,y-0.5) \leq (z-y) B(x+1,y-1).  
   \end{equation}
    
    \item For any $r > 0$, $b \geq 1$:
   \begin{equation}\label{ex-BetaNegBin:helper-ineq1}
        \sum_{y=1}^{\infty} \frac{\Gamma(y+r)}{y! \Gamma(r)} B(y,b+r) \leq \frac{r}{b-0.5}.
    \end{equation}
    
    \item For $b \geq 1$, for any $c > 0$, for any $K \geq c$:
    \begin{equation}\label{ex-BetaNegBin:helper-ineq3}
        \left| 1 - \frac{\Gamma(b)}{\Gamma(b+c/K)} \right| \leq \frac{c}{K} \left(2 + \ln b\right).
    \end{equation}
    
    \item For $b > 1, c > 0, K \geq 2c(\ln b + 2)$:
    \begin{equation}\label{ex-BetaNegBin:helper-ineq4}
        \left| c - \frac{K}{B(c/K, b)} \right| \leq \frac{c}{K} (3\ln b+8).
    \end{equation}
\end{itemize}

\begin{proofof}{\cref{ex-BetaNegBin:helper-ineq2}}
First we prove that for any $x \in [0,1)$:
\begin{equation*}
    \sqrt{1-x} \ln (1-x) + x \geq 0. 
\end{equation*}
Truly, let $g(x)$ be the function on the left hand side. Then its derivative is
\begin{equation*}
        g'(x) = \frac{2\sqrt{1-x} - \ln(1-x) - 2}{2\sqrt{1-x}}. 
\end{equation*}
Denote the numerator function by $h(x)$. Its derivative is
\begin{equation*}
        h'(x) = \frac{1}{1-x} - \frac{1}{\sqrt{1-x}} \geq 0,
\end{equation*}
since $x \in [0,1]$ meaning $h$ is monotone increasing. Since $h(0) = 0$, it means $h(x) \geq 0$. This means $g'(x) \geq 0$ i.e. $g$ itself is monotone increasing. Since $g(0) = 0$ it's true that $g(x) \geq 0$ for all $x \in [0,1)$. 

Second we prove that for all $x \in [0,1]$, for all $p \geq 0$:
\begin{equation}
	\label{ex-BetaNegBin:helper-ineq5}
	(1-x)^p + p \frac{x}{\sqrt{1-x}} - 1 \geq 0. 
\end{equation}
Truly, let $f(p) =  (1-x)^p + p \frac{x}{\sqrt{1-x}} - 1$.  Then $f'(p) = \ln(1-x)(1-x)^p + \frac{x}{\sqrt{1-x}}$. Also $f''(p) = (\ln(1-x))^2 (1-x)^p > 0$. So $f'(p)$ is monotone increasing. At $p = 0$, $f'(0) = \ln(1-x) + \frac{x}{\sqrt{1-x}} > 0$. Therefore $f'(p) \geq 0$ for all $p$. So $f(p)$ is increasing. Since $f(0) = 0$, it's true that $f(p) \geq 0$ for all $p$.

We finally prove the inequality about beta functions. 
\(
    B(x,y) - B(x,z) &= \int_{0}^{1} \theta^{x-1} (1-\theta)^{y-1}(1-(1-\theta)^{z-y}) d\theta \\
    &\leq \int_{0}^{1} \theta^{x-1} (1-\theta)^{y-1} (z-y) \theta (1-\theta)^{-0.5} d\theta \\
    &= (z-y) \int_{0}^{1} \theta^{x} (1-\theta)^{y-1.5}d\theta  = (z-y) B(x+1,y-0.5).
\)
where we have use $1-(1-\theta)^{z-y} \leq (z-y)\theta (1-\theta)^{-1/2}$ from \cref{ex-BetaNegBin:helper-ineq5}. As for $B(x+1,y-0.5) \leq B(x+1,y-1)$, it is because of the monotonicity of the beta function.
\end{proofof}

\begin{proofof}{\cref{ex-BetaNegBin:helper-ineq1}}
\(
    \sum_{y=1}^{\infty} \frac{\Gamma(y+r)}{y! \Gamma(r)} B(y,b+r) &= \int_{0}^{1} \sum_{y=1}^{\infty} \frac{\Gamma(y+r)}{y!\Gamma(r)}\theta^{y-1} (1-\theta)^{b+r-1}d\theta \\
       &=  \int_{0}^{1}  \theta^{-1}\left( \sum_{y=1}^{\infty} \frac{\Gamma(y+r)}{y! \
       \Gamma(r)}\theta^{y}\right)  (1-\theta)^{b+r-1}d\theta \\
       &=  \int_{0}^{1}  \left(\theta^{-1} \left(\frac{1}{(1-\theta)^r} -1\right)\right)  (1-\theta)^{b+r-1}d\theta \\
       &=  \int_{0}^{1}  \left(\theta^{-1} \left(1 - (1-\theta)^r\right)\right)  (1-\theta)^{b-1}d\theta \\
       &\leq  \int_{0}^{1} \theta^{-1} r \frac{\theta}{\sqrt{1-\theta}}(1-\theta)^{b-1}d\theta \\
       &= r  \int_0^1 (1-\theta)^{b-1.5}d\theta = \frac{r}{b-0.5},
\)
where the identity $\sum_{y=1}^{\infty} \frac{\Gamma(y+r)}{y! \
       \Gamma(r)}\theta^{y} = \frac{1}{(1-\theta)^r}-1$ is due to the normalization constant for negative binomial distributions, and we also used \cref{ex-BetaNegBin:helper-ineq5} on $1-(1-\theta)^r$. 
\end{proofof}

\begin{proofof}{\cref{ex-BetaNegBin:helper-ineq3}}
First we prove that: 
\begin{equation*}
	1 - \frac{\Gamma(b)}{\Gamma(b+c/K)} \leq \frac{c}{K} (2 + \ln b). 
\end{equation*}
The recursion defining $\Gamma(b)$ allows us to write: 
\begin{equation*}
    1 - \frac{\Gamma(b)}{\Gamma(b+c/K)} = 1 - \left(\prod_{i=1}^{\floor{b}-1} \frac{b-i}{b+c/K-i} \right) \frac{\Gamma(b-\floor{b}+1)}{\Gamma(b+c/K-\floor{b}+1)}.
\end{equation*}
The argument proceeds in one of two ways. If $\frac{\Gamma(b-\floor{b}+1)}{\Gamma(b+c/K-\floor{b}+1)} \geq 1$, then we have:
\(
     1 - \frac{\Gamma(b)}{\Gamma(b+c/K)} &\leq 1 - \prod_{i=1}^{\floor{b}-1} \frac{b-i}{b+c/K-i}  \\
     &= \left(1 - \frac{b-1}{b+c/K-1}\right) +  \frac{b-1}{b+c/K-1} - \left(\prod_{i=1}^{\floor{b}-1} \frac{b-i}{b+c/K-i} \right) \\
     &= \frac{c}{K} \frac{1}{b+c/K-1} + \frac{b-1}{b+c/K-1} \left(  1 - \prod_{i=2}^{\floor{b}-1} \frac{b-i}{b+c/K-i} \right) \\
     &\leq \frac{c}{K} \frac{1}{b-1} +  \left(  1 - \prod_{i=2}^{\floor{b}-1} \frac{b-i}{b+c/K-i} \right) \\
     &\leq ... \leq \frac{c}{K} \sum_{i=1}^{\floor{b}-1} \frac{1}{b-i} \leq \frac{c}{K} (\ln b+1). 
\)
Else, $\frac{\Gamma(b-\floor{b}+1)}{\Gamma(b+c/K-\floor{b}+1)} < 1$ and we write:
\(
        &1 - \frac{\Gamma(b)}{\Gamma(b+c/K)} \\
        &= 1 - \frac{\Gamma(b-\floor{b}+1)}{\Gamma(b+c/K-\floor{b}+1)} + \frac{\Gamma(b-\floor{b}+1)}{\Gamma(b+c/K-\floor{b}+1)} \left(1 - \prod_{i=1}^{\floor{b}-1} \frac{b-i}{b+c/K-i} \right) \\
        &\leq \left(1 -\frac{\Gamma(b-\floor{b}+1)}{\Gamma(b+c/K-\floor{b}+1)}\right) + \frac{c}{K}(\ln b+1). 
\)
We now argue that for all $x \in [1,2)$, for all $K \geq c$, $1-\frac{\Gamma(x)}{\Gamma(x+c/K)} \leq \frac{c}{K}$. By convexity of $\Gamma(x)$, we know that $\Gamma(x) \geq \Gamma(x+c/K) - \frac{c}{K}\Gamma'(x+c/K)$. Hence $\frac{\Gamma(x)}{\Gamma(x+1/K)} \geq 1 - \frac{c}{K} \frac{\Gamma'(x+c/K)}{\Gamma(x+c/K)}$. Since $x + c/K \in [1,3)$ and $\psi(y) = \frac{\Gamma'(y)}{\Gamma(y)}$, the digamma function, is a monotone increasing function (it is the derivative of a $\ln \Gamma(x)$, which is also convex), $\left|\frac{\Gamma'(x+c/K)}{\Gamma(x+c/K)}\right| \leq \left|\frac{\Gamma'(3)}{\Gamma(3)}\right| \leq 1$. Applying this to $x = b - \floor{b}+1$, we conclude that:
\begin{equation*}
     1 - \frac{\Gamma(b)}{\Gamma(b+c/K)} \leq \frac{c}{K} (2+\ln b). 
\end{equation*}

We now show that:
\begin{equation*}
    \frac{\Gamma(b)}{\Gamma(b+c/K)} - 1 \geq -\frac{c}{K} (\ln b + \ln 2). 
\end{equation*}
Convexity of $\Gamma(y)$ means that:
\begin{equation*}
        \Gamma(b) \geq \Gamma(b+c/K) - \frac{c}{K} \Gamma'(b+c/K) \xrightarrow{} \frac{\Gamma(b)}{\Gamma(b+c/K)} - 1 \geq -\frac{c}{K} \frac{\Gamma'(b+c/K)}{\Gamma(b+c/K)}. 
\end{equation*}
From \citet[Equation 2.2]{alzer1997some}, we know that $\psi(x) \leq \ln (x)$ for positive $x$. Therefore:
\begin{equation*}
-\frac{c}{K} \frac{\Gamma'(b+c/K)}{\Gamma(b+c/K)} \geq -\frac{c}{K} \ln (b+c/K) \geq -\frac{c}{K} (\ln b + \ln 2)
\end{equation*}
since $b + \frac{c}{K} \leq 2b$. 

We combine two sides of the inequality to conclude that the absolute value is at most $\frac{c}{K}(2+\ln b)$.
\end{proofof}

\begin{proofof}{\cref{ex-BetaNegBin:helper-ineq4}}
\(
    \left| c - \frac{K}{B(c/K,b)} \right| &= c\left|\frac{K/c}{\Gamma(c/K)} \frac{\Gamma(c/K+b)}{\Gamma(b)} - 1  \right| \\
    &= c \left| \frac{K/c}{\Gamma(c/K)} \left( \frac{\Gamma(c/K+b)}{\Gamma(b)}-1 \right) + \left( \frac{K/c}{\Gamma(c/K)}-1 \right) \right| \\
    &\leq c \left(\frac{K/c}{\Gamma(c/K)} \left| \frac{\Gamma(c/K+b)}{\Gamma(b)}-1 \right| + \left|\frac{K/c}{\Gamma(c/K)}-1 \right|\right). 
\)
On the one hand:
\begin{equation*}
    \frac{K/c}{\Gamma(c/K)} = \frac{\Gamma(1)}{\Gamma(1+c/K)}. 
\end{equation*}
From \cref{ex-BetaNegBin:helper-ineq3}, we know:
\begin{equation*}
    \left| \frac{\Gamma(1)}{\Gamma(1+c/K)} - 1 \right| \leq \frac{2c}{K}. 
\end{equation*}
On the other hand, let $y = {\Gamma(b)}/{\Gamma(c/K+b)}$. Then:
\begin{equation*}
        \left| \frac{\Gamma(c/K+b)}{\Gamma(b)}  - 1 \right| = \left| \frac{1}{y}-1 \right| = \frac{|1-y|}{y}.
\end{equation*}
Again using \cref{ex-BetaNegBin:helper-ineq3}, $|1-y| \leq \frac{c}{K}(2+\ln b)$. Since $K \geq 2c(\ln b + 2)$, $\frac{c}{K}(2+\ln b)$ is at most $0.5$, meaning $|1-y| \leq 0.5$ and $y \geq 0.5.$ Therefore
\begin{equation*}
	\left|   \frac{\Gamma(c/K+b)}{\Gamma(b)}  - 1 \right| \leq \frac{2c}{K}(2+\ln b).
\end{equation*}
In all:
\(
     \left| c - \frac{K}{B(c/K,b)} \right| &\leq c \left(  \left( 1 + \frac{2c}{K}  \right) 2 \frac{c}{K} (2+\ln b) + \frac{2c}{K} \right) \\
     &\leq \frac{c}{K}(3\ln b + 8).
\)
\end{proofof}

\section{Verification of upper bound's assumptions for additional examples} \label{app:more-verification}
Recall the definitions of $\targetcondLL$, $\approxcondLL$, and $M_{n,x}$ for exponential family CRM-likelihood in \cref{app:marginal}.

\subsection{Gamma--Poisson with zero discount}

First we write down the functions in \cref{condition:marginal-process} for non-power-law gamma--Poisson. This requires expressing the rate measure and likelihood in exponential-family form:
\begin{equation*}
	\traitLL(x \given \theta) = \frac{1}{x!} \theta^{x} \exp(-\theta), \hspace{10pt} \nu(d\theta) = \gamma \lambda \theta^{-1} \exp(-\lambda \theta), 
\end{equation*}
which means that $\kappa(x) = 1/x!, \phi(x) = x, \mu(\theta) = 0, A(\theta) = \theta$. This leads to the normalizer
\begin{equation*}
\normZ = \int_{0}^{\infty} \theta^{\xi} \exp(-\lambda \theta) d\theta = \Gamma(\xi+1) \lambda^{-(\xi+1)}.
\end{equation*}
Therefore, $\targetcondLL$ is
\(
\targetcondLL(x_{n} = x\given x_{1:(n-1)}) &= \frac{1}{x!} \frac{\Gamma(-1+\sum_{i=1}^{n-1}x_i + x + 1) (\lambda + n)^{-1+\sum_{i=1}^{n-1}x_i + x + 1} }{\Gamma(-1+\sum_{i=1}^{n-1}x_i + 1) (\lambda + n - 1)^{-1+\sum_{i=1}^{n-1}x_i + 1} } \\
&= \frac{1}{x!} \frac{\Gamma(\sum_{i=1}^{n-1}x_i + x )}{\Gamma(\sum_{i=1}^{n-1}x_i)} \left(\frac{1}{\lambda + n}\right)^{x} \left(1-\frac{1}{\lambda+n}\right)^{\sum_{i=1}^{n-1}x_i},
\)
and similarly $\approxcondLL$ is
\(
\approxcondLL(x_n = x \given  x_{1:(n-1)}) &= \frac{1}{x!} \frac{\Gamma(-1+\sum_{i=1}^{n-1}x_i + x + 1 + \gamma \lambda/K) (\lambda + n)^{-1+\sum_{i=1}^{n-1}x_i + x + 1 + \gamma \lambda/K} }{\Gamma(-1+\sum_{i=1}^{n-1}x_i + 1+\gamma \lambda/K) (\lambda + n - 1)^{-1+\sum_{i=1}^{n-1}x_i + 1 + \gamma \lambda/K} } \\
&= \frac{1}{x!} \frac{\Gamma(\sum_{i=1}^{n-1}x_i + x + \gamma \lambda/K)}{\Gamma(\sum_{i=1}^{n-1}x_i +\gamma \lambda/K)} \left(\frac{1}{\lambda + n}\right)^{x} \left(1-\frac{1}{\lambda+n}\right)^{\sum_{i=1}^{n-1}x_i+\gamma\lambda/K},
\)
and $M_{n,x}$ is
\begin{equation*}
	M_{n,x} = \gamma \lambda \frac{1}{x!} \Gamma(x) (\lambda + n)^{-x} = \frac{\gamma \lambda }{x(\lambda+n)^x}. 
\end{equation*}

Now, we state the constants so that gamma--Poisson satisfies \cref{condition:marginal-process}, and give the proof.

\bnprop [{Gamma--Poisson satisfies \cref{condition:marginal-process}}] \label{prop:gamma--Poisson} The following hold for arbitrary $\gamma,\lambda > 0$. For any $n$:
\begin{equation*}
\sum_{x=1}^{\infty} M_{n,x} \leq \frac{\gamma \lambda}{n-1+\lambda}.
\end{equation*}
\begin{equation*}
\sum_{x=1}^{\infty} \approxcondLL(x\given x_{1:(n-1)}=\zerovec{n}) \leq \frac{\gamma \lambda}{n-1+\lambda}.
\end{equation*}
For any $K$:
\begin{equation*}
\sum_{x=0}^{\infty} \left| \targetcondLL(x\given x_{1:(n-1)}) - \approxcondLL(x\given x_{1:(n-1)}) \right| \leq \frac{2\gamma \lambda}{K} \frac{1}{n-1+\lambda}.
\end{equation*}
For any $\approxlev \geq \gamma \lambda:$
\begin{equation*}
\sum_{x=1}^{\infty} \left| M_{n,x} - K \approxcondLL(x\given x_{1:(n-1)}=\zerovec{n})\right| \leq \frac{\gamma^2 \lambda + \euler \gamma^2 \lambda^2}{K} \frac{1}{n-1+\lambda}.
\end{equation*}
\enprop 

\begin{proofof}{\cref{prop:gamma--Poisson}}
	The growth rate condition of the target model is simple:
	\begin{equation*}
	\sum_{x=1}^{\infty} M_{n,x} = \gamma \lambda \sum_{x=1}^{\infty} \frac{1}{x(\lambda+n)^x} \leq  \gamma \lambda \sum_{x=1}^{\infty} \frac{1}{(\lambda+n)^x} =\frac{\gamma \lambda}{n - 1 + \lambda}.
	\end{equation*}
	
	The growth rate condition of the approximate model is also simple:
	\(
	\sum_{x=1}^{\infty} \approxcondLL(x\given x_{1:(n-1)}=\zerovec{n}) &= 1 - \approxcondLL(0\given x_{1:(n-1)}=\zerovec{n}) = 1 - \left( 1 - \frac{1}{\lambda+n}\right)^{\gamma \lambda/K} \\
	&\leq \frac{\gamma\lambda}{K} \frac{(\lambda+n)^{-1}}{1-(\lambda+n)^{-1}} = \frac{1}{K} \frac{\gamma \lambda}{n-1+\lambda},
	\)
	where we have used \cref{ex-GammaPoisson:helper-ineq1} with $p = \frac{\gamma \lambda}{K}, x = (\lambda+n)^{-1}$.
	
	For the total variation between $\targetcondLL$ and $\approxcondLL$ condition, observe that $\targetcondLL$ and $\approxcondLL$ are p.m.f's of negative binomial distributions, namely:
	\(
 	\targetcondLL(x \given x_{1:(n-1)}) &= \distNB \left(x \given \sum_{i=1}^{n-1}x_i, (\lambda+n)^{-1}\right), \\
 	 \approxcondLL(x\given x_{1:(n-1)}) &= \distNB \left(x \given \sum_{i=1}^{n-1}x_i+\gamma \lambda/K, (\lambda+n)^{-1}\right).
	\)
	The two negative binomial distributions have the same success probability and only differ in the number of trials. Hence using \cref{ex-GammaPoisson:helper-ineq2}, we have:
	\begin{equation*}
 	\sum_{x=0}^{\infty} \left| \targetcondLL(x\given x_{1:(n-1)}) - \approxcondLL(x\given x_{1:(n-1)}) \right|	\leq 2 \frac{\gamma\lambda}{K} \frac{(\lambda+n)^{-1}}{1-(\lambda+n)^{-1}} = \frac{2\gamma \lambda }{K} \frac{1}{n-1+\lambda},
	\end{equation*}
	where the factor $2$ reflects how total variation distance is ${1}/{2}$ the $L_1$ distance between p.m.f's.
	
	For the total variation between $M_{n,.}$ and $K\approxcondLL(\cdot \given 0)$ condition, 
	\(
	&\sum_{x=1}^{\infty} \left| M_{n,x} - K \approxcondLL(x\given x_{1:(n-1)}=\zerovec{n})\right| \\
	&= \sum_{x=1}^{\infty}  \frac{1}{(\lambda+n)^x} \left| \frac{ \gamma\lambda }{x} - K \frac{\Gamma(\gamma\lambda/K+x) }{\Gamma(\gamma\lambda/K)x!} \left(1-\frac{1}{\lambda+n}\right)^{\gamma \lambda/K} \right| \\
	&\leq \sum_{x=1}^{\infty} \frac{1}{(\lambda+n)^x}  \left( \left| \frac{\gamma \lambda}{x}  \left(1-\left(1-\frac{1}{\lambda+n}\right)^{\gamma \lambda/K}\right)  \right| +  \left| \frac{\gamma \lambda}{x} -  K \frac{\Gamma(\gamma\lambda/K+x) }{\Gamma(\gamma\lambda/K)x!}  \right| \right).
	\)
	Using \cref{ex-GammaPoisson:helper-ineq2} we can upper bound:
	\(
	1-\left(1-\frac{1}{\lambda+n}\right)^{\gamma \lambda/K}  &\leq \frac{\gamma\lambda}{K} \frac{1}{\lambda+n-1},
	\)
	while \cref{ex-GammaPoisson:helper-ineq3} gives the upper bound:
	\(
	\left| \frac{\gamma \lambda}{x} -  K \frac{\Gamma(\gamma\lambda/K+x) }{\Gamma(\gamma\lambda/K)x!} \right| &\leq \frac{\euler \gamma^2 \lambda^2}{K}.	
	\)
	This means:
	\(
	&\sum_{x=1}^{\infty} \left| M_{n,x} - K \approxcondLL(x\given x_{1:(n-1)}=\zerovec{n})\right| \\
	&\leq \sum_{x=1}^{\infty} \frac{1}{(\lambda+n)^x} \frac{\gamma \lambda}{x} \frac{\gamma\lambda}{K} \frac{1}{\lambda+n-1} + \sum_{x=1}^{\infty} \frac{1}{(\lambda+n)^x}  \frac{\euler \gamma^2 \lambda^2}{K} \\
	&\leq \frac{\gamma^2 \lambda^2}{K} \frac{1}{(\lambda+n-1)^2} + \frac{\euler \gamma^2\lambda^2}{K} \frac{1}{\lambda+n-1} \\
	&\leq \frac{\gamma^2\lambda + e \gamma^2 \lambda^2}{K} \frac{1}{n-1+\lambda}.
	\)
\end{proofof}

\subsection{Beta--negative binomial with zero discount}

First we write down the functions in \cref{condition:marginal-process} for non-power-law beta--negative binomial. This requires expressing the rate measure and likelihood in exponential-family form:
\(
\traitLL(x \given \theta) &= \frac{\Gamma(x+r)}{x!\Gamma(r)} \theta^{x} \exp(r \ln (1-\theta)) ,\\
\nu(d\theta) &= \gamma \alpha \theta^{-1} \exp( \ln (1-\theta) (\alpha-1)) \indict{\theta \leq 1}, 
\)
which means that $\kappa(x) = \Gamma(x+r)/\Gamma(r)x!, \phi(x) = x, \mu(\theta) = 0, A(\theta) = - r \ln (1-\theta)$. This leads to the normalizer:
\begin{equation*}
\normZ = \int_{0}^{1} \theta^{\xi} (1-\theta)^{r\lambda} d\theta = B(\xi+1, r\lambda+1). 
\end{equation*}
To match the parametrizations, we need to set $\lambda = \frac{\alpha-1}{r}$ i.e. $r\lambda = \alpha - 1$. Therefore, $\targetcondLL$ is
\begin{equation*}
\targetcondLL(x_{n} = x \given x_{1:(n-1)}) = \frac{\Gamma(x+r)}{x! \Gamma(r)} \frac{B(\sum_{i=1}^{n-1}x_i + x, rn + \alpha )}{B(\sum_{i=1}^{n-1}x_i, r(n-1)+\alpha)},
\end{equation*}
and $\approxcondLL$ is
\begin{equation*}
\approxcondLL(x_n = x \given  x_{1:(n-1)}) = \frac{\Gamma(x+r)}{x! \Gamma(r)} \frac{B(\gamma \alpha/K+\sum_{i=1}^{n-1}x_i + x, rn+\alpha)}{B(\gamma \alpha/K+\sum_{i=1}^{n-1}x_i, r(n-1)+\alpha)},
\end{equation*}
and $M_{n,x}$ is
\begin{equation*}
M_{n,x} = \gamma \alpha \frac{\Gamma(x+r)}{x!\Gamma(r)} B(x, rn + \alpha).
\end{equation*}

Now, we state the constants so that beta--negative binomial satisfies \cref{condition:marginal-process}, and give the proof.

\bnprop [{Beta--negative binomial satisfies \cref{condition:marginal-process}}] \label{prop:beta-NegBin} The following hold for any $\gamma > 0$ and $\alpha > 1$. For any $n$:
\begin{equation*}
\sum_{x=1}^{\infty}M_{n,x} \leq \frac{\gamma \alpha}{n-1+(\alpha-0.5)/r}.
\end{equation*}
For any $n$, any $K$:
\begin{equation*}
\sum_{x=1}^{\infty} \approxcondLL(x\given x_{1:(n-1)}=\zerovec{n}) \leq \frac{1}{K} \frac{4 \gamma \alpha}{n-1+(\alpha-0.5)/r}.
\end{equation*}
For any $K$:
\begin{equation*}
\sum_{x=0}^{\infty} \left| \targetcondLL(x\given x_{1:(n-1)}) - \approxcondLL(x\given x_{1:(n-1)}) \right| \leq 2\frac{\gamma \alpha}{K} \frac{1}{n-1+\alpha/r}.
\end{equation*}

For any $n$, for $K \geq \gamma \alpha (3 \ln (r(n-1)+\alpha)+8)$:
\(
&\sum_{x=1}^{\infty} \left| M_{n,x} - K \approxcondLL(x\given x_{1:(n-1)}=\zerovec{n})\right| \\
&\leq \frac{\gamma \alpha}{K} \frac{(4\gamma \alpha + 3) \ln (rn+\alpha+1) + (10+2r)\gamma \alpha + 24 }{n-1+(\alpha-0.5)/r}.
\)

\enprop 

\begin{proofof}{\cref{prop:beta-NegBin}}
	The growth rate condition for the target model is easy to verify:
	\begin{equation*}
	\sum_{x=1}^{\infty}M_{n,x} = \gamma \alpha \sum_{x=1}^{\infty} \frac{\Gamma(x+r)}{\Gamma(r) x!} B(x, rn+\alpha) \leq \gamma \alpha \frac{r}{r(n-1)+\alpha-0.5},
	\end{equation*}
	where we have used \cref{ex-BetaNegBin:helper-ineq1} with $b = r(n-1)+\alpha$. 
	
	As for the growth rate condition of the approximate model,
	\(
	\sum_{x=1}^{\infty} \approxcondLL(x\given x_{1:(n-1)}=\zerovec{n}) &= 1 - \approxcondLL(0\given x_{1:(n-1)}=\zerovec{n}) = 1 - \frac{B(\gamma \alpha/K, rn+\alpha)}{B(\gamma \alpha/K, r(n-1)+\alpha)} \\
	&=  \frac{B(\gamma \alpha/K, r(n-1)+\alpha)-B(\gamma \alpha/K, rn+\alpha)}{B(\gamma \alpha/K, r(n-1)+\alpha)}.
	\)
	The numerator is small because of \cref{ex-BetaNegBin:helper-ineq2} where $x = \gamma \alpha /K, y = r(n-1)+\alpha,  z=  rn + \alpha$:
	\begin{equation*}
	\begin{aligned}
			B(\gamma \alpha/K, r(n-1)+\alpha)-B(\gamma \alpha/K, rn+\alpha) &\leq r B(\gamma \alpha/K+1,r(n-1)+\alpha-0.5) \\
			&\leq rB(1,r(n-1)+\alpha-0.5) \\
			&= \frac{1}{n-1+(\alpha-0.5)/r}.
	\end{aligned}
	\end{equation*}
	The denominator is large because \cref{ex-BetaNegBin:helper-ineq4} with \cref{ex-BetaNegBin:helper-ineq4} with $c = \gamma \alpha, b = r(n-1)+\alpha$:
	\begin{equation*}
	\frac{1}{B(\gamma \alpha/K, r(n-1)+\alpha)} \leq \frac{4\gamma \alpha}{K}.	
	\end{equation*}
	Combining the two give yields
	\begin{equation*}
	\sum_{x=1}^{\infty} \approxcondLL(x\given x_{1:(n-1)}=\zerovec{n}) \leq \frac{1}{K} \frac{4 \gamma \alpha}{n-1+(\alpha-0.5)/r}.
	\end{equation*}
	
	For the total variation between $\targetcondLL$ and $\approxcondLL$ condition, we first discuss how each function can be expressed a p.m.f.\ of so-called beta negative binomial i.e., BNB \citep[Section 6.2.3]{johnson2005univariate} distribution. Let $A = \sum_{i=1}^{n-1}x_i$. Observe that:
	\begin{equation}\label{ex-BetaNegBin:transform-hc}
	\frac{\Gamma(x+r)}{\Gamma(r)x!} \frac{B(A+x,rn+\alpha)}{B(A,r(n-1)+\alpha)} = \frac{\Gamma(A+r)}{\Gamma(A)x!} \frac{B(r+x,A+r(n-1)+\alpha)}{B(r,r(n-1)+\alpha)}.
	\end{equation} 
	The random variable $V_1$ whose p.m.f at $x$ appears on the right hand side of \cref{ex-BetaNegBin:transform-hc} is the result of a two-step sampling procedure: 
	\begin{equation*}
	P \sim \distBeta(r,r(n-1)+\alpha), \hspace{10 pt} V_1\given P \sim \distNB(A; P). 
	\end{equation*}
	We denote such a distribution as $V_1 \sim \distBNB(A; r,r(n-1)+\alpha)$. An analogous argument applies to $\approxcondLL$:
	\begin{equation*}
	P \sim \distBeta(r,r(n-1)+\alpha), \hspace{10 pt} V_2\given P \sim \distNB \left(A + \frac{\gamma \alpha}{K}; P\right). 
	\end{equation*}
	 Therefore:
	\(
	\targetcondLL(x \given x_{1:(n-1)}) &= \distBNB \left(x \given A; r, r(n-1) + \alpha \right) \\
	\approxcondLL(x\given x_{1:(n-1)}) &= \distBNB \left(x \given A + \frac{\gamma \alpha}{K};r, r(n-1) + \alpha\right).
	\)
	We now bound the total variation between the $\distBNB$ distributions. Because they have a common mixing distribution, we can upper bound the distance with an integral using simple triangle inequalities:
	\(
	\dTV \left(\targetcondLL, \approxcondLL \right) &= \frac{1}{2} \sum_{x=0}^{\infty} |\Pr(V_1 = x) -\Pr(V_2=x) | \\
	&= \frac{1}{2} \sum_{x=0}^{\infty}  \left| \int_{0}^{1} (\Pr(V_1 = x \given  P = p) - \Pr(V_2 = x \given  P = p)) \Pr(P \in dp) \right| \\
	&\leq  \int_0^1 \left( \frac{1}{2}\sum_{x=0}^{\infty} \left|\Pr(V_1 = x \given  P = p) - \Pr(V_2 = x \given  P = p) \right| \right)  \Pr(P \in dp) \\
	&= \int_0^1 \dTV \left(\distNB(A,p),\distNB(A+\gamma\alpha/K,p)  \right) \Pr(P \in dp).
	\)
	For any $p$, we use \cref{ex-GammaPoisson:helper-ineq2} to upper bound the total variation distance between negative binomial distributions. Therefore:
	\(
	\dTV \left(\targetcondLL, \approxcondLL \right) 	&\leq \int_{0}^1 \frac{\gamma \alpha}{K} \frac{p}{1-p} \Pr(P \in dp) \\
	&= \frac{\gamma \alpha}{K} \frac{1}{B(r,r(n-1)+\alpha)} \int_0^1 p^{r} (1-p)^{r(n-1)+\alpha-2} dp \\ 
	&= \frac{\gamma \alpha}{K} \frac{B(r+1,r(n-1)+\alpha-1)}{B(r,r(n-1)+\alpha)} = \frac{\gamma \alpha}{K} \frac{1}{n-1+\alpha/r}. 
	\)
	
	Finally, we verify the condition between $K\approxcondLL$ and $M_{n,.}$, which is showing that the following sum is small:
	\begin{equation*}
	\sum_{x = 1}^{\infty} \frac{\Gamma(x+r)}{x!\Gamma(r)}\left|\gamma \alpha B(x, rn+\alpha)  - K \frac{B(\gamma \alpha/K + x, rn+\alpha)}{B(\gamma \alpha/K,  r(n-1)+\alpha)}  \right|.
	\end{equation*}
	We look at the summand for $x = 1$ and the summation from $x = 2$ through $\infty$ separately. For $x = 1$, we prove that:
	\begin{equation}
	 \label{ex-BetaNegBin:inter-ineq1}
 \frac{\Gamma(r+1)}{\Gamma(r)} \left|\gamma \alpha B(1, rn+\alpha)  - K \frac{B(\gamma \alpha/K + 1, rn+\alpha)}{B(\gamma \alpha/K, r(n-1)+\alpha)}  \right| \leq \frac{4r\gamma^2\alpha^2}{K} \frac{2+\ln(rn+\alpha+1)}{rn+\alpha}.
	\end{equation}
	Expanding gives:
	\begin{align}
	\label{ex-BetaNegBin:inter-fraction}
	&\left| \gamma \alpha B(1, rn+\alpha) - K \frac{B(1+\gamma \alpha/K, rn+\alpha)}{B(\gamma \alpha/K, r(n-1)+\alpha)} \right| \nonumber \\
	&= \frac{\left| \gamma \alpha B(1, rn+\alpha)B(\gamma \alpha/K,r(n-1)+\alpha) - K B(1+\gamma \alpha/K, rn+\alpha)\right| }{B(\gamma \alpha/K, r(n-1)+\alpha)}.
	\end{align}
	We look at the numerator of the right hand side in \cref{ex-BetaNegBin:inter-fraction}:
	\(
	&\left| \gamma \alpha B(1,rn+\alpha) \frac{\Gamma(\gamma \alpha/K) \Gamma(r(n-1)+\alpha)}{\Gamma(\gamma \alpha/K + r(n-1)+\alpha)} - K \frac{\Gamma(1+\gamma \alpha/K) \Gamma(rn+\alpha)}{\Gamma(1+\gamma \alpha/K + rn+\alpha)} \right| \\
	&= \gamma \alpha \Gamma(\gamma \alpha/K) \left| \frac{1}{rn+\alpha} \frac{\Gamma(r(n-1)+\alpha)}{\Gamma(\gamma \alpha/K+r(n-1)+\alpha)}  - \frac{\Gamma(rn+\alpha)}{\Gamma(\gamma \alpha/K+1+rn+\alpha)}\right| \\
	&= \frac{\gamma \alpha\Gamma(\gamma \alpha/K)}{rn+\alpha} \left| \frac{\Gamma(r(n-1)+\alpha)}{\Gamma(\gamma \alpha/K+r(n-1)+\alpha)} - \frac{\Gamma(rn+\alpha+1)}{\Gamma(\gamma \alpha/K+1+rn+\alpha)}  \right| \\ &\leq  \frac{\gamma \alpha\Gamma(\gamma \alpha/K)}{rn+\alpha} \left( \left|\frac{\Gamma(r(n-1)+\alpha)}{\Gamma(\gamma \alpha/K+r(n-1)+\alpha)} - 1 \right| + \left|\frac{\Gamma(rn+\alpha+1)}{\Gamma(\gamma \alpha/K+1+rn+\alpha)} - 1 \right| \right) \\
	&\leq \frac{\gamma \alpha\Gamma(\gamma \alpha/K)}{rn+\alpha}  \frac{2\gamma \alpha}{K} (2 + \ln (rn+\alpha+1)),
	\)
	where we have used \cref{ex-BetaNegBin:helper-ineq3} with $c = \gamma \alpha$ and $b = r(n-1)+\alpha$ or $b = rn+\alpha+1$. In all, \cref{ex-BetaNegBin:inter-fraction} is upper bounded by:
	\(
	&\frac{2\gamma^2\alpha^2}{rn+\alpha} \frac{2+\ln(rn+\alpha+1)}{K} \frac{\Gamma(\gamma\alpha/K)}{B(\gamma \alpha/K,r(n-1)+\alpha)} \\
	&=  \frac{2\gamma^2\alpha^2}{rn+\alpha} \frac{2+\ln(rn+\alpha+1)}{K} \frac{\Gamma(\gamma\alpha/K+r(n-1)+\alpha)}{\Gamma(r(n-1)+\alpha)} \\
	&\leq \frac{4\gamma^2\alpha^2}{K} \frac{2+\ln(rn+\alpha+1)}{rn+\alpha},
	\)
	since $\frac{\Gamma(r(n-1)+\alpha)}{\Gamma(r(n-1)+\alpha+\gamma\alpha/K)} \geq 1 - \frac{\gamma \alpha}{K}(2+\ln(r(n-1)+\alpha)) \geq 0.5$ with $K \ge 2 \gamma \alpha (2+\ln(r(n-1)+\alpha)$. Combining with ${\Gamma(r+1)}/{\Gamma(r)}  = r,$ this is the proof of \cref{ex-BetaNegBin:inter-ineq1}. 
	
	We now move onto the summands from $x = 2$ to $\infty$. By triangle inequality:
	\begin{equation*}
	\left|\gamma \alpha B(x, rn+\alpha)  - K \frac{B(\gamma \alpha/K + x, rn+\alpha)}{B(\gamma \alpha/K, r(n-1)+\alpha)} \right| \leq T_1(x) + T_2(x),
	\end{equation*}
	where:
	\(
	T_1(x) &\defined B(x, rn+\alpha) \left| \gamma \alpha - \frac{K}{B(\gamma \alpha/K,r(n-1)+\alpha)} \right|, \\
	T_2(x) &\defined K \frac{ \left|B(x, rn+\alpha) - B(\frac{\gamma \alpha}{K}+x,rn+\alpha)\right|}{B(\gamma \alpha/K, r(n-1)+\alpha)}.
	\)
	The helper inequalities we have proven once again are useful:
	\(
	\left| \gamma \alpha - \frac{K}{B(\gamma\alpha/K, r(n-1)+\alpha)} \right| &\leq \frac{\gamma \alpha}{K} (3\ln (r(n-1)+\alpha)+8) \\
	\frac{K}{B(\gamma\alpha/K, r(n-1)+\alpha)} &\leq \gamma \alpha + \frac{\gamma \alpha}{K} (3\ln (r(n-1)+\alpha)+8) \leq 2\gamma \alpha, \\
	\left| B(x,rn+\alpha) - B(\gamma \alpha/K+x, rn+\alpha)\right| &\leq \frac{\gamma \alpha}{K} B(x-1,rn+\alpha+1)
	\)
	since $K \geq \gamma \alpha (3 \ln (r(n-1)+\alpha)+8)$, we have applied \cref{ex-BetaNegBin:helper-ineq4} in the first and second inequality and \cref{ex-BetaNegBin:helper-ineq2} in the third one. So for each $x \geq 2$, each summand is at most
	\(
	& \frac{\gamma \alpha (3\ln(r(n-1)+\alpha)+8)}{K} \frac{\Gamma(x+r)}{x!\Gamma(r)}  B(x,rn+\alpha) \\
	&+ \frac{2\gamma^2\alpha^2}{K} \frac{\Gamma(x+r)}{x!\Gamma(r)} B(x-1, rn+\alpha+1).
	\)
	To upper bound the summation from $x = 2$ to $\infty$, it suffices to bound:
	\begin{equation*}
	\sum_{x=2}^{\infty} \frac{\Gamma(x+r)}{\Gamma(r)x!} B(x, rn+\alpha) \leq \sum_{x=1}^{\infty} \frac{\Gamma(x+r)}{\Gamma(r)x!} B(x, rn+\alpha)  \leq  \frac{r}{r(n-1)+\alpha-0.5},
	\end{equation*}
	and:
	\(
	\sum_{x=2}^{\infty} \frac{\Gamma(x+r)}{\Gamma(r)x!} B(x-1, rn+\alpha+1) &\leq r\sum_{x=2}^{\infty} \frac{\Gamma(x-1+r+1)}{\Gamma(r+1)(x-1)!} B(x-1, rn+\alpha+1) \\ &\leq r \sum_{z=1}^{\infty} \frac{\Gamma(z+r+1)}{\Gamma(r+1)z!} B(z, rn+\alpha+1) \\
	&\leq 
	\frac{r(r+1)}{r(n-1)+\alpha-0.5},
	\)
	where we have used \cref{ex-BetaNegBin:helper-ineq1} in each upper bound. 
	So the summation from $x = 2$ to $\infty$ is upper bounded by:
\begin{equation}\label{ex-BetaNegBin:inter-ineq2throughinf}
	\frac{\gamma \alpha (3\ln(r(n-1)+\alpha)+8)}{K} \frac{r}{r(n-1)+\alpha-0.5} + \frac{2\gamma^2\alpha^2}{K} \frac{r(r+1)}{r(n-1)+\alpha-0.5}
	\end{equation}
	\cref{ex-BetaNegBin:inter-ineq1,ex-BetaNegBin:inter-ineq2throughinf} combine to give:
	\(
	&\sum_{x=1}^{\infty} \left| M_{n,x} - K \approxcondLL(x\given x_{1:(n-1)}=\zerovec{n})\right| \\
	&\leq \frac{\gamma \alpha}{K} \frac{(4\gamma \alpha + 3) \ln (rn+\alpha+1) + (10+2r)\gamma \alpha + 24 }{n-1+(\alpha-0.5)/r}.
	\)
\end{proofof}

\section{Proofs of CRM bounds} \label{app:crm-proofs}
\subsection{Upper bound} \label{app:upperbound-proof}
\begin{proofof}{\cref{thm:CRM-upperbound}} \label{proof:crm-proof}
	
	We first give explicit formulas for the constants $C', C'', C''', C'''$.
 	Let $\beta$ be the smallest positive constant where $\beta^2/(1+\beta) \geq 4/C_1$. Such constant exists because $\beta^2/(1+\beta)$ is an increasing function. 
 	The constants are
 	\begin{equation*}
 		\begin{aligned}
 			C'&= (\beta+1) C_1 \ln(1+1/C_1) \left[ 4 C_1 \ln(1+1/C_1) + C_5 \right] \\
				&+ C_1^2 \psi_1(C_1) + \exp(2C_1(\psi(C_1) + 1)) \\
 			   &+ (\beta + 1) 2C_1 \ln (1+1/C_1) + C_2C_3 ,\\
 		\end{aligned}
 	\end{equation*}
	and
	\begin{equation*}
		\begin{aligned}
		C'' = &(\beta + 1) C_1(2C_1 + C_4) + [(\beta + 1) C_1 + C_2]/\ln 2 \\
	&+ (\beta + 1) \left[ C_1 (4 C_1 \ln(1+1/C_1) + C_5) + (2C_1+C_4) C_1 \ln(1+1/C_1) \right]/\ln 2  ,\\
	\end{aligned}
	\end{equation*}
	and
	\begin{equation*}
		\begin{aligned}
	C''' &= (\beta + 1)2C_1^2 \ln(1+1/C_1), \\
	C'''' &= (\beta + 1) 2C_1^2 \ln(1+1/C_1) +  (\beta + 1) C_1. 
\end{aligned}
\end{equation*}
 	By the end of the proof, the reasoning for these constants will be clear.
 	
 	We will focus on the case where the approximation level $K$ is $\Omega(\ln N)$:
	\begin{equation} \label{proof:K-atleast-lnN}
		K \geq \max \left\{\CRMtypical, C_2 (\ln N + C_3)  \right\},
	\end{equation} 
	where $C(N,\alpha)$ is the growth function from \cref{eq:growth-function}. 
	To see why it is sufficient, consider the case where $K < \max \left\{\CRMtypical, C_2 (\ln N + C_3)  \right\}$.
	This implies that $K$ is smaller than a sum
	\begin{equation*}
		\begin{aligned}
				K &< (\beta + 1) (C(N,C_1) + C(K, C_1)) + C_2 (\ln N + C_3) \\
				&\leq \left[ (\beta + 1) C_1 + C_2 \right] \ln N + (\beta + 1) C_1 \ln K + (\beta + 1) 2C_1 \ln (1+1/C_1) + C_2C_3
		\end{aligned}
	\end{equation*}
	where we have used upper bound on the growth function from \cref{lem:harmonic-like-sum}.
	Total variation distance is always upper bounded by $1$.
	Hence, $\dTV \left( P_{N,\infty}, P_{N,K} \right)$ is at most
	\begin{equation*}
			\frac{\left[ (\beta + 1) C_1 + C_2 \right] \ln N + (\beta + 1) C_1 \ln K + (\beta + 1) 2C_1 \ln (1+1/C_1) + C_2C_3}{K} \\
	\end{equation*}
	which is smaller than 
	\begin{equation} \label{proof:smallK}
			\frac{\hat{C}^{(0)} + \hat{C}^{(1)}  \ln^2 N + \hat{C}^{(2)} \ln K}{K}
	\end{equation}
	where
	\begin{equation*}
		\begin{aligned}
			\hat{C}^{(0)} &= (\beta + 1) 2C_1 \ln (1+1/C_1) + C_2C_3, \\ 
			\hat{C}^{(1)} &=  [(\beta + 1) C_1 + C_2]/\ln 2 , \\
			\hat{C}^{(2)} &= (\beta + 1) C_1.
		\end{aligned}
\end{equation*}
	
	In the sequel, we will only consider the situation in \cref{proof:K-atleast-lnN}.
	
	First, we argue that it suffices to bound the total variation distance between the \emph{trait-allocation matrices} coming from the target model and the approximate model. Given the latent measures $X_1,X_2,\ldots,X_N$ from the target model, we can read off the feature-allocation matrix $F$, which has $N$ rows and as many columns as there are unique atom locations among the $X_i$'s: 
	\benum[leftmargin=*]
		\item The $i$-th row of $F$ records the atom sizes of $X_i$. 
		\item Each column corresponds to an atom location: the locations are sorted first according to the index of the first measure $X_{i}$ to manifest it (counting from $1,2,\ldots$), and then its atom size in $X_{i}$. 
	\eenum
	For illustration, suppose $X_1 = 3\delta_{\psi_1} + 4\delta_{\psi_2} + 4\delta_{\psi_3}$, $X_2 = 2\delta_{\psi_1} + \delta_{\psi_3} + \delta_{\psi_4} + 2\delta_{\psi_5}$ and $X_3 = 6 \delta_{\psi_2} + 2\delta_{\psi_3} + \delta_{\psi_5} + 2\delta_{\psi_6} +3\delta_{\psi_7}.$ Then the associate trait-allocation matrix has $3$ rows and $7$ columns and has entries equal to
	\begin{equation} \label{eq:ex-feature-alloc}
			\begin{bmatrix}
			3 & 4 & 4 & 0 & 0 & 0 & 0 \\
			2 & 0 & 1 & 1 & 2 & 0 & 0 \\
			0 & 6 & 2 & 0 & 1 & 2 & 3
		\end{bmatrix}.
	\end{equation}	
	The marginal process that described the atom sizes of $X_{n}\given X_{n-1},X_{n-2},\ldots,X_1$ in \cref{prop:target-marginal} is also the description of how the rows of $F$ are generated. The joint distribution $X_{1},X_{2},\ldots,X_{n}$ can be two-step sampled. First, the trait-allocation matrix $F$ is sampled. Then, the atom locations are drawn iid from the base measure $H$: each column of $F$ is assigned an atom location, and the latent measure $X_i$ has atom size $F_{i,j}$ on the $j$th atom location. A similar two-step sampling generates $Z_1,Z_2,\ldots,Z_n$, the latent measures under the approximate model: the distribution over the feature-allocation matrix $F'$ follows \cref{prop:approx-marginal} instead of \cref{prop:target-marginal}, but conditioned on the feature-allocation matrix, the process generating atom locations and constructing latent measures is exactly the same. In other words, this implies that the conditional distributions $Y_{1:N}\given F$ and $W_{1:N}\given F'$ when $F = F'$ are the same, since both models have the same the observational likelihood $f$ given the latent measures $1$ through $N$. Denote $P_F$ to be the distribution of the feature-allocation matrix under the target model, and $P_{F'}$ the distribution of the feature-allocation matrix under the approximate model. \cref{lem:tv-prop-rule} implies that
	\begin{equation}\label{proof:tv-feature-alloc}
	\dTV \left( P_{N,\infty}, P_{N,K} \right) \leq \inf_{ F, F' \text{ coupling of } P_{F},P_{F'}} \Pr(F \neq F').
	\end{equation}

	Next, we parametrize the trait-allocation matrices in a way that is convenient for the analysis of total variation distance. Let $J$ be the number of columns of $F$. Our parametrization involves $d_{n,x}$, for $n \in [N]$ and $x \in \mathbb{N}$, and $s_j$, for $j \in [J]$: 
	\benum[leftmargin=*]
		\item For $n = 1,2,\ldots,N$:
		\benum[leftmargin=*]
			\item If $n = 1$, for each $x \in \mathbb{N}$, $d_{1,x}$ counts the number of columns $j$ where $F_{1,j} = x$. 
			\item For $n \geq 2$, for each $x \in \mathbb{N}$, let $J_n = \{j: \forall i < n, F_{i,j} = 0\}$ i.e. no observation before $n$ manifests the atom locations indexed by columns in $J_n$. For each $x \in \mathbb{N}$, $d_{n,x}$ counts the number of columns $j \in J_n$ where $F_{n,j} = x$.
		\eenum
		\item For $j = 1,2,\ldots,J$, let $I_j = \min \{i: F_{i,j} > 0 \}$ i.e. the first row to manifest the $j$-th atom location. Let $s_j = F_{I_{j}:N,j}$ i.e. the history of the $j$-th atom location. 
	\eenum
	In words, $d_{n,x}$ is the number of atom locations that is first instantiated by the individual $n$ and each atom has size $x$, while $s_j$ is the history of the $j$-th atom location. $\sum_{n=1}^{N}\sum_{x=1}^{\infty}d_{n,x}$ is exactly $J$, the number of columns. For the example in \cref{eq:ex-feature-alloc}:
	\benum[leftmargin=*]
	\item For $n = 1,2,\ldots,3$:
	\benum[leftmargin=*]
	\item For $n = 1$, $d_{1,1} = d_{1,2} = d_{1,j} = 0$ for $j > 4$. $d_{1,3} = 1$, $d_{1,4} = 2.$
	\item For $n = 2$, $d_{2,1} = 1$, $d_{2,2} = 1$, $d_{2,j} = 0$ for $j  > 2.$
	\item For $n = 3$, $d_{3,1} = 0$, $d_{3,2} = 1$, $d_{3,3} = 1$, $d_{3,j} = 0$ for $j  > 3.$
	\eenum
	\item For $j = 1,2,\ldots,7$, $s_1 = [3,2,0]$, $s_2 = [4,0,6]$, $s_3 = [4,1,2]$, $s_4 = [1,0]$, $s_5 = [2,1]$, $s_6 = [2]$, $s_7 = [3].$
	\eenum
	
	We use the short-hand $d$ to refer to the collection of $d_{n,x}$ and $s$ the collection of $s_j$. There is a one-to-one mapping between $(d,s)$ and the trait-allocation matrix $f$, since we can read-off $(d,s)$ from $f$ and use $(d,s)$ to reconstruct $f.$ Let $(D,S)$ be the distribution of $d$ and $s$ under the target model, while $(D',S')$ is the distribution under the approximate model. We have that
	\begin{equation*}
	\dTV \left( P_{N,\infty}, P_{N,K} \right) \leq \inf_{ (D,S), (D', S') \text{ coupling of } P_{D,S}, P_{D',S'}} \Pr ( (D,S) \neq (D',S')).
	\end{equation*}
	
	 To find an upper bound on $\dTV \left( P_{N,\infty}, P_{N,K} \right)$, we will demonstrate a joint distribution such that $\Pr( (D,S) \neq (D',S'))$ is small. The rest of the proof is dedicated to that end. To start, we only assume that $(D,S,D',S')$ is a proper coupling, in that marginally $(D,S) \sim P_{D,S}$ and $(D',S') \sim P_{D',S'}$. As we progress, gradually more structure is added to the joint distribution $(D,S,D',S')$ to control $\Pr( (D,S) \neq (D',S'))$.
	
	 We first decompose $\Pr( (D,S) \neq (D',S'))$ into other probabilistic quantities which can be analyzed using \cref{condition:marginal-process}. Define the \emph{typical} set:
	\begin{equation*}
	\mathcal{D}^* = \left\{d: \sum_{n=1}^{N} \sum_{x=1}^{\infty} d_{n,x} \leq \CRMtypical \right\}.
	\end{equation*}
	$d \in \mathcal{D}^*$ means that the trait-allocation matrix $f$ has a small number of columns. The claim is that: 
	\begin{equation}\label{proof:prob-diff-chain-rule}
	\Pr((D,S) \neq (D',S'))  \leq \Pr(D \neq D') + \Pr(S \neq S' \given  D = D', D \in \mathcal{D}^*) + \Pr(D \notin \mathcal{D}^*). 
	\end{equation}
	This is true from basic properties of probabilities and conditional probabilities:
	\(
	&\Pr((D,S) \neq (D',S')) \\
	&= \Pr(D \neq D') + \Pr(S \neq S', D = D') \\
	&= \Pr(D \neq D') + \Pr(S \neq S', D = D', D \in \mathcal{D}^*) + \Pr(S \neq S', D = D', D \notin \mathcal{D}^*) \\
	&\leq \Pr(D \neq D') + \Pr(S \neq S'\given D = D', D \in \mathcal{D}^*) + \Pr(D \notin \mathcal{D}^*),
	\)
	The three ideas behind this upper bound are the following. First, because of the growth condition, we can analyze the atypical set probability $\Pr(D \notin \mathcal{D}^*)$. Second, because of the total variation between $\targetcondLL$ and $\approxcondLL$, we can analyze $\Pr(S \neq S'\given D = D', D \in \mathcal{D}^*)$. Finally, we can analyze $\Pr(D \neq D')$ because of the total variation between $K\approxcondLL$ and $M_{n,.}$. In what follows we carry out the program.
	
	\textbf{Atypical set probability.} The $\Pr(D \notin \mathcal{D}^*)$ term in \cref{proof:prob-diff-chain-rule} is easiest to control. Under the target model \cref{prop:target-marginal}, the $D_{i,x}$'s are independent Poissons with mean $M_{i,x}$, so the sum $\sum_{i=1}^{N} \sum_{x=1}^{\infty} D_{i,x}$ is itself a Poisson with mean $M = \sum_{i=1}^{N} \sum_{x=1}^{\infty} M_{i,x}$. Because of \cref{lem:poisson-tail}, for any $x > 0$:
	\begin{equation*}
	\Pr \left(\sum_{i=1}^{N} \sum_{x=1}^{\infty} D_{i,x} >  M + x \right) \leq \exp\left(-\frac{x^2}{2(M+x)} \right).
	\end{equation*}
	For the event $\Pr(D \notin \mathcal{D}^*)$, $M+x = \CRMtypical $, $M \leq C(N,C_1)$ due to \cref{thm-ass:eCRM-tot}, so that $x \geq \beta \max(C(K,C_1),C(N,C_1))$. Therefore:
	\begin{equation}\label{proof:small-contribution}
	\Pr(D \notin \mathcal{D}^*) \leq  \exp \left( - \frac{\beta^2}{2(\beta+1)} \max(C(K,C_1),C(N,C_1)) \right).
	\end{equation}
	
	\textbf{Difference between histories.} To minimize the difference probability between the histories of atom sizes i.e. the $\Pr(S \neq S' \given  D = D', D \in \mathcal{D}^*)$ term in \cref{proof:prob-diff-chain-rule}, we will use \cref{thm-ass:old-location}. The claim is, there exists a coupling of $S'\given D'$ and $S\given D$ such that:
	\begin{equation} \label{proof:old-atom-tv}
	\Pr(S \neq S' \given  D = D', D \in \mathcal{D}^*) \leq \frac{\CRMtypical}{K} C(N,C_1). 
	\end{equation}
	Fix some $d \in \mathcal{D}^*$ --- since we are in the typical set, the number of columns in the trait-allocation matrix is at most $\CRMtypical$. Conditioned on $D = d$, there is a finite number of history variables $S$, one for each atom location; similar for conditioning of $S'$ on $D' = d$. For both the target and the approximate model, the density of the joint distribution factorizes:
	\(
	\Pr(S=s\given D=d) &= \prod_{j=1}^{J} \Pr(S_j = s_j\given D=d) \\
	\Pr(S'=s\given D'=d) &= \prod_{j=1}^{J} \Pr(S'_j = s_j\given D'=d),
	\)
	since in both marginal processes, the atom sizes for different atom locations are independent of each other. Each $S_j$ (or $S'_j$) only takes values from a countable set. Therefore, by \cref{lem:tv-prod-rule}, 
	\begin{equation*}
	\dTV (P_{S\given D=d}, P_{S'\given D'=d}) \leq \sum_{j=1}^{J} \dTV (P_{S_j\given D=d}, P_{S'_j\given D'=d}).
	\end{equation*}
	We inspect each $\dTV (P_{S_j\given D=d}, P_{S'_j\given D'=d})$. Fixing $d$ also fixes $I_j$, the first row to manifest the $j$-th atom location. The history $s_j$ is then a $N-I_j+1$ dimensional integer vector, whose $t$th entry is the atom size over the $j$the atom location of the $t+I_j-1$ row. Because of \cref{thm-ass:old-location}, we know that conditioned on the same partial history $S_j(1:(t-1)) = S'_j(1:(t-1)) = s$, the distributions $S_j(t)$ and $S'_j(t)$ are very similar. The conditional distribution $S_j(t)\given D=d, S_j(1:(t-1)) = s$ is governed by $\targetcondLL$ \cref{prop:target-marginal} while $S'_j(t)\given D'=d, S'_j(1:(t-1)) = s$ is governed by $\approxcondLL$ \cref{prop:approx-marginal}. Hence: 
	\begin{equation*}
	\dTV \left( P_{S_j(t)\given D=d, S_j(1:(t-1)) = s} , P_{S'_j(t)\given D'=d, S'_j(1:(t-1)) = s} \right) \leq 2 \frac{1}{K} \frac{C_1}{t+I_j-2+C_1},
	\end{equation*}
	for any partial history $s$. To use this conditional bound, we repeatedly use \cref{lem:tv-chain-rule} to compare the joint $S_j = (S_j(1), S_j(2), \ldots, S_j(N-I_j+1))$ with the joint $S'_j = (S'_j(1), S'_j(2), \ldots, S'_j(N-I_j+1))$, peeling off one layer of random variables (indexed by $t$) at a time. 
	\(
	& \dTV (P_{S_j\given D=d}, P_{S'_j\given D'=d}) \\
	&\leq \sum_{t=1} ^{N-I_j+1} \max_{s} \dTV \left( P_{S_j(t)\given D=d, S_j(1:(t-1)) = s}  , P_{S'_j(t)\given D'=d, S'_j(1:(t-1)) = s} \right) \\
	&\leq \sum_{t=1}^{N-I_j+1} 2 \frac{1}{K} \frac{C_1}{t+I_j-2+C_1} \\
	&\leq 2 \frac{C(N,C_1)}{K}. 
	\)
	Multiplying the right hand side by $\CRMtypical$, the upper bound on $J$, we arrive at the same upper bound for the total variation between $P_{S\given D=d}$ and $P_{S'\given D'=d}$ in \cref{proof:old-atom-tv}. Furthermore, our analysis of the total variation can be back-tracked to construct the coupling between the conditional distributions $S\given D=d$ and $S'\given D'=d$ which attains that small probability of difference because all the distributions being analyzed are discrete. Since the choice of conditioning $d \in \mathcal{D}^*$ was arbitrary, we have actually shown \cref{proof:old-atom-tv}. 
	
	\textbf{Difference between new atom sizes.} Finally, to control the difference probability for the distribution over new atom sizes i.e. the $\Pr(D \neq D')$ term in \cref{proof:prob-diff-chain-rule}, we will utilize \cref{thm-ass:IFA-tot,thm-ass:new-location}. For each $n$, define the short-hand $d_{1:n}$ to refer to the collection $d_{i,x}$ for $i \in [n]$, $x \in \mathbb{N}$, and the typical sets:
	\begin{equation*}
	\mathcal{D}_{n}^* = \left\{d_{1:n}: \sum_{i=1}^{n} \sum_{x=1}^{\infty} d_{i,x} \leq \CRMtypical \right\}.
	\end{equation*}
	The type of expansion performed in \cref{proof:prob-diff-chain-rule} can be done once here to see that:
	\(
	&\Pr(D \neq D') \\
	&= \Pr((D_{1:(N-1},D_N) \neq (D'_{1:(N-1)},D'_N)) \\
	&\leq \Pr(D_{1:(N-1)} \neq D'_{1:(N-1)}) \\
	&\phantom{=~} + \Pr(D_{N} \neq D'_{N} \given  D_{1:(N-1)} = D'_{1:(N-1)}, D_{1:(N-1)} \in \mathcal{D}_{n-1}^*) \\
	&\phantom{=~} + \Pr(D_{1:(N-1)} \notin \mathcal{D}_{n-1}^*). 
	\)
	Apply the expansion once more to $\Pr(D_{1:(N-1)} \neq D'_{1:(N-1)})$, then to $\Pr(D_{1:(N-2)} \neq D'_{1:(N-2)})$. If we define:
	\begin{equation*}
	B_j = \Pr(D_{j} \neq D'_{j} \given  D_{1:(j-1)} = D'_{1:(j-1)}, D_{1:(j-1)} \in \mathcal{D}_{j-1}^*),
	\end{equation*}
	with the special case $B_1$ simply being $\Pr(D_1 \neq D_1')$, then:
	\begin{equation} \label{proof:decompose-DneqD'}
	\Pr(D \neq D') \leq \sum_{j=1}^{N} B_j + \sum_{j=2}^{N} \Pr(D_{1:(j-1)} \notin \mathcal{D}_{j-1}^*). 
	\end{equation}
	
	The second summation in \cref{proof:decompose-DneqD'}, comprising of only atypical probabilities, is easier to control. For any $j$, since $\sum_{i=1}^{j-1} \sum_{x=1}^{\infty} D_{i,x} \leq \sum_{i=1}^{N} \sum_{x=1}^{\infty} D_{i,x}$, $\Pr(D_{1:(j-1)} \notin \mathcal{D}_{j-1}^*) \leq \Pr( D \notin \mathcal{D}^*)$,	so a generous upper bound for the contribution of all the atypical probabilities including the first one from \cref{proof:small-contribution} is
	\(
	\Pr( D \notin \mathcal{D}^*) &+ \sum_{j=2}^{N} \Pr(D_{1:(j-1)} \notin \mathcal{D}_{j-1}^*) &\\
	&\leq \exp \left(- \left( \frac{\beta^2}{2(\beta+1)} \max(C(K,C_1),C(N,C_1)) - \ln N \right) \right).
	\)
	By \cref{lem:harmonic-like-sum}, $\max(C(K,C_1),C(N,C_1)) \geq C_1 (\max(\ln N, \ln K) - C_1 (\psi(C_1)+1))$. Since we have set $\beta$ so that $\frac{\beta^2}{\beta+1} C_1 = 4$, we have
	\begin{equation*}
		\begin{aligned}
			\frac{\beta^2}{2(\beta+1)} \max(C(K,C_1),C(N,C_1)) - \ln N &\geq 2 \max(\ln N, \ln K) - 2C_1 (\psi(C_1) + 1)  - \ln N \\
			&\geq \ln K - 2C_1 (\psi(C_1) + 1).
		\end{aligned}
	\end{equation*}
	meaning the overall atypical probabilities is at most
	\begin{equation} \label{proof:total-atypical-probs}
		\Pr( D \notin \mathcal{D}^*) + \sum_{j=2}^{N} \Pr(D_{1:(j-1)} \notin \mathcal{D}_{j-1}^*) \leq \frac{\exp(2C_1 (\psi(C_1) + 1)) }{K}.
	\end{equation}
	
	As for the first summation in \cref{proof:decompose-DneqD'}, we look at the individual $B_j$'s. For any fixed $d_{1:(j-1)} \in \mathcal{D}^*_{j-1}$ , we claim that there exists a coupling between the conditionals $D_j \given  D_{1:(j-1)} = d_{1:(j-1)}$ and $D'_j \given  D'_{1:(j-1)} = d_{1:(j-1)}$ such that $\Pr(D_j \neq D_j' \given  D_{1:(j-1)} = D'_{1:(j-1)} = d_{1:(j-1)})$ is at most
	\begin{equation} \label{proof:Bj-upper}
	\frac{C_1^2}{K} \frac{1}{(j-1+C_1)^2} + \left[ C_4 \ln j + C_5 + (\beta + 1) \max(C(K,C_1), C(N,C_1)) \right] \frac{1}{j-1+C_1}.
	\end{equation}
	Because the upper bound holds for arbitrary values $d_{1:(j-1)}$, the coupling actually ensures that, as long as $D_{1:(j-1)} = D'_{1:(j-1)}$ for some value in $\mathcal{D}^*_{j-1}$, the probability of difference between $D_j$ and $D_j'$ is small i.e. $B_j$ is at most the right hand side. 
	
	We demonstrate the existence of a distribution $U = \{ U_x\}_{x=1}^{\infty}$ of independent Poisson random variables, such that both the total variation between $P_{D_j \given  D_{1:(j-1)} = d_{1:(j-1)}}$ and $P_U$ and the total variation between $P_{D'_j \given  D'_{1:(j-1)} = d_{1:(j-1)}}$ and $P_U$ are small. Here, each $U_x$ has mean:
	\begin{equation*}
	\mathbb{E}(U_x) = \left( K - \sum_{i=1}^{j-1} \sum_{y=1}^{\infty}d_{i,y} \right) \approxcondLL(x\given x_{1:(j-1)}=0).
	\end{equation*}
	
	On the one hand, conditioned on $D'_{1:(j-1)} = d_{1:(j-1)}$, $D'_j = \{D'_{j,x}\}_{x=1}^{\infty}$ is the joint distribution of types of successes of type $x$, where there are $K - \sum_{i=1}^{j-1} \sum_{x=1}^{\infty} d_{i,x}$ independent trials and types $x$ success has probability $\approxcondLL(x\given x_{1:(j-1)}=0)$ by \cref{prop:approx-marginal}. Because of \cref{lem:multi-poiss-approx} and \cref{thm-ass:IFA-tot}:
	\begin{align} \label{proof:IFA-almost-poisson}
	\Pr(D'_j \neq U \given D'_{1:(j-1)} = d_{1:(j-1)}) &\leq \left( K - \sum_{i=1}^{j-1} \sum_{y=1}^{\infty}d_{i,y} \right) \left(\sum_{x=1}^{\infty}\approxcondLL(x\given x_{1:(j-1)}=0) \right)^2 \nonumber \\
	&\leq K \left( \frac{1}{K} \frac{C_1}{j-1+C_1} \right)^2 \nonumber \\
	&\leq \frac{C_1^2}{K} \frac{1}{(j-1+C_1)^2}.
	\end{align}
	On the other hand, conditioned on $D_{1:(j-1)}$, $D_j = \{D_{j,x}\}_{x=1}^{\infty}$ consists of independent Poissons, where the mean of $D_{j,x}$ is $M_{j,x}$ by \cref{prop:target-marginal}. We show that there exists a coupling of $P_U$ and $P_{D_j}$ such that
	\begin{equation} \label{proof:bound-by-sum}
		\Pr(U \neq D_j) \leq \sum_{x=1}^{\infty}  \dTV(P_{U_x}, P_{D_{j,x}}).
	\end{equation}
	For each $x \geq 1$, let $O_x$ be the maximal coupling distribution between $P_{U_x}$ and $P_{D_{j,x}}$ i.e.\ for $(A, B) \sim O_x$, $\Pr(A \neq B) = \dTV(P_{U_x}, P_{D_{j,x}}).$ Such $O_x$ exists because both $P_{U_x}$ and $P_{D_{j,x}}$ are Poisson (hence discrete) distributions. Furthermore, since $O_x$ is itself a discrete distribution, the conditional distributions $D_{j,x} \given U_x$ exists. 	
	 Denote the natural zig-zag bijection from $\{\mathbb{N} \cup 0\}^2$ to $\mathbb{N}$ to be $L.$\footnote{$L(0,0) = 1$, $L(0,1) = 2$, $L(1,0) = 3$, $L(2,0) = 4$, $L(1,1) = 5$, $L(0,2) = 6$ and so on.} Denote by $F_x$ the cdf of the distribution of $L(A,B)$ for $(A, B) \sim O_x.$ To generate samples from $O_x$, it suffices to generate samples from $F_x$ and transform using the inverse of $L.$ Consider the following coupling of $P_U$ and $P_{D_j}$:
	\begin{itemize}
		\item Generate i.i.d uniform random random variables $V_1,V_2,\ldots$ 
		\item For $x \geq 1$, let $(U_x, D_{j,x}) = L^{-1}(F^{-1}_x(V_x)). $
	\end{itemize}
	Marginally, each $U_x$ (or $D_{j,x}$) is Poisson with the right mean, and across $x$, the $U_x$ (or $D_{j,x}$) are independent of each other because we use i.i.d uniform r.v's. Alternatively, the conditional distribution of $D_{j} \given U$ implied by this joint distribution is as follows:
	\begin{itemize}
		\item For $x \geq 1$, sample $U_x \given D_{j,x}$ from the conditional distribution implied by the maximal coupling $O_x.$
	\end{itemize}
	If $U$ is different from $D_j$, it must be that for at least one $x$, $U_x \neq D_{j,x}$.
	Therefore
	\begin{equation*}
		\Pr(U \neq D_j) \leq \sum_{x=1}^{\infty} \Pr(U_x \neq D_{j,x}).
	\end{equation*}
	Since the coupling $(U_x,D_{j,x})$ attains the $\dTV(P_{U_x}, P_{D_{j,x}})$, we are done.	
	From \cref{lem:poisson-tv}, we know 
	\begin{align} \label{proof:poissons-close}
		 &\sum_{x=1}^{\infty} \dTV(P_{U_x}, P_{D_{j,x}}) \nonumber\\
		 &\leq \sum_{x=1}^{\infty} \left|M_{j,x} - \left( K - \sum_{i=1}^{j-1} \sum_{y=1}^{\infty}d_{i,y} \right) \approxcondLL(x|x_{1:(j-1)}=0) \right| \nonumber \\
		 &\leq \sum_{x=1}^{\infty} \left( |M_{j,x} - K\approxcondLL(x\given x_{1:(j-1)}=0)| +  \sum_{i=1}^{j-1} \sum_{y=1}^{\infty}d_{i,y} \approxcondLL(x\given x_{1:(j-1)}=0)  \right) \nonumber \\
		 &\leq \sum_{x=1}^{\infty} |M_{j,x} - K\approxcondLL(x\given x_{1:(j-1)}=0)| + \left(  \sum_{i=1}^{j-1} \sum_{y=1}^{\infty}d_{i,y} \right) \left(  \sum_{x=1}^{\infty}  \approxcondLL(x\given x_{1:(j-1)}=0)\right). 
	\end{align}
	The first term is upper bounded by \cref{thm-ass:new-location}. Regarding the second term, since we are in the typical set, $\sum_{i=1}^{j-1} \sum_{y=1}^{\infty}d_{i,y}$ is small and we also use \cref{thm-ass:IFA-tot}. Therefore the overall bound on the second term is:
	\begin{equation*}
		\CRMtypical \frac{1}{K}\frac{C_1}{j-1+C_1}. 
	\end{equation*}
	Combining the two bounds and \cref{proof:bound-by-sum} give the following bound on $\Pr(U \neq D_j)$:
	\begin{equation} \label{proof:PUclosePDj}
		\Pr(U \neq D_j) \leq	\frac{1}{K} \frac{C_4\ln j + C_5}{j-1+C_1} +  \CRMtypical \frac{1}{K}\frac{C_1}{j-1+C_1}.
	\end{equation}

	We now show how the combination of \cref{proof:PUclosePDj,proof:IFA-almost-poisson} imply \cref{proof:Bj-upper}. From \cref{proof:PUclosePDj}, there exists a coupling of $P_U$ and $P_{D_j}$ such that the difference probability is small. From \cref{proof:IFA-almost-poisson}, there exists a coupling of $P_U$ and $P_{D'_j \given D'_{1:(j-1)} = d_{1:(j-1)}}$ such that the difference probability is small. In both cases, we can sample from the conditional distribution based on $U$. $D_j \given U$ exists because of the discussion after \cref{proof:bound-by-sum}, while $D'_j \given D'_{1:(j-1)} = d_{1:(j-1)}, U$ exists because of \cref{lem:multi-poiss-approx}. Therefore, we can glue the two couplings together, by first sampling $U$, and then sample from the appropriate conditional distributions. By taking expectations of the simple triangle inequality for the discrete metric i.e.\
	\begin{equation*}
		\indict{D_j \neq D'_j} \leq \indict{D_j \neq U} + \indict{D'_j \neq U},
	\end{equation*}
	we reach \cref{proof:Bj-upper}. 
	
	We sum of the right hand side of \cref{proof:Bj-upper} across $j$. 
	This shows that $\sum_{j=1}^{N} B_j$ is at most
	\begin{equation*}
		\begin{aligned} 
	\frac{C_1^2}{K} \left( \sum_{j=1}^{N} \frac{1}{(j-1 + C_1)^2} \right) &+ \frac{\CRMtypical}{K} C(N, C_1) \\
	&+\frac{C_4 \ln N + C_5}{K} C(N, C_1).
		\end{aligned}
	\end{equation*}
	The first term is upper bounded by the trigamma function $\psi_1(\cdot)$:
	\begin{equation*}
		\frac{C_1^2}{K} \sum_{j=1}^{N} \frac{1}{(j-1 + C_1)^2} \leq \frac{C_1^2 \psi_1(C_1)}{K}. 
	\end{equation*}
	This means, an upper bound on $\sum_{j=1}^{N} B_j$ is 
	\begin{equation} \label{proof:totalBj}
		\frac{C_1^2 \psi_1(C_1)}{K} + \frac{\beta+1}{K} C(N,C_1) \left[ (C_1 + C_4) \ln N + C_1 \ln K + 2C_1 \ln (1+1/C_1) + C_5 \right].
	\end{equation}
	Because of \cref{proof:decompose-DneqD',proof:total-atypical-probs,proof:totalBj}, we can couple $D$ and $D'$ such that $\Pr(D \neq D') + 
	\Pr(D \notin  \mathcal{D}^*)$ is at most
	\begin{equation}\label{proof:coupleDD'}
		\begin{aligned} 
		&\frac{C_1^2 \psi_1(C_1) + \exp(2C_1(\psi(C_1) + 1))}{K} \\
		&+\frac{\beta+1}{K} C(N,C_1) \left[ (C_1 + C_4) \ln N + C_1 \ln K + 2C_1 \ln (1+1/C_1) + C_5 \right].
		\end{aligned}
	\end{equation}
	Aggregating the results from \cref{proof:coupleDD',proof:old-atom-tv}, we have that $\dTV \left( P_{N,\infty}, P_{N,K} \right)$ is at most
	\begin{equation*}
		\begin{aligned}
			&\frac{C_1^2 \psi_1(C_1) + \exp(2C_1(\psi(C_1) + 1))}{K}\\
			&+\frac{\beta+1}{K} C(N,C_1) \left[ \max(C(K,C_1), C(N,C_1)) + (C_1 + C_4) \ln N \right]\\
			&+\frac{\beta+1}{K} C(N,C_1) \left[ C_1 \ln K + 2C_1 \ln (1+1/C_1) + C_5 \right].
		\end{aligned}
	\end{equation*}
	We expand the sum of the last two term by upper bounding $\max(C(K,C_1), C(N,C_1))$ by $C(K,C_1) + C(N,C_1)$ and using the upper bound \cref{lem:harmonic-like-sum}.
	The end result is 
	\begin{equation*}
		\frac{\tilde{C}^{(0)} + \tilde{C}^{(1)} \ln K + \tilde{C}^{(2)} \ln N + \tilde{C}^{(3)} \ln N \ln K +  \tilde{C}^{(4)} \ln^2 N }{K}
	\end{equation*}
	where $\tilde{C}^{(0)}$ is equal to
	\begin{equation*}
		(\beta+1) C_1 \ln(1+1/C_1) \left[ 4 C_1 \ln(1+1/C_1) + C_5 \right] + C_1^2 \psi_1(C_1) + \exp(2C_1(\psi(C_1) + 1)), \\ 
	\end{equation*}
	and 
	\begin{equation*}
		\begin{aligned}
			\tilde{C}^{(1)} &= (\beta + 1) 2C_1^2 \ln(1+1/C_1), \\
			\tilde{C}^{(2)} &= (\beta + 1) \left[ C_1 (4 C_1 \ln(1+1/C_1) + C_5) + (2C_1+C_4) C_1 \ln(1+1/C_1) \right], \\
			\tilde{C}^{(3)} &= (\beta + 1)2C_1^2 \ln(1+1/C_1), \\
			\tilde{C}^{(4)} &= (\beta + 1) C_1(2C_1 + C_4).
		\end{aligned}
	\end{equation*}
	Since $N$ is a natural number, $N \geq 1$ we can write $\ln N \leq (1/\ln 2) \ln^2 N$, to simplify the upper bound on total variation as
	\begin{equation} \label{proof:largeK}
		\frac{\tilde{C}^{(0)} +  \left( \tilde{C}^{(4)} + \tilde{C}^{(2)}/\ln 2 \right) \ln^2 N + \tilde{C}^{(3)} \ln N  \ln K  + \tilde{C}^{(1)}  \ln K}{K}.
	\end{equation}
	Taking the sum of individual coefficients in front of $\ln^2 N$ (et cetera) between \cref{proof:largeK} and \cref{proof:smallK} yields the constants at the beginning of the proof.
\end{proofof}

In applications, the observational likelihood $f$ and the ground measure $H$ might be random rather than fixed quantities.
For instance, in linear--Gaussian beta--Bernoulli processes without good prior information, probabilistic models put priors on the variances of the Gaussian features as well as the noise in observed data. 
In such cases, the AIFAs remain the same as the in \cref{thm:general-CRM-convergence} (or \cref{cor:expCRM-d-zero}) since the rate measure $\nu$ is still fixed.  
The above proof of \cref{thm:CRM-upperbound} can be easily extended to the case where $f$ and $H$ are random, because the argument leading to \cref{proof:tv-feature-alloc} retains validity when $f$ and $H$ have the same distribution under the target and the approximate model. For completeness, we state the error bound in such cases where hyper-priors are used. 

\bncor [Upper bound for hyper-priors] \label{cor:hyper-prior-bound}
Let $\mathcal{H}$ be a prior distribution for ground measures $H$ and $\mathcal{F}$ be a prior distribution for observational likelihoods $f.$
Suppose the target model is 
\(
\begin{split}
	H &\sim \mathcal{H}(.), \\
	f &\sim \mathcal{F}(.), \\  
	\Theta \given H &\sim \distCRM(H, \nu), \\
	X_n \mid \Theta &\distiid \distLP(\traitLL, \Theta), \quad n = 1, 2, \ldots, N, \\
	Y_n \mid f, X_n &\distind f(\cdot \given X_n), \quad n = 1, 2, \ldots, N.
\end{split}
\)
The approximate model, with $\nu_K$ as in \cref{thm:general-CRM-convergence} (or \cref{cor:expCRM-d-zero}), is 
\(
\begin{split}
	H &\sim \mathcal{H}(.), \\
	f &\sim \mathcal{F}(.), \\  
	\Theta_{\approxlev} \given H &\sim \distIFA_\approxlev(H, \nu_\approxlev),  \\
	Z_n \mid \Theta_{\approxlev} &\distiid \distLP(\traitLL, \Theta_\approxlev), \quad n = 1, 2, \ldots, N, \\
	W_n \mid f, Z_n &\distind f(\cdot \mid Z_n),  \,\,~\quad n = 1, 2, \ldots, N. \\
\end{split}	
\)
	If \cref{assume:rate-measure-near-0} and \cref{condition:marginal-process} hold, then there exist positive constants $C',C'',C'''$ depending only on $\{C_i\}_{i=1}^{5}$ such that 
	\begin{equation*}
		\dTV \left( P_{Y_{1:N}}, P_{W_{1:N}} \right) \leq \frac{C' + C''\ln^2 N + C''' \ln N \ln K}{K}.
	\end{equation*}
\encor

The upper bound in \cref{cor:hyper-prior-bound} is visually identical to \cref{thm:CRM-upperbound}, and has no dependence on the hyper-priors $\mathcal{H}$ or $\mathcal{F}.$

\subsection{Lower bound} \label{app:lowerbound-proof}

\begin{proofof}{\cref{thm:betaBer-lnN-necessary}} \label{crm-proof:betaBer-lnN-necessary}
	First we mention which probability kernel $f$ results in the large
	total variation distance: the pathological $f$ is the Dirac measure i.e., $f(\cdot \given X) \defined \delta_X(.)$. With this conditional likelihood $\targetLatent_{n} = \targetObs_{n}$ and $\approxLatent_n = \approxObs_n$, meaning:
	\begin{equation*}
		\dTV(\BPproc{N}{\infty}, \BPproc{N}{K}) = \dTV(P_{\targetLatent_{1:N}},P_{\approxLatent_{1:N}}).	
	\end{equation*}
	
	Now we discuss why the total variation is lower bounded by the function of $N$. Let $\mathcal{A}$ be the event that there are at least $\frac{1}{2} \gamma C(N,\alpha)$ unique atom locations in among the latent states:
	\begin{equation*}
		\mathcal{A} \defined \left\{x_{1:N}: \# \text{unique atom locations} \geq \frac{1}{2} \gamma C(N,\alpha) \right\}.	
	\end{equation*}
	
	The probabilities assigned to this event by the approximate and the target models are
	very different from each other.	On the one hand, since $K < \frac{\gamma C(N,\alpha)}{2}$, under $\AIFA{\approxlev}$, $\mathcal{A}$ has measure zero:
	\begin{equation} \label{crm-proofs:zero-mass-approx}
		\Pr_{\approxLatent_{1:N}}(\mathcal{A}) = 0.
	\end{equation}
	On the other hand, under beta--Bernoulli, the number of unique atom locations drawn is a Poisson random variable with mean exactly $\gamma C(N,\alpha)$ --- see \cref{prop:target-marginal} and \cref{prop:approx-marginal}. The complement of $\mathcal{A}$ is a lower tail event. By \cref{lem:poisson-tail} with $\lambda = \gamma C(N,\alpha)$ and $x = \frac{1}{2} \gamma C(N,\alpha)$:
	\begin{equation}\label{crm-proofs:large-mass-target}
		\Pr_{\targetLatent_{1:N}}(\mathcal{A}) \geq 1 - \exp \left(-\frac{\gamma C(N,\alpha)}{8} \right).
	\end{equation}
	Because of \cref{lem:harmonic-like-sum}, we can lower bound $C(N,\alpha)$ by a multiple of $\ln N$:
	\begin{equation*}
		\exp \left(-\frac{\gamma C(N,\alpha)}{8} \right) \leq \exp \left( -\frac{\gamma \alpha \ln N}{8} + \frac{\alpha \gamma (\psi(\alpha)+1)}{8} \right) = \frac{\text{constant}}{N^{\gamma \alpha/8}}.
	\end{equation*}
	
	We now combine \cref{crm-proofs:zero-mass-approx,crm-proofs:large-mass-target} and recall that total variation is the maximum over discrepancy in probabilistic masses.
\end{proofof}

The proof of \cref{thm:betaBer-1/K-lowerbound} relies on the ability to compute a lower bound on the total variation distance between a binomial distribution and a Poisson distribution.

\bnprop [Lower bound on total variation between binomial and Poisson] \label{prop:binom-poiss-lowerbound}
For all $K$, it is true that
\begin{equation*}
	\dTV \left( \distPoisson\left( \gamma \right), \distBinom \left( K, \frac{\gamma/K}{\gamma/K+1} \right) \right) \geq C(\gamma) K \left(\frac{\gamma/K}{\gamma/K+1}  \right)^2,  
\end{equation*}
where
\begin{equation*}
C(\gamma) = \frac{1}{8} \frac{1}{\gamma + \exp(-1)(\gamma+1) \max(12 \gamma^2, 48 \gamma, 28)}. \end{equation*}
\enprop

\begin{proofof}{\cref{prop:binom-poiss-lowerbound}}
	
	We adapt the proof of \cite[Theorem 2]{barbour1984rate} to our setting. The $\distPoisson(\gamma)$ distribution satisfies the functional equality:
	\begin{equation} \label{lb:poisson-stein-eq}
		\mathbb{E}[\gamma y(Z+1) - Zy(Z)] = 0,
	\end{equation}
	where $y$ is any real-valued function and $Z \sim \distPoisson(\gamma)$. 
	
	Denote $\gamma_K = \frac{\gamma}{\gamma/K+1}$. For $m \in \mathbb{N}$, let
	\begin{equation*}
	x(m) = m \exp \left( -\frac{m^2}{\gamma_K \theta}\right),
	\end{equation*}
	where $\theta$ is a constant which will be specified later. $x(m)$ serves as a test function to lower bound the total variation distance between $\distPoisson(\gamma)$ and $\distBinom \left( K, \gamma_K/K \right)$. Let $X_i \sim \text{Ber}(\frac{\gamma_K}{K})$, independently across $i$ from $1$ to $K$, and $W = \sum_{i=1}^{K}$. Then $W \sim \text{Binomial} \left( K, \gamma_K/K\right)$. The following identity is adapted from \cite[Equation 2.1]{barbour1984rate}:
	\begin{equation} \label{eq:functional}
	\mathbb{E}[\gamma_K x(W+1) - Wx(W)] = \left( \frac{\gamma_K}{K} \right)^2 \sum_{i=1}^{K} \mathbb{E}[x(W_i+2) - x(W_i+1)],
	\end{equation}
	where $W_i = W - X_i$. 
	
	We first argue that the right hand side is not too small i.e.\ for any $i$,
	\begin{equation} \label{eq:lower-bound-one-step-diff}
	\mathbb{E}[x(W_i+2) - x(W_i+1)] \geq 1 - \frac{3\gamma_K^2 + 12 \gamma_K+7}{\theta \gamma_K}. 
	\end{equation}
	Consider the derivative of $x(m)$:
	\begin{equation*}
	\frac{d}{dm} x(m) = \exp \left( -\frac{m^2}{\gamma_K \theta}\right) \left( 1 - \frac{2m^2}{\gamma_K \theta} \right) \geq 1 - \frac{3m^2}{\theta \gamma_K},
	\end{equation*}
	because of the easy-to-verify inequality $e^{-x}(1-2x) \geq 1 - 3x$ for $x \geq 0$. This means that
	\begin{equation*}
	x(W_i+2) - x(W_i+1) \geq \int_{W_i+1}^{W_i+2} \left(1 - \frac{3m^2}{\theta \gamma_K}\right) dm = 1 - \frac{1}{\theta \gamma_K} (3W_i^2 + 9W_i + 7). 
	\end{equation*}
	Taking expectations, noting that $\mathbb{E}(W_i) \leq \gamma_K$ and $\mathbb{E}(W_i^2) = \text{Var}(W_i) + [\mathbb{E}(W_i)]^2 \leq \sum_{j=1}^{K} \frac{\gamma_K}{K} + (\gamma_K)^2 = \gamma_K^2 + \gamma_K$ we have proven \cref{eq:lower-bound-one-step-diff}. 
	
	Now, because of positivity of $x$, and that $\gamma \geq \gamma_K$, we trivially have
	\begin{equation}\label{eq:positivity-of-x}
	\mathbb{E}[\gamma x(W+1) - Wx(W)] \geq \mathbb{E}[\gamma_K x(W+1) - Wx(W)]. 
	\end{equation} 
	Combining \cref{eq:functional}, \cref{eq:lower-bound-one-step-diff} and \cref{eq:positivity-of-x} we have that
	\begin{equation*}
	\mathbb{E}[\gamma x(W+1) - Wx(W)] \geq K\left(\frac{\gamma_K}{K}\right)^2 \left( 1 - \frac{3\gamma_K^2 + 12 \gamma_K+7}{\theta \gamma_K}\right). 
	\end{equation*}
	
	Recalling \cref{lb:poisson-stein-eq}, for any coupling $(W,Z)$ such that $W \sim \distBinom \left( K, \frac{\gamma/K}{\gamma/K+1} \right)$ and $Z \sim \distPoisson(\gamma)$:
	\begin{equation*}
	\mathbb{E}[\gamma(x(W+1)-x(Z+1)) + Zx(Z) - Wx(W)] \geq \frac{\gamma^2_K}{K}\left( 1 - \frac{3\gamma_K^2 + 12 \gamma_K+7}{\theta \gamma_K}\right). 
	\end{equation*}
	Suppose $(W,Z)$ is the maximal coupling attaining the total variation distance between $P_W$ and $P_Z$ i.e. $\Pr(W \neq Z) = \dTV(P_Y,P_Z)$. Clearly,
	\(
	&\gamma(x(W+1)-x(Z+1)) + Zx(Z) - Wx(W) 
	\\ &\leq \indict{W \neq Z} \sup_{m_1,m_2} | (\gamma x(m_1+1) - m_1 x(m_1)) - (\gamma x(m_2+1) - m_2 x(m_2)) | \\
	&\leq 2\indict{W \neq Z} \sup_{m} | (\gamma x(m+1) - m x(m) |. 
	\)
	Taking expectations on both sides, we conclude that
	\begin{equation} \label{eq:almost-tv-bound}
	2\dTV(P_W, P_Z) \times \sup_{m} |\gamma x(m+1) - m x(m)| \geq \frac{\gamma_K^2}{K} \left( 1 - \frac{3\gamma_K^2 + 12 \gamma_K+7}{\theta \gamma_K}\right).
	\end{equation}
	
	It remains to upper bound $\sup_{m} |\gamma x(m+1) - m x(m)|$. Recall that the derivative of $x$ is $\exp \left( -\frac{m^2}{\gamma_K \theta}\right) \left( 1 - \frac{2m^2}{\gamma_K \theta} \right)$, taking values in $[-2e^{-3/2},1]$. This means for any $m$, $-2e^{-3/2} \leq x(m+1) - x(m) \leq 1$. Hence:
	\begin{align} \label{eq:upper-bound-integrand}
	|\gamma x(m+1) - m x(m)| &= | \gamma(x(m+1)-x(m)) + (\gamma - m) x(m)| \nonumber \\
	&\leq \gamma + (m+\gamma) m \exp\left(-\frac{m^2}{\gamma_K \theta}\right) \nonumber \\
	&\leq \gamma + (\gamma+1) m^2 \exp\left(-\frac{m^2}{\gamma_K \theta}\right) \nonumber \\
	&\leq \gamma + \theta \gamma_K (\gamma+1)\exp(-1).
	\end{align}
	where the last inequality owes to the easy-to-verify $x \exp(-x) \leq \exp(-1)$. Combining \cref{eq:upper-bound-integrand} and \cref{eq:almost-tv-bound} we have that
	\begin{equation*}
	\dTV \left(\text{Binomial} \left( K, \frac{\gamma/K}{\gamma/K+1} \right), \text{Poisson}(\gamma)\right) \geq \frac{1}{2} \frac{ 1 - \frac{3\gamma_K^2 + 12 \gamma_K+7}{\theta \gamma_K}}{\gamma + (\gamma+1)\theta \gamma_K \exp(-1)} K\left(\frac{\gamma_K}{K}\right)^2 . 
	\end{equation*}
	
	Finally, we calibrate $\theta$. By selecting $\theta = \max \left(12 \gamma_K, \frac{28}{\gamma_K}, 48 \right)$ we have that the numerator of the unwieldy fraction is at least $\frac{1}{4}$ and its denominator is at most $\gamma + \exp(-1)(\gamma+1) \max(12 \gamma^2, 48 \gamma, 28)$, because $\gamma_K < \gamma$. This completes the proof. 
\end{proofof}

\begin{proofof}{\cref{thm:betaBer-1/K-lowerbound}} \label{crm-proof:betaBer-1/K-lowerbound}
	
	The constant $C$ in the theorem statement is 
	\begin{equation*}
		C \coloneqq  \gamma^2 /\left(\gamma + \exp(-1)(\gamma+1) \max(12 \gamma^2, 48 \gamma, 28)\right),
	\end{equation*}
	which is equal to $\gamma^2 C(\gamma)$, with $C(\gamma)$ from \cref{prop:binom-poiss-lowerbound}.
	
	First we mention which probability kernel $f$ results in the large total variation distance. For any discrete measure $\sum_{i=1}^{M} \delta_{\psi_i}$, $f$ is the Dirac measure sitting on $M$, the number of atoms.
	\begin{equation}\label{lb:likelihood}
	f(.\given \sum_{i=1}^{M} \delta_{\psi_i}) \defined \delta_{M}(.).
	\end{equation}
	
	Now we show that under such $f$, the total variation distance is lower bounded.	From \cref{lem:tv-redux-rule}, we know that
	\begin{equation*}
	\dTV(\BPproc{N}{\infty}, \BPproc{N}{K}) = \dTV(P_{Y_{1:N}}, P_{W_{1:N}}) \geq \dTV(P_{Y_1}, P_{W_1}). 
	\end{equation*}
	Hence it suffices to show:
	\begin{equation*}
	\dTV(P_{Y_1}, P_{W_1}) \geq C(\gamma) \frac{\gamma^2}{K} \frac{1}{(1+\gamma/K)^2}. 
	\end{equation*}
	
	Recall the generative process defining $P_{Y_1}$ and $P_{W_1}$. $Y_1$ is an observation from the target beta--Bernoulli model, and the functions $h,\widetilde{h},$ and $M_{n,x}$ are given in \cref{exa:beta-Ber}. By \cref{prop:target-marginal},
	\begin{equation*}
	N_T \sim \distPoisson(\gamma), \hspace{10pt} \psi_k \distiid H, \hspace{10pt} X_1 = \sum_{i=1}^{N_T} \delta_{\psi_k}, \hspace{10pt} Y_1 \sim f(.\given X_1).
	\end{equation*}
	$W_1$ is an observation from the approximate model, so by \cref{prop:approx-marginal},
	\begin{equation*}
	N_A \sim \distBinom \left(K, \frac{\gamma/K}{1+\gamma/K} \right), \hspace{10pt} \phi_k \distiid H, \hspace{10pt} Z_1 = \sum_{i=1}^{N_A} \delta_{\phi_k}, \hspace{10pt} W_1 \sim f(.\given Z_1). 
	\end{equation*}
	Because of the choice of $f$, $Y_1 = N_T$ and $W_1 = N_A$. Hence, by \cref{prop:binom-poiss-lowerbound}, 
	\(
	\dTV \left( P_{Y_1}, P_{W_1} \right)	&= \dTV \left( P_{N_T}, P_{N_A} \right) \\
	&\geq C(\gamma) \frac{\gamma^2}{K} \frac{1}{(1+\gamma/K)^2}. 
	\)
\end{proofof}

\section{DPMM results} \label{app:dp-results}
We consider Dirichlet process mixture models \citep{Antoniak:1974}
\begin{equation}
	\begin{split}
		\Theta &\sim \distDP(\alpha, H), \\
		\targetLatent_{n} \given \Theta &\distiid \Theta, \quad n = 1, 2, \ldots, N, \\
		\targetObs_n \given \targetLatent_n &\distind f(\cdot \given \targetLatent_n), \quad n = 1, 2, \ldots, N.
	\end{split}  \label{dpmm:target}
\end{equation}
with corresponding approximation 
\begin{equation}
	\begin{split}
		\Theta_K &\sim \distFSD_K(\alpha, H), \\
		\approxLatent_n \given \Theta_K &\distiid \Theta_K, \quad n = 1, 2, \ldots, N, \\
		\approxObs_n \given \approxLatent_n &\distind f(\cdot \given \approxLatent_n), \quad n = 1, 2, \ldots, N,.
	\end{split} 
\label{dpmm:approx}
\end{equation}
Let $P_{N,\infty}$ be the distribution of the observations $Y_{1:N}$. Let $P_{N,K}$ be the distribution of the observations $W_{1:N}$. 
\subsection{Upper bound}
 \label{sub-sec:dp-upper}
Upper bounds on the error made by $\distFSD_{\approxlev}$ can be used to determine the sufficient $\approxlev$ to approximate the target process for a given $N$ and accuracy level. We upper bound $\dTV \left(P_{N,\infty}, P_{N,K}\right)$ in \cref{thm:DPMM-upperbound}. 

\bnthm[Upper bound for DPMM] \label{thm:DPMM-upperbound}
	For some constants $C',C'',C''',C''''$ that only depend on $\alpha$, 
	\begin{equation*}
		\dTV \left(P_{N,\infty}, P_{N,K}\right) \leq \frac{C' + C''\ln^2 N + C'''\ln N \ln K + C''''\ln K}{K}.
	\end{equation*}
\enthm

The proof and explicit values of the constants are given in \cref{app-proof:DPMM-upperbound}. \cref{thm:DPMM-upperbound} is  similar to \cref{thm:CRM-upperbound}, although the exact values of the constants $C', C'', C''', C''''$ are different. The $O(\ln^2 N)$ growth of the bound for fixed $N$ can likely be reduced to $O(\ln N)$, the inherent growth rate of DP mixture models \citep[Section 5.2]{arratia2003logarithmic}. The $O ( {\ln K}/{K} )$ rate of decrease to zero is tight because of a ${1}/{K}$ lower bound on the approximation error. 
\cref{thm:DPMM-upperbound} is an improvement over the existing theory for $\distFSD_K$, in the sense that \citet[Theorem 4]{Ishwaran:2002} provide an upper bound on $\dTV \left(P_{N,\infty}, P_{N,K}\right)$ that lacks an explicit dependence on $K$ or $N$ --- that bound cannot be inverted to determine the sufficient $K$ to approximate the target to a given accuracy, while it is simple to determine using \cref{thm:DPMM-upperbound}.

\subsection{Lower bounds}
\label{sub-sec:dp-lower}

As \cref{thm:DPMM-upperbound} is only an upper bound, we now investigate the tightness of the inequality in terms of $N$ and $K$. We first look at the dependence of the error bound in terms of $\ln N$. \cref{thm:DPMM-lnN-necessary} shows that finite approximations cannot be accurate if the approximation level is too small compared to the growth rate $\ln N$. 

\bnthm [$\ln N$ is necessary] \label{thm:DPMM-lnN-necessary}
There exists a probability kernel $f(\cdot)$, independent of $K,N$, such that for any $N \geq 2$, if $K \leq \frac{1}{2}C(N,\alpha)$, then
\begin{equation*}
\dTV \left(P_{N,\infty}, P_{N,K}\right)\geq 1-\frac{C'}{N^{\alpha/8}}
\end{equation*}
where $C'$ is a constant only dependent on $\alpha$.
\enthm 

See \cref{app-proof:DPMM-lnN-necessary} for the proof.  \cref{thm:DPMM-lnN-necessary} implies that as $N$ grows, if the approximation level $K$ fails to surpass the ${C(N,\alpha)}/{2}$ threshold, then the total variation between the approximate and the target model remains bounded from zero --- in fact, the error tends to one. Recall that $C(N,\alpha) = \Omega(\ln N)$, so the necessary approximation level is $\Omega(\ln N)$. \cref{thm:DPMM-lnN-necessary} is the analog of \cref{thm:betaBer-lnN-necessary}.

We also investigate the tightness of \cref{thm:DPMM-upperbound} in terms of $K$. In \cref{thm:DPMM-1/K-lowerbound}, our lower bound indicates that the ${1}/{K}$ factor in \cref{thm:DPMM-upperbound} is tight (up to log factors). 

\bnthm[${1}/{K}$ lower bound] \label{thm:DPMM-1/K-lowerbound}
	There exists a probability kernel $f(\cdot)$, independent of $K,N$, such that for any $N \geq 2$, 
	\begin{equation*}
	\dTV \left(P_{N,\infty}, P_{N,K}\right) \geq \frac{\alpha}{1+\alpha}\frac{1}{K}.
	\end{equation*}
\enthm

See \cref{app-proof:DPMM-1/K-lowerbound} for the proof. While \cref{thm:DPMM-upperbound} implies that the normalized AIFA with $K = O \left(\text{poly}(\ln N)/\epsilon\right)$ atoms suffices in approximating the DP mixture model to less than $\epsilon$ error, \cref{thm:DPMM-1/K-lowerbound} implies that a normalized AIFA with $K = \Omega \left( 1/\epsilon\right)$ atoms is \emph{necessary} in the worst case. This worst-case behavior is analogous to \cref{thm:betaBer-1/K-lowerbound} for DP-based models.

The ${1}/{\epsilon}$ dependence means that AIFAs are worse than TFAs in theory. It is known that small TFA models are already excellent approximations of the $\distDP$. \cref{exa:TSB} is a very well-known finite approximation whose error is upper bounded in \cref{prop:TFA-DP-upperbound}.

\bnprop {\citep[Theorem 2]{Ishwaran:2001}} \label{prop:TFA-DP-upperbound}
Let $\Xi_K \sim \distTSB_K(\alpha, H)$, $R_n \given \Xi_K \distiid \Xi_K, T_n \given R_n \distind f(\cdot \given R_n)$ with $N$ observations. Let $Q_{N,K}$ be the distribution of the observations $T_{1:N}$. Then: $\dTV \left(P_{N,\infty}, Q_{N,K}\right) \leq 2N \exp \left( -\frac{K-1}{\alpha} \right).$
\enprop 

\cref{prop:TFA-DP-upperbound} implies that a TFA with $K = O \left( \ln \left( N/\epsilon \right) \right)$ atoms suffices in approximating the DP mixture model to less than $\epsilon$ error. Modulo log factors, comparing the necessary ${1}/{\epsilon}$ level for AIFA and the sufficient $\ln \left({1}/{\epsilon}\right)$ level for TFA, we conclude that the necessary size for normalized IFA is exponentially larger than the sufficient size for TFA, in the worst case.

\section{Proofs of DP bounds} \label{app:dp-proofs}
Our technique to analyze the error made by $\distFSD_{\approxlev}$ follows a similar vein to the technique in \cref{app:crm-proofs}. We compare the joint distribution of the latents $\targetLatent_{1:N}$ and $\approxLatent_{1:N}$ (with the underlying $\Theta$ or $\Theta_{\approxlev}$ marginalized out) using the conditional distributions $\targetLatent_{n} \given \targetLatent_{1:(n-1)}$ and $\approxLatent_{n} \given \approxLatent_{1:(n-1)}$. Before going into the proofs, we give the form of the conditionals.

The conditional $\targetLatent_{1:N} \given \targetLatent_{1:(n-1)}$ is the well-known Blackwell-MacQueen prediction rule.

\bnprop \label{app:DP-urn} \citet{blackwell1973ferguson}
For $n = 1$, $X_1 \sim H$. For $n \geq 2$, 
\begin{equation*}
	X_n \given X_{n-1},X_{n-2},\ldots, X_1 \sim \frac{\alpha}{n-1+\alpha} H + \sum_{j} \frac{n_j}{n-1+\alpha} \delta_{\psi_j},
\end{equation*}
where $\{\psi_j\}$ is the set of unique values among $X_{n-1},X_{n-2},\ldots, X_1$ and $n_j$ is the cardinality of the set $\{i: 1 \leq i \leq n-1, X_i = \psi_j\}$.
\enprop 

The conditionals $\approxLatent_{n} \given \approxLatent_{1:(n-1)}$ are related to the Blackwell-MacQueen prediction rule. 
\bnprop \citet{pitman1996some} \label{app:FSD-urn}
For $n = 1$, $Z_1 \sim H$. For $n \geq 2$, let $\{\psi_j\}_{j=1}^{J_n}$ be the set of unique values among $Z_{n-1},Z_{n-2},\ldots, Z_1$ and $n_j$ is the cardinality of the set $\{i: 1 \leq i \leq n-1, Z_i = \psi_j\}$. If $J_n < K$:
\begin{equation*}
Z_n \given Z_{n-1},Z_{n-2},\ldots, Z_1 \sim \frac{(K-J_n) \alpha/K}{n-1+\alpha} H + \sum_{j=1}^{J_n} \frac{n_j + \alpha/K}{n-1+\alpha} \delta_{\psi_j},
\end{equation*}
Otherwise, if $J_n = K$, there is zero probability of drawing a fresh component from $H$ i.e. $Z_n$ comes only from $\{\psi_j\}_{j=1}{J_n}$:
\begin{equation*}
Z_n \given Z_{n-1},Z_{n-2},\ldots, Z_1 \sim \sum_{j=1}^{J_n} \frac{n_j + \alpha/K}{n-1+\alpha} \delta_{\psi_j}.
\end{equation*}
$J_n \leq K$ is an invariant of these of prediction rules: once $J_{n} = K$, all subsequent $J_{m}$ for $m \geq n$ is also equal to $K$.
\enprop 

\subsection{Upper bounds}

\begin{proofof}{\cref{thm:DPMM-upperbound}} \label{app-proof:DPMM-upperbound}
	
	The constants $C', C'', C''', C''''$ are as follows
	\begin{equation} \label{proof:dp-cdef}
		\begin{aligned}
			C' &= \exp(\alpha(\psi(\alpha) + 1)) + 2 \alpha^2 \ln^2 (1 + 1/\alpha) ,\\
			C'' &= \alpha^2 + \frac{3 \alpha^2 \ln(1+1/\alpha) }{\ln2} ,\\
			C''' &=  \alpha^2,\\
			C'''' &=  \alpha^2 \ln(1+1/\alpha).\\
		\end{aligned}
	\end{equation}
	The reasoning for these constants will be clear by the end of the proof.
	
 	To begin, observe that the conditional distributions of the observations given the latent variables are the same across target and approximate models: $P_{Y_{1:N}|X_{1:N}}$ is the same as $P_{W_{1:N}|Z_{1:N}}$ if $X_{1:N} = Z_{1:N}$. Therefore, using \cref{lem:tv-prop-rule}, we want to show that there exists a coupling of $P_{\targetLatent_{1:N}}$ and $P_{\approxLatent_{1:N}}$ that has small difference probability. 
 	
 	First, we construct a coupling of $P_{\targetLatent_{1:N}}$ and $P_{\approxLatent_{1:N}}$ such that, for any $n \geq 1$, for any $x_{1:(n-1)}$ such that $J_n$ is the number of unique atom locations among $x_{1:(n-1)}$ is at most $K$,
	\begin{equation} \label{app-dpmm:nice-coupling}
	\Pr(\targetLatent_{n} \neq \approxLatent_{n} \given \targetLatent_{1:(n-1)} = \approxLatent_{1:(n-1)} = x_{1:(n-1)}) \leq \frac{\alpha}{K} \frac{J_n}{n-1+\alpha}.	
	\end{equation}
	The case where $n = 1$ reads that $\Pr(\targetLatent_{1} \neq \approxLatent_{1}) = 0$.
	Such a coupling exists because the total variation distance between the prediction rules $\targetLatent_{n} \given \targetLatent_{1:(n-1)}$ and $\approxLatent_{n} \given \approxLatent_{1:(n-1)}$ is small.
	Let $\{\psi_j\}_{j=1}^{J_n}$ be the unique atom locations in $x_{1:(n-1)}$ and $n_j$ be the number of latents $x_i$ that manifest atom location $\psi_j.$ The distribution $\targetLatent_{n} \given \targetLatent_{1:(n-1)}$ can be sampled from in two steps:
	\begin{itemize}
		\item Sample $I_1$ from the categorical distribution over $J_n+1$ elements where, for $1 \leq j \leq J_n$, $\Pr(I_1 = j) = {n_j} / {(n-1+\alpha)}$ and $\Pr(I_1 = J_n + 1) = {\alpha} / {(n-1+\alpha)}.$
		\item If $I_1 = j$ for $1 \leq j \leq J_n$, set $\targetLatent_n = \delta_{\psi_j}$. If $I_1 = J_{n}+1$, draw a fresh atom from $H$, label $\psi_{J_{n}+1}$ and set $\targetLatent_n = \delta_{\psi_{J_{n}+1}}.$
	\end{itemize} 
	Similarly, we can generate $\approxLatent_{n} \given \approxLatent_{1:(n-1)}$ in two steps:
	\begin{itemize}
		\item Sample $I_2$ from the categorical distribution over $J_n+1$ elements where, for $1 \leq j \leq J_n$, $\Pr(I_2 = j) = \frac{n_j+\alpha/K}{n-1+\alpha}$ and $\Pr(I_2 = J_n + 1) = \frac{\alpha(1-J_n/K)}{n-1+\alpha}.$
		\item If $I_2 = j$ for $1 \leq j \leq J_n$, set $\approxLatent_n = \delta_{\psi_j}$. If $I_2 = J_{n}+1$, draw a fresh atom from $H$, label $\psi_{J_{n}+1}$ and set $\approxLatent_n = \delta_{\psi_{J_{n}+1}}.$
	\end{itemize} 
	Still conditioning on $\targetLatent_{1:(n-1)}$ and $\approxLatent_{1:(n-1)}$, we observe that the distribution of $\targetLatent_n \given I_1$ is the same as $\approxLatent_{n} \given I_2.$ Hence, using the propagation argument from \cref{lem:tv-prop-rule}, it suffices to couple $I_1$ and $I_2$ so that
	\begin{equation*}
	\Pr(I_1 \neq I_2 \given \targetLatent_{1:(n-1)} = \approxLatent_{1:(n-1)} = x_{1:(n-1)})
	\end{equation*}
	is small. 
	Since $I_1$ and $I_2$ are categorical distributions, the minimum of the difference probability is the total variation distance between the two distributions, which equals ${1}/{2}$ the $L_1$ distance between marginals
	\begin{equation*}
		\sum_{j=1}^{J_n} \left| 
		\frac{n_j+\alpha/K}{n-1+\alpha} - \frac{n_j}{n-1+\alpha}\right| + \left| \frac{\alpha}{n-1+\alpha} - \frac{\alpha(1-J_n/K)}{n-1+\alpha} \right| = 2 \frac{\alpha}{K} \frac{J_n}{n-1+\alpha}.
	\end{equation*}
	Dividing the last equation by $2$ gives \cref{app-dpmm:nice-coupling}. The joint coupling of  $P_{\targetLatent_{1:N}}$ and $P_{\approxLatent_{1:N}}$ is the natural gluing of the couplings $P_{\targetLatent_{n} \given \targetLatent_{1:(n-1)}}$ and $P_{\approxLatent_{n} \given \approxLatent_{1:(n-1)}}.$

	We now show that for the coupling satisfying \cref{app-dpmm:nice-coupling}, the overall probability of difference $\Pr(\targetLatent_{1:N} \neq \approxLatent_{1:N})$ is small. Recall the growth function from \cref{eq:growth-function}.
	We will use the notation of a \emph{typical} set in the rest of the proof:
	\begin{equation*}
	\mathcal{D}_{n} \defined \left\{x_{1:(n-1)}: J_n \leq (1+\delta) \max(C(N,\alpha),C(K,\alpha) )  \right\}.	
	\end{equation*}
	In other words, the number of unique values among the $x_{1:(n-1)}$ is small.
	The constant $\delta$ satisfies $\frac{\delta^2}{2+\delta} \alpha = 2$: such $\delta$ always exists and is unique.
	 The following decomposition is used to investigate the difference probability on the typical set:
	\begin{align}\label{app-dpmm:proto-expansion}
\Pr(\targetLatent_{1:N} \neq \approxLatent_{1:N}) &= \Pr((\targetLatent_{1:(N-1)},\targetLatent_{N}) \neq (\approxLatent_{1:(N-1)},\approxLatent_{N})) \nonumber \\
	&= \Pr(\targetLatent_{1:(N-1)} \neq \approxLatent_{1:(N-1)}) + \Pr(\targetLatent_{N} \neq \approxLatent_{N}, \targetLatent_{1:(N-1)} = \approxLatent_{1:(N-1)}).
	\end{align}
	The second term can be further expanded:
	\(
	 \Pr(\targetLatent_{N} \neq \approxLatent_{N}, &\targetLatent_{1:(N-1)} = \approxLatent_{1:(N-1)}, \targetLatent_{1:(N-1)} \in 	\mathcal{D}_{N}) \\
	  &+ \Pr(\targetLatent_{N} \neq \approxLatent_{N}, \targetLatent_{1:(N-1)} = \approxLatent_{1:(N-1)}, \targetLatent_{1:(N-1)} \notin 	\mathcal{D}_{N}).
	\)
	The former term is at most
	\begin{equation*}
	 \Pr(\targetLatent_{N} \neq \approxLatent_{N} \given \targetLatent_{1:(N-1)} = \approxLatent_{1:(N-1)}, \targetLatent_{1:(N-1)} \in \mathcal{D}_{N}),
	\end{equation*}
	while the latter term is at most
	\begin{equation*}
	\Pr(\targetLatent_{1:(N-1)} \notin 	\mathcal{D}_{N}). 
	\end{equation*}
	To recap, we can bound $\Pr(\targetLatent_{1:N} \neq \approxLatent_{1:N})$ by bounding three quantities:
	\benum[leftmargin=*]
	\item The difference probability of a shorter process $\Pr(\targetLatent_{1:(N-1)} \neq \approxLatent_{1:(N-1)})$. 
	\item The difference probability of the prediction rule on typical sets $\Pr(\targetLatent_{N} \neq \approxLatent_{N} \given \targetLatent_{1:(N-1)} = \approxLatent_{1:(N-1)}, \targetLatent_{1:(N-1)} \in \mathcal{D}_{N})$.
	\item The probability of the atypical set $\Pr(\targetLatent_{1:(N-1)} \notin \mathcal{D}_{N})$. 
	\eenum 
	By recursively applying the expansion initiated in \cref{app-dpmm:proto-expansion} to $\Pr(\targetLatent_{1:(N-1)} \neq \approxLatent_{1:(N-1)})$, we actually only need to bound difference probability of the different prediction rules on typical sets and the atypical set probabilities.
	
	 Regarding difference probability of the different prediction rules, being in the typical set allows us to control $J_n$ in \cref{app-dpmm:nice-coupling}. Summation across $n = 1$ through $N$ gives the overall bound of
	\begin{equation}\label{app-dpmm:prediction-differences}
	\frac{\alpha}{K} (1+\delta) \max(C(N,\alpha),C(K,\alpha)) C(N,\alpha).
	\end{equation}
	Regarding the atypical set probabilities, because $J_{n-1}$ is stochastically dominated by $J_n$ i.e., the number of unique values at time $n$ is at least the number at time $n-1$, all the atypical set probabilities are upper bounded by the last one i.e. $\Pr(X_{1:(N-1)} \notin \mathcal{D}_N)$. When $N > 1$, $J_{N}$ is the sum of independent Poisson trials, with an overall mean equaling exactly $C(N-1,\alpha)$ and $J_1$ is defined to be $0$. Therefore, the atypical event has small probability because of \cref{lem:upper-Chernoff}:
	\begin{equation*}
		\begin{aligned}
			\Pr(J_{N} > (1+\delta) \max(C(N-1,\alpha),C(K,\alpha)) &\leq \Pr(J_{N} > (1+\delta) \max(C(N,\alpha),C(K,\alpha))  \\ 
			&\leq \exp \left( -\frac{\delta^2}{2+\delta}  \max(C(N,\alpha),C(K,\alpha) \right).	
		\end{aligned}
	\end{equation*}
	
	Even accounting for all $N$ atypical events through union bound, the total probability is still small small:
	\begin{equation*}
	\exp \left(- \left( \frac{\delta^2}{2+\delta}  \max(C(N,\alpha),C(K,\alpha) - \ln N \right) \right).
	\end{equation*}
	By \cref{lem:harmonic-like-sum}, $\max(C(N,\alpha),C(K,\alpha) \geq \alpha \max (\ln N, \ln K -  \alpha (\psi(\alpha) + 1)$. we have
	\begin{equation*}
	\frac{\delta^2}{2+\delta}  \max(C(N,\alpha),C(K,\alpha) - \ln N \geq \ln K - \alpha (\psi(\alpha) + 1) ,
	\end{equation*}
	meaning the overall atypical probabilities is at most
 	\begin{equation}\label{app-dpmm:atypical}
	\frac{\exp(\alpha(\psi(\alpha) + 1))}{K}.
	\end{equation}
	The overall total variation bound combines \cref{app-dpmm:prediction-differences,app-dpmm:atypical}.
	We first upper bound $C(N,\alpha)$ using \cref{lem:harmonic-like-sum} and upper bound $\max(C(N,\alpha),C(K,\alpha))$ by the sum of the two constituent terms. 
	We also upper bound $\ln N \leq \ln^2 N / \ln 2$ to remove the dependence on the sole $\ln N$ factor.
	After the algebraic manipulations, we arrive at the constants in \cref{proof:dp-cdef}s.
	
\end{proofof}

\begin{proofof}{\cref{thm:mHDP-upperbound}} \label{app-proof:mHDP-upperbound}
	
	The constants $C',C'',C''',C''''$ are as follows:
	\begin{equation*}
		\begin{aligned}
			C' &= \exp(\omega(\psi(\omega) + 1)) + 2 \omega^2 \ln^2 (1 + 1/\omega) ,\\
			C'' &= \omega^2 + \frac{3 \omega^2 \ln(1+1/\omega) }{\ln2} ,\\
			C''' &=  \omega^2,\\
			C'''' &=  \omega^2 \ln(1+1/\omega).\\
		\end{aligned}
	\end{equation*}
	
	The main idea is reducing to the Dirichlet process mixture model. We do this in two steps.

	First, the conditional distribution of the observations $W \given H_{1:D}$ of the target model is the same as the conditional distribution $Z \given F_{1:D}$ of the approximate model if $H_{1:D} = F_{1:D}$. 
	Second, there exists latent variables $\Lambda$ and $\Phi$ such that the conditional distribution of $H_{1:D} \given \Lambda $ and the conditional $F_{1:D} \given \Phi$ are the same when $\Lambda = \Phi$. Recall the construction of the $F_d$ in terms of atom locations $\phi_{d,j}$ and stick-breaking weights $\gamma_{d,j}$:
	\(
	G_K &\sim \distFSD_{\approxlev}(\omega,H), \\
	\phi_{dj} \given G_K &\distiid G_K(.) &\text{ across } d,j, \\
	\gamma_{dj} &\distiid \text{Beta}(1,\alpha) &\text{ across } d,j \text{ (except } \gamma_{dT} = 1 ),\\
	F_d \given \phi_{d,.}, \gamma_{d,.} &= \sum_{i=1}^{T} \left(\gamma_{di} \prod_{j < i} \left( 1 - \gamma_{dj} \right)\right) \delta_{\phi_{dj}}.
	\)
	Similarly $H_d$ is also constructed in terms of atom locations $\lambda_{d,j}$ and stick-breaking weights $\eta_{d,j}$:
	\(
	G &\sim \distDP(\omega,H), \\
	\lambda_{dj} \given G &\distiid G(.) &\text{ across } d,j, \\
	\eta_{dj} &\distiid \text{Beta}(1,\alpha) &\text{ across } d,j \text{ (except } \eta_{dT} = 1 ),\\
	H_d \given \lambda_{d,.}, \eta_{d,.} &= \sum_{i=1}^{T} \left(\eta_{di} \prod_{j < i} \left( 1 - \eta_{dj} \right)\right) \delta_{\lambda_{dj}}.
	\)
	Therefore, if we set $\Lambda = \{\lambda_{dj}\}_{d,j}$ and $\Phi = \{\phi_{dj}\}_{d,j}$, then $H_{1:D}\given\Lambda$ is the same as the conditional $F_{1:D}\given\Phi$ if $\Lambda = \Phi$. 
	
	Overall, this means that $W \given \Lambda$ is the same as $Z \given \Phi$. Again by \cref{lem:tv-prop-rule}, we only need to demonstrate a coupling between $P_{\Lambda}$ and $P_{\Phi}$ such that the difference probability is small. 
	
	From the proof of \cref{thm:CRM-upperbound} in \cref{app-proof:DPMM-upperbound}, we already know how to couple $P_{\Lambda}$ and $P_{\Phi}$. On the one hand, since $\lambda_{dj}$ are conditionally iid given $G$ across $d,j$, the joint distribution of $\lambda_{dj}$ is from a DPMM (probability kernel $f$ being Dirac $f(\cdot \given x) = \delta_{x}(\cdot)$) where the underlying $\distDP$ has concentration $\omega$. On the other hand, since $\phi_{dj}$ are conditionally iid given $G_K$ across $d,j$, the joint distribution  $\phi_{dj}$ comes from the finite mixture with $\distFSD_{\approxlev}$. Each observational process has cardinality $DT$. Therefore, we can couple $P_{\Lambda}$ and $P_{\Phi}$ such that
	\begin{equation*}
	\Pr(\Lambda \neq \Phi) \leq \frac{C' + C'' \ln^2(DT) + C''' \ln (DT) \ln K + C'''' \ln K}{K},
	\end{equation*}
	where the constants have been given at the beginning of this proof. 
\end{proofof}

\subsection{Lower bounds}
\begin{proofof}{\cref{thm:DPMM-lnN-necessary}} \label{app-proof:DPMM-lnN-necessary}
	First we mention which probability kernel $f$ results in the large total variation distance: the pathological $f$ is the Dirac measure i.e., $f(\cdot \given x) = \delta_x(.)$. With this conditional likelihood $\targetLatent_{n} = \targetObs_{n}$ and $\approxLatent_n = \approxObs_n$, meaning:
	\begin{equation*}
	\dTV(P_{N,\infty}, P_{N,K}) = \dTV(P_{\targetLatent_{1:N}},P_{\approxLatent_{1:N}}).	
	\end{equation*}
	
	Now we discuss why the total variation is lower bounded by the function of $N$. 	
	Let $\mathcal{A}$ be the event that there are at least $\frac{1}{2} C(N,\alpha)$ unique components in among the latent states:
	\begin{equation*}
	\mathcal{A} \defined \left\{x_{1:N}: \# \text{unique values} \geq \frac{1}{2} C(N,\alpha) \right\}.	
	\end{equation*}
	
	The probabilities assigned to this event by the approximate and the target models are very different from each other.	On the one hand, since $K < \frac{C(N,\alpha)}{2}$, under $\distFSD_{\approxlev}$, $\mathcal{A}$ has measure zero:
	\begin{equation} \label{dp-proofs:zero-mass-approx}
	\Pr_{\approxLatent_{1:N}}(\mathcal{A}) = 0.
	\end{equation}
	On the other hand, under $\distDP$, the number of unique atoms drawn is the sum of Poisson trials with expectation exactly $C(N,\alpha)$. The complement of $\mathcal{A}$ is a lower tail event. Hence by \cref{lem:lower-Chernoff} with $\delta = 1/2, \mu = C(N,\alpha)$, we have:
	\begin{equation}\label{dp-proofs:large-mass-target}
	\Pr_{\targetLatent_{1:N}}(\mathcal{A}) \geq 1 - \exp \left(-\frac{C(N,\alpha)}{8} \right)
	\end{equation} 
	Because of \cref{lem:harmonic-like-sum}, we can lower bound $C(N,\alpha)$ by a multiple of $\ln N$:
	\begin{equation*}
	\exp \left(-\frac{C(N,\alpha)}{8} \right) \leq \exp \left( -\frac{\alpha \ln N}{8} + \frac{\alpha(\psi(\alpha)+1)}{8} \right) = \frac{\text{constant}}{N^{\alpha/8}}.
	\end{equation*}
	
	We now combine \cref{dp-proofs:zero-mass-approx,dp-proofs:large-mass-target} and recall that total variation is the maximum over probability discrepancies.
\end{proofof}

\begin{proofof}{\cref{thm:DPMM-1/K-lowerbound}} \label{app-proof:DPMM-1/K-lowerbound}
	First we mention which probability kernel $f$ results in the large total variation distance: the pathological $f$ is the Dirac measure i.e., $f(\cdot \given x) = \delta_x(.)$.
	
	Now we show that under such f, the total variation distance is lower bounded. Observe that it suffices to understand the total variation between $P_{Y_1,Y_2}$ and $P_{W_1,W_2}$, because \cref{lem:tv-redux-rule} already implies
	\begin{equation*}
	\dTV(P_{N,\infty}, P_{N,K}) \geq \dTV(P_{Y_1,Y_2}, P_{W_1,W_2}).
	\end{equation*}
	Since $f$ is Dirac, $\targetLatent_{n} = \targetObs_{n}$ and $\approxLatent_n = \approxObs_n$ and we have:
	\begin{equation*}
	\dTV(P_{Y_1,Y_2}, P_{W_1,W_2}) = \dTV(P_{\targetLatent_1,\targetLatent_2}, P_{\approxLatent_1,\approxLatent_2}).
	\end{equation*}

	Consider the event that the two latent states are equal. Under the target model,
	\begin{equation*}
		\Pr(\targetLatent_2 = \targetLatent_1) = \frac{1}{1+\alpha}, 
	\end{equation*}
	while under the approximate one,
	\begin{equation*}
		\Pr(\approxLatent_2 = \approxLatent_1) = \frac{1+\alpha/K}{1+\alpha}.
	\end{equation*}
	They are simple consequences of the prediction rules in \cref{app:DP-urn,app:FSD-urn}. 
	Therefore, there exists a measurable event where the probability mass assigned by the target and approximate models differ by
	\begin{equation}
		\frac{1+\alpha/K}{1+\alpha} - \frac{1}{1+\alpha} = \frac{\alpha}{1+\alpha} \frac{1}{K},
	\end{equation}
	meaning $\dTV(P_{\targetLatent_1,\targetLatent_2}, P_{\approxLatent_1,\approxLatent_2}) \geq \frac{\alpha}{1+\alpha} \frac{1}{K}.$

\end{proofof}

\section{More ease-of-use results} \label{app:conceptual-proofs}
\subsection{Conceptual results (continued.)}

We begin by stating the log density of the optimal $q_{\approxrate}^*$ under general priors.
\bnprop [{Optimal distribution over atom rates}] \label{prop:optim-q1}
Define the normalization constant $C := \int_\rho \exp \left( \ln \Pr(\rho) + \sum_{n,k} \EE_{x_{n,k} \sim q_{x}} \ln \Pr(x_{n,k} \mid \rho_k) \right) d\rho$
where $x_{n,k} \sim q_{x}$ denote the marginal distribution of $x_{n,k}$ under $q_{x}(x)$ (which is a distribution over the whole set $(x_{n,k})_{n,k}$).
Then
\begin{equation} \label{eq:logq1-eq}
	q_{\approxrate}^*(\rho) = - \ln C + \ln \Pr(\rho) + \sum_{n,k} \EE_{x_{n,k} \sim q_{x}} \ln \traitLL(x_{n,k} \mid \rho_k).
\end{equation}
\enprop
The proof of \cref{prop:optim-q1} is given in \cref{app-proof:optim-q1}.

Knowing the log density \cref{eq:logq1-eq} does not mean that drawing inference is easy.
By drawing inference, we mean computing posterior expectations of important integrands.
Polynomials (such as $\rho_k$) are natural integrands.
In addition, we also need to compute quantities like $\EE_{\rho_k \sim q_{\approxrate}}\{\ln \traitLL(x_{n,k} \given \rho_k)\}$ to derive the optimal distribution for $q_{x}(x)$.
\bnprop \label{prop:necssary-ElogPr} 
Suppose that the variational distribution $q_{x}(x)$ factorizes as $ q_{x}(x) = \prod_{n,k} f_{n,k}(x_{n,k})$.
For a particular $n,k$, let $f_{n,k}^*$ be the optimal distribution over $(n,k)$ trait count with all other variational distributions being fixed i.e.\
\begin{equation*}
	f_{n,k}^* :=\arg \min_{f_{n,k}} \text{KL} \left( q_{\approxrate} q_{\psi}  f_{n,k} \prod_{(n',k') \neq (n,k)} f_{n',k'}  \mid \mid \bar{P}  \right),
\end{equation*}
where $\bar{P}$ denotes the posterior $\Pr(\cdot,\cdot,\cdot \mid y)$.
Then, the p.m.f.\ of $f_{n,k}^*$ at $x_{n,k}$ is equal to
\begin{equation*}
	- \ln C + \EE_{\rho_k \sim q_{\approxrate}} \ln \traitLL(x_{n,k} \mid \rho_k) + \EE_{\psi \sim q_{\psi}, x_{n,-k} \sim f_{n,-k}} \ln \Pr(y_n \mid x_{n,.}, \psi). 
\end{equation*}
for some positive constant $C$.
\enprop 
See \cref{proof:necssary-ElogPr} for the proof of this proposition.

Under TFAs such as \cref{exa:stick-breaking-beta}, since we cannot identify the log density in \cref{eq:logq1-eq} with a well-known distribution, we do not have formulas for expectations.
For \cref{exa:stick-breaking-beta}, strategies to make computing expectations more tractable include introducing auxiliary round indicator variables $r_k$, replacing the product $\prod_{l=1}^{i-1}(1-V_{i,j}^{(l)})$ with a more succinct representation and fixing the functional form $q_{\approxrate}$ rather than using optimality conditions \citep[Section 3.2]{paisley2011variational}. 
However, \citet[Section 3.3]{paisley2011variational} still runs into intractability issues when evaluating $\EE_{\rho_k \sim q_{\approxrate}}\{\ln \traitLL(x_{n,k} \given \rho_k)\}$ in the beta--Bernoulli process, and additional approximations such as Taylor series expansion are needed.

In our second TFA example, the complete conditional of the atom sizes can be sampled without auxiliary variables, but important expectations are not analytically tractable. 

\bexa [Bondesson approximation \citep{doshi-velez2009variational,Teh:2007}] \label{exa:Bondesson-beta-alpha=1} When $\alpha = 1$, the Bondesson approximation in \cref{exa:Bondesson-of-beta} becomes
\[
\Theta_K &= \sum_{i=1}^{K} \approxrate_{i} \delta_{\psi_{i}}, &
\approxrate_{i} &=   \prod_{j=1}^{i} p_j, & 
p_j &\distiid \text{Beta}(\gamma,1), &
\psi_{i} \distiid H.
\]
The atom sizes are dependent because they jointly depend on $p_{1},\dots,p_{K}$, but the complete conditional of atom sizes $\Pr(\approxrate\given x)$ admits a density proportional to
\begin{equation*}
	\indict{0 \leq \approxrate_K \leq \approxrate_{K-1} \leq \ldots \leq \approxrate_1 \leq 1} \prod_{j=1}^{K}\approxrate_{j}^{\gamma \mathbf{1}\{j=K\}+ \sum_{n=1}^{N}x_{n,j}-1} (1-\approxrate_{j})^{N-\sum_{n=1}^{N}x_{n,j}}.
\end{equation*}
The conditional distributions $\Pr(\approxrate_i\given \approxrate_{-i},x)$ are truncated betas, so adaptive rejection sampling \citep{gilks1992adaptive} can be used as a sub-routine to sample each $\Pr(\approxrate_i\given \approxrate_{-i},x)$ and then sweep over all atom sizes. 
However, for this exponential family, expectations of the sufficient statistics are not tractable.
The optimal $q_{\approxrate}^*$ in the sense of \cref{eq:q1-as-optim} has a density proportional to
\begin{equation*}
	\indict{0 \leq \approxrate_K \leq \approxrate_{K-1} \leq \ldots \leq \approxrate_1 \leq 1} \prod_{j=1}^{K}\approxrate_{j}^{\gamma \mathbf{1}\{j=K\}+ \sum_{n=1}^{N}\EE_{q_{x}} x_{n,j}-1} (1-\approxrate_{j})^{N-\sum_{n=1}^{N} \EE_{q_{x}}x_{n,j}}.
\end{equation*}
We do not know closed-form formulas for $\EE\{\ln(\approxrate_i)\}$ or $\EE\{\ln (1-\approxrate_i)\}$.
Rather than using the $q_{\approxrate}^*$ which comes from optimality arguments, \citet{doshi-velez2009variational} fixes the functional form of the variational distribution.
Even then, further approximations such as Taylor series expansion are necessary to approximate $\EE\{\ln(\approxrate_i)\}$ or $\EE\{\ln (1-\approxrate_i)\}$.
\eexa 

Other series-based approximations, like thinning or rejection sampling \citep{campbell2019truncated}, are characterized by even less tractable dependencies between atom sizes in both the prior and the conditional $\Pr(\approxrate\given x)$. 

\subsection{Proofs}

\begin{proofof}{\cref{lem:IFA-conditional-conjugacy}} \label{app-proof:conjugacy}
	Because of the Markov blanket, conditioning on $x,\psi,y$ is the same as conditioning on $x$:
	\begin{equation*}
		\Pr(\approxrate\mid x,\psi,y) =	\Pr(\approxrate\mid x).
	\end{equation*}
	Conditioned on the atom rates, the trait counts are independent across the atoms.
	In the prior over atom rates, the atom rates are independent across the atoms.
	These facts mean that the posterior also factorizes across the atoms
	\begin{equation*}
		\Pr(\approxrate\mid x) = \prod_{k=1}^{K} \Pr(\approxrate_k \mid x_{.,k})
	\end{equation*}
	We look at each factor $\Pr(\approxrate_k \mid x_{.,k})$.
	This is the posterior for $\rho_k$ after observing $N$ observations $(x_{n,k})_{n=1}^{N}$.
	Since the AIFA prior over $\rho_k$ is the conjugate prior of the trait count likelihood, the posterior is in the same exponential family, with updated parameters based on the sufficient statistics and the log partition function.
\end{proofof}

\begin{proofof}{\cref{prop:optim-q1}} \label{app-proof:optim-q1}
	Minimizing the KL divergence is equivalent to maximizing the evidence lower bound (ELBO):
	\begin{equation} \label{eq:ELBO}
		\text{ELBO}(q) :=\EE_{(\rho, \psi, x) \sim q} \ln \Pr( y, \rho, \psi, x) - \EE_{(\rho, \psi, x) \sim q} \ln q( \rho, \psi, x). 
	\end{equation}
	The log joint probability $\Pr( y, \rho, \psi, x)$, regardless of the prior over $\rho$, decomposes as
	\begin{equation} \label{eq:logJoint}
		\begin{aligned}
			\ln \Pr( y, \rho, \psi, x) = \ln \Pr(\rho) &+ \sum_{k} \ln \Pr(\psi_k)  \\
			&+ \sum_{n,k} \ln \Pr(x_{n,k} \mid \rho_k) + \sum_{n} \ln \Pr(y_n \mid x_{n,.}, \psi).
		\end{aligned}
	\end{equation}
	
	Recall that the variational distribution factorizes like as $q(\rho, \psi, x) = q_{\approxrate}(\rho) q_{\psi}(\psi) q_{x}(x)$.
	Therefore, for fixed $q_{\psi}(\psi)$ and $q_{x}(x)$, the ELBO from \cref{eq:ELBO} depends on $q_{\approxrate}(\rho)$ only through 
	\begin{equation*} \label{eq:f}
		f(q_{\approxrate}) := \EE_{\rho \sim q_{\approxrate}} \ln \Pr(\rho) + \sum_{n,k} \EE_{x_{n,k} \sim q_{x}, \rho_k \sim q_{\approxrate}} \ln \Pr(x_{n,k} \mid \rho_k) - \EE_{\rho \sim q_{\approxrate}} \ln q_{\approxrate}(\rho).
	\end{equation*}
	Here, the notation $\rho_k \sim q_{\approxrate}$ means the marginal distribution of $\rho_k$ under $ q_{\approxrate}$.
	Using Fubin's theorem, we rewrite the last integral as
	\begin{equation*}
		f(q_{\approxrate}) = -\EE_{\rho \sim q_{\approxrate}} \ln \frac{ q_{\approxrate}(\rho)}{\Pr(\rho) \times \exp( \sum_{n,k} \EE_{x_{n,k} \sim q_{x}} \ln \Pr(x_{n,k} \mid \rho_k)) }
	\end{equation*}
	The denominator $\Pr(\rho) \times \exp( \sum_{n,k} \EE_{x_{n,k} \sim q_{x}} \ln \Pr(x_{n,k} \mid \rho_k))$ is exactly equal to $C q_0(\rho)$ where  
	$\ln q_0(\rho) = - \ln C + \ln \Pr(\rho) + \sum_{n,k} \EE_{x_{n,k} \sim q_{x}} \ln \traitLL(x_{n,k} \mid \rho_k)$.
	Therefore
	\begin{equation*}
		f(q_{\approxrate}) = -\text{KL}(q_{\approxrate} || q_0) + \ln C.
	\end{equation*}
	This means that the unique maximizer of $f(q_{\approxrate})$ is $q_{\approxrate} = q_0$ i.e.\ the log density of $q_{\approxrate}^*$ is as given in \cref{eq:logq1-eq}.
\end{proofof}

\begin{proofof}{\cref{cor:AIFA-q1}} \label{app-proof:AIFA-q1}
	
	We specialize the formula in \cref{eq:logq1-eq} to the AIFA prior.
	
	Recall the exponential-family form of $\traitLL(x_{n,k} \mid \rho_k)$:
	\begin{equation} \label{eq:exp-fam}
		\traitLL (x_{n,k} \mid \rho_k) = \ln \kappa(x_{n,k}) + \phi(x_{n,k}) \ln \rho_k + \langle \mu(\rho_k), t(x_{n,k}) \rangle - A(\rho_k).
	\end{equation}
	Next, observe that $\EE_{x_{n,k} \sim q_{x}} \ln \traitLL(x_{n,k} \mid \rho_k)$ is equal to
	\begin{equation} \label{eq:factorize}
			\EE_{x_{n,k} \sim q_{x}}  \ln \kappa(x_{n,k})   +  \EE_{x_{n,k} \sim q_{x}} \phi (x_{n,k}) \times \ln \rho_k
			+ \langle \mu(\rho_k), \EE_{x_{n,k} \sim q_{x}} t(x_{n,k}) \rangle - A(\rho_k).
	\end{equation}
	
	Recall that AIFA prior over $\rho_k$ is the conjugate prior for the likelihood in \cref{eq:exp-fam}:
	\begin{equation} \label{eq:AIFA}
		\ln \Pr (\rho_k) = (c/K - 1) \ln \rho_k + \langle \begin{bmatrix}
			\psi \\ \lambda
		\end{bmatrix}, \begin{bmatrix}
			\mu(\rho_k) \\ -A(\rho_k)
		\end{bmatrix}  \rangle - \ln Z (c/K-1, \begin{bmatrix}
			\psi \\ \lambda
		\end{bmatrix}),
	\end{equation}
	and the prior factorizes across atoms:
	\begin{equation*}
		\ln \Pr(\rho) = \sum_{k} \ln \Pr(\rho_k)
	\end{equation*}
	
	Putting \cref{eq:factorize} and \cref{eq:AIFA} together, We have 
	\begin{equation*}
		\ln \Pr(\rho) + \sum_{n,k} \EE_{x_{n,k} \sim q_{x}} \ln \traitLL(x_{n,k} \mid \rho_k) = \sum_{k} T_k(\rho_k)
	\end{equation*}
	where $T_k(\rho_k)$ is equal to
	\begin{equation*}
			(c/K + \sum_n \EE_{x_{n,k} \sim q_{x}} \phi (x_{n,k})  - 1) \ln \rho_k) \\
		+ \left\langle  \begin{bmatrix}
			\psi+ \sum_{n} \EE_{x_{n,k} \sim q_{x}} t(x_{n,k}) \\
			\lambda + N
		\end{bmatrix},  \begin{bmatrix}
			\mu(\rho_k) \\ - A(\rho_k)
		\end{bmatrix}  \right\rangle 
	\end{equation*}
	Accounting for the normalization constant $Z_k$ for each dimension $k$, we arrive at \cref{eq:AIFA-q1}.
\end{proofof}

\begin{proofof}{\cref{prop:necssary-ElogPr}} \label{proof:necssary-ElogPr}
The argument is the same as \cref{app-proof:optim-q1}.
In the overall ELBO, the only terms that depend on $f_{n,k}$ is
\begin{equation*}
	\begin{aligned}
			\EE_{x_{n,k} \sim f_{n,k}, \rho_k \sim q_{\approxrate}} \ln \traitLL(x_{n,k} \mid \rho_k) &+ \EE_{x_{n,k} \sim f_{n,k}, x_{n,-k} \sim f_{n,-k}, \psi \sim q_{\psi}} \ln \Pr(y_n \mid x_{n,.}, \psi) \\
			 &- \EE_{x_{n,k} \sim f_{n,k}} \ln f_{n,k}(x_{n,k}).
	\end{aligned}
\end{equation*}
We use Fubini to express the last integral as a negative KL-like quantity, and use optimality of KL arguments to derive the p.m.f.\ of the minimizer.
\end{proofof}

\section{Experimental setup}
In this section, the notation for atom sizes, atom locations, latent trait counts and observed data follow that of \cref{sec:conceptual} i.e.\ $(\approxrate_k)_{k=1}^{K}$ denotes the collection of atom sizes, 
$(\psi_k)_{k=1}^{K}$ denotes the collection of atom locations, 
$(x_{n,k})_{k=1,n=1}^{K,N}$ denotes the latent trait counts of each observation, 
and $(y_n)_{n=1}^{N}$ denotes the observed data. 

\subsection{Image denoising using the beta--Bernoulli process} \label{app:image-setup}

\paragraph{Data.} 
We obtain the ``clean'' house image from \url{http://sipi.usc.edu/database/}. We downscale the original $512 \times 512$ image to $256 \times 256$ and convert colors to gray scale.
We add iid Gaussian noise to the pixels of the clean image, resulting in the noisy input image. 
We follow \citet{zhou2009nonparametric} in extracting the patches.
We use patches of size $8 \times 8$, and flatten each observed patch $y_i$ into a vector in $\mathbb{R}^{64}$.

\paragraph{Finite approximations.} We use finite approximations that target the beta--Bernoulli process with $\distBetaP(1,1,0)$ i.e.\ $\gamma = 1, \alpha = 1, d = 0.$ \citet{zhou2009nonparametric} remark that the denoising performance is not sensitive to the choice of $\gamma$ and $\alpha$. 
Therefore, we pick $\gamma = \alpha = 1$ for computational convenience, since the beta process with $\alpha = 1$ has the simple TFA in \cref{exa:Bondesson-beta-alpha=1}. 
To be explicit, the TFA for the given beta--Bernoulli process is
\begin{equation} \label{eq:img-TFA}
	\begin{aligned}
		v_j &\distiid \distBeta(1,1), \hspace{10pt} i = 1,2,\ldots,K, \\
		\approxrate_i &= \prod_{j=1}^{i} v_j, \hspace{10pt} i = 1,2,\ldots,K, \\
		x_{n,i} \given \approxrate_i &\distind \distBer(\approxrate_i), \hspace{10pt} \text{across } n,i. \\
	\end{aligned}
\end{equation}
while the corresponding AIFA is
\begin{equation} \label{eq:img-IFA}
	\begin{aligned}
		\approxrate_i &\distiid \distBeta\left(\frac{1}{K}, 1\right), \hspace{10pt} i = 1,2,\ldots, K, \\
		x_{n,i} \given \approxrate_i &\distind \distBer(\approxrate_i), \hspace{10pt} \text{across } n,i. \\
	\end{aligned}
\end{equation}
We report the performance for $K$'s between $10$ and $100$ with spacing $10$. 

\paragraph{Ground measure and observational likelihood.}  
Following \citet{zhou2009nonparametric}, we fix the ground measure but put a hyper-prior (in the sense of \cref{cor:hyper-prior-bound}) on the observational likelihood.  
The ground measure is a fixed Gaussian distribution:
\begin{equation} \label{eq:img-H}
\psi_i \distiid \mathcal{N}\left( 0, \frac{1}{64}\mathbf{I}_{64}\right), \hspace{10pt} i = 1,2,\ldots,K. 
\end{equation}
The observational likelihood involves two Gaussian distributions with random variances:
\begin{equation} \label{eq:img-F}
\begin{aligned}
\gamma_w &\sim \distGamma(10^{-6}, 10^{-6}),\\
\gamma_e &\sim \distGamma(10^{-6}, 10^{-6}), \\
w_{n,i} \given \gamma_w  &\distiid \mathcal{N}(0, \gamma_w^{-1}), \hspace{10pt} \text{across }i,n,\\
y_n \given x_{n,.},  w_{n,.} \psi, \gamma_e &\distind \mathcal{N}(\sum_{i=1}^{K} x_{n,i} w_{n,i} \psi_{i}, \gamma_e^{-1} \mathbf{I}_{64}), \hspace{10pt} \text{across }n.\\
\end{aligned}
\end{equation}
We use the (shape,rate) parametrization of the gamma distribution. 
The weights $w_{n,i}$ enable an observation to manifest a non-integer (and potentially negative) scaled version of the $i$-th basis element. 
The precision $\gamma_w$ determines the scale of these weights. 
The precision $\gamma_e$ determines the noise variance of the observations. 
We are uninformative about the precisions by choosing the $\distGamma(10^{-6}, 10^{-6}$) priors. 

In sum, the full finite models combine either \cref{eq:img-IFA,eq:img-H,eq:img-F} (for AIFA) or \cref{eq:img-TFA,eq:img-H,eq:img-F} (for TFA). 

\paragraph{Approximate inference.}
We use Gibbs sampling to traverse the posterior over all the latent variables --- the ones that are most important for denoising are $x,w,\psi$. 
The chosen ground measure and observational likelihood have the right conditional conjugacies so that blocked Gibbs sampling is conceptually simple for most of the latent variables. 
The only difference between AIFA and TFA is the step to sample the feature proportions $\approxrate$: TFA updates are much more involved compared to AIFA (see \cref{sec:conceptual}). 
The order in which Gibbs sampler scans through the blocks of variables does not affect the denoising quality. 
To generate the PSNR in \cref{sub-fig:house256-similar-metric}, after finishing the gradual introduction of all patches, we run $150$ Gibbs sweeps. 
We use the final state of the latent variables at the end of these Gibbs sweep as the warm-start configurations in \cref{sub-fig:house256-TFA-similar-posterior,sub-fig:house256-IFA-similar-posterior}.  

\paragraph{Evaluation metric.}
We discuss how iterates from Gibbs sampling define output images. 
Each configuration of $x,w,\psi$ defines each patch's ``noiseless'' value:
\begin{equation*}
	\widetilde{y}_n = \sum_{i=1}^{K}x_{n,i} w_{n,i}\psi_i.
\end{equation*}
Each pixel in the overall image is covered by a small number of patches. 
The ``noiseless'' value of each pixel is the average of the pixel value suggested by the various patches that cover that pixel.
We aggregate the output images across Gibbs sweeps by a simple weighted averaging mechanism.
We report the PSNR of the output image with the original image following the formulas from \cite{hore2010image}. 

\subsection{Topic modelling with the modified HDP} \label{app:topic-setup}

\paragraph{Data.} We download and pre-process into bags-of-words about one million random Wikipedia documents, following \citet{hoffman2010online}.

\paragraph{Finite models.} 
We fix the ground measure to be a Dirichlet distribution and the observational likelihood to be a categorical distribution i.e.\ no hyper-priors. The AIFA is
	\(
	G_0 &\sim \distFSD_K(\omega, \distDir(\eta \mathbf{1}_V)), \\
	G_{d} \mid G_0 &\distiid \distTSB_T(\alpha, G_0), & \text{across } d, \\
	\beta_{dn} \mid G_d &\distind  G_d(\cdot), &\text{across } d,n, \\
	w_{dn} \given \beta_{dn} &\distind \distCat(\beta_{dn}), &\text{across } d,n.
	\)
while the TFA is
	\(
	G_0 &\sim \distTSB_K(\omega, \distDir(\eta \mathbf{1}_V)), \\
	G_{d} \mid G_0 &\distiid \distTSB_T(\alpha, G_0), & \text{across } d, \\
	\beta_{dn} \mid G_d &\distind  G_d(\cdot), &\text{across } d,n, \\
	w_{dn} \given \beta_{dn} &\distind \distCat(\beta_{dn}), &\text{across } d,n.
	\)
We set the hyperparameters $\eta, \alpha, \omega,$ and $T$ following \citet{wang2011online}, in that $\eta = 0.01, \alpha = 1.0, \omega = 1.0, T = 20.$ We report the performance for $K$'s between $20$ and $300$ with spacing $40$. 

\paragraph{Approximate inference.} We approximate the posterior in each model using stochastic variational inference \citep{Hoffman:2013}. Both models have conditional conjugacies that enable the use of exponential family variational distributions and closed-form expectation equations for all update types. The batch size is $500$. We use the learning rate $(t + \tau)^{-\kappa}$, where $t$ is the number of data mini-batches. For cold-start experiments, we set $\tau = 1.0$ and $\kappa = 0.9.$ To generate the results of \cref{sub-fig:wiki1m-similar-metric}, we process $4000$ mini-batches of documents. 
We obtain the warm-start initializations in \cref{sub-fig:wiki1m-TFA-similar-posterior,sub-fig:wiki1m-IFA-similar-posterior} by processing $512$ mini-batches of documents.
When training from warm-start initialization, to reflect the fact that the initial topics are the results of a training period, we change $\tau = 512$, but use the same $\kappa$ as cold start. 

\paragraph{Evaluation metrics.} We compute held-out log-likelihood following \citet{Hoffman:2013}. Each test document $d'$ is separated into two parts $w_{ho}$ and $w_{obs}$\footnote{How each document is separated into these two parts can have an impact on the range of test log-likelihood values encountered. For instance, if the first (in order of appearance in the document) $x\%$ of words were the observed words and the last $(100-x)\%$ words were unseen, then the test log-likelihood is low, presumably since predicting future words using only past words and without any filtering is challenging. Randomly assigning words to be observed and unseen gives better test log-likelihood.}, with no common words between the two. In our experiments, we set $75\%$ of words to be observed, the remaining $25\%$ unseen. The predictive distribution of each word $w_{new}$ in the $w_{ho}$ is exactly equal to:
\begin{equation*}
		p(w_{new} \given \mathcal{D}, w_{obs}) = \int_{\theta_{d'}, \beta} p(w_{new} \given \theta_{d'}, \beta) p(\theta_{d'}, \beta \given \mathcal{D}, w_{obs}) d\theta_{d'} d\beta.
\end{equation*}
This is an intractable computation as the posterior $p(\theta_{d'}, \beta \given \mathcal{D}, w_{obs})$ is not analytical. We approximate it with a factorized distribution:
\begin{equation*}
p(\theta_{d'}, \beta \given \mathcal{D}, w_{obs}) \approx q(\beta \given \mathcal{D}) q(\theta_{d'}), 
\end{equation*}
where $q(\beta\given\mathcal{D})$ is fixed to be the variational approximation found during training and $q(\theta_{d'})$ minimizes the KL between the variational distribution and the posterior. Operationally, we do an E-step for the document $d'$ based on the variational distribution of $\beta$ and the observed words $w_{obs}$, and discard the distribution over $z_{d',.}$, the per-word topic assignments because of the mean-field assumption. Using those approximations, the predictive approximation is approximately: 
\begin{equation*}
p(w_{new} \given \mathcal{D}, w_{obs}) \approx \widetilde{p}(w_{new} \given \mathcal{D}, w_{obs}) = \sum_{k=1}^{K} \mathbb{E}_q (\theta_{d'}(k)) \mathbb{E}_q(\beta_{k}(w_{new})),
\end{equation*}
and the final number we report for document $d'$ is:
\begin{equation*}
\frac{1}{|w_{ho}|} \sum_{w \in w_{ho}} \ln  \widetilde{p}(w \given \mathcal{D}, w_{obs}).
\end{equation*}

\subsection{Comparing IFAs} \label{app:synthetic-setup}

\paragraph{Data.} For the AIFA versus BFRY IFA comparison i.e. \cref{sub-fig:bfry-v-aifa}, we generate synthetic data $\{y_n\}_{n=1}^{2000}$ from a power-law beta--Bernoulli process $\distBetaP(2,0,0.6)$.
\begin{equation*}
	\begin{aligned}
		\sum_{i=1}^{\infty} \theta_i \psi_i  &\sim \distBP(2,0,0.6; \mathcal{N}(0, 5I_5)), \\
		x_{n,i} \given \theta_i &\distind \text{Ber}(\theta_i), & \text{across }n,i, \\
		y_{n} \given x_{n,.}, \psi &\distind \mathcal{N}(\sum_{i}x_{n,i}\psi_i,I_5), &\text{across } n. 
	\end{aligned}
\end{equation*}
For the AIFA vs \genpar{} IFA comparison i.e. \cref{sub-fig:genpar-v-aifa}, we use the same generative process except the beta process is $\distBetaP(2,1.0,0.6)$.
We marginalize out the feature proportions $\theta_i$ and sample the assignment matrix $X = \{x_{n,i}\}$ from the power-law Indian buffet process \citep{Teh:2009}.
The feature means are Gaussian distributed, with prior mean $0$ and prior covariance $5I_5$. Conditioned on the feature combination, the observations are Gaussian with noise variance $I_5$. 
Since the data is exchangeable, without loss of generality, we use $y_{1:1500}$ for training and $y_{1501:2000}$ for evaluation. 

\paragraph{Finite approximations.} 
We use finite approximations that have exact knowledge of the beta process hyperparameters.
For instance, for the AIFA versus BFRY IFA comparison, we use $K$-atom AIFA prior with densities
\begin{equation} \label{eq:IFA-nuK}
	\nu_{\text{AIFA}}(d\theta) \coloneqq \frac{\mathbf{1} \{0 \leq \theta \leq 1\}}{Z_K} \theta^{-1+c/K-0.6S(\theta-1/K)}(1-\theta)^{-0.4}d\theta,
\end{equation}
where $c \coloneqq \frac{2}{B(0.6,0.4)}$ and $S(\theta) = \begin{cases}
	\exp\left(\frac{-1}{1-K^2(\theta-1/K)^{2}}+1\right) & \text{if } \theta \in (0,1/K) \\
	\indict{\theta > 0} & \text{otherwise.}
\end{cases}$, and $Z_K$ is the suitable normalization constant. 

In all, the approximation to the beta--Bernoulli part of the generative process is
\begin{equation} \label{eq:synthetic-beta-Ber}
\begin{aligned}
	\approxrate_i &\distiid \widetilde{\nu}(.) &\text{ for } i \in [K], \\
	x_{n,i} \given \approxrate_i &\distind \distBer(\approxrate_i) & \text{across }, n, i,
\end{aligned}
\end{equation}
where $\widetilde{\nu}(.)$ is either $\nu_{\text{AIFA}}$, $\nu_{\text{BFRY}}$ or $\nu_{\text{\genpar{}}}$.
We report the performance for $K$ from $2$ to $100$.

\paragraph{Ground measure and observational likelihood.} 
We use hyper-priors in the sense of \cref{cor:hyper-prior-bound}.
The ground measure is random because the we do not fix the variance of the feature means.
\begin{equation} \label{eq:synthetic-H}
	\begin{aligned}
		\sigma_g &\sim \text{Gamma}(5,5), \\
		\psi_i &\distiid \mathcal{N}(0, \sigma_g^2 I_5) \text{ for } i \in [K]. \\ 
	\end{aligned}
\end{equation}

The observational likelihood is also random because we do not fix the noise variance of the observed data.
\begin{equation} \label{eq:synthetic-F}
	\begin{aligned}
		\sigma_c &\sim \text{Gamma}(5,5), \\
		y_{n} \given x_{n,.}, \psi, \sigma_c &\distind \mathcal{N}(\sum_{i}x_{n,i}\psi_i,\sigma_c^2 I_5).
	\end{aligned}
\end{equation}
In \cref{eq:synthetic-H,eq:synthetic-F}, we use the (shape, rate) parametrization of the gamma distribution.
The full finite models are described by \cref{eq:synthetic-beta-Ber,eq:synthetic-H,eq:synthetic-F}.\footnote{During inference, we add a small tolerance of $10^{-3}$ to the standard deviations $\sigma_c, \sigma_g, \zeta_i$ in the model to avoid singular covariance matrices, although this is not strictly necessary if we clip gradients.}

\paragraph{Approximate inference.}
We use mean-field variational inference to approximate the posterior. We pick the variational distribution $q(\sigma_c,\sigma_g,\approxrate,\psi,x)$ with the following factorization structure:
\begin{equation*}
	q(\sigma_c) q(\sigma_g) \prod_{i} q(\approxrate_i) \prod_{i}q(\psi_i) \prod_{i,n} q(x_{n,i}).
\end{equation*}
Each variation distribution is the natural exponential family. Specifically, we have $q(\sigma_c) = \text{Gamma}(\nu_c(0), \nu_c(1))$, $q(\sigma_g) = \text{Gamma}(\nu_g(0), \nu_g(1))$, $q(\psi_i) = \mathcal{N}(\tau_i, \zeta_i),$ $q(\approxrate_i) = \text{Beta}(\kappa_i(0),\kappa_i(1))$, $q(x_{n,i}) = \text{Ber}(\phi_{n,i}).$
We set the initial variational parameters using using the latent features, feature assignment matrix, and the variances of the features prior and the observations around the feature combination. 
We use the ADAM optimizer in Pyro (learning rate $0.001$, $\beta_1 = 0.9$, clipping gradients if their norms exceed $40$) to minimize the KL divergence between the approximation and exact posterior. 
We sub-sample $50$ data points at a time to form the objective for stochastic variational inference. 
We terminate training after processing $5{,}000$ mini-batches of data.

\paragraph{Evaluation metrics.}
We use the following definition of predictive likelihood: 
\begin{equation}  \label{eq:sum-of-log-individuals}
	\sum_{i=1}^{m} \ln  \Pr(y_{n+i}\given y_{1:n}),
\end{equation}
where $y_{1:n}$ are the training data and $\{y_{n+i}\}_{i=1}^{m}$ are the held-out data points. 

We estimate $\Pr(y_{n+i}\given y_{1:n})$ using Monte Carlo samples, since the predictive likelihood is an integral of the posterior over training data:
\begin{equation*}
	\Pr(y_{n+i}\given y_{1:n}) = \int_{x_{n+i}, \sigma, \psi, \approxrate} \Pr(y_{n+i}\given x_{n+i}, \psi, \sigma) \Pr(x_{n+i}, \psi, \sigma, \approxrate\given y_{1:n}),
\end{equation*}
where $x_{n+i}$ is the assignment vector of the $n+i$ test point. 
Define the $S$ Monte Carlo samples of the variational approximation to the posterior as $(x_{(n+1):(n+m),.}^{s},\approxrate^s, \psi^{s}, \sigma^{s})_{s=1}^{S}$.
We jointly estimate $\Pr(y_{n+i}\given y_{1:n})$ across test points $y_{n+i}$ using the $S$ Monte Carlo samples:
\begin{equation*}
	\Pr(y_{n+i}\given y_{1:n}) \approx \frac{1}{S}\sum_{s=1}^{S} \Pr(y_{n+i}\given x^s_{n+i}, \psi^s, \sigma^s).
\end{equation*}
We use $S = 1{,}000$ samples from the (approximate) posterior to estimate the average log test-likelihood in \cref{eq:sum-of-log-individuals}.

\subsection{Beta process hyperparameter estimation}\label{app:discount-setup}

\paragraph{Data.}
In this experiment, the number of observations, or the number of rows in the feature matrix, is $N = 1000$.
For discount estimation, we generate $50$ matrices from the corresponding IBP~\citep{Teh:2009} with for mass $\gamma = 3.0$, concentration $\alpha = 1.0$ and discount varying from $0$ through $0.5$.
For mass estimation, we generate $50$ matrices from the IBP with concentration $\alpha = 1.0$, discount $d=0.25$ and mass varying from $1.0$ through $5.0$.
For the concentration estimation, we generate $50$ matrices from the IBP with mass $\alpha = 3.0$, discount $d=0.25$ and concentration varying from $0$ through $5.0$.

\paragraph{AIFA marginal likelihood.}
The $K$-atom AIFA rates define a generative process over feature matrices with $N$ rows and $K$ columns:
\begin{equation*}
	\begin{aligned}
		\theta_k &\distiid \AIFA{K} & \text{ across } k, \\
		x_{n,k} \mid \theta_k &\distind \distBer(\theta_k) &\text{ across } n, k.
	\end{aligned}
\end{equation*}
$x_{n,k}$ is the entry in the $n$th row and $k$th column of the feature matrix.
Treating the beta process hyperparameters $\gamma, \alpha, d$ as unknowns, we compute the probability of observing a particular feature matrix $\{x_{n,k}\}$ (integrating out the AIFA rates) as a function of $\gamma, \alpha, d$. 
By symmetry and independence among the columns $x_{.,k}$, it suffices to compute the probability of observing just one column, say  $\{x_{n,1}\}_{n=1}^{N}$.
Conditioned on $\theta_1$, the probability of observing  $\{x_{n,1}\}_{n=1}^{N}$ is exactly 
\begin{equation*}
	\prod_{n=1}^{N} \theta_1^{x_{n,1}}  (1-\theta_1)^{1-x_{n,1}}
\end{equation*}
We integrate out $\theta_1$ to compute the marginal likelihood.
Recall that $c(\gamma, \alpha, d) = \gamma /B(\alpha + d, 1 - d)$ for the beta process AIFA.
The marginal likelihood of observing the first column $\{x_{n,1}\}_{n=1}^{N}$ is
\begin{equation*}
	\begin{aligned}
	\mathbb{E}_{\theta \sim \AIFA{K}} &\left[\prod_{n=1}^{N} \theta_1^{x_{n,1}}(1-\theta_1)^{1-x_{n,1}} \right] \\
	&= \frac{\int_{0}^{1} \theta^{-1 + c(\gamma, \alpha, d)/K + \sum_n x_{n,1} - d S_{1/K} (\theta - 1/K)} (1 - \theta)^{\alpha + d + N - \sum_{n}x_{n,1} - 1} d\theta }{\int_{0}^{1} \theta^{-1 + c(\gamma, \alpha, d)/K - d S_{1/K} (\theta - 1/K)} (1 - \theta)^{\alpha + d - 1} d\theta }.
	\end{aligned}
\end{equation*}
In all, if we denote 
\begin{equation*}
	Z_K(\gamma, \alpha, d; x, y) := \int_{0}^{1} \theta^{-1 + c(\gamma, \alpha, d)/K + x - d S_{1/K} (\theta - 1/K)} (1 - \theta)^{\alpha + d + (y - x) - 1} d\theta,
\end{equation*}
then the marginal probability of observing a particular binary matrix $\{x_{n,k}\}$, as a function of $\gamma, \alpha, d$, is
\begin{equation} \label{eq:margLik}
	\prod_{k=1}^{K} \frac{Z_K(\gamma, \alpha, d; \sum_{n} x_{n,k}, N)}{Z_K(\gamma, \alpha, d; 0, 0)}.
\end{equation}
For feature matrices coming from an IBP, the number of columns $\widehat{K}$ is random, and usually (much) smaller than the number of atoms in the approximation.
In this section, the approximation level is $K = 100{,}000$: the distribution of the number of active features in the finite model (for $d \in [0,0.5])$ has no noticeable change between $K = 100{,}000$ and $K > 100{,}000$.
The fact that $K - \widehat{K}$ columns are missing is the same as $K - \widehat{K}$ columns being identically zero; hence, when evaluating the marginal probability of matrices that have less than $K$ columns, we simple pad the missing columns with zeros.

It remains to show how to compute \cref{eq:margLik} using numerical methods.
The bottleneck is computing $Z_K(\gamma, \alpha, d; x, y)$.
We split the integral into two disjoint domains.
The first domain is $(0,1/K)$: on this domain, the integral is an incomplete beta integral, which is implemented in libraries such as \citet{virtanen2020scipy}.
The second domain is $[1/K,1]$.
On this domain, we first compute $m^*$, the maximum value of the integrand $\theta^{-1 + c(\gamma, \alpha, d)/K + x - d S_{1/K} (\theta - 1/K)} (1 - \theta)^{\alpha + d + (y - x) - 1}$.
We then use numerical integration to integrate $\theta^{-1 + c(\gamma, \alpha, d)/K + x - d S_{1/K} (\theta - 1/K)} (1 - \theta)^{\alpha + d + (y - x) - 1}/m^*$.
We divide by $m^*$ to avoid the integrand getting too small, which happens if $x$ or $y$ are large.
The last integrand is well-behaved (bounded and smooth), and we expect numerical integration to be accurate.

\paragraph{Marginal likelihood under BFRY IFA (or \genpar{} IFA) are challenging to estimate.}
In theory, for the BFRY IFA, it is also possible to express the marginal likelihood (as a function of $\gamma, \alpha, d$) 
for an observed feature matrix $x_{n,k}$ under the BFRY IFA prior as a ratio between normalization constants.
However, we run into numerical issues (divergence errors) computing the BFRY IFA normalization constants that are not present in computing the AIFA normalization constants.
For completeness, the BFRY IFA normalization constants are of the kind
\begin{equation} \label{eq:hard-integral}
	Z_{\text{BFRY}}(\gamma, d; x, y) = \int_0^1 \frac{\gamma/K}{B(d, 1-d)} \theta^{x - d - 1} (1-\theta)^{y - x + d - 1} \left[1  - \exp\left( - (Kd/\gamma)^{1/d} \frac{\theta}{1-\theta}   \right)  \right]. 
\end{equation}
Whether this integral has a closed-form solution is unknown: the closed-formed marginal likelihoods from \citet{lee2016finite} apply to clustering models from normalized CRMs rather than feature-allocation models from unnormalized CRMs.
Numerical integration struggles with \Cref{eq:hard-integral} for $x = 0$.
$(Kd/\gamma)^{1/d}$ is typically very large: when $\gamma = 1, d = 0.1$, even $K = 100$ leads to $(Kd/\gamma)^{1/d}$ being on the order of $10^{20}$. 
As a result, under standard floating point precision, $1-\exp\left( - (Kd/\gamma)^{1/d} \frac{\theta}{1-\theta} \right)$ evaluates to $1$ on all points of the quadrature grid: this leads to divergent behavior, as the factor $\theta^{-d - 1}$ by itself grows too fast near $0$.

We resort to Monte Carlo to estimate the normalization constant.
In each Monte Carlo batch, we draw $K$ random variables $\theta_1, \theta_2,\ldots,\theta_K$ from the BFRY density \cref{eq:BFRY-nuK1}, and estimate the log of $Z_{\text{BFRY}}(\gamma, d; x, y)$ with 
\begin{equation*}
 \text{logsumexp} \bigg\{ [(x-d-1) \ln \theta_k  + (y - x + d - 1) \ln(1-\theta_k) - \ln K ]\bigg\}_{k=1}^{K}.
\end{equation*} 
In the left panel of \cref{sub-fig:likelihoodCurve}, we first generate an feature matrix from IBP with mass $3.0$, concentration $0.0$ and discount $0.25$.
We then plot the estimate of the marginal likelihood under BFRY IFA for this feature matrix as a function of $d$ for mass fixed at $3.0$ and discount fixed at $0.0$.

\genpar{} IFA faces similar problems as BFRY IFA.
We are not aware of a closed-form formula for the marginal likelihood. 
Namely, we are not able to show that \cref{eq:Pareto_IFA} is a conjugate prior for the Bernoulli likelihood: when we observe an observation $X = 1$ from the model $X \sim \text{Ber}(\theta), \theta \sim \nu_{\text{\genpar{}}}$, the posterior density for $\theta$ is proportional to
\begin{equation*}
	 \frac{\theta^{-d}(1-\theta)^{\alpha + d - 1} }{B(1-d, \alpha + d)} \left(1  - \frac{1}{\left( \theta \left[\left( 1 + \frac{K d}{\gamma \alpha } \right)^{1/d} -1 \right] + 1   \right)^{\alpha} } \right)    \mathbf{1} \{0 \leq \theta \leq 1\}.
\end{equation*}
This new density is not in the same family as the original generalized Pareto variate. 
Default schemes to numerically integrate $\Pr(0 \mid \theta_k)$ against the generalized Pareto prior for $\theta_k$ fail because of overflow issues associated with the magnitude of the term $\left( 1 + \frac{K d}{\gamma \alpha } \right)^{1/d}$.
In the left panel of \cref{sub-fig:likelihoodCurve}, we first generate an feature matrix from IBP with mass $3.0$, concentration $1.0$ and discount $0.25$.
We then plot the estimate of the marginal likelihood under BFRY IFA for this feature matrix as a function of $d$ for mass fixed at $3.0$ and discount fixed at $0.0$.

\paragraph{Optimization.}
For AIFA i.e.\ left panel of \cref{sub-fig:discountEstimation}, to estimate the beta process hyperparameters given an observed feature matrix, we maximize the marginal probability in \cref{eq:margLik} with respect to $\gamma, \alpha, d$, by doing a grid search with a fine resolution.
The base grid for the triplet $\gamma, \alpha, d$ is the Cartesian product of three lists: $[1.0, 2.0, 3.0, 4.0, 5.0], [0.5, 1.0, 1.5, 2.0, 2.5], [0.0, 0.1, 0.2, 0.3, 0.4, 0.5]$. 
We refine the base grid around the true hyperparameters.
For example, in the discount estimation experiment, a true configuration is $(3.0, 1.0, 0.4)$
The refinement here is the Cartesian product of three lists $[2.6, 2.8, 3.0, 3.2, 3.4]$, $[0.8, 0.9, 1.0, 1.1, 1.2]$, $[0.36, 0.38, 0.4, 0.42, 0.44]$.
We append the refinement to the base grid by looping through all the configurations.
We propose the best hyperparameters by evaluating the marginal likelihood (\cref{eq:margLik}) at all points on the grid, and reporting the maximizer.

For the nonparametric process i.e.\ right panel of \cref{sub-fig:discountEstimation}, the probability of observing a particular feature matrix under the IBP prior over $N$ rows is given in \citet[Equation 7]{broderick2013feature}.
We maximize this function with respect to $\gamma, \alpha, d$ using differential evolution techniques \citep{storn1997differential,virtanen2020scipy}. 

\subsection{Dispersion estimation}\label{app:dispersion-setup}

\paragraph{Generative model.} 
The probabilistic model is 
\begin{equation} \label{eq:eG-CMP}
	\begin{aligned}
		\lambda_k &\distiid \distXGamma(\alpha/K, c, \tau, T)  &\text{across } k, \\
		\phi_{k} &\distiid \distDir(a_\phi \mathbf{1}_V)  &\text{across } k, \\
		z_{n,k} \mid \lambda_k &\distind  \distCMP(\lambda_k, \tau) &\text{across } k,n, \\
		x_{n,v} \mid z_{n,:},\phi &\distind  \distPoisson(\sum_{k} z_{n,k} \phi_{k,v}) , &\text{across } v,n.
	\end{aligned}
\end{equation}
Recall the definition of the Xgamma variate from \cref{def:Xgamma}.
The observed data is the count matrix $x_{n,v}$, the number of times document $n$ manifests vocab word $v$.
The hyperparameters are $\alpha,c,\tau,T$ and $a_\phi$.
To draw data for \cref{eq:eG-CMP}, we need to sample from $\distXGamma$ and $\distCMP$, two distributions that are not implemented in standard numerical libraries.
The only bottleneck in drawing $\distCMP(\theta, \tau)$ is computing $Z_{\tau}(\theta)$.
We approximate the infinite sum $\sum_{y=0}^{\infty} \frac{\theta^y}{(y!)^{\tau}}$ with a truncation $\sum_{y=0}^{L} \frac{\theta^y}{(y!)^{\tau}}$, using the bounds from \citet{minka2001computing} to make sure the contribution of the left-out terms is small.
To draw from $\distXGamma$, whose unnormalized density has a contribution from $Z_{\tau}^{-c}(\theta)$, we use the above approximation of $Z_{\tau}(\theta)$ and slice sampling on the approximation of the unnormalized density.

When generating synthetic data, we draw $N = 600$ documents, over a vocabulary of size $100$, from a model with $K = 500$.
The under-dispersed case and the over-dispersed case have the same following hyperparameters: $\alpha = 20$, $c = 1$, $T = 1000$, $a_{\phi} = 0.01$. 
For underdispersion, $\tau = 1.5$, while for overdispersion, $\tau = 0.7$.
Our primary goal of inference is estimating the topics and the shape $\tau$.
As such, during posterior inference, we fix the hyperparameters $\alpha$, $c$, $T$, and $a_\phi$ at the data-generating values, and sample the remaining latent variables ($\lambda, \phi, z$) and shape $\tau$.
We put a uniform $(0, 100]$ prior on the shape $\tau$: $\tau$ is always positive, and there is no noticeable difference in amount of dispersion (ratio of variance over mean) between $\tau = 100$ and $\tau > 100$.
Furthermore, during sampling, the values of $\tau$ are much smaller than $100$, indicating that inference would have remained the same for different choices of the uniform's upper limit.

\paragraph{Gibbs sampling.}
During sampling, following \citet[Section 4]{zhou2012beta}, we augment the original model by introducing three additional families of latent variables: $s$, $u$ and $q$. 
Conditioned on $z$ and $\phi$, the \emph{pseudocount} $s_{n,k,v}$ is distributed as Poisson
\begin{equation*}
	\begin{aligned}
	s_{n,k,v} \given z, \phi &\distind \distPoisson(z_{n,k} \phi_{k,v}), &\text{across } n,k,v,
	\end{aligned}
\end{equation*}
and the $s_{n,k,v}$ add up to be $x_{n,v}$ in the following way
\begin{equation*}
	x_{n,v} = \sum_{k} s_{n,k,v}. 
\end{equation*}
Summing up the pseudocounts across words, we have
\begin{equation*}
	u_{n,k} := \sum_{v} s_{n,k,v}.
\end{equation*}
It is true that
\begin{equation} \label{eq:augment}
	\begin{aligned}
		u_{n,k} \mid z_{n,k} &\distind \distPoisson(z_{n,k}), &\text{ across } n, k, \\
		\{ s_{n,k,v} \}_{v=1}^V \mid u_{n,k}, \phi_k &\distind \distMulti(u_{n,k}; \phi_k), &\text{ across } n, k.
	\end{aligned}
\end{equation}
Summing up the pseudocounts across documents, we have
\begin{equation*} 
	q_{k,v} := \sum_{n} s_{n,k,v}.
\end{equation*} 

We use a blocked Gibbs sampling strategy. 
The variable blocks variables are $\phi$, $\lambda$, $s$ (which determines $u$ and $q$), $z$, $\tau$.
First, we compute the Gibbs conditional of the \emph{topics} $\phi$.
Since $u$ is determined by $s$ (\cref{eq:augment}), conditioned on $s$, $\phi$ is independent of the remaining latent variables:
\begin{equation*}
\Pr( \phi \mid x, \lambda, s, z, \tau) = \Pr( \phi \mid s) \propto \Pr(\phi) \Pr(s \mid u, \phi) = \prod_{k=1}^{K} \distDir(\phi_k \mathbf{1}_V \mid [a_\phi +  q_{k,v} ]_{v=1}^{V}).
\end{equation*}
We compute the Gibbs conditionals of the \emph{rates} $\lambda$.
Conditioned on the trait counts $z$ and shape $\tau$, $\lambda$ is independent of the remaining latent variables:
\begin{equation*}
	\begin{aligned}
		\Pr(\lambda \mid x, \phi, s, z, \tau) &= \Pr(\lambda \mid z, \tau) = \prod_{k=1}^{K} \Pr(\lambda_k \mid z_{.,k}, \tau) \\
		&= \prod_{k=1}^{K} \distXGamma \left(\lambda_k \given \frac{\alpha}{K} + \sum_{n}z_{n,k}, c + N, \tau, T\right).
	\end{aligned}
\end{equation*}
We use the scheme discussed after \cref{eq:eG-CMP} to sample these $\distXGamma$ variates.  
The Gibbs conditionals of the \emph{trait counts} $z$ are
\begin{equation*}
	\begin{aligned}
		\Pr(z \mid x, \phi, \lambda, s, \tau) &= \Pr(z \mid \lambda, s, \tau) = \prod_{n,k} \Pr(z_{n,k} \mid \lambda_k, u_{n,k}, \tau) \\
		&\propto \prod_{n,k} \distPoisson( u_{n,k} \mid z_{n,k} ) \distCMP(z_{n,k} \mid \lambda_k, \tau). 
	\end{aligned}
\end{equation*}
To draw from the distribution whose p.m.f.\ at $z \in \naturals \cup \{0\}$ is proportional to $\distPoisson( u_{n,k} \mid z ) \distCMP(z \mid \lambda_k, \tau)$, the only bottleneck is computing the normalization constant
\begin{equation*}
	\sum_{z=0}^{\infty} \exp(-z) \frac{z^{u_{n,k}}}{u_{n,k}!} Z_{\tau}^{-1}(\lambda_k) \frac{\lambda_k^z}{(z!)^{\tau}}
\end{equation*}
The multiplicative factors that don't depend on $z$ can be taken out of the sum: we only need to compute $\sum_{z=0}^{\infty} \frac{(\lambda_k/e)^z z^{u_{n,k}}}{(z!)^{\tau}}$. 
Similar to the computation of $Z_{\tau}(\theta)$, we approximate the above infinite sum with a finite truncation, making sure the left-out terms have a small contribution.
The Gibbs conditionals of the \emph{pseudocounts} $s$ are
\begin{equation*}
	\begin{aligned}
		\Pr(s \mid x, \phi, \lambda, z, \tau) &= \Pr(s \mid x, \phi, z) \\
			&=\prod_{n,v} \distMulti( \{ s_{n,k,v} \}_{k=1}^{K} \mid x_{n,v}; [z_{n,k} \phi_{k,v}/\sum_{k'} z_{n,k'} \phi_{k',v}]).
	\end{aligned}
\end{equation*}
Finally, the Gibbs conditionals of the \emph{shape} $\tau$ are
\begin{equation*}
	\Pr(\tau \mid x, \phi, \lambda, s, z) = \Pr(\tau \mid z, \lambda) \propto \Pr(z \mid \tau, \lambda) \Pr(\lambda \mid \tau) \Pr(\tau).
\end{equation*}
In implementations, we omit the contribution from $\Pr(\lambda \mid \tau)$, since it contributes a very small amount (less than $0.1\%$) to the overall value of $\ln \Pr(z \mid \tau, \lambda) + \ln \Pr(\lambda \mid \tau) + \ln \Pr(\tau)$, but takes up more time to evaluate than the other two components.
In other words, the unnormalized log density of $\tau$ conditioned on the other variables is just
\begin{equation*}
	\ln \Pr(z \given \tau, \lambda)  + \ln \Pr(\tau) = \sum_{n,k} \ln \distCMP(z_{n,k} \mid \lambda_k, \tau)  + \ln \indict{\tau \in (0, 100]}.
\end{equation*}
We use slice sampling to draw from this distribution. 

\paragraph{MCMC results.}
We run $40$ chains, each for $50{,}000$ iterations.
By discarding the first $25{,}000$ iterations, all chains have $\widehat{R}$ diagnostic \citep{gelman1992inference} smaller than $1.01$.
To combat the serial correlation, we thin samples after burn-in, selecting only one draw after $2{,}000$ iterations. 
The effective number of samples remaining after burn-in and thinning is about $1{,}000$.

\section{Additional experiments} \label{app:more-experiments}
\subsection{Denoising other images} \label{app:more-denoising}
\begin{figure}[t]
	\begin{subfigure}[b]{.24\linewidth}
		\centering
		\includegraphics[width=\linewidth]{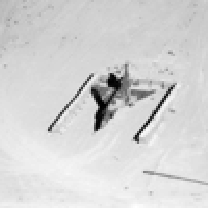}
		\caption{Original}\label{sub-fig:plane-original}
	\end{subfigure}%
	\begin{subfigure}[b]{.24\linewidth}
		\centering
		\includegraphics[width=\linewidth]{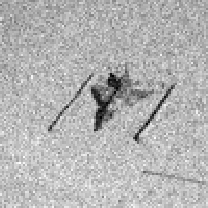}
		\caption{Input, $24.68$ dB}\label{sub-fig:plane-corrupted}
	\end{subfigure}%
	\begin{subfigure}[b]{.24\linewidth}
		\includegraphics[width=\linewidth]{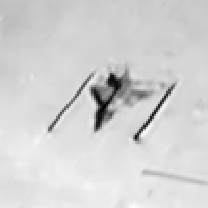}
		\caption{AIFA, $34.62$ dB}\label{sub-fig:plane-IFA}
	\end{subfigure}%
	\begin{subfigure}[b]{.24\linewidth}
		\includegraphics[width=\linewidth]{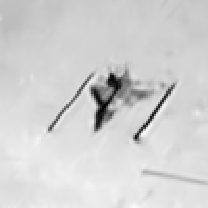}
		\caption{TFA, $34.76$ dB}\label{sub-fig:plane-TFA}
	\end{subfigure}
	\caption{Sample AIFA and TFA denoised images have comparable quality. \textbf{(a)} shows the noiseless image. \textbf{(b)} shows the corrupted image. \textbf{(c,d)} are sample denoised images from finite models with $K = 60$. PSNR (in dB) is computed with respect to the noiseless image.}
	\label{app-fig:example-plane}
\end{figure}

\begin{figure}[t]
	\begin{subfigure}[b]{.33\linewidth}
		\centering
		\includegraphics[width=\linewidth]{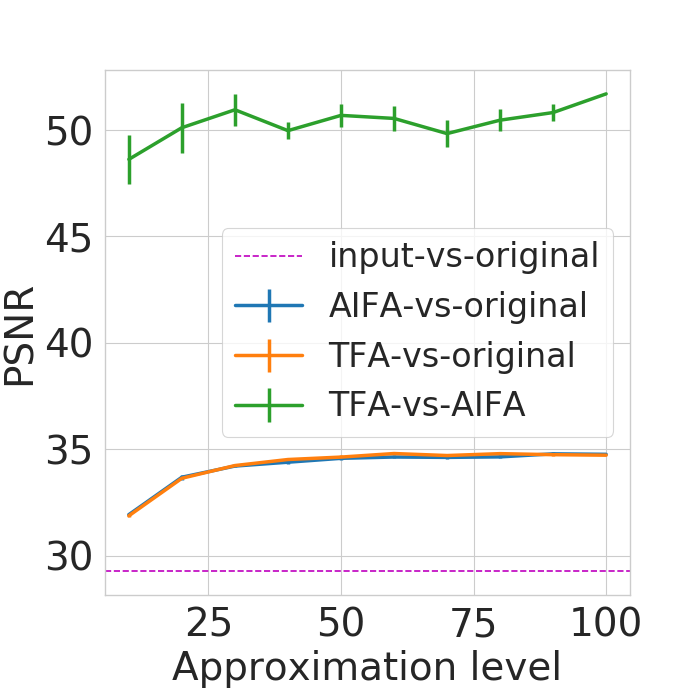}
		\caption{Performance versus $K$}\label{sub-fig:plane256-across-K}
	\end{subfigure}%
	\begin{subfigure}[b]{.33\linewidth}
		\centering
		\includegraphics[width=\linewidth]{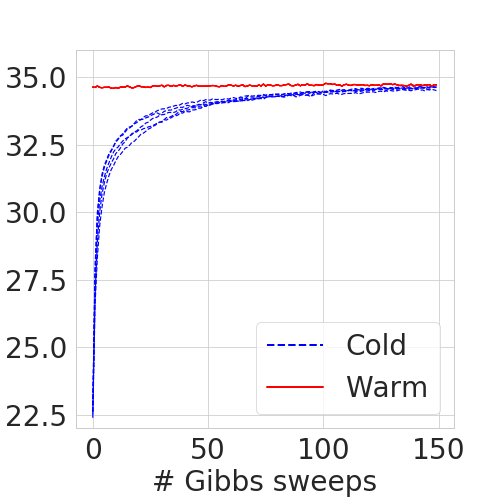}
		\caption{TFA training}\label{sub-fig:plane256-TFA-similar-posterior}
	\end{subfigure}%
	\begin{subfigure}[b]{.33\linewidth}
		\includegraphics[width=\linewidth]{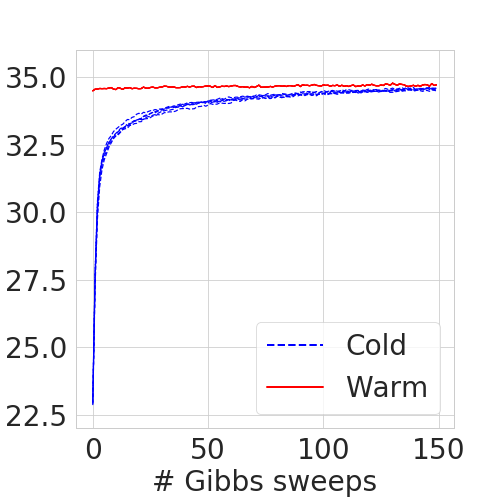}
		\caption{AIFA training}\label{sub-fig:plane256-IFA-similar-posterior}
	\end{subfigure}
	\caption{\textbf{(a)} Peak signal-to-noise ratio (PNSR) as a function of approximation level $K$.
		The error bars reflect randomness in both initialization and simulation of the conditionals across $5$ trials. 
		AIFA denoising quality improves as $K$ increases, and the performance is similar to TFA across approximation levels. 
		Moreover, the TFA- and AIFA-denoised images are very similar: the PSNR $\approx 50$ for TFA versus AIFA, whereas PSNR $< 35$
		for TFA or AIFA  versus the original image.
		\textbf{(b,c)} Show how PSNR evolves during inference. 
		The ``warm-start'' lines in indicate that the AIFA-inferred (respectively, TFA-inferred) parameters are excellent initializations for 
		TFA (respectively, AIFA) inference.}
	\label{fig:plane256-similar-posterior}
\end{figure}

Similar to the house image, the clean plane image was obtained from \url{http://sipi.usc.edu/database/}. 
The clean, the corrupted, and the example denoised images from AIFA/TFA for plane images are  given in \cref{app-fig:example-plane}.
In \cref{sub-fig:plane256-IFA-similar-posterior,sub-fig:plane256-TFA-similar-posterior}, the approximation level is $K = 60$. 

\begin{figure}[t]
	\begin{subfigure}[b]{.24\linewidth}
		\centering
		\includegraphics[width=\linewidth]{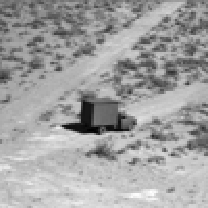}
		\caption{Original}\label{sub-fig:truck-original}
	\end{subfigure}%
	\begin{subfigure}[b]{.24\linewidth}
		\centering
		\includegraphics[width=\linewidth]{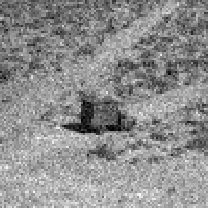}
		\caption{Input, $24.69$ dB}\label{sub-fig:truck-corrupted}
	\end{subfigure}%
	\begin{subfigure}[b]{.24\linewidth}
		\includegraphics[width=\linewidth]{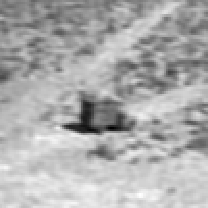}
		\caption{AIFA, $30.06$ dB}\label{sub-fig:truck-IFA}
	\end{subfigure}%
	\begin{subfigure}[b]{.24\linewidth}
		\includegraphics[width=\linewidth]{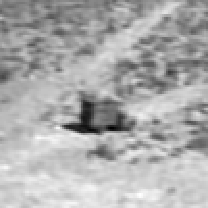}
		\caption{TFA, $30.24$ dB}\label{sub-fig:truck-TFA}
	\end{subfigure}
	\caption{Sample AIFA and TFA denoised images have comparable quality. \textbf{(a)} shows the noiseless image. \textbf{(b)} shows the corrupted image. \textbf{(c,d)} are sample denoised images from finite models with $K = 60$. PSNR (in dB) is computed with respect to the noiseless image.}
	\label{app-fig:example-truck}
\end{figure}

\begin{figure}[t]
	\begin{subfigure}[b]{.33\linewidth}
		\centering
		\includegraphics[width=\linewidth]{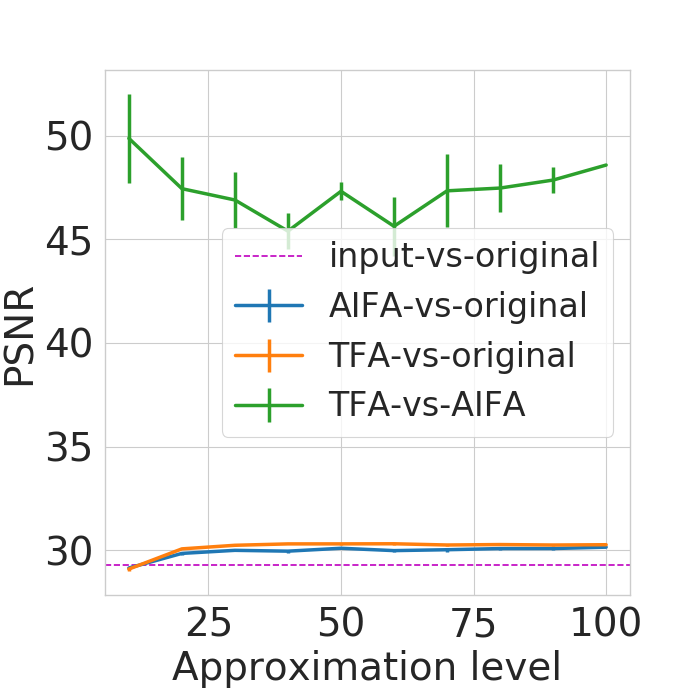}
		\caption{Performance versus $K$.}\label{sub-fig:truck256-across-K}
	\end{subfigure}%
	\begin{subfigure}[b]{.33\linewidth}
		\centering
		\includegraphics[width=\linewidth]{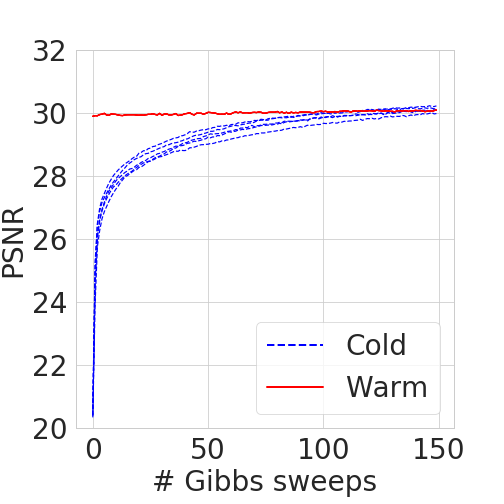}
		\caption{TFA training}\label{sub-fig:truck256-TFA-similar-posterior}
	\end{subfigure}%
	\begin{subfigure}[b]{.33\linewidth}
		\includegraphics[width=\linewidth]{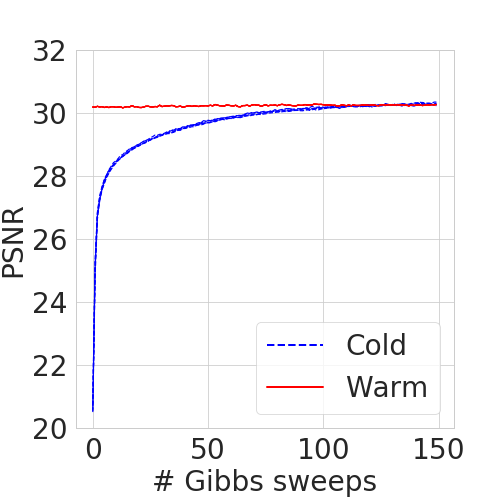}
		\caption{AIFA training}\label{sub-fig:truck256-IFA-similar-posterior}
	\end{subfigure}
	\caption{\textbf{(a)} Peak signal-to-noise ratio (PNSR) as a function of approximation level $K$.
		The error bars reflect randomness in both initialization and simulation of the conditionals across $5$ trials. 
		AIFA denoising quality improves as $K$ increases, and the performance is similar to TFA across approximation levels. 
		Moreover, the TFA- and AIFA-denoised images are very similar: the PSNR $\approx 47$ for TFA versus AIFA, whereas PSNR $< 31$
		for TFA or AIFA versus the original image.
		\textbf{(b,c)} Show how PSNR evolves during inference. 
		The ``warm-start'' lines in indicate that the AIFA-inferred (respectively, TFA-inferred) parameters are excellent initializations for 
		TFA (respectively, AIFA) inference.}
	\label{fig:truck256-similar-posterior}
\end{figure}

Similar to the house image, the clean truck image was obtained from \url{http://sipi.usc.edu/database/}. 
The clean, the corrupted, and the example denoised images from AIFA/TFA for truck images are  given in \cref{app-fig:example-truck}.
In \cref{sub-fig:truck256-IFA-similar-posterior,sub-fig:truck256-TFA-similar-posterior}, the approximation level is $K = 60$. 

\subsection{Effect of AIFA tuning hyperparamters} \label{app:ab-impact}
We investigate the impact of $a$ and $b_K$,which are two tunable parameters in the more general definition of AIFA from \cref{thm:general-CRM-convergence}.
Other than the setting of $a$ and $b_K$, the experimental set up is the same as \cref{app:synthetic-setup}.

\begin{figure}[t]
	\begin{subfigure}[b]{.45\linewidth}
		\centering
		\includegraphics[width=\linewidth]{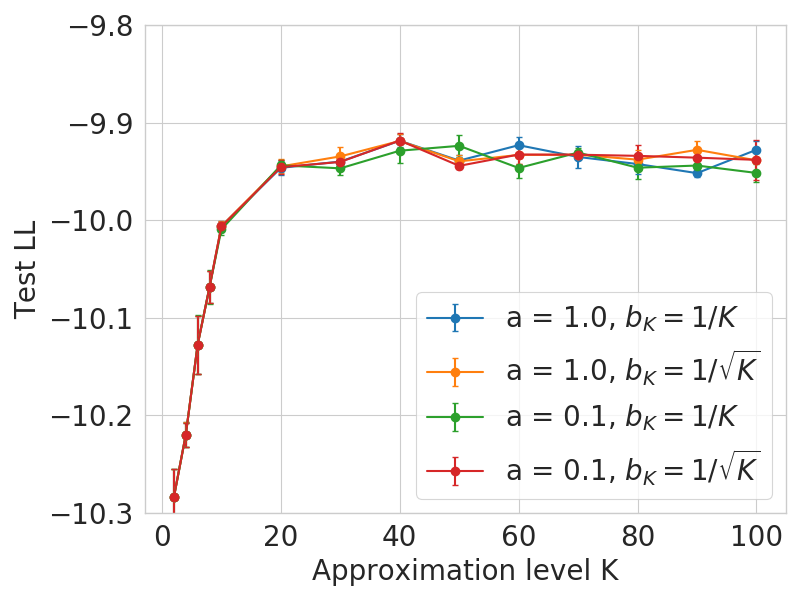}
		\caption{Average}\label{sub-fig:ifa-comparison-avg}
	\end{subfigure}\hspace{5pt} %
	\begin{subfigure}[b]{.45\linewidth}
		\centering
		\includegraphics[width=\linewidth]{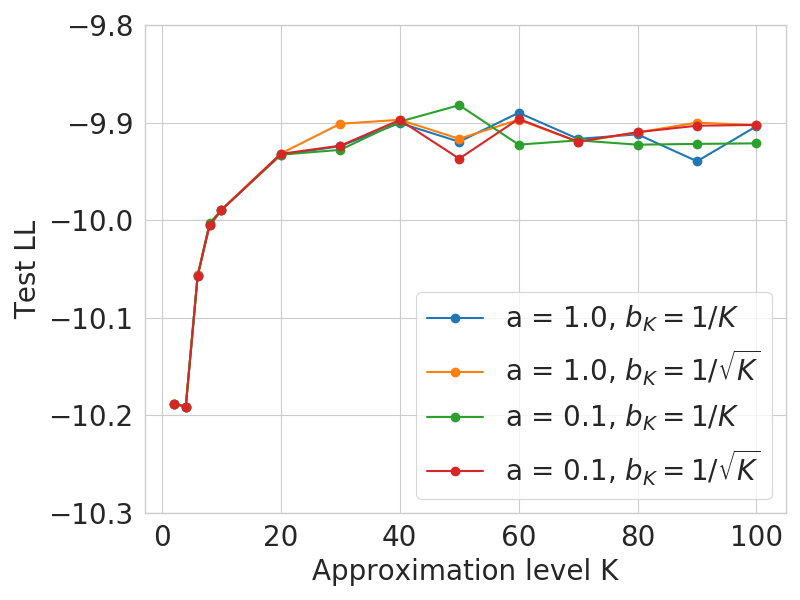}
		\caption{Best}\label{sub-fig:ifa-comparison-best}
	\end{subfigure}%
	\caption{The predictive log-likelihood of AIFA is not sensitive to different settings of $a$ and $b_K$. Each color corresponds to a combination of $a$ and $b_K$. \textbf{(a)} is the average across 5 trials with different random seeds for the stochastic optimizer, while \textbf{(b)} is the best across the same trials.}
	\label{fig:ifa-comparison}
\end{figure}

From \cref{fig:ifa-comparison}, we see that the setting of $a$ and $b_K$ do not have a big impact on the performance of the IFA from \cref{thm:general-CRM-convergence}.
We report results for a combination of $a \in \{0.1, 1\}$ and $b_K = {1}/{\sqrt{K}}$ or  $b_K = {1}/{K}.$

\subsection{Estimation of mass and concentration} \label{app:mass-concentration}
\begin{figure}[t]
	\begin{subfigure}[b]{.5\linewidth}
		\centering
		\includegraphics[width=\linewidth]{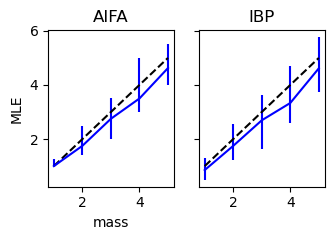}
		\caption{Estimation of mass $\gamma$}\label{sub-fig:IBP-mass}
	\end{subfigure}%
	\begin{subfigure}[b]{.5\linewidth}
		\centering
		\includegraphics[width=\linewidth]{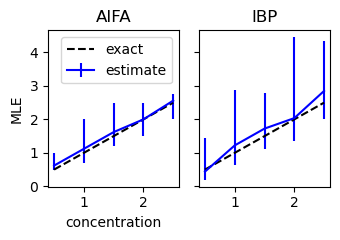}
		\caption{Estimation of concentration $\alpha$}\label{sub-fig:IBP-concentration}
	\end{subfigure}%
	\\
	\caption{In \cref{sub-fig:IBP-mass}, we estimate the mass by maximizing the marginal likelihood of the AIFA (left panel) or the full process (right panel).
		The solid blue line is the median of the estimated masses, while the lower and upper bounds of the error bars are the $20\%$ and $80\%$ quantiles. The black dashed line is the ideal value of the estimated mass, equal to the ground-truth mass. The key for  \cref{sub-fig:IBP-concentration} is the same, but for concentration instead of mass.}
	\label{fig:IBP-mass-concentration}
\end{figure}

\cref{fig:IBP-mass-concentration} shows that we can use an AIFA to estimate the underlying mass and concentration for a variety of ground-truth masses and concentrations.
The experimental setup is from \cref{app:discount-setup}.
Since the error bars in the left and right panels are comparable, we conclude that the AIFA yields comparable inference to the full nonparametric process.

\end{document}